\documentstyle[12pt]{article}

\large
\newcommand{\be}{\begin{eqnarray}}
\newcommand{\ee}{\end{eqnarray}}

\newcommand\ub{\underline}

\begin{document}
\setlength{\baselineskip}{21pt}
\pagestyle{empty}
\vfill
\eject
\begin{flushright}
SUNY-NTG-94-57
\end{flushright}

\vskip 1.0cm
\centerline{\bf A Master Formula for Chiral Symmetry Breaking}
\vskip 1.0 cm
\centerline{Hidenaga Yamagishi}
\vskip .1cm
\centerline{4 Chome 11-16-502}
\centerline{Shimomeguro, Meguro, Tokyo 153, Japan}
\vskip .1cm
\centerline{and}
\vskip .1cm
\centerline{Ismail Zahed}
\vskip .1cm
\centerline{Department of Physics}
\centerline{SUNY, Stony Brook, New York 11794-3800, USA}
\vskip 1.5cm

\centerline{\bf Abstract}
\vskip .20cm

We derive a master formula for chiral $SU(2)\times SU(2)$
breaking, based on the Veltman-Bell equations and the Peierls-Dyson relation.
Our approach does not rely on the use of the soft pion limit or an expansion
around the chiral limit, and yields exact results for on-shell pions.
Threshold theorems for
$\pi N\rightarrow \pi N$, $\gamma N\rightarrow \pi N$,
$\pi N\rightarrow\pi\pi N$, $\gamma N\rightarrow \gamma\pi N$,
$\gamma N\rightarrow \pi\pi N$ and $\pi N\rightarrow \pi\gamma N$
are recovered, and corrections to them are given.
The reactions $\pi\rightarrow e\nu \gamma$,
$\pi\rightarrow e\nu e^+e^-$, $\gamma\pi\rightarrow \gamma\pi$
and $\gamma\gamma\rightarrow \pi\pi$ are also discussed.
A general formula for $\pi\pi$ scattering and a new one loop effective action
are obtained. The new effective action reproduces the KSFR relation, and yields
specific estimates for the pion  polarisabilities.
A detailed comparison with baryon-free chiral perturbation
theory to one loop is made. An extension of our effective action to two loops
is outlined.

\vfill
\noindent
\begin{flushleft}
SUNY-NTG-94-57\\
March   1995
\end{flushleft}
\eject
\pagestyle{plain}

{\bf 1. Introduction}

At low energies, chiral symmetry offers a powerful method for dealing with
hadronic processes involving pions.
Beginning with the original work of Nambu and coworkers \cite{nambu}, various
approaches have been formulated to assess
the consequences of chiral symmetry and
its breaking, such as the infinite momentum limit \cite{fubini}, effective
Lagrangians \cite{weinberg1,coleman,weinberg2}, chiral perturbation theory
\cite{dashen,pagels,leutwyler}, and gauge-covariant divergence equations
\cite{veltman}.

In the infinite momentum limit of Fubini and Furlan, the role of the pion as a
Nambu-Goldstone boson is not emphasized. Instead, the multiplet aspect of an
approximate symmetry is used. The approach relies on the algebraic structure of
chiral symmetry through the use of sum rules.

The effective Lagrangian formulation introduced by Weinberg \cite{weinberg2}
stems from soft pion theorems, and emphasizes the character
of the pion as a Goldstone boson. The idea is that since only chiral
symmetry is relevant, then any choice of an effective Lagrangian for
the pion as a Golstone mode that accounts for this symmetry and its
soft breaking, as well as the general precepts of unitarity and causality,
should yield the same low energy predictions. In principle, by calculating and
renormalizing all loop diagrams using infinitely many counterterms one has the
most general result compatible with the underlying symmetries. In practice,
this task can only be carried out using chiral perturbation theory.

In chiral perturbation theory, amplitudes are systematically expanded in the
square of the pion momenta. To leading order, chiral symmetry and
covariance uniquely specify the character of the effective Lagrangian. To
maintain unitarity, loop diagrams have to be included. The latter generate
a growing set of divergences at higher momentum that require the introduction
of new counterterms. As such, the effective Lagrangian  is
nonrenormalizable. Nevertheless, to any order in chiral perturbation theory,
the counterterms are usually finite in number, and their coefficients
may be extracted from the data. The power of this construction is that it
yields relationships between various processes at threshold.
In a series of papers, Gasser and Leutwyler have systematically
used this scheme to one loop \cite{leutwyler}.
Their result is an empirically determined
effective Lagrangian to fourth order in the pion momentum.
Their work has rekindled interest in the subject and triggered a flurry of
papers in this direction \cite{CPT}.

In the approach pioneered by Veltman and Bell \cite{veltman},
the gauge covariant divergence equations as opposed to an effective Lagrangian,
are taken as fundamental. In retrospect, this approach appears to be
relatively unknown and thus unexploited. The purpose of this paper is to
develop the method to the point where it gives results comparable to other
methods. In the process, we also derive  new results in the form of
general formula for $\pi\pi$ scattering, $\pi\rightarrow e\nu\gamma$ decay,
$\gamma\gamma\rightarrow \pi\pi$ fusion, $\pi N$ reactions, $\gamma N$
reactions, and finally a new one-loop effective action.
Through chiral reduction formulas our approach allows an exact rewriting of
scattering amplitudes in terms of correlation functions and form factors, some
of which are directly measurable. The results embody chiral symmetry and
unitarity to all orders in the pion momentum, and as such provide important
insight beyond the loop expansion or specific models.

The organization of the paper is as follows. In section
2, we spell out our conventions.
In section 3, we derive master formulas based on the Veltman-Bell equations
\cite{veltman} and the Peierls-Dyson formula \cite{peierls} in the case where
the pion is massless. In section 4, these formulas are extended to the massive
case. In sections 5-17, the chiral reduction scheme is
applied to $\pi\pi\rightarrow \pi\pi$, $\pi\rightarrow e\nu\gamma$,
$\gamma\pi\rightarrow \gamma\pi$, $\gamma\gamma\rightarrow \pi\pi$,
$\pi N\rightarrow \pi N$, $\gamma N\rightarrow \pi N$,
$\pi N\rightarrow \pi\pi N$, $\gamma N\rightarrow \gamma\pi N$,
$\gamma N\rightarrow \pi\pi N$ and $\pi N\rightarrow \pi\gamma N$
processes. Our treatment of the electromagnetic interaction is only of first
order in the electric charge.
In section 18 the new one-loop effective action is derived, and shown to
reproduce the KSFR relation. This effective action is used to estimate
$\pi\rightarrow e\nu\gamma$, $\pi\rightarrow e\nu e^+ e^-$, $\pi\pi\rightarrow
\pi\pi$, $\gamma\gamma \rightarrow \pi\pi$, $\gamma\pi\rightarrow \gamma\pi$
and the pion  polarisabilities. In section 19, comparison is made with
other approaches, and some open issues are discussed.
In Appendix A, we give some other applications of the Peierls-Dyson formula and
related equations.
In Appendix B, we introduce some definitions in relation to the Dirac
constraint problem. In Appendix C, we discuss the Gell-Mann algebra in our
context.
In Appendix D, we establish some consistency checks and reformulate the master
formula.
In Appendix E, we give a general analysis of one-loop effects.
In Appendix F, we summarise some useful one-loop integrals.
In Appendix G, we discuss some unsolved technical issues
in chiral perturbation theory as formulated by Gasser and Leutwyler
\cite{leutwyler}. In Appendix H, we outline the calculational framework for a
two-loop calculation.

\vskip .5cm
{\bf 2. Conventions}

Throughout, isospin indices will be often suppressed using matrix notation and
\be
A\cdot B = A^a B^a\qquad\qquad
\ub A^{ac} =\epsilon^{abc} A^b\qquad\qquad
{\bf 1}^{ab} =\delta^{ab}.
\label{A1}
\ee
Green's functions are all normalized to $1/(k^2-m^2)$ in momentum space, so
that
\be
(-\Box -m^2 ) \Delta_{R} (x) =
(-\Box -m^2 ) \Delta_{A} (x) =
(-\Box -m^2 ) \Delta_{F} (x) = \delta^4 (x).
\label{A2}
\ee
The retarded and advanced Green's functions satisfy
\be
\Delta_R (x-y) = \Delta_A (y-x) = 0 \qquad\qquad  ( x\preceq y ),
\label{A3}
\ee
where $x\preceq y$ means that $x$ is in the past light-cone of $y$ or
spacelike with
respect to $y$. The Jordan-Pauli distribution $\Delta =\Delta_R -\Delta_A$ is
related to free (incoming) pion fields through
\be
\bigg[\pi_{\rm in}^a (x) , \pi_{\rm in}^b (y) \bigg] =i\delta^{ab}\Delta (x-y )
=\delta^{ab}\frac 1{(2\pi )^3}\int \frac {d^3k}{2k^0}
\bigg(e^{-ik\cdot (x-y)} -e^{+ik\cdot (x-y)}\bigg),
\label{A4}
\ee
where the pion energy is on-shell $k^0=\sqrt{{\vec k}^2 + m_{\pi}^2}$.
The propagator is given by
\be
\Big<0\,{\rm in} |T\Big( \pi_{\rm in}^a (x)\pi_{\rm in}^b (y)\Big) |0\,{\rm in}
\Big>=
i\delta^{ab}\Delta_F (x-y ).
\label{A5}
\ee
The corresponding distributions for the massless case are denoted as usual with
$D$. Creation and annihilation operators are defined as
\be
\pi_{\rm in}^a (x) =
\frac 1{(2\pi )^3}\int \frac {d^3k}{2k^0}
\bigg(a_{\rm in}^a (k) e^{-ik\cdot x} +
a_{\rm in}^{a\dagger} (k) e^{+ik\cdot x}\bigg)
\label{A6}
\ee
so that one-pion states are covariantly normalized.

\vskip .5cm
{\bf 3. The Master Formula : Massless Case}

\vskip .3cm
{\it 3.1. Veltman-Bell Equations}
\vskip .15cm

Our starting point is an action ${\bf I}$, invariant under local
$SU(2)\times SU(2)$ gauged with external c-number vector and axial vector
fields $v_{\mu}^a $ and $a_{\mu}^a$. Examples are massless QCD with two
flavors
\be
{\bf I} = -\frac 1{2g^2}\int d^4x {\rm Tr}_C \bigg( G_{\mu\nu}G^{\mu\nu}\bigg)
+\int d^4x \overline{q}\gamma^{\mu}
\bigg(i\partial_{\mu} + G_{\mu} + v_{\mu}^a\frac {\tau^a}2 +
a_{\mu}^a\frac {\tau^a}2\gamma_5 \bigg) q,
\label{B1}
\ee
where the gluon field strength is given by
\be
G_{\mu\nu} = \partial_{\mu} G_{\nu} -\partial_{\nu} G_{\mu }
-i [G_{\mu} , G_{\nu } ]\,\,
\nonumber
\ee
and the gauged nonlinear sigma model
\be
{\bf I} = \frac{f_{\pi}^2}4 \int d^4x {\rm Tr} \bigg[ &&
\bigg(i\partial_{\mu} U + v_{\mu}^a \bigg[ \frac{\tau^a}2 , U \bigg] +
a_{\mu}^a \bigg[ \frac{\tau^a}2 , U \bigg]_+\bigg)\nonumber\\&&
\bigg(-i\partial^{\mu} U^{\dagger} - v^{\mu b} \bigg[ \frac{\tau^b}2 ,
U^{\dagger} \bigg] +
a^{\mu b}\bigg[ \frac{\tau^b}2 , U^{\dagger} \bigg]_+\bigg) \,\,\,\,\,\bigg],
\label{B2}
\ee
where $U$ is a chiral field. We will assume that the external fields
$v_{\mu}^a$ and $a_{\mu}^a$ are smooth and fall off rapidly at infinity, so
that various operations such as partial integration and functional
differentiation are allowed.

By Noether's theorem, the currents ${\bf V}$ and ${\bf A}$ as defined through
\be
{\bf V}^{\mu a} (x) =\frac {\delta {\bf I} }{\delta v_{\mu}^a (x) }\qquad\qquad
{\bf A}^{\mu a} (x) =\frac {\delta {\bf I} }{\delta a_{\mu}^a (x) }
\nonumber
\ee
obey the Veltman-Bell equations \cite{veltman}
\be
\nabla^{\mu}{\bf V}_{\mu} +\ub a^{\mu} {\bf A}_{\mu} = 0
\label{B3}
\ee
\be
\nabla^{\mu}{\bf A}_{\mu} +\ub a^{\mu} {\bf V}_{\mu} = 0,
\label{B4}
\ee
where $\nabla_{\mu} =\partial_{\mu }\bf{1} +\ub v_{\mu}$ is the vector
covariant
derivative. In writing (\ref{B3}) and (\ref{B4}) we have used the fact that the
Bardeen anomaly and the Wess-Zumino term vanish for $SU(2)\times SU(2)$.
Actually, we should also consider the possibility of global anomalies
associated with
the fact that $\Pi_4 (SU (2 )) = {\bf Z}_2$ \cite{witten}. However,
this issue will be ignored in this paper,
and only gauge transformations connected to the identity will be considered.

Introducing the S-matrix $\cal S$ for fixed incoming fields and using the
Schwinger action principle yield
\be
<\beta\, {\rm in} |\delta {\cal S} |\alpha\, {\rm in} > = i
<\beta \,{\rm in} | {\cal S}\int d^4x \bigg( {\bf V}^{\mu a}\delta v_{\mu}^a +
{\bf A}^{\mu a} \delta a_{\mu}^a \bigg) |\alpha \,{\rm in} > .
\label{B6}
\ee
The completeness of asymptotic states leads to the Peierls-Dyson formula
\cite{peierls} \footnote{See the footnote on p. 147 in particular.}
\be
{\bf V}^{\mu a } (x) = -i {\cal S}^{\dagger} \frac{\delta{\cal S}}{\delta
v^a_{\mu} (x) }
\label{B7}
\ee
\be
{\bf A}^{\mu a } (x) = -i {\cal S}^{\dagger} \frac{\delta{\cal S}}{\delta
a^a_{\mu} (x) }.
\label{B8}
\ee
The Veltman-Bell equations (\ref{B3}-\ref{B4}) are then equivalent to
\be
{\bf X}_V^a (x) {\cal S} = 0
\label{B9}
\ee
\be
{\bf X}_A^a (x) {\cal S} = 0,
\label{B10}
\ee
where we have introduced
\be
{\bf X}_V^a (x) = \nabla_{\mu}^{ac}\frac{\delta}{\delta v_{\mu}^c (x)} +
                  {\ub a}_{\mu}^{ac} (x) \frac{\delta}{\delta a_{\mu}^c (x)}
\label{B11}
\ee
\be
{\bf X}_A^a (x) = \nabla_{\mu}^{ac}\frac{\delta}{\delta a_{\mu}^c (x)} +
                  {\ub a}_{\mu}^{ac} (x) \frac{\delta}{\delta v_{\mu}^c (x)}.
\label{B12}
\ee
Some further related issues to the Peierls-Dyson formula are discussed in
Appendix A.

\vskip .3cm
{\it 3.2. Algebraization}
\vskip .15cm

The relations (\ref{B9}-\ref{B10}) simply express the invariance of the
S-matrix under gauge transformations connected with the identity, since
(\ref{B11}-\ref{B12}) are the generators of the local $SU(2)\times SU (2)$
symmetry:
\be
\bigg[ {\bf X}_V^a (x) , {\bf X}_V^b (y ) \bigg] =
\bigg[ {\bf X}_A^a (x) , {\bf X}_A^b (y ) \bigg] =
-\epsilon^{abc} {\bf X}^c_V (y ) \delta^4 (x-y).
\label{B13}
\ee
\be
\bigg[ {\bf X}_V^a (x) , {\bf X}_A^b (y ) \bigg] =
\bigg[ {\bf X}_A^a (x) , {\bf X}_V^b (y ) \bigg] =
-\epsilon^{abc} {\bf X}^c_A (y )\delta^4 (x-y).
\label{B14}
\ee

We will assume that ${\cal S}$ obeys the Bogoliubov causality condition
\cite{bogoliubov}
\be
\frac{\delta}{\delta v_{\nu}^b (y)}\bigg(\frac{\delta{\cal S}}{\delta v_{\mu}^a
(x) }{\cal S}^{\dagger}\bigg) =
\frac{\delta}{\delta a_{\nu}^b (y)}\bigg(\frac{\delta{\cal S}}{\delta v_{\mu}^a
(x) }{\cal S}^{\dagger}\bigg) = 0 \qquad  ( x\succeq y )
\label{B15}
\ee
\be
\frac{\delta}{\delta v_{\nu}^b (y)}\bigg(\frac{\delta{\cal S}}{\delta a_{\mu}^a
(x) }{\cal S}^{\dagger}\bigg) =
\frac{\delta}{\delta a_{\nu}^b (y)}\bigg(\frac{\delta{\cal S}}{\delta a_{\mu}^a
(x) }{\cal S}^{\dagger}\bigg) = 0 \qquad  ( x\succeq y ),
\label{B16}
\ee
which can be rewritten more conveniently in the following form:
\be
\frac{\delta}{\delta v_{\nu}^b (y )}{\bf V}_{\mu}^a (x) =
\frac{\delta}{\delta a_{\nu}^b (y )}{\bf V}_{\mu}^a (x) = 0 \qquad
( y\succeq x )
\label{B17}
\ee
\be
\frac{\delta}{\delta v_{\nu}^b (y )}{\bf A}_{\mu}^a (x) =
\frac{\delta}{\delta a_{\nu}^b (y )}{\bf A}_{\mu}^a (x) = 0 \qquad
( y\succeq x ).
\label{B18}
\ee
In other words, for fixed incoming fields, the current at x depends on the
external fields only in the past light-cone of x. The Maurer-Cartan equations
\be
\frac{\delta}{\delta v_{\nu}^b (y )}{\bf V}^{\mu a} (x) -
\frac{\delta}{\delta v_{\mu}^a (x )}{\bf V}^{\nu b} (y) =
i\bigg[ {\bf V}^{\mu a } (x) , {\bf V}^{\nu b} (y) \bigg]
\label{B19}
\ee
\be
\frac{\delta}{\delta a_{\nu}^b (y )}{\bf V}^{\mu a} (x) -
\frac{\delta}{\delta v_{\mu}^a (x )}{\bf A}^{\nu b} (y) =
i\bigg[ {\bf V}^{\mu a } (x) , {\bf A}^{\nu b} (y) \bigg]
\label{B20}
\ee
\be
\frac{\delta}{\delta a_{\nu}^b (y )}{\bf A}^{\mu a} (x) -
\frac{\delta}{\delta a_{\mu}^a (x )}{\bf A}^{\nu b} (y) =
i\bigg[ {\bf A}^{\mu a } (x) , {\bf A}^{\nu b} (y) \bigg]
\label{B21}
\ee
then imply that the currents are local
\be
\bigg[ {\bf V}^{\mu a } (x) , {\bf V}^{\nu b} (y) \bigg] =
\bigg[ {\bf V}^{\mu a } (x) , {\bf A}^{\nu b} (y) \bigg] =
\bigg[ {\bf A}^{\mu a } (x) , {\bf A}^{\nu b} (y) \bigg] = 0
\qquad  ( x\sim y ).
\label{B22}
\ee

\vskip .3cm
{\it 3.3. Ward-Identities}
\vskip .15cm

The relations (\ref{B7}-\ref{B8}) and (\ref{B17}-\ref{B21}) allow us to define
time-ordered and retarded products  by
\be
&&{\rm T^*} \bigg( {\bf V}^{\mu a} (x_1 ) ... {\bf V}^{\nu b} (x_m)
{\bf A}^{\rho c} (x_{m+1} ) ...{\bf A}^{\sigma d} (x_n )\bigg) =\nonumber\\&&
(-i)^n {\cal S}^{\dagger}
\frac{\delta}{\delta v_{\mu}^a (x_1 )} ...
\frac{\delta}{\delta v_{\nu}^b (x_m )}
\frac{\delta}{\delta a_{\rho}^c (x_{m+1} )} ...
\frac{\delta}{\delta a_{\sigma}^d (x_n )} {\cal S}
\label{B23}
\ee
\be
&&{\rm R^*} \bigg[ {\bf V}^{\alpha e} (x),
 {\bf V}^{\mu a} (x_1 ) ... {\bf V}^{\nu b} (x_m)
{\bf A}^{\rho c} (x_{m+1} ) ...{\bf A}^{\sigma d} (x_n )\bigg] =\nonumber\\&&
(-i)^n
\frac{\delta}{\delta v_{\mu}^a (x_1 )} ...
\frac{\delta}{\delta v_{\nu}^b (x_m )}
\frac{\delta}{\delta a_{\rho}^c (x_{m+1} )} ...
\frac{\delta}{\delta a_{\sigma}^d (x_n )} {\bf V}^{\alpha e} (x)
\label{B24}
\ee
\be
&&{\rm R^*} \bigg[ {\bf A}^{\alpha e} (x),
 {\bf V}^{\mu a} (x_1 ) ... {\bf V}^{\nu b} (x_m)
{\bf A}^{\rho c} (x_{m+1} ) ...{\bf A}^{\sigma d} (x_n )\bigg] =\nonumber\\&&
(-i)^n
\frac{\delta}{\delta v_{\mu}^a (x_1 )} ...
\frac{\delta}{\delta v_{\nu}^b (x_m )}
\frac{\delta}{\delta a_{\rho}^c (x_{m+1} )} ...
\frac{\delta}{\delta a_{\sigma}^d (x_n )} {\bf A}^{\alpha e} (x).
\label{B25}
\ee
Differentiating (\ref{B3},\ref{B4},\ref{B9},\ref{B10}) then yields the
Ward identities for (\ref{B23}-{\ref{B25}), $e.g.$
\be
&&\nabla_{\mu}^{ac} (x)
{\rm T}^* \bigg( {\bf V}^{\mu c} (x) {\bf V}^{\nu b} (y )\bigg)
+ {\ub a}_{\mu}^{ac} (x)
{\rm T}^* \bigg( {\bf A}^{\mu c} (x) {\bf V}^{\nu b} (y )\bigg) =\nonumber\\
&&\nabla_{\mu}^{ac} (x)
{\rm R}^* \bigg[ {\bf V}^{\mu c} (x) ,{\bf V}^{\nu b} (y )\bigg]
+ {\ub a}_{\mu}^{ac} (x)
{\rm R}^* \bigg[ {\bf A}^{\mu c} (x) ,{\bf V}^{\nu b} (y )\bigg] =\nonumber\\
&&\nabla_{\mu}^{ac} (x)
{\rm T}^* \bigg( {\bf A}^{\mu c} (x) {\bf A}^{\nu b} (y )\bigg)
+ {\ub a}_{\mu}^{ac} (x)
{\rm T}^* \bigg( {\bf V}^{\mu c} (x) {\bf A}^{\nu b} (y )\bigg) =\nonumber\\
&&\nabla_{\mu}^{ac} (x)
{\rm R}^* \bigg[ {\bf A}^{\mu c} (x) , {\bf A}^{\nu b} (y )\bigg]
+ {\ub a}_{\mu}^{ac} (x)
{\rm R}^* \bigg[ {\bf V}^{\mu c} (x) , {\bf A}^{\nu b} (y )\bigg] =\nonumber\\
&&i\epsilon^{abc}\delta^4 (x-y) {\bf V}^{\nu c} (y)
\label{B26}
\ee
\be
&&\nabla_{\mu}^{ac} (x)
{\rm T}^* \bigg( {\bf V}^{\mu c} (x) {\bf A}^{\nu b} (y )\bigg)
+ {\ub a}_{\mu}^{ac} (x)
{\rm T}^* \bigg( {\bf A}^{\mu c} (x) {\bf A}^{\nu b} (y )\bigg) =\nonumber\\
&&\nabla_{\mu}^{ac} (x)
{\rm R}^* \bigg[ {\bf V}^{\mu c} (x) ,{\bf A}^{\nu b} (y )\bigg]
+ {\ub a}_{\mu}^{ac} (x)
{\rm R}^* \bigg[ {\bf A}^{\mu c} (x) ,{\bf A}^{\nu b} (y )\bigg] =\nonumber\\
&&\nabla_{\mu}^{ac} (x)
{\rm T}^* \bigg( {\bf A}^{\mu c} (x) {\bf V}^{\nu b} (y )\bigg)
+ {\ub a}_{\mu}^{ac} (x)
{\rm T}^* \bigg( {\bf V}^{\mu c} (x) {\bf V}^{\nu b} (y )\bigg) =\nonumber\\
&&\nabla_{\mu}^{ac} (x)
{\rm R}^* \bigg[ {\bf A}^{\mu c} (x) , {\bf V}^{\nu b} (y )\bigg]
+ {\ub a}_{\mu}^{ac} (x)
{\rm R}^* \bigg[ {\bf V}^{\mu c} (x) , {\bf V}^{\nu b} (y )\bigg] =\nonumber\\
&&i\epsilon^{abc}\delta^4 (x-y) {\bf A}^{\nu c} (y).
\label{B27}
\ee

\vskip .3cm
{\it 3.4. Asymptotic Condition}
\vskip .15cm

So far the discussion has been independent of whether chiral symmetry is
manifest  or spontaneously broken. The spontaneous breaking of chiral
symmetry can be expressed in terms of the  asymptotic conditions
\be
{\bf A}_{\mu}^a (x)\rightarrow -f_{\pi} \partial_{\mu} \pi_{\rm in, out}^a (x)
\qquad\qquad x^0\rightarrow\mp\infty,
\label{B28}
\ee
where $\pi_{\rm in}$ and $\pi_{\rm out}$ are free incoming and outgoing
massless pions, and $f_{\pi}$ is the pion decay constant. In writing
(\ref{B28}) we have also assumed the absence of stable axial vector mesons
or other stable pseudoscalars.
Analysis of the gauged nonlinear sigma model and experience with ordinary
scalar field theories \cite{yamagishi} indicates that the condition
(\ref{B28}) can be conveniently incorporated into the present analysis if we
impose the constraint
\be
f_{\pi}\nabla^{\mu} a_{\mu} - J =0
\label{B29}
\ee
Here $J$ is some
new function. Some useful concepts regarding the use of (\ref{B29}) as a Dirac
constraint may be found in Appendix B.

With the above in mind, let us introduce a modified action,
\be
\hat{\bf I} = {\bf I} -
\frac {f_{\pi}^2}2 \int d^4x \,\,a_{\mu}(x) \cdot a^{\mu} (x) -
\int d^4x \bigg(f_{\pi} \nabla^{\mu} a_{\mu} - J \bigg) (x) \cdot \pi (x),
\label{B39}
\ee
where $\pi$ is some suitable pion field. The new S-matrix $\hat{\cal S}$ is
related to the old S-matrix ${\cal S}$ subject to (\ref{B29}) as
\be
\hat{\cal S} \approx {\cal S} {\rm exp}\bigg({-i\frac{f_{\pi}^2}2 \int d^4x
a_{\mu}(x) \cdot a^{\mu} (x)}\bigg).
\label{B40}
\ee
In general, the new currents
\be
{\bf j}_V^{\mu a} (x) = \frac{\delta\hat{\bf I}}{\delta v_{\mu}^a (x)}
\qquad\qquad
{\bf j}_A^{\mu a} (x) = \frac{\delta\hat{\bf I}}{\delta a_{\mu}^a (x)}
\nonumber
\ee
have no simple relation with the old currents, ${\bf V}_{\mu}^a$
and ${\bf A}_{\mu}^a$, since the quark or pion fields in ${\bf I}$ and
$\hat{\bf I}$ obey different equations of motion. However, they can be weakly
related as
\be
{\bf j}_{V\mu}^a (x) = {\bf V}_{\mu}^a (x) - f_{\pi} {\ub a}_{\mu}^{ac}
(x)\pi^c (x)
\label{B41}
\ee
\be
{\bf j}_{A\mu}^a (x) \approx {\bf A}_{\mu}^a (x) - f^2_{\pi} { a}_{\mu}^{a}
(x) + f_{\pi} (\nabla_{\mu} \pi )^a (x).
\label{B42}
\ee
We also observe that the pion field $\pi$ follows from $\hat{\bf I}$ through
\be
\pi^a (x) = \frac{\delta\hat{\bf I}}{\delta J^a (x)}.
\label{B43}
\ee
Since (\ref{B41}-\ref{B42}) remain invariant under the analog of (\ref{B38}),
\be
{\bf j}_{V}^{\mu a} (x)  &&\rightarrow
{\bf j}_{V}^{\mu a} (x) + f_{\pi} ({\ub a}^{\mu} \Lambda )^a (x )
\nonumber\\
{\bf j}_{A}^{\mu a} (x)  &&\rightarrow
{\bf j}_{A}^{\mu a} (x) -f_{\pi} (\nabla^{\mu} \Lambda )^a (x )
\nonumber\\
\pi^a (x) &&\rightarrow  \pi^a (x) -\Lambda^a (x ),
\label{B44}
\ee
we may assume without loss of generality that the asymptotic one-pion
components (\ref{B28}) are contained entirely in the field ${\pi}$
and not in the new axial current ${\bf j}_A$ or $\Lambda$.
As a result, the transformations (\ref{B44}) may be regarded as the analog of
the Nishijima-Gursey  transformations \cite{nishijima} or
reparametrizations of the chiral field $U$ in the nonlinear sigma model.

\vskip .3cm
{\it 3.5. Master Formula}
\vskip .15cm

Upon substitution of (\ref{B41}-\ref{B42}) into the Veltman-Bell equations
(\ref{B3}-\ref{B4}), we obtain the weak relations
\be
\nabla^{\mu} {\bf j}_{V\mu} + {\ub a}^{\mu} {\bf j}_{A\mu} +  {\ub J} \pi
\approx 0
\label{B45}
\ee
\be
\bigg( -\nabla^{\mu}\nabla_{\mu} +{\ub a}^{\mu}{\ub a}_{\mu} \bigg) \pi \approx
-J - \frac 1{f_{\pi}}
\bigg( \nabla^{\mu} {\bf j}_{A\mu} +{\ub a}^{\mu} {\bf j}_{V \mu}\bigg)
\label{B46}
\ee
The first equation is simply the weak version of the vector conservation law
for $\hat{\bf I}$, namely
\be
\nabla^{\mu} {\bf j}_{V\mu} + {\ub a}^{\mu} {\bf j}_{A\mu} + {\ub J} \pi
=0.
\label{B47}
\ee

To solve the second equation, we will introduce the retarded and advanced
Green's functions
\be
\bigg( -\nabla^{\mu}\nabla_{\mu} +{\ub a}^{\mu}{\ub a}_{\mu} \bigg)
\,G_{R,A} (x, y ) = \delta^4 (x-y) \,\,{\bf 1}
\label{B48}
\ee
and
\be
{\bf K} = \nabla^{\mu}\nabla_{\mu} -{\ub a}^{\mu}{\ub a}_{\mu} -\Box =
2{\ub v}^{\mu}\partial_{\mu} + (\partial^{\mu}{\ub v}_{\mu}) +
{\ub v}^{\mu}{\ub v}_{\mu} -{\ub a}^{\mu}{\ub a}_{\mu}.
\label{B49}
\ee
The combination (\ref{B48}-\ref{B49}) obeys a standard identity
\be
G_{R,A} = D_{R, A} + D_{R, A} {\bf K} G_{R, A} =
D_{R, A} + G_{R, A} {\bf K} D_{R, A}
\label{B50}
\ee
where we have introduced a condensed notation with space-time integration
absorbed into matrix multiplication. Also,
\be
G_{R}^{ab} (x, y ) = G_{A}^{ba} (y, x )
\label{B51}
\ee
since the operator $-\nabla^{\mu}\nabla_{\mu} +{\ub a}^{\mu}{\ub a}_{\mu}$
is self-adjoint, being associated with the quadratic action
\be
{\bf I}_Q^0 = \frac 12 \int d^4x \bigg(
(\nabla^{\mu}\pi )\cdot (\nabla_{\mu}\pi ) -({\ub a}^{\mu}\pi )\cdot
({\ub a}_{\mu}\pi )\bigg).
\label{B52}
\ee

Solving (\ref{B46}) under the asymptotic conditions (\ref{B28}) gives
\be
\pi \approx &&\bigg( 1 +G_R {\bf K} \bigg) \pi_{\rm in} - G_R J
-\frac 1{f_{\pi}} G_R \bigg( \nabla^{\mu}{\bf j}_{A \mu} +{\ub a}^{\mu}
{\bf j}_{V\mu }\bigg) \nonumber\\\approx &&
\bigg( 1 +G_A {\bf K} \bigg) \pi_{\rm out} -  G_A J
-\frac 1{f_{\pi}} G_A \bigg( \nabla^{\mu}{\bf j}_{A \mu} +{\ub a}^{\mu}
{\bf j}_{V\mu }\bigg).
\label{B53}
\ee
Using the Peierls-Dyson formula
\be
&&{\bf j}_{V}^{\mu a } (x)  = -i\hat{\cal S}^{\dagger}
\frac{\delta \hat {\cal S}}{\delta v_{\mu}^a (x)}\nonumber\\ &&
{\bf j}_{A}^{\mu a } (x)  = -i\hat{\cal S}^{\dagger}
\frac{\delta \hat {\cal S}}{\delta a_{\mu}^a (x)}\nonumber\\ &&
\pi^a (x)  = - i\hat{\cal S}^{\dagger}
\frac{\delta \hat {\cal S}}{\delta J^a (x)}
\label{B54}
\ee
and the Yang-Feldman relation $\pi_{\rm out} = \hat{\cal S}^{\dagger}\pi_{\rm
in}\hat{\cal S}$, we may rewrite (\ref{B53}) into the following form
\be
\bigg( \frac{\delta}{\delta J} + \frac 1{f_{\pi}}
G_R {\bf X}_A \bigg) \hat{\cal S} \approx
-i G_R J \hat{\cal S} + i \hat{\cal S}
\bigg( 1 + G_R {\bf K}\bigg) \pi_{\rm in}
\label{B55}
\ee
\be
\bigg( \frac{\delta}{\delta J} + \frac 1{f_{\pi}}
G_A {\bf X}_A \bigg) \hat{\cal S} \approx
-i G_A J \hat{\cal S} +
i\bigg( 1 + G_A {\bf K}\bigg) \pi_{\rm in} \hat{\cal S}.
\label{B56}
\ee
{}From (\ref{B47}) we also have
\be
{\bf T}_V \,\hat{\cal S} \equiv
\bigg( {\bf X}_V +{\ub J}\frac {\delta}{\delta J} \bigg)
\hat{\cal S} = 0.
\label{B57}
\ee
Equations (\ref{B55}-\ref{B57}) constitute the $master$ $formulas$ in the
chiral limit, replacing (\ref{B9}-\ref{B10}). The differential
operators on the left-hand side of (\ref{B55}-\ref{B56}) may be checked to be
tangent vectors, so the system is meaningful as a weak relation.

Some insight into these equations may be obtained by expanding the gauged
nonlinear sigma model action (\ref{B2})
\be
{\bf I} =\frac{f_{\pi}^2}2 \int d^4 x\, a^{\mu} (x)\cdot a_{\mu} (x) +
f_{\pi} \int d^4x (\nabla^{\mu} a_{\mu} ) (x) \cdot \pi (x) +
{\bf I}_Q^0 + {\cal O} \bigg( \frac{\pi^3}{f_{\pi}}\bigg)
\label{B58}
\ee
\be
\hat{\bf I}  = \int d^4x J(x)\cdot \pi (x)  +{\bf I}_Q^0  +
{\cal O} \bigg( \frac{\pi^3}{f_{\pi}}\bigg),
\label{B59}
\ee
where ${\bf I}_Q^0$ is the quadratic action (\ref{B52}). We note that the above
expansion up to and including quadratic terms is invariant under redefinitions
of the pion field, $\pi\rightarrow \pi' =\pi + {\cal O} (\pi^3 )$, which are
independent of the external fields and conserve G-parity. It is worth noting
that the system (\ref{B55}-\ref{B57}) is linear in the extended S-matrix
$\hat{\cal S}$. This is in contrast with the unitarity and causality conditions
which are quadratic in $\hat{\cal S}$. As shown in Appendix D, this allows us
to associate characteristic curves with the system. The equations
(\ref{B55}-\ref{B57}) may be used to derive various results in the chiral
limit. However, we will proceed to the more realistic case with massive pions.

\vskip .5cm
{\bf 4. The Master Formula : Massive Case}

\vskip .3cm
{\it 4.1. Asymptotic Condition}
\vskip .15cm

As before, we will assume the asymptotic conditions
\be
{\bf A}_{\mu}^a (x)\rightarrow -f_{\pi} \partial_{\mu} \pi_{\rm in, out}^a (x)
\qquad\qquad x^0\rightarrow\mp\infty,
\label{B60}
\ee
which also imply that
\be
\partial^{\mu}{\bf A}_{\mu}^a (x)\rightarrow +f_{\pi} m_{\pi}^2
\pi_{\rm in, out}^a (x)
\qquad\qquad x^0\rightarrow\mp\infty.
\label{B61}
\ee
For definiteness, we will take the symmetry breaking term to be a
scalar-isoscalar that transforms as the $(2,2)$ representation of
$SU(2)\times SU(2)$. Specifically, for QCD we will take
\be
-\hat{m}\int d^4x \overline{q} q
\label{B62}
\ee
while for the nonlinear sigma model we will take
\be
+\frac{f_{\pi}^2m_{\pi}^2}4 \int d^4 x {\rm Tr} \bigg( U + U^{\dagger} \bigg).
\label{B63}
\ee
We may further couple external scalar and pseudoscalar sources $s$ and $p^a$,
through
\be
-\frac{\hat{m}}{m_{\pi}^2}\int d^4x \overline{q}\bigg(
m_{\pi}^2 + s -i\gamma_5 \tau^a p^a \bigg) q
\label{B64}
\ee
\be
+\frac{f_{\pi}^2}4 \int d^4x {\rm Tr}\bigg(
(m_{\pi}^2 + s- i\tau^a p^a ) U + U^{\dagger}
(m_{\pi}^2 + s +i\tau^a p^a )  \bigg)
\label{B65}
\ee
and assume that the sources are again smooth and fall off rapidly enough at
infinity.

\vskip .3cm
{\it 4.2. Master Formula}
\vskip .15cm

{}From here on, the action ${\bf I}$ will be understood as the sum of
(\ref{B1}) and (\ref{B64}) for QCD, and (\ref{B2}) and (\ref{B65}) for the
nonlinear sigma model. The associated currents and S-matrix will be denoted by
${\bf V}_{\mu}^a$, ${\bf A}_{\mu}^a$, and ${\cal S}$ as before.
We will define the scalar and pseudoscalar densities as
\be
\sigma (x ) = \frac 1{f_{\pi}} \frac{\delta{\bf I}}{\delta s (x)} = -
\frac{i}{f_{\pi}} {\cal S}^{\dagger}\frac{\delta{\cal S}}{\delta s (x )}
\label{B66}
\ee
\be
\pi^a  (x ) = \frac 1{f_{\pi}} \frac{\delta{\bf I}}{\delta p^a (x)} = -
\frac{i}{f_{\pi}} {\cal S}^{\dagger}\frac{\delta{\cal S}}{\delta p^a (x )}.
\label{B67}
\ee
Explicitly, these densities for QCD are
\be
\sigma = -\frac{\hat m}{f_{\pi}m_{\pi}^2} \overline{q} q\qquad\qquad
\pi^a =+ \frac {\hat m}{f_{\pi} m_{\pi}^2}\overline q i\gamma_5 \tau^a q
\label{B68}
\ee
and for the nonlinear sigma model are
\be
\sigma = +\frac{f_{\pi}}4 {\rm Tr}\bigg( U + U^{\dagger} \bigg) \qquad\qquad
\pi^a= -i\frac{f_{\pi}}4 {\rm Tr}\bigg( \tau^a (U - U^{\dagger}) \bigg).
\label{B69}
\ee
The Veltman-Bell equations (\ref{B3}-\ref{B4}) then read
\be
\nabla^{\mu}{\bf V}_{\mu} +\ub a^{\mu} {\bf A}_{\mu} + f_{\pi} {\ub p} \pi = 0
\label{B70}
\ee
\be
\nabla^{\mu}{\bf A}_{\mu} +\ub a^{\mu} {\bf V}_{\mu} = f_{\pi} (m_{\pi}^2 + s )
\pi -f_{\pi} p \sigma,
\label{B71}
\ee
where we have used the fact that no anomaly is generated by adding the
densities (\ref{B68}) to the QCD action. For the nonlinear sigma model, we may
take (\ref{B70}-\ref{B71}) as part of the definition of the theory. By the
Peierls-Dyson formula (\ref{B7}-\ref{B8}, \ref{B66}-\ref{B67}), the
Veltman-Bell equations (\ref{B70}-\ref{B71}) are equivalent to
\be
\bigg( {\bf X}_V +{\ub p}\frac {\delta}{\delta p} \bigg) {\cal S} = 0
\label{B72}
\ee
\be
\bigg( {\bf X}_A -(m_{\pi}^2 +s)\frac {\delta}{\delta p} +
p\frac {\delta}{\delta s} \bigg) {\cal S} = 0.
\label{B73}
\ee
Differentiation of these equations yields Ward identities as for the massless
case before.

To incorporate the asymptotic conditions (\ref{B60}-\ref{B61}) more
efficiently, we introduce as before a modified action $\hat{\bf I}$
\be
\hat{\bf I} = {\bf I} -f_{\pi}^2 \int d^4x \bigg( s (x) +
\frac 12 a^{\mu} (x) \cdot a_{\mu} (x) \bigg).
\label{B74}
\ee
The new S-matrix $\hat{\cal S}$ relates to the old S-matrix ${\cal S}$ through
\be
\hat{\cal S} = {\cal S} {\rm exp}\bigg({-if_{\pi}^2\int d^4x
\bigg( s (x) + \frac 12 a^{\mu} (x) \cdot a_{\mu} (x) \bigg)}\bigg).
\label{B75}
\ee
We further set $p=J/f_{\pi}-\nabla^{\mu} a_{\mu}$ and treat
$\phi$ = ($v_{\mu}^a$, $a_{\mu}^a$, $s$ ,$J^a$) as $independent$ variables.
Modified currents and densities may then be introduced as
\be
{\bf j}_{V}^{\mu a}(x ) &&= \frac{\delta\hat{\bf I}}{\delta v_{\mu}^a (x)} = -
i \hat{\cal S}^{\dagger}\frac{\delta\hat{\cal S}}{\delta v_{\mu}^a (x )}
\nonumber\\
{\bf j}_{A}^{\mu a}(x ) &&= \frac{\delta\hat{\bf I}}{\delta a_{\mu}^a (x)} = -
i \hat{\cal S}^{\dagger}\frac{\delta\hat{\cal S}}{\delta a_{\mu}^a (x )}
\nonumber\\
\hat{\sigma }(x ) &&= \frac 1{f_{\pi}} \frac{\delta\hat{\bf I}}{\delta s (x)} =
-
\frac{i}{f_{\pi}} \hat{\cal S}^{\dagger}\frac{\delta\hat{\cal S}}{\delta s (x
)}
\nonumber\\
\hat\pi^a  (x ) &&=  \frac{\delta\hat{\bf I}}{\delta J^a (x)} = -
i \hat{\cal S}^{\dagger}\frac{\delta\hat{\cal S}}{\delta J^a (x )}.
\label{B76}
\ee
Applying the chain rule
\be
\frac{\delta }{\delta v_{\mu}^a (x) }|_p =
\frac{\delta }{\delta v_{\mu}^a (x) }|_J + f_{\pi}
{\ub a}^{\mu ac} (x) \frac{\delta
}{\delta J^c (x)}
\label{B77}
\ee
\be
\frac{\delta }{\delta a_{\mu}^a (x) }|_p =
\frac{\delta }{\delta a_{\mu}^a (x) }|_J -f_{\pi}
\nabla^{\mu ac} (x) \frac{\delta
}{\delta J^c (x)}
\label{B78}
\ee
and so on to (\ref{B74}-\ref{B75}) yields
\be
{\bf V}_{\mu}^a (x) = {\bf j}_{V\mu}^a (x) + f_{\pi} {\ub a}_{\mu}^{ac} (x)
\hat{\pi}^c (x)
\label{B79}
\ee
\be
{\bf A}_{\mu}^a (x) = {\bf j}_{A\mu}^a (x) + f_{\pi}^2 a_{\mu}^a (x)
-f_{\pi} (\nabla_{\mu} \hat\pi )^a (x)
\label{B80}
\ee
\be
\sigma (x ) =\hat{\sigma } (x) + f_{\pi}
\label{B81}
\ee
\be
\pi^a(x) = \hat{\pi }^a (x).
\label{B82}
\ee
We observe that ${\bf j}_{A\mu}^a$ is again free of the asymptotic one-pion
component by (\ref{B60}, \ref{B61}, \ref{B71}, \ref{B80} ).

Further differentiation also gives
\be
T^*\bigg( {\bf V}_{\mu}^a (x) {\bf V}_{\nu}^b (y) \bigg) = &&
T^*\bigg( {\bf j}_{V\mu}^a (x) {\bf j}_{V\nu}^b (y) \bigg) +
f_{\pi}{\ub a}_{\mu}^{ac} (x)
T^*\bigg( \pi^c (x) {\bf j}_{V\nu}^b (y) \bigg)  \nonumber \\&&+
f_{\pi}{\ub a}_{\nu}^{bd} (y)
T^*\bigg( {\bf j}_{V\mu}^a (x) \pi^d (y) \bigg)
+ f_{\pi}^2 {\ub a}_{\mu}^{ac} (x) {\ub a}_{\nu}^{bd} (y)
T^*\bigg( \pi^c (x) \pi^d (y )\bigg)\nonumber\\
\label{B83}
\ee
\be
T^*\bigg( {\bf V}_{\mu}^a (x) {\bf A}_{\nu}^b (y) \bigg) = &&
T^*\bigg( {\bf j}_{V\mu}^a (x) {\bf j}_{A\nu}^b (y) \bigg) \nonumber\\&&+
f_{\pi}^2 a_{\nu}^b (y) {\bf j}_{V\mu}^a (x) + f_{\pi}^3 a_{\nu}^b (y) {\ub
a}_{\mu}^{ac} (x) \pi^c (x) \nonumber\\ &&-
if_{\pi}\epsilon^{abc} g_{\mu\nu} \delta^4 (x-y) \pi^c (x) +
f_{\pi}{\ub a}_{\mu}^{ac} (x)
T^*\bigg( \pi^c (x) {\bf j}_{A\nu}^b (y) \bigg)  \nonumber \\&&-
f_{\pi}\nabla_{\nu}^{bd} (y)
T^*\bigg( {\bf j}_{V\mu}^a (x) \pi^d (y) \bigg)
-f_{\pi}^2 {\ub a}_{\mu}^{ac} (x) \nabla_{\nu}^{bd} (y)
T^*\bigg( \pi^c (x) \pi^d (y )\bigg)\nonumber\\
\label{B84}
\ee
\be
T^*\bigg( {\bf A}_{\mu}^a (x) {\bf A}_{\nu}^b (y) \bigg) = &&
-if_{\pi}^2 \delta^4 (x-y) g_{\mu\nu} \delta^{ab} \nonumber\\ &&+
f_{\pi}^4 a_{\mu}^a (x) a_{\nu}^b (y) + f_{\pi}^2 a_{\mu}^a (x) {\bf
j}_{A\nu}^b (y) + f_{\pi}^2 {\bf j}_{A\mu}^a (x) a_{\nu}^b (y) \nonumber\\ &&+
T^*\bigg( {\bf j}_{A\mu}^a (x) {\bf j}_{A\nu}^b (y) \bigg) -
f_{\pi}^3 a_{\mu}^a (x) \nabla_{\nu}^{bd} (y) \pi^d (y) \nonumber\\ &&-
f_{\pi}^3 \nabla_{\mu}^{ac} (x) \pi^c (x) a_{\nu}^b (y) -
f_{\pi}\nabla_{\mu}^{ac} (x) T^*\bigg( \pi^c (x) {\bf j}_{A\nu}^b (y)\bigg)
\nonumber\\&&-
f_{\pi}\nabla_{\nu}^{bd} (y) T^*\bigg( {\bf j}_{A\mu}^a (x) \pi^d (y)\bigg) +
f_{\pi}^2 \nabla_{\mu}^{ac} (x)\nabla_{\nu}^{bd} (y)
T^*\bigg( \pi^c (x) \pi^d (y) \bigg)\nonumber\\
\label{B85}
\ee
\be
T^*\bigg( {\bf V}_{\mu}^a (x) \pi^b (y) \bigg) =
T^*\bigg( {\bf j}_{V\mu}^a (x) \pi^b (y)\bigg) +
f_{\pi} {\ub a}_{\mu}^{ac} (x) T^*\bigg( \pi^c (x) \pi^b (y )\bigg)
\label{B86}
\ee
\be
T^*\bigg( {\bf A}_{\mu}^a (x) \pi^b (y) \bigg) =
T^*\bigg( {\bf j}_{A\mu}^a (x) \pi^b (y)\bigg) +
f_{\pi}^2 a_{\mu}^a (x) \pi^b (y) -
f_{\pi} \nabla_{\mu}^{ac} (x) T^*\bigg( \pi^c (x) \pi^b (y )\bigg).
\label{B87}
\ee

Substitution of (\ref{B79}-\ref{B82}) into (\ref{B70}) gives (\ref{B47})
as before for the vector current and hence (\ref{B57}).
For the axial current, we find
\be
\nabla^{\mu} {\bf j}_{A\mu} + {\ub a}^{\mu} {\bf j}_{V\mu} =&&
-f_{\pi}^2 \nabla^{\mu}a_{\mu} + f_{\pi} \nabla^{\mu}\nabla_{\mu} \pi -f_{\pi}
{\ub a}^{\mu}{\ub a}_{\mu} \pi \nonumber\\&&+
f_{\pi} (m_{\pi}^2 + s ) \pi - (J-f_{\pi}\nabla^{\mu} a_{\mu} ) (\hat\sigma
+f_{\pi} ).
\label{B88}
\ee
As before, we may introduce the Green's functions
\be
\bigg( -\nabla^{\mu}\nabla_{\mu} +{\ub a}^{\mu}{\ub a}_{\mu}
-m_{\pi}^2 -s \bigg)
\,G_{R,A} (x, y ) = \delta^4 (x-y)\,\, {\bf 1}
\label{B89}
\ee
and
\be
{\bf K} = \nabla^{\mu}\nabla_{\mu} -{\ub a}^{\mu}{\ub a}_{\mu}+s -\Box =
2{\ub v}^{\mu}\partial_{\mu} + (\partial^{\mu}{\ub v}_{\mu}) +
{\ub v}^{\mu}{\ub v}_{\mu} -{\ub a}^{\mu}{\ub a}_{\mu} +s
\label{B90}
\ee
associated with the quadratic action
\be
{\bf I}_Q = \frac 12 \int d^4x \bigg(
(\nabla^{\mu}\pi )\cdot (\nabla_{\mu}\pi ) -({\ub a}^{\mu}\pi )\cdot
({\ub a}_{\mu}\pi ) -(m_{\pi}^2 +s ) \pi\cdot\pi \bigg).
\label{B91}
\ee
The Green's functions obey
\be
G_{R,A} = \Delta_{R, A} + \Delta_{R, A} {\bf K} G_{R, A} =
\Delta_{R, A} + G_{R, A} {\bf K} \Delta_{R, A}
\label{B92}
\ee
and (\ref{B51}) as before.
Integration of (\ref{B88}) under the boundary conditions (\ref{B60}-\ref{B61})
gives
\be
\pi =&&\bigg( 1 +G_R {\bf K} \bigg) \pi_{\rm in} - G_R J
+G_R \bigg( \nabla^{\mu} a_{\mu} - J/f_{\pi} \bigg) \hat\sigma
-\frac 1{f_{\pi}} G_R \bigg( \nabla^{\mu}{\bf j}_{A \mu} +{\ub a}^{\mu}
{\bf j}_{V\mu }\bigg) \nonumber\\ =&&
\bigg( 1 +G_A {\bf K} \bigg) \pi_{\rm out} - G_A J
+G_A \bigg( \nabla^{\mu} a_{\mu}- J/f_{\pi} \bigg) \hat\sigma
-\frac 1{f_{\pi}} G_A \bigg( \nabla^{\mu}{\bf j}_{A \mu} +{\ub a}^{\mu}
{\bf j}_{V\mu }\bigg).\nonumber\\
\label{B93}
\ee
The Peierls-Dyson formula and the Yang-Feldman relation may be used as before
to convert (\ref{B93}) into
\be
\frac{\delta}{\delta J} \hat{\cal S} =&& -iG_R J \hat{\cal S} +
i \hat{\cal S} \bigg( 1+ G_R {\bf K}\bigg) \pi_{in} \nonumber\\&& +
\frac 1{f_{\pi}}
G_R \bigg( \nabla^{\mu} a_{\mu} - J/f_{\pi} \bigg) \frac{\delta \hat{\cal
S}}{\delta
s}-\frac 1{f_{\pi}}
G_R {\bf X}_A \hat{\cal S}\nonumber\\ =&&
-i G_A J \hat{\cal S} +
i   \bigg( 1+ G_A {\bf K}\bigg) \pi_{\rm in} \hat{\cal S}\nonumber\\&& +
\frac 1{f_{\pi}}G_A \bigg( \nabla^{\mu} a_{\mu} - J/f_{\pi} \bigg)
\frac{\delta \hat{\cal S}}{\delta
s}-\frac 1{f_{\pi}}G_A {\bf X}_A \hat{\cal S}.
\label{B94}
\ee
This equation together with  (\ref{B57}) constitute
the desired master formulas
for $SU(2)\times SU(2)$ symmetry with massive pions.

\vskip .3cm
{\it 4.3. Chronological Products}
\vskip .15cm

The equation (\ref{B94}) allows us to express the Green's functions involving
$\pi$ fields
 in terms of ${\bf j}_{V\mu}^a$, ${\bf j}_{A\mu}^a$, $\hat\sigma$ and
$\pi_{\rm in}$. Using the infinitesimal variations
\be
\delta G_{R, A} = G_{R, A} \bigg( \delta{\bf K}\bigg) G_{R, A}
\label{B95}
\ee
we find,
\be
T^*\bigg( \pi^a (x) {\bf j}_{A\beta}^b (y)\bigg) =&& -\hat{\cal
S}^{\dagger} \frac{\delta^2\hat{\cal S}}{\delta J^a (x) \delta a^{\beta b }
(y)}\nonumber\\ =&&
+iG_R^{ac} (x, y) {\ub a}_{\beta}^{cd} (y) \epsilon^{dbe} \pi^e (y) +
i\nabla_{\beta}^{bc} (y) \bigg( G_R^{ac} (x, y) \hat\sigma (y
)\bigg)\nonumber\\&&+
\frac i{f_{\pi}} G_R^{ac} (x , y ) \epsilon^{cbd} \bigg( {\bf j}_{V\beta}^d (y)
+f_{\pi} {\ub a}_{\beta}^{de} (y ) \pi^e (y) \bigg) \nonumber\\&&+
{\bf j}_{A\beta }^b (y)\int d^4 x' \bigg( 1 + G_R {\bf K} \bigg)^{ac} (x, x' )
\pi_{\rm in}^c (x' )\nonumber\\&&-
{\bf j}_{A\beta}^b (y) \int d^4 x' G_R^{ac} (x, x') J^c (x' )
\nonumber\\&&+
\int d^4x' G_R^{ac} (x, x') (\nabla^{\mu}a_{\mu} -J/f_{\pi})^c (x')
T^*\bigg( \hat\sigma (x' ) {\bf j}_{A\beta}^b (y )\bigg)\nonumber\\&&-
\frac 1{f_{\pi}}\int d^4x' G_R^{ac} (x,x')
\bigg(\nabla^{\alpha cd} (x' ) T^* \bigg({\bf j}_{A\alpha}^d (x' ){\bf
j}_{A\beta}^b (y)\bigg) \nonumber\\&&+
{\ub a}^{\alpha cd} (x ' ) T^*\bigg({\bf j}_{V\alpha}^d
(x') {\bf j}_{A\beta}^b (y )\bigg)\bigg)
\label{B96}
\ee
\be
T^*\bigg( \pi^a (x) {\bf j}_{V\beta}^b (y)\bigg) =&& - \hat{\cal
S}^{\dagger} \frac{\delta^2\hat{\cal S}}{\delta J^a (x) \delta v^{\beta b }
(y)}\nonumber\\ =&&
+iG_R^{ac} (x, y) {\ub a}_{\beta}^{cb} (y)  \hat{\sigma} (y)
\nonumber\\&&+
iG_R^{ac} (x , y ) \bigg(
\frac{
\stackrel\leftarrow
\partial}{\partial y^{\beta}}\epsilon^{cbd} -{\ub v}^{ce}_{\beta} (y)
\epsilon^{ebd} \bigg) \pi^d (y )\nonumber\\&&+
\frac i{f_{\pi}} G_R^{ac} (x , y ) \epsilon^{cbd} \bigg( {\bf j}_{A\beta}^d (y)
-f_{\pi} \nabla_{\beta}^{de} (y ) \pi^e (y) \bigg) \nonumber\\&&+
{\bf j}_{V\beta }^b (y)\int d^4 x' \bigg( 1 + G_R {\bf K} \bigg)^{ac} (x, x' )
\pi_{\rm in}^c (x' )\nonumber\\&&-
{\bf j}_{V\beta}^b (y) \int d^4 x' G_R^{ac} (x, x') J^c (x' )
\nonumber\\&&+
\int d^4x' G_R^{ac} (x, x') (\nabla^{\mu}a_{\mu} -J/f_{\pi})^c (x')
T^*\bigg( \hat\sigma (x' ) {\bf j}_{V\beta}^b (y )\bigg)\nonumber\\&&-
\frac 1{f_{\pi}}\int d^4x' G_R^{ac} (x,x')
\bigg(\nabla^{\alpha cd} (x' ) T^* \bigg({\bf j}_{A\alpha}^d (x' ){\bf
j}_{V\beta}^b (y)\bigg) \nonumber\\&&+
{\ub a}^{\alpha cd} (x ' ) T^*\bigg({\bf j}_{V\alpha}^d
(x') {\bf j}_{V\beta}^b (y )\bigg)\bigg)
\label{B97}
\ee
\be
T^*\bigg( \pi^a (x) \hat\sigma (y)\bigg) =&& -\frac 1{f_{\pi}} \hat{\cal
S}^{\dagger} \frac{\delta^2\hat{\cal S}}{\delta J^a (x) \delta s (y) }
\nonumber\\ =&& -\frac{i}{f_{\pi}} G_R^{ac} (x, y) \pi^c (y) \nonumber\\&&+
\hat\sigma (y)\int d^4 x' \bigg( 1 + G_R {\bf K} \bigg)^{ac} (x, x' )
\pi_{\rm in}^c (x' )\nonumber\\&&-
 \hat\sigma (y) \int d^4 x' G_R^{ac} (x, x') J^c (x' )
\nonumber\\&&+
\int d^4x' G_R^{ac} (x, x') (\nabla^{\mu}a_{\mu} -J/f_{\pi})^c (x')
T^*\bigg( \hat\sigma (x' ) \hat\sigma (y )\bigg)\nonumber\\&&-
\frac 1{f_{\pi}}\int d^4x' G_R^{ac} (x,x')
\bigg(\nabla^{\alpha cd} (x' ) T^* \bigg(
{\bf j}_{A\alpha}^d (x') \hat\sigma (y )\bigg) \nonumber\\&&+
{\ub a}^{\alpha cd} (x ' )
T^*\bigg({\bf j}_{V\alpha}^d (x') \hat\sigma (y )\bigg)\bigg)
\label{B98}
\ee
\be
T^*\bigg( \pi^a (x) \pi^b (y)\bigg) =&& - \hat{\cal
S}^{\dagger} \frac{\delta^2\hat{\cal S}}{\delta J^a (x) \delta J^b (y) }
\nonumber\\ =&& +iG_R^{ab} (x , y )+
\frac{i}{f_{\pi}} G_R^{ab} (x, y) \hat\sigma (y) \nonumber\\&&+
\pi^b (y)\int d^4 x' \bigg( 1 + G_R {\bf K} \bigg)^{ac} (x, x' )
\pi_{\rm in}^c (x' )\nonumber\\&&-
\pi^b (y) \int d^4 x' G_R^{ac} (x, x') J^c (x' ) \nonumber\\&&+
\int d^4x' G_R^{ac} (x, x') (\nabla^{\mu}a_{\mu} -J/f_{\pi})^c (x')
T^*\bigg( \hat\sigma (x' ) \pi^b (y )\bigg)\nonumber\\&&-
\frac 1{f_{\pi}}\int d^4x' G_R^{ac} (x,x')
\bigg(\nabla^{\alpha cd} (x' ) T^* \bigg(
{\bf j}_{A\alpha}^d (x') \pi^b (y )\bigg) \nonumber\\&&+
{\ub a}^{\alpha cd} (x ' )
T^*\bigg({\bf j}_{V\alpha}^d (x') \pi^b (y )\bigg)\bigg).
\label{B99}
\ee

\vskip .3cm
{\it 4.4. Ward Identities}
\vskip .15cm

Ward identities may also be derived. For instance,
\be
&&\bigg(-\Box_x -m_{\pi}^2 -{\bf K} (x) \bigg)^{ac}
\bigg(-\Box_y -m_{\pi}^2 -{\bf K} (y) \bigg)^{bd}
T^*\bigg( \pi^c (x )\pi^d (y)\bigg) = J^a (x) J^b (y) \nonumber\\&&-
J^a (x) \bigg(\nabla^{\nu}a_{\nu} -J/f_{\pi}\bigg)^b (y) \hat{\sigma} (y) -
J^b (y) \bigg(\nabla^{\mu}a_{\mu} -J/f_{\pi}\bigg)^a (x) \hat{\sigma} (y)
\nonumber\\&&+
\frac 1{f_{\pi}}J^a (x) \bigg(\nabla^{\beta} {\bf j}_{A\beta } +{\ub
a}^{\beta}{\bf j}_{V\beta }
\bigg)^b (y) +
\frac 1{f_{\pi}}J^b (y) \bigg(\nabla^{\beta} {\bf j}_{A\beta } +{\ub
a}^{\beta}{\bf j}_{V\beta }
\bigg)^a (x) \nonumber\\&&+
i\bigg( -\Box_y - m_{\pi}^2 -{\bf K} (y) \bigg)^{ba}\delta^4 (x-y)
\bigg( 1+\frac{\hat\sigma (x)}{f_{\pi}} \bigg) \nonumber\\&&+
\frac i{f_{\pi}}\bigg(-\nabla^{\alpha}\nabla_{\alpha} +{\ub a}^{\alpha}{\ub
a}_{\alpha} \bigg)^{ab} (x) \delta^4 (x-y) \hat\sigma (y)\nonumber\\&&-
\frac i{f_{\pi}}
\bigg(\nabla^{\mu}a_{\mu} - J/f_{\pi}\bigg)^a (x) \pi^b (x) \delta^4 (x-y
)\nonumber\\&&-
\frac i{f_{\pi}}\nabla^{\alpha ac} (x) \delta^4 (x-y){\ub a}_{\alpha}^{bd} (y)
\epsilon^{dce} \pi^e (y )\nonumber\\&&-
\frac i{f^2_{\pi}}\nabla^{\alpha ac} (x) \delta^4 (x-y)
\epsilon^{bce}
\bigg({\bf j}_{V\alpha}^e (y) + f_{\pi} {\ub a}_{\alpha}^{ef} (y) \pi^f (y)
\bigg) \nonumber\\&&-
\frac i{f_{\pi}}{\ub a}^{\alpha ac} (x)
\bigg( \delta^4 (x-y) \bigg(\frac{\stackrel\leftarrow{\partial}}{\partial
x^{\alpha}}\epsilon^{bce}-{\ub v}^{bd}_{\alpha} (x)\epsilon^{dce}\bigg) \pi^e
(x)\bigg) \nonumber\\&&-
\frac i{f^2_{\pi}}{\ub a}^{\alpha ac} (x) \delta^4 (x-y)\epsilon^{bce}
\bigg({\bf j}_{A\alpha}^e (x) -f_{\pi} \nabla_{\alpha}^{ef} (x) \pi^f (x)
\bigg) \nonumber\\&&+
(\nabla^{\mu}a_{\mu} -J/f_{\pi})^a (x) (\nabla^{\nu}a_{\nu} -J/f_{\pi})^b (y)
T^*\bigg(\hat{\sigma} (x) \hat\sigma (y )\bigg) \nonumber\\&&-
\frac 1{f_{\pi}} (\nabla^{\mu}a_{\mu} -J/f_{\pi})^a (x)
\bigg(\nabla^{\alpha bd} (y) T^* \bigg( \hat{\sigma} (x) {\bf j}_{A\alpha}^d
(y) \bigg) + {\ub a}^{\alpha bd} (y) T^*\bigg( \hat\sigma (x) {\bf
j}_{V\alpha}^d
(y) \bigg)\bigg)\nonumber\\&&-
\frac 1{f_{\pi}} (\nabla^{\mu}a_{\mu} -J/f_{\pi})^b (y)
\bigg(\nabla^{\alpha ac} (x) T^* \bigg(  {\bf j}_{A\alpha}^c (x)
\hat\sigma (y) \bigg) + {\ub a}^{\alpha ac} (x)
T^*\bigg( {\bf j}_{V\alpha}^c (x) \hat\sigma (y) \bigg)\bigg)\nonumber\\&&+
\frac 1{f_{\pi}^2}\bigg(
\nabla^{\alpha ac} (x) \nabla^{\beta bd } (y) T^*\bigg(
{\bf j}_{A\alpha}^c (x) {\bf j}_{A\beta}^d (y) \bigg) +
\nabla^{\alpha ac} (x) {\ub a}^{\beta bd} (y) T^*\bigg(
{\bf j}_{A\alpha}^c (x) {\bf j}_{V\beta}^d (y) \bigg) \nonumber\\&&+
{\ub a}^{\alpha ac} (x) \nabla^{\beta bd } (y) T^*\bigg(
{\bf j}_{V\alpha}^c (x) {\bf j}_{A\beta}^d (y) \bigg) +
{\ub a}^{\alpha ac} (x) {\ub a}^{\beta bd} (y) T^*\bigg(
{\bf j}_{V\alpha}^c (x) {\bf j}_{V\beta}^d (y) \bigg) \bigg).
\label{B100}
\ee
Further results will be given in the next section and in Appendix D.

\vskip .5cm
{\bf 5. The Goldberger-Treiman Relation}
\vskip .3cm

{}From (\ref{B80}) we have
\be
<N (p_2) | {\bf A}_{\mu}^a (x) |N (p_1) > =&& -i f_{\pi} (p_2-p_1)_{\mu}
<N(p_2) |\pi^a (x) |N(p_1) > \nonumber\\&&
+ <N(p_2) | {\bf j}_{A\mu}^a (x) |N(p_1) >
\label{GT1}
\ee
Introducing the nucleon axial form factor
\be
<N(p_2) | {\bf A}^a_{\mu} (x) |N(p_1) > =
\overline{u} (p_2) \bigg(\gamma_{\mu} \gamma_5 \,G_1 (t) +
(p_2-p_1)_{\mu} \gamma_5 \,G_2 (t) \bigg) \frac {\tau^a}2 \,\,u(p_1)
e^{i(p_2-p_1)\cdot x}
\label{GT2}
\ee
with $t=(p_1-p_2)^2$, we have
\be
<N(p_2) | {\bf j}_{A\mu}^a (x) |N (p_1) > =
\overline{u} (p_2) \bigg(\gamma_{\mu} \gamma_5 \,G_1 (t) +
(p_2-p_1)_{\mu} \gamma_5 \,\overline{G}_2 (t) \bigg) \frac {\tau^a}2 \,\,u(p_1)
e^{i(p_2-p_1)\cdot x}
\label{GT3}
\ee
where $G_1$ and $\overline{G}_2$ are free of pion poles.

On the other hand, the master formula (\ref{B93}) gives
\be
<N(p_2) | \pi^a (x) |N (p_1) > =&&+
<N(p_2) | \pi_{\rm in}^a (x) |N (p_1) >\nonumber\\&&
-\frac 1{f_{\pi}} \int d^4y \,\Delta_R (x-y)
<N(p_2) | \partial^{\mu}{\bf j}_{A\mu}^a (y) |N (p_1) >\nonumber\\=&&
-\frac 1{f_{\pi}}\frac 1{t-m_{\pi}^2}
\overline{u} (p_2)\bigg( 2Mi\gamma_5 G_1 (t) + it\gamma_5 \overline{G}_2
(t)\bigg)\, \frac {\tau^a}2 \, u(p_1) \, e^{i(p_2-p_1)\cdot x}
\nonumber\\
\label{GT4}
\ee
This is by definition
\be
-g_{\pi NN} (t) \frac 1{t-m_{\pi}^2} \overline{u} (p_2) i\gamma_5 \tau^a u
(p_1)\, e^{i (p_2-p_1)\cdot x}
\label{GT5}
\ee
where the minus sign corresponds to the coupling
${\cal L} = g_{\pi NN} \overline{N} i\gamma_5 \tau^a N \, \pi^a$ in the
effective Lagrangian.
Hence,
\be
f_{\pi} g_{\pi NN} (0)  = M g_A
\label{GT6}
\ee
where $g_A = G_1 (0)$ is the nucleon axial charge.
Extrapolating $g_{\pi NN}$ to the physical value $g_{\pi
NN} =g_{\pi NN} (m_{\pi}^2)$ yields the Goldberger-Treiman relation.

\vskip .25cm
{\bf 6. Chiral Reduction Formula}

\vskip .3cm
{\it 6.1. Single Commutator}
\vskip .15cm

To extract results from the master formula, we multiply (\ref{B94}) by
$(1+G_A{\bf K} )^{-1} = 1 -\Delta_A {\bf K}$ to obtain
\be
\bigg[ \pi_{\rm in} , \hat{\cal S} \bigg]
= &&+\hat{\cal S} \Delta \bigg( 1 + {\bf
K} G_R \bigg) {\bf K} \pi_{\rm in} - \Delta
\bigg(1 +{\bf K} G_R \bigg) J\hat{\cal
S} \nonumber\\&&-
\frac i{f_{\pi}} \Delta \bigg(1 + {\bf K} G_R \bigg)
\bigg(\nabla^{\mu}a_{\mu} - J/f_{\pi}\bigg)
\frac{\delta\hat{\cal S}}{\delta s} +\frac i{f_{\pi}}\Delta
\bigg(1 +{\bf K} G_R \bigg) {\bf X}_A \hat{\cal S}.
\label{C1}
\ee
The Fourier decompositions (\ref{A4}, \ref{A6}) then give
\be
\bigg[ a_{\rm in}^a (k) , \hat{\cal S} \bigg] = &&
-i \hat{\cal S} \int d^4y d^4z e^{ik\cdot y}
\bigg( 1 + {\bf K} G_R \bigg)^{ac} (y, z)
({\bf K} \pi_{\rm in})^c (z) \nonumber\\&&+
i\int d^4y d^4z e^{ik\cdot y}
\bigg(1 +{\bf K} G_R \bigg)^{ac} (y,z) J^c (z) \hat{\cal
S} \nonumber\\&&-
\frac 1{f_{\pi}} \int d^4y d^4z e^{ik\cdot y}
\bigg(1 + {\bf K} G_R \bigg)^{ac} (y,z)
\bigg(\nabla^{\mu}a_{\mu} - J/f_{\pi}\bigg)^c (z)
\frac{\delta\hat{\cal S}}{\delta s (z)} \nonumber\\&&+
\frac 1{f_{\pi}}\int d^4y d^4z e^{ik\cdot y}
\bigg(1 +{\bf K} G_R \bigg)^{ac} (y, z)  {\bf X}^c_A(z) \hat{\cal S}
\label{C2}
\ee
\be
\bigg[\hat{\cal S} ,  a^{a \dagger }_{\rm in} (k)  \bigg] = &&
-i \hat{\cal S} \int d^4y d^4z e^{-ik\cdot y}
\bigg( 1 + {\bf K} G_R \bigg)^{ac} (y, z)
({\bf K} \pi_{\rm in})^c (z) \nonumber\\&&+
i\int d^4y d^4z e^{-ik\cdot y}
\bigg(1 +{\bf K} G_R \bigg)^{ac} (y,z) J^c (z) \hat{\cal
S} \nonumber\\&&-
\frac 1{f_{\pi}} \int d^4y d^4z e^{-ik\cdot y}
\bigg(1 + {\bf K} G_R \bigg)^{ac} (y,z)
\bigg(\nabla^{\mu}a_{\mu} - J/f_{\pi}\bigg)^c (z)
\frac{\delta\hat{\cal S}}{\delta s (z)} \nonumber\\&&+
\frac 1{f_{\pi}}\int d^4y d^4z e^{-ik\cdot y}
\bigg(1 +{\bf K} G_R \bigg)^{ac} (y, z)  {\bf X}^c_A(z) \hat{\cal S}.
\label{C3}
\ee
The master equations in Fourier form have the structure of reduction formulas.
In particular, to lowest order in the external fields
$\phi = (v, a, J, s)$,
\be
<\beta\,\,{\rm in} | \bigg[ a_{\rm in}^a (k) , {\bf S} \bigg]  |\alpha
\,\,{\rm in} >= +\frac 1{f_{\pi}} k^{\alpha}\int d^4y e^{+ik\cdot y}
<\beta\,\,{\rm in} | {\bf S}\, {\bf j}_{A\alpha }^a (y) |\alpha\,\, {\rm in} >
\label{C4}
\ee
\be
<\beta\,\,{\rm in} | \bigg[ {\bf S}, a_{\rm in}^{a\dagger} (k) \bigg]  |\alpha
\,\,{\rm in} >= -\frac 1{f_{\pi}} k^{\alpha}\int d^4y e^{-ik\cdot y}
<\beta\,\,{\rm in} | {\bf S}\, {\bf j}_{A\alpha }^a (y) |\alpha\,\, {\rm in} >,
\label{C5}
\ee
where ${\bf S} ={\cal S}|_{\phi =0} =\hat{\cal S}|_{\phi =0}$ is the on-shell
S-matrix. Equations (\ref{C4}-\ref{C5}) define an off-shell extension
of ${\bf S}$. The
Adler consistency condition and crossing symmetry are manifest.

\vskip .3cm
{\it 6.2. Double Commutator}
\vskip .15cm

For the double commutator
\be
\bigg[ a_{\rm in}^b (k_2 ), \bigg[ {\bf S} , a_{\rm in }^{a\dagger}
(k_1)\bigg]\bigg] = +\frac i{f_{\pi}} k_{1\alpha } \int d^4 y e^{-ik_1\cdot y}
\bigg[ a_{\rm in}^b (k_2 ), \frac{\delta\hat{\cal S}}{\delta a_{\alpha}^a
(y)} \bigg]
\label{C6}
\ee
we need the result
\be
\frac{\delta}{\delta a^{a}_{\alpha}  (y)} \bigg[ a_{\rm in}^b (k_2 ) ,
\hat{\cal S} \bigg] =&&-\frac 1{f_{\pi}} e^{ik_2\cdot y}
\delta^{ab} \bigg( -\frac{\partial}{\partial y^{\alpha}} -
\frac{\stackrel\leftarrow\partial}{\partial y^{\alpha}}\bigg)
\frac{\delta\hat{\cal S}}{\delta s
(y)} \nonumber\\&&+
\frac 1{f_{\pi}}\int d^4 y_2 e^{ik_2\cdot y_2 }\frac{\partial}{\partial
y_2^{\beta}}\frac{\delta^2\hat{\cal S}}{\delta a_{\beta}^b (y_2 ) \delta
a_{\alpha}^a (y)} +\frac 1{f_{\pi}} e^{ik_2\cdot y}\epsilon^{bae}
\frac{\delta\hat{\cal S}}{\delta v_{\alpha}^e (y)}
\nonumber\\ = &&
+e^{ik_2\cdot y}
\delta^{ab} \bigg( {\partial^{\alpha}} +ik_2^{\alpha}\bigg)
i{\bf S} \hat\sigma (y )
\nonumber\\&&
-\frac i{f_{\pi}}k_{2\beta}
\int d^4 y_2 e^{ik_2\cdot y_2 }
\frac{\delta^2\hat{\cal S}}{\delta a_{\beta}^b (y_2 ) \delta
a_{\alpha}^a (y)} +\frac 1{f_{\pi}} e^{ik_2\cdot y}\epsilon^{bae}
i{\bf S} {\bf V}^{\alpha e} (y ).
\label{C7}
\ee
to leading order in the external fields.
Thus (\ref{C6}) reduces to
\be
&&\bigg[ a_{\rm in}^b (k_2 ), \bigg[ {\bf S} , a_{\rm in }^{a\dagger}
(k_1)\bigg]\bigg] =\nonumber \\&&
-\frac i{f_{\pi}} {m_{\pi}^2}\int d^4y e^{-i(k_1-k_2)\cdot y}
{\bf S}\hat\sigma (y ) \delta^{ab} +
\frac 1{f^2_{\pi}} k_1^{\alpha} \int d^4y e^{-i(k_1-k_2)\cdot y}
\epsilon^{abe} {\bf S} {\bf V}^{\alpha e} (y ) \nonumber\\ &&+
\frac 1{f^2_{\pi}}k_{1\alpha}k_{2\beta}
\int d^4 y_1 d^4 y_2 e^{-ik_1\cdot y_1 +ik_2\cdot y_2 }
\frac{\delta^2\hat{\cal S}}{\delta a_{\alpha}^a (y_1 ) \delta
a_{\beta}^b (y_2)}.
\label{C8}
\ee
For later use, we quote the crossed version of (\ref{C7})
\be
&&\frac{\delta}{\delta a^{b}_{\beta} (y)} \bigg[ \hat{\cal S},
a_{\rm in}^{a\dagger} (k_1 )\bigg]
= e^{-ik_1\cdot y}
\delta^{ab} \bigg( {\partial^{\beta}} -ik_1^{\beta}\bigg)
i{\bf S} \hat\sigma (y )
\nonumber\\&& +
\frac i{f_{\pi}}k_{1\alpha}
\int d^4 y_1 e^{-ik_1\cdot y_1 }
\frac{\delta^2\hat{\cal S}}{\delta a_{\alpha}^a (y_1 ) \delta
a_{\beta}^b (y)} +\frac 1{f_{\pi}} e^{-ik_1\cdot y}\epsilon^{abe}
i{\bf S} {\bf V}^{\beta e} (y ).
\label{C9}
\ee
The fact that ${\bf j}_{A\mu}^a$ is free of one-pion components also implies
\be
&&0=i<0 | \bigg[ a_{\rm in}^b (k_2 ), {\bf S} {\bf j}_{A\alpha}^a (y) \bigg]
|0>=
\frac{\delta}{\delta a^{\alpha a} (y)}
<0 | \bigg[ a_{\rm in}^b (k_2 ) , \hat{\cal S} \bigg] | 0 > =\nonumber\\ &&
-e^{ik_2\cdot y} \delta^{ab} k_{2\alpha} <0 | \hat\sigma | 0 > +
\frac i{f_{\pi}} k_2^{\beta} e^{ik_2 \cdot y} \int d^4y_2 e^{i k_2\cdot y_2}
<0 | T^*\bigg( {\bf j}_{A\alpha}^a (0) {\bf j}_{A\beta }^b (y_2)\bigg) | 0 >
\label{C10}
\ee
where we have made explicit use of (\ref{C7}) and translational invariance of
vacuum expectation values. The crossed result reads
\be
&&0=i<0 | \bigg[  {\bf S} {\bf j}_{A\beta}^b (y),
a_{\rm in}^{a \dagger} (k_1 ) \bigg] |0>=
\nonumber\\ &&
+e^{-ik_1\cdot y} \delta^{ab} k_{1\beta} <0 | \hat\sigma | 0 > -
\frac i{f_{\pi}} k_1^{\alpha} e^{-ik_1 \cdot y} \int d^4y_1 e^{-i k_1\cdot y_1}
<0 | T^*\bigg( {\bf j}_{A\alpha}^a (y_1) {\bf j}_{A\beta }^b (0)\bigg) | 0 >
\label{C11}\nonumber\\
\ee
Energy-momentum conservation and (\ref{C10}-\ref{C11}) suffice to guarantee the
stability of the one pion state $[{\bf S}, a_{\rm in}^{a\dagger} (k) ] |0> =0$.

\vskip .5cm
{\bf 7. Pion-Nucleon Scattering :
{$\pi N\rightarrow \pi N$}}
\vskip .3cm

The equation (\ref{C8}) may be applied to $\pi N$ scattering. Indeed,
sandwiched between one-nucleon states, it becomes
\be
&&<N (p_2 ) | a_{\rm in}^b (k_2 ) ({\bf S} - {\bf 1}) a_{\rm in}^{a\dagger}
(k_1) |N(p_1)>=
<N(p_2 ) | \bigg[a_{\rm in}^b (k_2) ,\bigg[ {\bf S} , a_{\rm
in}^{a\dagger}(k_1)\bigg]\bigg] |N(p_1 ) > =\nonumber\\&&
-\frac i{f_{\pi}} m_{\pi}^2 \delta^{ab} \int d^4y e^{-i(k_1-k_2)\cdot y}
<N(p_2 ) | {\hat \sigma} (y ) | N(p_1) > \nonumber\\&&
-\frac 1{f_{\pi}^2} k_1^{\alpha} k_2^{\beta} \int d^4y_1 d^4 y_2 e^{-ik_1\cdot
y_1 + ik_2\cdot y_1} <N(p_2) | T^*\bigg(
{\bf j}_{A\alpha}^a (y_1) {\bf j}_{A\beta}^b (y_2 ) \bigg) | N(p_1)
>\nonumber\\&&
+\frac 1{f_{\pi}^2} k_1^{\alpha} \int d^4y e^{-i (k_1-k_2)\cdot y}
\epsilon^{abe} <N(p_2) | {\bf V}_{\alpha}^e (y) | N(p_1) >
\label{C12}
\ee
In (\ref{C12}),
 we have assumed that $a_{\rm in}^b (k_2) |N(p_1) > =0$ as well
as the stability of the one-nucleon state ${\bf S} |N(p_1) > =
|N (p_1) >$. As a check
of (\ref{C12}), we note that the same result is obtained from the conventional
WLSZ reduction formula\footnote{W stands for Watanabe \cite{watanabe}.}
and the Ward identity (\ref{B100}). We also
note that the disconnected part
\be
&&-\frac{i}{f_{\pi}} m_{\pi}^2 \delta^{ab} \int d^4 y e^{-i(k_1-k_2)\cdot y}
<0|\hat{\sigma} (y) |0 > <N(p_2 )| N(p_1) >\nonumber\\&&
-\frac{1}{f_{\pi}^2} k_1^{\alpha}k_2^{\beta} \int d^4y_1 d^4y_2
e^{-i(k_1-k_2)\cdot y_1}
<0| T^*\bigg( {\bf j}_{A\alpha}^a (y_1) {\bf j}_{A\beta}^b (y_2)\bigg) |0>
<N(p_2) |
N(p_1 )>
\label{C13}\nonumber\\
\ee
vanishes by (\ref{C10}-\ref{C11}).

Eq. (\ref{C12}) implies that the part of the amplitude that is even under
$a \leftrightarrow b$ is quadratic in the momenta $k\sim m_{\pi}$, while the
odd part is linear in the momenta. Since ${\bf V}_{\mu}^a$ reduces to the
isospin operator in the limit of vanishing momentum transfer
$k_1-k_2\rightarrow 0$, the amplitude evaluated at threshold gives
the Tomozawa-Weinberg relation \cite{tomozawa,weinberg1}
for the S-wave scattering lengths $a_{ab}$,
\be
a_{ab} =\frac {-i}{\pi} \bigg( \frac{M}{M+m_{\pi}}\bigg)
\frac {m_{\pi}}{f_{\pi}^2} \,\epsilon^{abc} \, <N |\tau^c | N>
\label{C14}
\ee
where $M$ is the nucleon mass.
Specifically, if we denote by $i{\cal T}^{ab} (\omega )$ the forward on-shell
$\pi N$ scattering amplitude (\ref{C12}) for a nucleon at rest,
\be
<N (p ) | a_{\rm in}^b (k) ({\bf S} - {\bf 1})
   a_{\rm in}^{a\dagger} (k) |N(p)> = (2\pi )^{4} \delta^4 (p+k-p-k)
i{\cal T}^{ab} (\omega =k^0)
\label{C144}
\ee
then by isospin symmetry
\be
i{\cal T}^{ab} (\omega ) =
\overline{u} (p) \bigg(
 i{\cal T}^+ (\omega ) \delta^{ba}
+i{\cal T}^- (\omega ) \,\,i\epsilon^{bac} \tau^c \bigg) u(p)
\label{XC140}
\ee
The even-odd S-wave scattering lengths $a^{\pm}$ are defined to be
\be
{\cal T}^{\pm } ( m_{\pi}) = 4\pi \bigg( 1 +\frac {m_{\pi}}M \bigg) \,\,a^{\pm}
\label{XC141}
\ee
To lowest order $a^+ = 0$ and
$a^- = 8.76 \,10^{-2}\,m_{\pi}^{-1}$ in good agreement
with $a^+ (\rm exp ) = (-0.83 \pm 0.38 )\times 10^{-2} m_{\pi}^{-1}$
and $a^- (\rm exp ) = (+9.17 \pm 0.17 )\times 10^{-2} m_{\pi}^{-1}$. The
empirical results are extracted from forward $\pi N$ scattering and pionic
atom data.

Combining (\ref{C14}) with the Goldberger-Miyazawa-Oehme sum rule
\cite{miyazawa}
\be
\frac 23 \bigg( \frac 1{m_{\pi}} +\frac 1{M} \bigg)
\bigg( a_{1/2} - a_{3/2}\bigg) =&&+
\frac 1{\pi} \bigg(\frac {g_{\pi NN}}{2M}\bigg)^2
\bigg(1-\frac{m_{\pi}^2}{4M^2}\bigg)^{-1} \nonumber\\&&
-\frac 1{2\pi^2} \int_{m_{\pi}}^{\infty}
\frac {d\nu}{\sqrt{\nu^2-m_{\pi}^2}}
\bigg(\sigma_{\pi^+ p} (\nu ) - \sigma_{\pi^- p} (\nu )\bigg)
\label{C15}
\ee
and using the Goldberger-Treiman relation $g_A= f_{\pi} g_{\pi NN}/ M$,
give
\be
\frac 1{g_A^2} = \bigg(1-\frac{m_{\pi}^2}{4M^2}\bigg)^{-1}
+\frac 1{2\pi}\bigg(\frac {2M}{g_{\pi NN}}\bigg)^2 \int_{m_{\pi}}^{\infty}
\frac {d\nu}{\sqrt{\nu^2-m_{\pi}^2}}
\bigg(\sigma_{\pi^- p} (\nu ) - \sigma_{\pi^+ p} (\nu )\bigg)
\label{C16}
\ee
which is one form of the Adler-Weisberger sum rule \cite{weisberger}.
Here $\sigma_{\pi^{\pm}p}$ are the total cross sections for the
elastic  $\pi^{\pm} p\rightarrow \pi^{\pm} p$ reactions.
The sum rule (\ref{C16}) can be verified to a good accuracy
using the Karlsruhe-Helsinki data for the elastic cross sections.

\vskip .5cm
{\bf 8. Pion Photo- and Electro-Production : $\gamma N\rightarrow \pi N$}
\vskip .3cm

Another process of interest is the  photo- or electro-production reaction.
The isovector part of the amplitude is given by
\be
&&\int d^4y_1 e^{-ik_1\cdot y_1}
<N (p_2) | a_{\rm in}^b (k_2) {\bf S}{\bf V}_{\alpha}^a (y_1) |N(p_1)> =
\nonumber\\&&
+i k_{2\alpha} \epsilon^{baf}\int d^4y e^{-i(k_1-k_2)\cdot y}
<N(p_2) | \pi^f (y) |N(p_1) > \nonumber\\&& +
\frac 1{f_{\pi}}
\epsilon^{baf}\int d^4y e^{-i(k_1-k_2)\cdot y}
<N(p_2) | {\bf A}^f_{\alpha} (y) |N(p_1) >\nonumber\\&&
+\frac 1{f_{\pi}}  k^{\beta}_2
\int d^4y_1 d^4y_2  e^{-ik_1\cdot y_1+ik_2\cdot y_2}
<N(p_2) | T^*\bigg( {\bf V}_{\alpha}^a (y_1) {\bf j}_{A\beta}^b (y_2)\bigg)
|N(p_1) >
\label{C68}
\ee
where we have used (\ref{C29}). We may evaluate (\ref{C68}) for the isospin
$3/2$ channel, where there is no contribution from the isoscalar part of
the current. The third term involving ${\bf j}_{A\mu}^a $ is one power of
momentum higher than the second term with ${\bf A}_{\mu}^a$ and will be
ignored. The first term with $\pi$ also carries an extra power of momentum, but
it cannot be dropped since it contains a pole.

Using the axial form factor (\ref{GT2})
and factoring out $(2\pi)^4 \delta^4 (k_1+p_1-k_2-p_2)$, give
\be
&&+ik_{2\alpha} \epsilon^{baf} <N(p_2) | \pi^f (0) | N(p_1) > \nonumber\\&&
+\frac 1{f_{\pi}} \epsilon^{baf}
\overline{u} (p_2) \bigg(\gamma_{\alpha} \gamma_5 \,G_1 (t) +
(k_1-k_2)_{\alpha} \gamma_5 \,G_2 (t) \bigg) \frac {\tau^f}2 \,\,u(p_1)
\label{CA69}
\ee
At threshold $t=-m_{\pi}^2 + k_1^2 + {\cal O} (m_{\pi}/M )$, we may replace
$<N(p_2) |\pi^f (0) |N(p_1) >$ and $G_2 (t)$ by their pole terms
\be
<N(p_2) |\pi^f (0) |N(p_1) >\rightarrow -g_{\pi NN} \frac 1{t-m_{\pi}^2}
\overline u (p_2) i\gamma_5 \tau^f u (p_1)
\label{CB69}
\ee
\be
G_2 (t) \rightarrow -2f_{\pi} g_{\pi NN} \frac 1{t-m_{\pi}^2}
\label{CC69}
\ee
and $G_1 (t)$ by the axial coupling $G_1 (0) = g_A =1.25$, to obtain
\be
\bigg( 2ik_{2\alpha} -ik_{1\alpha} \bigg) \epsilon^{baf}
\frac {g_{\pi NN}}{2m_{\pi}^2-k_1^2}\overline u (p_2) i\gamma_5 \tau^f u (p_1)
+\frac 1{f_{\pi}} \epsilon^{baf} g_A
\overline u (p_2) \gamma_{\alpha}\gamma_5 \frac{\tau^f}2 u (p_1)
\label{CD69}
\ee
We note that the pole term simply corresponds to Fig. 1a.
Also, it may be checked that current conservation is satisfied by virtue of the
Goldberger-Treiman relation and
\be
\overline u (p_2) k_2\cdot\gamma \gamma_5 \tau^f u (p_1) =
{\cal O} (m_{\pi}^2/M)\,\, u^{\dagger} (p_2)\, u (p_1)
\label{CE69}
\ee
The result (\ref{CD69}) is consistent with the
Kroll-Ruderman theorem  for photoproduction \cite{kroll}, and shows
that the isospin $3/2$ amplitudes $\gamma p\rightarrow \pi^+ n$
and $\gamma n\rightarrow \pi^- n$ are totally fixed at threshold by chiral
symmetry. This is not the case for the
isospin $1/2$ channel that is $\gamma p\rightarrow \pi^0 p$, for which the
Kroll-Ruderman term drops. This process also receives contribution from the
isoscalar part of the electromagnetic current and thus requires a full
$SU(3)\times SU(3)$ calculation. This will be discussed elsewhere \cite{yaza}.
We note that since our treatment of electromagnetism is first order in
the electromagnetic charge, the Primakoff term  in (\ref{CD69}) is absent.

Finally, we would like to note that the neutral pion photoproduction
experiments carried out at Saclay \cite{saclay} and Mainz \cite{mainz}
on proton targets, $\gamma p\rightarrow \pi^0 p$, have questioned some
current algebra results at threshold, although the new data from
Mainz \cite{newmainz} for $\pi^0$ electroproduction seems to indicate
otherwise. Further data on these issues are expected from the SAL
collaboration at Saskatoon. Given the experimental interest in pion photo-
and electro-production near threshold,
there has been some recent theoretical activity
in this direction using heavy-baryon chiral perturbation theory \cite{theo}.
While our results provide exact constraints at threshold, our formalism
has to be extended to baryon intermediate states to account properly for
near-threshold effects. We will report on some of these issues elsewhere
\cite{yaza}.

\vskip .5cm
{\bf 9. Two-Pion Pion-Production : $\pi N\rightarrow \pi\pi N$}
\vskip .3cm

We may also consider the two-pion production  $\pi N\rightarrow \pi\pi
N$ amplitude
\be
&&<N(p_2) | a_{\rm in}^c (k_3) a_{\rm in}^b (k_2) {\bf S}
a_{\rm in}^{a\dagger} (k_1) |N (p_1) > =\nonumber\\&&
-\frac i{f_{\pi}} m_{\pi}^2 \delta^{ab}
\int d^4y e^{-i (k_1-k_2)\cdot y} <N(p_2)| a_{\rm in}^c (k_3) {\bf S}
\hat\sigma (y) |N(p_1)> \,\,+\,\,2\,\, {\rm perm.}\nonumber\\&&
+\frac 1{f_{\pi}^2} k_2^{\alpha} \epsilon^{abe}
\int d^4y e^{-i (k_1-k_2)\cdot y} <N(p_2)| a_{\rm in}^c (k_3) {\bf S}
{\bf V}_{\alpha}^e (y) |N(p_1)> + \,\,{\rm {2\,\, perm}} \nonumber\\&&
-\frac i{f^2_{\pi}} m_{\pi}^2 \delta^{ac}
\int d^4y e^{-i (k_1-k_2-k_3)\cdot y} <N(p_2)| \pi^b (y) |N(p_1)> \nonumber\\&&
-\frac i{f^2_{\pi}} m_{\pi}^2 \delta^{bc}
\int d^4y e^{-i (k_1-k_2-k_3)\cdot y} <N(p_2)| \pi^a (y) |N(p_1)> \nonumber\\&&
+\frac 1{f_{\pi}^3} k_1^{\alpha} \epsilon^{cae}\epsilon^{bef}
\int d^4y e^{-i (k_1-k_2-k_3)\cdot y}
<N(p_2)| {\bf A}_{\alpha}^f (y) |N(p_1)> \nonumber\\&&
-\frac 1{f_{\pi}^3} k_2^{\beta} \epsilon^{cbe}\epsilon^{aef}
\int d^4y e^{-i (k_1-k_2-k_3)\cdot y}
<N(p_2)| {\bf A}_{\beta}^f (y) |N(p_1)> \nonumber\\&&
-\frac 1{f_{\pi}^3} k_1^{\alpha} k_2^{\beta} k_3^{\gamma}
\int d^4y_1 d^4y_2 d^4y_3 e^{-ik_1\cdot y_1 +ik_2\cdot y_2 +ik_3\cdot y_3}
\nonumber\\&&
\times <N(p_2)| T^*\bigg({\bf j}_{A\alpha}^a (y_1) {\bf j}_{A\beta}^b (y_2)
{\bf j}_{A\gamma}^c (y_3)\bigg) |N(p_1)>
\nonumber\\
\label{C72}
\ee
where we have used the crossed version of (\ref{C31}). The term with three
${\bf j}_{A\mu}^a$'s is three pion reduced and cubic in the momenta, so it will
be ignored. In the isospin $3/2$ channel, the term with $\hat\sigma$
does not contribute. In the same channel, the contribution involving ${\bf
V}_{\mu}^a $ is quadratic in the momenta, since ${\bf V}_{\mu}^a$ at zero
momentum is proportional to the isospin operator, which does not change the
total isospin of the nucleon $1/2$. Hence, the dominant contribution at
low momentum is given by the terms with ${\bf A}_{\mu}^a $ and $\pi$.
Using (\ref{GT2},\ref{CB69},\ref{CC69}) and $G_1 (t)\rightarrow g_A$ as before
and
factoring out the momentum delta function, we obtain for the isospin $3/2$
channel of (\ref{C72}) at threshold
\be
&&-\frac {i}{4f_{\pi}^2}\delta^{ac} g_{\pi NN}
\,\,\overline u (p_2) i\gamma_5 \tau^b u  (p_1) \nonumber\\&&
-\frac 1{f^3_{\pi}}\delta^{bc} g_{A} k_1^{\alpha}
\,\,\overline u (p_2) \gamma_{\alpha} \gamma_5 \frac{\tau^a}2 u(p_1)
\nonumber\\&&
+\frac 1{f^3_{\pi}}\delta^{ab} g_{A} k_1^{\alpha}
\,\,\overline u (p_2) \gamma_{\alpha} \gamma_5 \frac{\tau^c}2 u(p_1)
\nonumber\\&&
+\frac i{4m_{\pi}^2f_{\pi}^2}\delta^{bc} g_{\pi NN} k_1\cdot (p_2-p_1)
\,\,\overline u (p_2) i\gamma_5 \tau^a u  (p_1) \nonumber\\&&
-\frac i{4m_{\pi}^2f_{\pi}^2}\delta^{ab} g_{\pi NN} k_1\cdot (p_2-p_1)
\,\,\overline u (p_2) i\gamma_5 \tau^c u  (p_1) \nonumber\\&&
+\bigg( k_1\,\,\,a \leftrightarrow -k_2\,\,\,b\bigg)
\label{C73}
\ee
which may be checked to be totally symmetric by the Goldberger-Treiman
relation.

The process has also been treated by Chang \cite{chang} in the soft pion limit.
A chiral calculation has been made by Beringer \cite{beringer}, and
heavy-baryon chiral perturbation theory has been applied by Bernard, Kaiser and
Meissner \cite{bernard}.

New data on $\pi^+ p\rightarrow \pi^+ \pi^+ n$ and $\pi^- p\rightarrow
\pi^0\pi^0 n$ reactions are now available in the threshold region, with
pion energies as large as 200 MeV in the lab frame \cite{newdata}. These
new measurements complement the high statistics measurements of $\pi^-
p\rightarrow \pi^-\pi^+ n$ at 17.2 GeV carried out at CERN \cite{grayer},
and the $\pi^+ n\rightarrow \pi^+\pi^- p$ measurement at 5.98 GeV and
11.85 GeV on polarised deuterons also carried out at CERN \cite{lesquen}.
We will report on these calculations and data in the context of our
framework including nucleon intermediate states elsewhere \cite{yaza}.

We note that since the reaction $\pi N\rightarrow \pi\pi N$
has been conventionally used to extract the S-wave $\pi\pi$ scattering
length \cite{newdata}, we may ask to what extent this extraction depends on the
model used. Indeed our result (\ref{C72}) shows exactly how this amplitude is
related to other pion-nucleon amplitudes under the dictates of chiral symmetry
and unitarity. From (\ref{C72}) it appears that the S-wave $\pi\pi$ scattering
amplitude (see below) does not factorize out of $\pi N\rightarrow \pi\pi N$
in the t-channel. This means that the present extractions
can only be performed in a scheme dependent way. In fact these extractions have
been recently questioned by Svec \cite{svec}.

Finally, it has been suggested that the two experiments
$\pi^- p_{ \uparrow} \rightarrow \pi^-\pi^+ n$
at 17.2 GeV, and $\pi^+ n_{\uparrow}\rightarrow
\pi^+\pi^- p$ at 5.98 GeV and 11.85 GeV on polarized targets, show evidence for
a low lying scalar state $0^{++} (750)$ with a width of 250 MeV \cite{svec}.
It would be interesting to see whether this can follow from the general
principles we have presently used when nucleon intermediate states are
also included.

\vskip .5cm
{\bf 11. Pion Emission in Compton Scattering :
$\gamma N\rightarrow \gamma\pi N$}
\vskip .3cm

The amplitude for the reaction
$\gamma N\rightarrow \gamma\pi N $ may be written as
\be
&&+\int d^4y_1 d^4y_2 e^{-iq_1\cdot y_1 +iq_2\cdot y_2}
<N(p_2) | a_{\rm in}^a (k) {\bf S} T^*
\bigg({\bf V}_{\mu}^c (y_1) {\bf V}_{\nu}^d (y_2)\bigg) |N (p_1)> \nonumber\\
=&&-
\int d^4y_1 d^4y_2 e^{-iq_1\cdot y_1 +iq_2\cdot y_2}
<N(p_2) | \frac {\delta^2}{\delta v^{\mu c} (y_1) \delta v^{\nu d} (y_2)}
\bigg[ a_{\rm in}^a (k), \hat{\cal S}\bigg] |N(p_1)>
\label{ADD1}
\ee
in the isospin $3/2$ channel. To evaluate (\ref{ADD1}) we note that
differentiating the one-pion reduction formula
\be
\bigg[ a_{\rm in}^a (k) , \hat{\cal S}\bigg] =
\frac 1{f_{\pi}}\int d^4y d^4z e^{ik\cdot y}
\bigg( 1 +{\bf K}{\bf G}_R\bigg)^{ae} (y,z)\,\, {\bf X}_A^e (z)
\hat{\cal S} +{\cal O}\bigg( J,a,\pi_{\rm in} \bigg)
\label{ADD2}
\ee
yields
\be
&&\frac {\delta^2}{\delta v^{\mu c} (y_1) \delta v^{\nu d} (y_2)}
\bigg[ a_{\rm in}^a (k), \hat{\cal S}\bigg] =\nonumber\\&&
+\frac 1{f_{\pi}}e^{ik\cdot y_1}g_{\mu\nu}\delta^4
(y_1-y_2)\epsilon^{ace}\epsilon^{edf}
\int d^4z \Delta_R (y_1-z) \frac{\partial}{\partial z_{\alpha}}
i{\bf S}{\bf j}_{A\alpha}^f (z)\nonumber\\&&
+\frac 1{f_{\pi}}e^{ik\cdot y_1}
\bigg(\frac{\partial}{\partial y_{1}^{\mu}} -ik_{\mu}\bigg)
\epsilon^{ace}\Delta_R (y_1-y_2)
\bigg(\frac{\partial}{\partial y_{2}^{\nu}} -
\frac{\stackrel\leftarrow\partial}{\partial y_{2}^{\nu}}\bigg)
\epsilon^{edf}\nonumber\\&&\times
\int d^4z \Delta_R (y_1-z) \frac{\partial}{\partial z_{\alpha}}
i{\bf S}{\bf j}_{A\alpha}^f (z)\nonumber\\&&
+\frac 1{f_{\pi}}e^{ik\cdot y_1}
\bigg(\frac{\partial}{\partial y_{1}^{\mu}} -ik_{\mu}\bigg)
\epsilon^{ace}\Delta_R (y_1-y_2)
\epsilon^{edf}i{\bf S}{\bf j}_{A\nu}^f (y_2)\nonumber\\&&
-\frac 1{f_{\pi}}e^{ik\cdot y_1}
\bigg(\frac{\partial}{\partial y_{1}^{\mu}} -ik_{\mu}\bigg)
\epsilon^{ace}\int d^4z \Delta_R (y_1-z)
\frac{\partial}{\partial z_{\alpha}}
{\bf S}T^*\bigg( {\bf j}_{A\alpha}^e (z){\bf V}_{\nu}^d (y_2)\bigg)
\nonumber\\&&
-\frac 1{f_{\pi}}e^{ik\cdot y_1}
\epsilon^{ace}
{\bf S}T^*\bigg( {\bf j}_{A\mu}^e (y_1){\bf V}_{\nu}^d (y_2)\bigg)
\nonumber\\&&
+\bigg( y_1\,\,\,  c\,\,\, \mu\leftrightarrow y_2\,\,\,d\,\,\,\nu\bigg)
\nonumber\\&&
-\frac 1{f_{\pi}}\int d^4y e^{ik\cdot y}
\frac{\partial}{\partial y_{\alpha}}
i{\bf S}T^*\bigg( {\bf j}_{A\alpha}^a (y){\bf V}_{\mu}^c (y_1)
{\bf V}_{\nu}^d (y_2)\bigg)
\nonumber\\&&
+{\cal O}\bigg( J, a, \pi_{\rm in} \bigg)
\label{ADD3}
\ee
Counting $\partial$ as ${\cal O} (k)$ and $\Delta$ as ${\cal O} (k^{-2})$,
the low momentum behaviour of (\ref{ADD1}) follows from the first three terms
of (\ref{ADD3}). Using (\ref{GT3}) and dividing out by
$(2\pi)^4\delta^4(p_1+q_1-p_2-q_2-k)$ gives
\be
&&+\frac {g_A}{f_{\pi}}g_{\mu\nu}\epsilon^{ace}\epsilon^{edf}
\frac 1{(k+q_2-q_1)^2-m_{\pi}^2}(p_2-p_1)^{\alpha}\,\,
\overline{u} (p_2)\gamma_{\alpha}\gamma_5\frac{\tau^f}2 u(p_1)\nonumber\\&&
-\frac {g_A}{f_{\pi}}
(2k_{\mu}-q_{1\mu})(2k_{\nu}-2q_{1\nu}+q_{2\nu})\epsilon^{ace}\epsilon^{edf}
\frac 1{(k-q_1)^2-m_{\pi}^2}\nonumber\\&&\,\,\,\,\times
\frac 1{(k+q_2-q_1)^2-m_{\pi}^2}(p_2-p_1)^{\alpha}\,\,
\overline{u} (p_2)\gamma_{\alpha}\gamma_5\frac{\tau^f}2 u(p_1)\nonumber\\&&
-\frac {g_A}{f_{\pi}}
(2k_{\mu}-q_{1\mu})\epsilon^{ace}\epsilon^{edf}
\frac 1{(k-q_1)^2-m_{\pi}^2}\,\,
\overline{u} (p_2)\gamma_{\nu}\gamma_5\frac{\tau^f}2 u(p_1)\nonumber\\&&
+\bigg( q_1\,\,\,c\,\,\,\mu\leftrightarrow -q_2\,\,\,d\,\,\,\nu\bigg)
\label{ADD5}
\ee
The threshold contributions (\ref{ADD5}) are shown in Fig. 2. We note that
the (strong) radiative corrections are subleading effects at threshold.
The reaction $\gamma p\rightarrow \gamma\pi^+ n$ has been analysed at Lebedev
\cite{lebedev} in the kinematical regime where the t-channel pion pole
dominates. The experiment at Lebedev was used to extract
the pion polarisabilities (see below).

\vskip .5cm
{\bf 11. Two-Pion Photo- and Electro-Production :
$\gamma N\rightarrow \pi\pi N$}
\vskip .3cm

Finally, consider the  two-pion photo- and electro-production
$\gamma N\rightarrow \pi\pi N$ reaction. The isovector part of the amplitude
is given by
\be
&&\int d^4y e^{-iq\cdot y}
<N(p_2) | a_{\rm in}^a (k_1) a_{\rm in}^b (k_2) {\bf S} {\bf V}^{a }_{\mu}
(y) | N(p_1) >_{\rm conn.} =\nonumber\\ &&
<N(p_2) | (-i)\int d^4y e^{-iq\cdot y} \frac{\delta}{\delta v^{\mu c} (y)}
\bigg[a_{\rm in}^a (k_1) ,\bigg[ a_{\rm in}^b (k_2) , \hat{\cal S} \bigg]\bigg]
|N(p_1)>_{\rm conn.}
\label{T1}
\ee
As before, we consider the isospin $3/2$ channel where the isoscalar part of
the current does not contribute. Iteration of (\ref{C2}) gives
\be
&&\bigg[a_{\rm in}^a (k_1) ,\bigg[ a_{\rm in}^b (k_2) ,
\hat{\cal S} \bigg]\bigg]=\nonumber\\
&&-i \bigg[ a_{\rm in}^a (k_1), \hat{\cal S} \bigg]
\int d^4y_2 d^4z_2 e^{ik_2\cdot y_2} \bigg( 1 +{\bf K}{\bf G}_R\bigg)^{bf}
(y_2,z_2) \bigg({\bf K}{\pi_{\rm in}}\bigg)^f (z_2)\nonumber\\&&
-i\hat{\cal S}
\int d^4y_2 d^4z_2 e^{ik_2\cdot y_2} \bigg( 1 +{\bf K}{\bf G}_R\bigg)^{bf}
(y_2,z_2) \bigg[ a_{\rm in}^a (k_1), \bigg( {\bf K}\pi_{\rm in}\bigg)^f
(z_2)\bigg]\nonumber\\&&
+\frac 1{f_{\pi}}
\int d^4y_2 d^4z_2 e^{ik_2\cdot y_2} \bigg( 1 +{\bf K}{\bf G}_R\bigg)^{bf}
(y_2,z_2) {\bf X}_A^f (z_2) \bigg[ a_{\rm in}^a (k_1), \hat{\cal S}\bigg]
\nonumber\\&& +{\cal O} (J, a )
\label{T2}
\ee
and then
\be
&&\bigg[a_{\rm in}^a (k_1) ,\bigg[ a_{\rm in}^b (k_2) ,
\hat{\cal S} \bigg]\bigg]=\nonumber\\&&
-\hat{\cal S}\int d^4y_1 d^4z_1 e^{ik_1\cdot y_1}
\bigg( 1 +{\bf K}{\bf G}_R\bigg)^{ae}
(y_1,z_1) \bigg({\bf K}{\pi_{\rm in}}\bigg)^e (z_1)\nonumber\\&&
\times
\int d^4y_2 d^4z_2 e^{ik_2\cdot y_2} \bigg( 1 +{\bf K}{\bf G}_R\bigg)^{bf}
(y_2,z_2) \bigg({\bf K}{\pi_{\rm in}}\bigg)^f (z_2)\nonumber\\&&
-\frac i{f_{\pi}}
\int d^4y_1 d^4z_1 e^{ik_1\cdot y_1}
\bigg( 1 +{\bf K}{\bf G}_R\bigg)^{ae}
(y_1,z_1) \bigg({\bf X}_A^e (z_1)\hat{\cal S}\bigg) \nonumber\\&&
\times
\int d^4y_2 d^4z_2 e^{ik_2\cdot y_2}
\bigg( 1 +{\bf K}{\bf G}_R\bigg)^{bf}
(y_2,z_2) \bigg({\bf K}{\pi_{\rm in}}\bigg)^f (z_2)\nonumber\\&&
-i\hat{\cal S}
\int d^4y_2 d^4z_2 e^{ik_2\cdot y_2} \bigg( 1 +{\bf K}{\bf G}_R\bigg)^{bf}
(y_2,z_2) {\bf K}^{fa} (z_2)  \nonumber\\&&
-i\frac 1{f_{\pi}}
\int d^4y_2 d^4z_2 e^{ik_2\cdot y_2} \bigg( 1 +{\bf K}{\bf G}_R\bigg)^{bf}
(y_2,z_2) \bigg({\bf X}_A^f (z_2) \hat{\cal S}\bigg) \nonumber\\&&
\times \int d^4y_1 d^4z_1 e^{ik_1\cdot y_1}
\bigg( 1 +{\bf K}{\bf G}_R\bigg)^{ae}
(y_1,z_1) \bigg({\bf K}{\pi_{\rm in}}\bigg)^e (z_1)\nonumber\\&&
-\frac 1{f_{\pi}^2}
\int d^4y_2 d^4z_2 e^{ik_2\cdot y_2} \bigg( 1 +{\bf K}{\bf G}_R\bigg)^{bf}
(y_2,z_2) \nonumber\\&&
\times \int d^4y_1 d^4z_1 e^{ik_1\cdot y_1}
\bigg( 1 +{\bf K}{\bf G}_R\bigg)^{ae}
(y_1,z_1) \,\,
\bigg[ {\bf X}_A^f (z_2) (\nabla^{\mu} a_{\mu})^e (z_1) \bigg]
\frac{\delta\hat{\cal S}}{\delta s (z_1)}\nonumber\\&&
+\frac 1{f_{\pi}^2}
\int d^4y_2 d^4z_2 e^{ik_2\cdot y_2} \bigg( 1 +{\bf K}{\bf G}_R\bigg)^{bf}
(y_2,z_2) \nonumber\\&&
\times \int d^4y_1 d^4z_1 e^{ik_1\cdot y_1}
\bigg( 1 +{\bf K}{\bf G}_R\bigg)^{ae}
(y_1,z_1) {\bf X}_A^f (z_2)\bigg({\bf X}_A^e (z_1)\hat{\cal
S}\bigg)\nonumber\\&&
+{\cal O} (J, a)\,\,\, ,
\label{T3}
\ee
where we have used ${\bf X}_A {\bf K} = {\cal O} (a)$. We also note that the
sixth term of (\ref{T3}) is essentially symmetric since
\be
\bigg[ {\bf X}_A^f (z_2), {\bf X}_A^e (z_1) \bigg]\hat{\cal S} =
{\cal O} (J)\,\,\,
\label{T4}
\ee
by (\ref{B13},\ref{B57}). The fifth term may be reduced by
\be
&&+\frac 1{f_{\pi}^2}
\int d^4y_1 d^4z_1 e^{ik_1\cdot y_1}
\bigg( 1 +{\bf K}{\bf G}_R\bigg)^{ae} (y_1,z_1) \frac{\delta\hat{\cal S}}
{\delta s(z_1)}\nonumber\\&&
\times \bigg(\nabla^{\mu}\nabla_{\mu}\bigg)^{ef} (z_1)
\int d^4y_2 e^{ik_2\cdot y_2} \bigg( 1 +{\bf K}{\bf G}_R\bigg)^{bf}
(y_2,z_1) +{\cal O} (a)  \nonumber\\ = &&
-\frac 1{f_{\pi}^2}m_{\pi}^2
\int d^4y_1 d^4z_1 e^{ik_1\cdot y_1}
\bigg( 1 +{\bf K}{\bf G}_R\bigg)^{ae} (y_1,z_1) \frac{\delta\hat{\cal S}}
{\delta s(z_1)}\nonumber\\&&
\times \int d^4y_2 e^{ik_2\cdot y_2} \bigg( 1 +{\bf K}{\bf G}_R\bigg)^{bf}
(y_2,z_1) +{\cal O} (a, s)
\label{T5}
\ee
where we have used the fact that $\delta\hat{\cal S}/\delta s(z)$ has no
asymptotic component when integrating by parts.

The sixth term gives two components
\be
&&+\frac 1{f_{\pi}^2}
\int d^4y_2 d^4z_1 e^{ik_2\cdot y_2}
\bigg( 1 +{\bf K}{\bf G}_R\bigg)^{bf} (y_2,z_1) \nabla^{fh}_{\beta} (z_1)
\nonumber\\&&
\times \int d^4y_1 e^{ik_1\cdot y_1} \bigg( 1 +{\bf K}{\bf G}_R\bigg)^{ae}
(y_1,z_1) \epsilon^{ehg}
\frac {\delta\hat{\cal S}}{\delta v^{g}_{\beta} (z_1)}
\label{T6}
\ee
and
\be
&&+\frac 1{f_{\pi}^2}
\int d^4y_2 d^4z_2 e^{ik_2\cdot y_2}
\bigg( 1 +{\bf K}{\bf G}_R\bigg)^{bf} (y_2,z_2) \nonumber\\&&
\times \int d^4y_1 d^4z_1 e^{ik_1\cdot y_1}
\bigg( 1 +{\bf K}{\bf G}_R\bigg)^{ae}
(y_1,z_1) \nonumber\\&&
\times\nabla^{fh}_{\beta} (z_2) \nabla^{eg}_{\alpha} (z_1)
\frac{\delta^2\hat{\cal S}}{\delta a_{\alpha}^g (z_1)
\delta a_{\beta}^h (z_2) }
+{\cal O} (a)
\label{T7}
\ee

Applying
\be
-i\int d^4y \,\,e^{-iq\cdot y}\frac{\delta}{\delta v^{\mu c} (y)}
\label{T8}
\ee
the first term in (\ref{T3}) gives no contribution.  The second and fourth
terms give
\be
&&-\bigg[ a_{\rm in}^a (k_1), {\bf S} \bigg]
\int d^4y e^{-iq\cdot y}
\bigg[ e^{ik_2\cdot y}\bigg( \frac{\partial}{\partial y_{\mu}} -
\frac{\stackrel\leftarrow\partial}{\partial y_{\mu}}\bigg)
\epsilon^{bcf} \pi_{\rm in}^f (y) \bigg] +
\bigg( k_1\,\, a\leftrightarrow k_2\,\, b\bigg)=\nonumber\\
&&+ i\bigg[ a_{\rm in}^a (k_1), {\bf S} \bigg] \bigg( 2k_{2\mu}-q_{\mu}\bigg)
\epsilon^{bcf} \int d^4y e^{+i(k_2-q)\cdot y}
\pi_{\rm in}^f (y) +
\bigg( k_1\,\, a\leftrightarrow k_2\,\, b\bigg)
\label{T9}
\ee
For photoproduction $q^2=0$ with $q\neq 0$, so that (\ref{T9}) vanishes by
energy-momentum conservation. For electroproduction $q^2 <0$, we may choose a
frame with $q^0=0$ so that $k^0-q^0$ is positive. Only the annihilation part of
$\pi_{\rm in} (y)$ is then picked up, and its contribution vanishes when
sandwiched between nucleon states.

The third term gives
\be
-{\bf S}\int d^4y e^{-iq\cdot y}
\bigg[ e^{ik_2\cdot y}\epsilon^{bca}
\bigg( \frac{\partial}{\partial y_{\mu}} -
\frac{\stackrel\leftarrow\partial}{\partial y_{\mu}}\bigg)\,\,e^{ik_1\cdot y}
\bigg]
\label{T10}
\ee
which contributes only to the disconnected part.

(\ref{T5}) and (\ref{T6}) give respectively
\be
&&+\frac i{f_{\pi}^2} m_{\pi}^2\int d^4y\,\, e^{-iq\cdot y}\int d^4z_1
\bigg[ e^{ik_1\cdot y}\bigg(\frac{\partial}{\partial y^{\mu}} -
\frac{\stackrel\leftarrow\partial}{\partial y^{\mu}}\bigg)\epsilon^{ace}
\Delta_R (y-z_1) \bigg]\frac{\delta\hat{\cal S}}{\delta s(z_1)}
e^{ik_2\cdot z_1}\delta^{be}\nonumber\\&&
+\bigg( k_1\,\, a \leftrightarrow k_2\,\, b\bigg)\nonumber\\
&&+\frac i{f_{\pi}^2} m_{\pi}^2\int d^4y\,\, e^{-iq\cdot y}\int d^4z
e^{i(k_1+k_2)\cdot z}\delta^{ab}
\frac{\delta^2\hat{\cal S}}{\delta s(z)\delta v^{\mu c} (y)}\nonumber\\
=&&+\frac i{f_{\pi}} m_{{\pi}}^2  (2k_{1\mu} -q_{\mu})\epsilon^{acb}
\frac 1{(k_1-q)^2-m_{\pi}^2}\int d^4z
e^{i(k_1+k_2-q)\cdot z} {\bf S}\hat\sigma (z)\nonumber\\&&
+\bigg( k_1\,\, a \leftrightarrow k_2\,\, b\bigg)\nonumber\\
&&-\frac i{f_{\pi}} m_{\pi}^2\delta^{ab}
\int d^4y d^4z e^{-iq\cdot y+i(k_1+k_2)\cdot z}
{\bf S}T^*\bigg( {\bf V}_{\mu}^c (z) \hat\sigma (z)\bigg)
\label{T11}
\ee
and
\be
&&-\frac i{f_{\pi}^2}\int d^4y\,\, e^{-iq\cdot y}\int d^4z
\bigg[ e^{ik_2\cdot y}\bigg(\frac{\partial}{\partial y^{\mu}} -
\frac{\stackrel\leftarrow\partial}{\partial y^{\mu}}\bigg)\epsilon^{bcf}
\Delta_R (y-z) \bigg]\nonumber\\
&&\times\frac{\partial}{\partial z_{\beta}}
\bigg(e^{ik_1\cdot z}\epsilon^{afg} i{\bf S}{\bf V}_{\beta}^g (z)\bigg)
\nonumber\\
&&-\frac i{f_{\pi}^2} \int d^4y\,\, e^{i(k_1+k_2-q)\cdot
y}\epsilon^{bch}\epsilon^{ahg} i{\bf S}{\bf V}_{\mu}^g (y)\nonumber\\
&&-\frac i{f_{\pi}^2}\int d^4y\,\, e^{-iq\cdot y}\int d^4z e^{ik_2\cdot z}
\nonumber\\&&
\times\frac {\partial}{\partial z_{\beta}}\bigg(
\bigg[ e^{ik_1\cdot y}\bigg(\frac{\partial}{\partial y^{\mu}} -
\frac{\stackrel\leftarrow\partial}{\partial y^{\mu}}\bigg)
\Delta_R (y-z) \bigg]\epsilon^{ace}\epsilon^{beg}
i{\bf S}{\bf V}_{\beta}^g (z) \bigg)\nonumber\\
&&+\frac i{f_{\pi}^2}\int d^4y\,\, e^{-iq\cdot y}\int d^4z e^{ik_2\cdot z}
\nonumber\\&&
\times\frac {\partial}{\partial z_{\beta}}
\bigg[ e^{ik_1\cdot z}\epsilon^{abg} {\bf S}T^*\bigg({\bf V}_{\mu}^c (y)
{\bf V}_{\beta}^g (z)\bigg)\bigg]\nonumber\\
=&&+\frac 1{f_{\pi}^2} (2k_{2\mu}-q_{\mu})
k_1^{\beta}\epsilon^{bcf}\epsilon^{afg}
\frac 1{(k_2-q)^2-m_{\pi}^2}
\int d^4z e^{i(k_1+k_2-q)\cdot z} {\bf S}{\bf V}_{\beta}^g (z)
\nonumber\\&& +\bigg( k_1\,\, a\leftrightarrow k_2\,\, b\bigg)\nonumber\\&&
-\frac 1{2f_{\pi}^2} (k_1^{\beta}-k_2^{\beta})\epsilon^{abg}
\int d^4y d^4z e^{-iq\cdot y+(k_1+k_2)\cdot z}
{\bf S}T^*\bigg( {\bf V}_{\mu}^c (y) {\bf V}_{\beta}^g (z) \bigg)
\label{T12}
\ee
where we have used the Ward identity
\be
\frac{\partial}{\partial z_{\beta}}{\bf S}
T^*\bigg( {\bf V}_{\mu}^c (y) {\bf V}_{\beta}^g (z) \bigg) =
i\epsilon^{gce} \delta^4 (z-y) \,{\bf S}{\bf V}_{\mu}^e (z)\,\,\,.
\label{T13}
\ee

(\ref{T7}) gives
\be
&&+\frac i{f_{\pi}^2}\int d^4y\,\, e^{-iq\cdot y}\int d^4z_2
\bigg[ e^{ik_2\cdot y}\bigg(\frac{\partial}{\partial y^{\mu}} -
\frac{\stackrel\leftarrow\partial}{\partial y^{\mu}}\bigg)\epsilon^{bcf}
\Delta_R (y-z_2) \bigg]\nonumber\\
&&\times\int d^4z_1 e^{ik_1\cdot z_1}\frac{\partial}{\partial z_{1\alpha}}
\frac{\partial}{\partial z_{2\beta}} {\bf S}T^*
\bigg({\bf j}_{A\alpha}^a (z_1){\bf j}_{A\beta}^f (z_2)\bigg)\nonumber\\
&&+\frac i{f_{\pi}^2}\int d^4y\,\, e^{-iq\cdot y}\int d^4z_1
e^{ik_1\cdot z_1 +ik_2\cdot y}\epsilon^{bch}
\frac{\partial}{\partial z_{1\alpha}} {\bf S}T^*
\bigg({\bf j}_{A\alpha}^a (z_1) {\bf j}_{A\mu}^h (y)\bigg)\nonumber\\
&&+\bigg( k_1\,\, a\leftrightarrow k_2\,\, b\bigg)\nonumber\\
&&-\frac i{f_{\pi}^2}\int d^4y\,\, e^{-iq\cdot y}\int d^4z_2
e^{ik_2\cdot z_2}\int d^4z_1 e^{ik_1\cdot z_1}\nonumber\\&&
\times\frac{\partial}{\partial z_{1\alpha}}
\frac{\partial}{\partial z_{2\beta}} i^3 {\bf S}T^*
\bigg({\bf j}_{A\alpha}^a (z_1) {\bf j}_{A\beta}^b (z_2)
{\bf V}_{\mu}^c (y)\bigg)\nonumber\\
=&&-\frac 1{f_{\pi}^2} (2k_{2\mu}-q_{\mu}) k_{1\alpha} (k_1^{\beta}-q^{\beta})
\epsilon^{bcf}\frac 1{(k_2-q)^2-m_{\pi}^2}\nonumber\\&&
\times\int d^4z_1d^4z_2 e^{ik_1\cdot z_1 + i(k_2-q)\cdot z_2}
{\bf S} T^*\bigg({\bf j}_{A\alpha}^a (z_1){\bf j}_{A\beta}^f (z_2)\bigg)
\nonumber\\
&&+\frac 1{f_{\pi}^2} k_1^{\alpha} \epsilon^{bch}\int d^4y d^4z_1
e^{i(k_2-q)\cdot y+ik_1\cdot z_1}
\,\,{\bf S} T^*\bigg({\bf j}_{A\alpha}^a (z_1){\bf j}_{A\mu}^h (y)\bigg)
\nonumber\\&&
+\bigg( k_1\,\, a\leftrightarrow k_2\,\, b\bigg)\nonumber\\
&&+\frac 1{f_{\pi}^2} k_1^{\alpha} k_2^{\beta}
\int d^4y d^4z_1 d^4z_2 e^{-iq\cdot y+ik_1\cdot z_1 + ik_2\cdot z_2}
\bigg({\bf j}_{A\alpha}^a (z_1) {\bf j}_{A\beta}^b (z_2)
{\bf V}_{\mu}^c (y)\bigg)
\label{T14}
\ee

To lowest order in the momentum, the above results yield for the isovector part
of the two-pion photo- and electro-production amplitude
\be
&&\int d^4y e^{-iq\cdot y}
<N(p_2) | a_{\rm in}^a (k_1) a_{\rm in}^b (k_2) {\bf S} {\bf V}^{a }_{\mu}
(y) | N(p_1) >_{\rm conn.} =\nonumber\\ &&
+\frac 1{f_{\pi}^2} (2k_{2\mu}-q_{\mu}) k_1^{\beta}
\epsilon^{bcf}\epsilon^{afg}
\frac 1{(k_2-q)^2-m_{\pi}^2}
\int d^4z e^{i(k_1+k_2-q)\cdot z}
<N(p_2)|{\bf V}_{\beta}^g (z) |N(p_1)>\nonumber\\
&&+\frac 1{2f_{\pi}^2}  \epsilon^{bch}\epsilon^{ahg}
\int d^4z e^{i(k_1+k_2-q)\cdot z}
<N(p_2)|{\bf V}_{\mu}^g (z) |N(p_1)>
+\bigg( k_1\,\, a\leftrightarrow k_2\,\, b\bigg)\nonumber\\
&&-\frac 1{2f_{\pi}^2} (k_1^{\beta}-k_2^{\beta})\epsilon^{abg}
\int d^4y d^4z e^{-iq\cdot y+i(k_1+k_2)\cdot z}
<N(p_2)|T^*\bigg({\bf V}_{\mu}^c (y){\bf V}_{\beta}^g (z)\bigg) |N(p_1)>
\label{T15}
\ee
where the last term is necessary to maintain current conservation.
Fig. 3 illustrates the various contributions to (\ref{T15}). As before,
the result applies to the isospin $3/2$ channel where the isoscalar part of the
current does not contribute. Our result is in agreement with part of the
original result derived by Carruthers and Huang \cite{huang} in the soft
pion limit.

Since the
isovector part of the electromagnetic current is G-parity even, the
anomalous process $\gamma \rightarrow \pi^+\pi^0\pi^-$
as a t-channel pole is absent from (\ref{T15}).
The anomalous process is expected in the isospin $1/2$ channel
through the isoscalar part of the electromagnetic current.
The latter requires a full $SU(3)\times SU(3)$ analysis.
The extraction of the $\gamma \rightarrow \pi^+\pi^0\pi^-$ amplitude from the
$\gamma N\rightarrow \pi\pi N$ reaction, provide an alternative
to the conventional collider experiments using
$e^+e^-\rightarrow \pi^0\pi^+\pi^-$.

In the threshold region, the reaction $\gamma N\rightarrow \pi\pi N$ has been
studied in the isobar regime with a particular emphasis on
$\gamma N\rightarrow \pi\Delta\rightarrow \pi\pi N$ \cite{laget}. Since our
formulation does not account for baryon dynamics, our results would not be
accurate in the resonance region. However, since the isobar excitation
energy is about $M_{\Delta}-M-2m_{\pi} \sim 14 $ MeV, it would be interesting
to see how our results compare with the data within this energy range, although
the width of the isobar and the nearness of the $N^* (1520)$
may cause some problems.

Finally, we note that a set of interesting and new experiments
by the MAMI collaboration are presently being carried out at MAINZ,
for two-pion photoproduction on proton targets, measuring directly
$\gamma p\rightarrow \pi^+\pi^- p$, $\gamma p\rightarrow \pi^0\pi^0$
and $\gamma^* p\rightarrow \pi^+\pi^0 n$. Also the reaction $\gamma
p\rightarrow
\pi^+ \pi^0 n$ will be measured at CEBAF using tagged photons with special
emphasis on the extraction of $\gamma \pi\rightarrow \pi\pi^0$ from the
t-channel pole. In light of this and the more recent
calculations in the context of heavy-baryon chiral
perturbation theory \cite{dahm,meissner}, it would be interesting to extend
the present analysis to baryon dynamics beyond threshold.

\vskip .5cm
{{\bf 12. Radiative Pion-Nucleon Scattering :} $\pi N\rightarrow \pi  \gamma
N$}
\vskip .3cm

The radiative pion-nucleon transition amplitude  involves the isovector current
in the form
\be
&&+\epsilon^{\mu}(q)\int d^4y e^{iq\cdot y}
<N(p_2) |a_{\rm in}^b (k_2) {\bf S} {\bf V}_{\mu}^c (y) a_{\rm in}^{a\dagger}
(k_1) |N(p_1)> = \nonumber\\&&
-i\epsilon^{\mu}(q)\int d^4y e^{iq\cdot y}
<N(p_2) | \frac{\delta}{\delta v^{\mu c} (y)}
\bigg[\bigg[ a_{\rm in}^b (k_2) ,\hat{\cal S}, a_{\rm in}^{a\dagger}
(k_1)\bigg]\bigg] |N(p_1)>
\label{ADD7}
\ee
This amplitude follows from $\gamma N\rightarrow \pi\pi N$ by crossing $k_1,
q\rightarrow -k_1, -q$. The result is the sum of three terms
\be
&&-\frac {2i}{f_{\pi}} m_{\pi}^2 \epsilon (q)\cdot k_1\epsilon^{acb}
\frac 1{(k_1-q)^2-m_{\pi}^2}\nonumber\\
&&\,\,\,\times \int d^4z e^{i(-k_1+k_2+q)\cdot z}
<N(p_2) |\hat\sigma (z) |N(p_1) >\nonumber\\
&&+\bigg(-k_1\,\,\,a\leftrightarrow k_2\,\,\, b\bigg)\nonumber\\
&&-\frac i{f_{\pi}} m_{\pi}^2 \epsilon^{\mu} (q) \delta^{ab}
\int d^4y d^4z e^{iq\cdot y+i(-k_1+k_2)\cdot z}
<N(p_2)| T^*\bigg({\bf V}_{\mu}^c (y) \hat\sigma (z) \bigg) |N (p_1)>
\label{ADD8}
\ee
\be
&&-\frac 2{f_{\pi}^2} \epsilon(q)\cdot k_2 k_1^{\beta}
\epsilon^{bcf}\epsilon^{afg}
\frac 1{(k_2+q)^2-m_{\pi}^2}
\int d^4z e^{i(-k_1+k_2+q)\cdot z}
<N(p_2)|{\bf V}_{\beta}^g (z) |N(p_1)>\nonumber\\
&&+\frac 1{2f_{\pi}^2}  \epsilon^{\mu} (q) \epsilon^{bch}\epsilon^{ahg}
\int d^4z e^{i(-k_1+k_2+q)\cdot z}
<N(p_2)|{\bf V}_{\mu}^g (z) |N(p_1)>
+\bigg(- k_1\,\, a\leftrightarrow k_2\,\, b\bigg)\nonumber\\
&&+\frac 1{2f_{\pi}^2} \epsilon^{\mu}
(q)(k_1^{\beta}+k_2^{\beta})\epsilon^{abg}
\int d^4y d^4z e^{+iq\cdot y+i(-k_1+k_2)\cdot z}
<N(p_2)|T^*\bigg({\bf V}_{\mu}^c (y){\bf V}_{\beta}^g (z)\bigg) |N(p_1)>
\nonumber\\
\label{ADD9}
\ee
\be
&&-\frac 2{f_{\pi}^2} \epsilon (q)\cdot k_2 k_1^{\alpha}
(k_1^{\beta}-q^{\beta}) \frac 1{(k_2+q)^2-m_{\pi}^2}\epsilon^{bcf}
\nonumber\\
&&\,\,\,\times
\int d^4z_1 d^4z_2 e^{-ik_1\cdot z_1 +i (k_2 +q)\cdot z_2}
<N(p_2) | T^*\bigg( {\bf j}_{A\alpha}^a (z_1) {\bf j}_{A\beta }^f (z_2)\bigg)
|N(p_1)>\nonumber\\
&&-\frac 1{f_{\pi}^2} \epsilon^{\mu} (q) \epsilon^{bch}
\int d^4y d^4z_1 e^{-ik_1\cdot z_1 +i (k_2 +q)\cdot y}
<N(p_2) | T^*\bigg( {\bf j}_{A\alpha}^a (z_1) {\bf j}_{A\mu }^h (y)
\bigg)|N(p_1)>\nonumber\\
&&+\bigg(-k_1\,\,\,a\leftrightarrow k_2 \,\,\, b\bigg)\nonumber\\
&&-\frac 1{f_{\pi}^2} k_1^{\alpha} k_2^{\beta}\epsilon^{\mu} (q)
\int d^4y d^4z_1 d^4z_2 e^{+iq\cdot y-ik_1\cdot z_1 +i k_2\cdot z_2}
\nonumber\\
&&\,\,\,\times
<N(p_2) | T^*\bigg( {\bf j}_{A\alpha}^a (z_1) {\bf j}_{A\beta }^b (z_2)
{\bf V}_{\mu}^c (y)\bigg) |N(p_1)>\nonumber\\
\label{ADD10}
\ee
The low momentum result is given by (\ref{ADD9}), and illustrated in Fig. 4.

Radiative pion-nucleus scattering $\pi^- A\rightarrow \pi^- \gamma A$ has been
studied experimentally at Serpukhov \cite{serp}, using 40 GeV pions
with final $\pi^-\gamma$ detected in coincidence. In the kinematical
regime $t\sim 0$ the pion scatters off a virtual photon in the Coulomb field of
a fixed nucleus A. The reaction is of the Primakoff type and is equivalent to
$\gamma\pi^-\rightarrow \gamma\pi^-$ on a fixed $\pi^-$ target,
with $\gamma$'s in the $60-600$ MeV
range. This experiment was used to determine the pion
electric polarizability as we will discuss below. Here we would like to note
that since we are treating the electromagnetic interaction to first order in
the electric charge, the Primakoff term is absent from our result
(\ref{ADD10}). The inclusion of electromagnetic effects beyond leading order
requires the $U(1)$ gauging of (\ref{B1}-\ref{B2}), and a rerun of the above
arguments. This point will be discussed elsewhere.

\vskip .5cm
{\bf 13. Pion-Pion Scattering : $\pi\pi\rightarrow \pi\pi$}
\vskip .3cm

We now proceed to the $\pi\pi$-scattering amplitude.
\be
&&<0| a_{\rm in}^d (p_2)a_{\rm in}^b (k_2) \bigg( {\bf S} -{\bf 1}\bigg)
a_{\rm in}^{a\dagger} (k_1)a_{\rm in}^{c\dagger} (p_1) |0> \nonumber\\ =&&
<0| \bigg[a_{\rm in}^d (p_2) ,\bigg[\bigg[
a_{\rm in}^b (k_2), \bigg[ {\bf S} , a_{\rm in}^{a\dagger} (k_1)
\bigg]\bigg], a_{\rm in}^{c\dagger} (p_1)\bigg]\bigg] |0> \nonumber\\ = &&
(2\pi )^4 \delta^4 (k_1+ p_1 -k_2-p_2 )
i{\cal T} \bigg(p_2 d, k_2 b \leftarrow k_1 a, p_1 c \bigg).
\label{C17}
\ee
Here we have used the stability of the vacuum and one-particle states, and
anticipated energy and momentum conservation.

\vskip .15cm
{\it 13.1. Reduction}

For the triple commutator
\be
\bigg[\bigg[ a_{\rm in}^b (k_2) , \bigg[ {\bf S}, a_{\rm in}^{a\dagger}
(k_1)\bigg]\bigg], a_{\rm in}^{c\dagger} (p_1 )\bigg]
\label{C18}
\ee
we need to evaluate
\be
&&\frac{\delta}{\delta a_{\alpha}^a (y_1 )}
\frac{\delta}{\delta a_{\beta}^b  (y_2)}
\bigg[ \hat{\cal S}, a_{\rm in}^{c\dagger} (p_1 )\bigg] =\nonumber\\&&
+i{\bf S} e^{-ip_1\cdot y_1} \bigg(\epsilon^{cae}\epsilon^{ebf}
+\epsilon^{cbe}\epsilon^{eaf}\bigg) \delta^4 (y_1-y_2) g^{\alpha\beta} \pi^f
(y_1)\nonumber\\ &&
+\frac{1}{f_{\pi}} e^{-ip_1\cdot y_1} \delta^{ac}
(\partial_1^{\alpha}-ip_1^{\alpha})
\frac{\delta^2\hat{\cal S}}{\delta s (y_1) \delta a_{\beta}^b (y_2)}
+\frac{1}{f_{\pi}} e^{-ip_1\cdot y_2} \delta^{bc}
(\partial_2^{\beta}-ip_2^{\beta})
\frac{\delta^2\hat{\cal S}}{\delta s (y_1) \delta a_{\alpha}^a (y_1)}
\nonumber\\&&
+\frac{1}{f_{\pi}} e^{-ip_1\cdot y_1} \epsilon^{cae}
\frac{\delta^2\hat{\cal S}}{\delta v_{\alpha}^e (y_1) \delta a_{\beta}^b (y_2)}
+\frac{1}{f_{\pi}} e^{-ip_1\cdot y_2} \epsilon^{cbe}
\frac{\delta^2\hat{\cal S}}{\delta v_{\beta}^e (y_2) \delta a_{\alpha}^a (y_1)}
\nonumber\\&&
+\frac{i}{f_{\pi}} p_{1\gamma}\int d^4 y_3
e^{-ip_1\cdot y_3} \frac{\delta^3\hat{\cal S}}
{\delta a_{\alpha}^a (y_1) \delta a_{\beta}^b (y_2)\delta a_{\gamma}^c (y_3)}.
\label{C19}
\ee
Hence (\ref{C18}) consists of the following five terms
\be
&&-\frac{i}{f_{\pi}} m_{\pi}^2 \int d^4y e^{-i (k_1-k_2)\cdot y} \delta^{ab}
\bigg[ {\bf S} \hat{\sigma} (y) , a_{\rm in}^{c\dagger} (p_1 )
\bigg]\nonumber\\&&
+\frac{1}{f^2_{\pi}} k_{1}^{\alpha} \int d^4y e^{-i (k_1-k_2)\cdot y}
\epsilon^{abe}
\bigg[ {\bf S} {\bf V}_{\alpha}^e (y) , a_{\rm in}^{c\dagger} (p_1 )
\bigg]
\label{C20}
\ee
\be
+\frac{i}{f_{\pi}^2} {\bf S} k_1\cdot k_2
\int d^4y e^{-i (k_1-k_2 + p_1)\cdot y} \bigg(\epsilon^{cae}\epsilon^{ebf}+
\epsilon^{cbe}\epsilon^{eaf}\bigg) \pi^f (y)
\label{C21}
\ee
\be
&&+\frac{i}{f^3_{\pi}} m_{\pi}^2 k_{2\beta} \delta^{ac}
\int d^4y_1 d^4y_2  e^{-i (k_1+p_1)\cdot y_1 + ik_2\cdot y_2}
\frac{\delta^2\hat{\cal S}}{\delta s(y_1)\delta a_{\beta}^b (y_2)}
\nonumber\\&&
-\frac{i}{f^3_{\pi}} m_{\pi}^2 k_{1\alpha} \delta^{bc}
\int d^4y_1 d^4y_2  e^{-i k_1\cdot y_1 +i (k_2-p_1)\cdot y_2}
\frac{\delta^2\hat{\cal S}}{\delta s(y_2)\delta a_{\alpha}^a (y_1)}
\label{C22}
\ee
\be
+\frac{i}{f^3_{\pi}} k_{1\alpha} k_{2\beta} p_{1\gamma}
\int d^4y_1 d^4y_2  d^4y_3 e^{-i k_1\cdot y_1 +ik_2\cdot y_2 -ip_1\cdot y_3}
\frac{\delta^3\hat{\cal S}}{\delta a_{\alpha}^a (y_1)\delta a_{\beta}^b (y_2)
\delta a_{\gamma}^c (y_3)}
\label{C23}
\ee
\be
&&+\frac{1}{f^3_{\pi}} k_{1\alpha} k_{2\beta} \epsilon^{cae}
\int d^4y_1 d^4y_2   e^{-i (k_1 +p_1)\cdot y_1 +ik_2\cdot y_2}
\frac{\delta^2\hat{\cal S}}{\delta v_{\alpha}^e (y_1)\delta a_{\beta}^b (y_2)}
\nonumber\\&&
+\frac{1}{f^3_{\pi}} k_{1\alpha} k_{2\beta} \epsilon^{cbe}
\int d^4y_1 d^4y_2   e^{-i k_1\cdot y_1 +i(k_2-p_1)\cdot y_2}
\frac{\delta^2\hat{\cal S}}{\delta a_{\alpha}^a (y_1)\delta v_{\beta}^e (y_2)}.
\label{C24}
\ee
To obtain a more symmetric representation, we note that
\be
&&\bigg[ {\bf S}\hat{\sigma} (y) , a_{\rm in}^{a\dagger} (k_1) \bigg] =
-\frac{i}{f_{\pi}} \frac{\delta}{\delta s (y)} \bigg[\hat{\cal S} ,
a_{\rm in}^{a\dagger} (k_1 )\bigg] =\nonumber\\&&
-\frac 1{f_{\pi}} {\bf S} e^{-ik_1\cdot y} \pi_{\rm in}^a (y) -
\frac i{f_{\pi}^2} e^{-ik_1\cdot y} \int d^4 z \Delta_R (y-z) \partial_{\beta}
\frac{\delta\hat{\cal S}}{\delta a_{\beta}^a (z)}
\nonumber\\&&
-\frac i{f_{\pi}^2}\int d^4 y_1 e^{-ik_1\cdot y_1} \frac{\partial}{\partial
y_1^{\alpha}}
\frac{\delta^2\hat{\cal S}}{\delta a_{\alpha}^a (y_1) \delta s
(y)}  =\nonumber\\&&
-\frac 1{f_{\pi}} {\bf S} e^{-ik_1\cdot y} \pi^a (y) +
\frac 1{f_{\pi}^2} k_{1\alpha} \int d^4y_1 e^{-ik_1\cdot y_1}
\frac{\delta^2\hat{\cal S}}{\delta a_{\alpha}^a (y_1) \delta s(y)}
\label{C25}
\ee
Similarly
\be
\bigg[ a_{\rm in}^b (k_2) , {\bf S} \hat{\sigma} (y) \bigg] =
-\frac 1{f_{\pi}} {\bf S} e^{ik_2\cdot y} \pi^b (y) -
\frac 1{f_{\pi}^2} k_{2\beta} \int d^4y_2 e^{ik_2\cdot y_1}
\frac{\delta^2\hat{\cal S}}{\delta a_{\beta}^b (y_2)\delta s
(y)}
\label{C26}
\ee
Hence (\ref{C22}) is equal to
\be
&&-\frac i{f_{\pi}} m_{\pi}^2 \delta^{bc}\int d^4y e^{i(k_2-p_1)\cdot y}
\bigg[ {\bf S} \hat\sigma (y) , a_{\rm in}^{a\dagger} (k_1) \bigg]\nonumber\\&&
-\frac i{f_{\pi}^2} m_{\pi}^2 \delta^{bc}\int d^4y e^{i(k_2-p_1-k_1)\cdot y}
{\bf S} \pi^a (y)\nonumber\\&&
-\frac i{f_{\pi}} m_{\pi}^2 \delta^{ac}\int d^4y e^{-i(k_1+p_1)\cdot y}
\bigg[ a_{\rm in}^{b} (k_2), {\bf S} \hat\sigma (y) \bigg]\nonumber\\&&
-\frac i{f_{\pi}^2} m_{\pi}^2 \delta^{ac}\int d^4y e^{-i(k_1-k_2+p_1)\cdot y}
{\bf S} \pi^b (y)
\label{C27}
\ee

In the same manner we have
\be
&&\bigg[{\bf S}{\bf V}^{\gamma e} (y) , a_{\rm in}^{a\dagger} (k_1 ) \bigg] =
-i\frac {\delta}{\delta v_{\gamma}^e (y) } \bigg[ \hat{\cal S}, a_{\rm
in}^{a\dagger} (k_1) \bigg] =\nonumber\\&&
-{\bf S} e^{-ik_1\cdot y} \epsilon^{aef}
\bigg(\frac{\partial}{\partial y_{\gamma}} -
\frac{\stackrel\leftarrow\partial}{\partial
y_{\gamma}}\bigg) \pi_{\rm in}^f (y) \nonumber\\ &&
-\frac{i}{f_{\pi}} e^{-ik_1\cdot y} \epsilon^{aef}
\bigg(\frac{\partial}{\partial y_{\gamma}} -
\frac{\stackrel\leftarrow\partial}{\partial
y_{\gamma}}\bigg) \int d^4z \Delta_R (y-z) \partial_{\mu}
\frac{\delta\hat{\cal S}}{\delta a_{\mu}^f (z)}
\nonumber\\ &&
-\frac {i}{f_{\pi}} e^{-ik_1\cdot y}\epsilon^{aef}
\frac{\delta\hat{\cal S}}{\delta
a_{\gamma}^f (y)}\nonumber\\&&
-\frac {i}{f_{\pi}} \int d^4y_1 e^{-ik_1\cdot y_1}
\frac{\partial}{\partial y_1^{\alpha}}
\frac{\delta^2\hat{\cal S}}
{\delta a_{\alpha}^a (y_1) \delta v_{\gamma}^e (y)}
\nonumber\\&&
=-ik_{1}^{\gamma} e^{-ik_1\cdot y}\epsilon^{aef} {\bf S} \pi^f (y) +\frac
1{f_{\pi}} e^{-ik_1\cdot y} \epsilon^{aef} {\bf S} {\bf A}^{\gamma f}
(y)\nonumber\\&&
+\frac 1{f_{\pi}} k_{1\alpha} \int d^4 y_1 e^{-ik_1\cdot y_1}
\frac{\delta^2\hat{\cal S}}{\delta a_{\alpha}^a (y_1) \delta v_{\gamma}^e (y)}
\label{C28}
\ee
and
\be
&&\bigg[ a_{\rm in}^b (k_2) , {\bf S} {\bf V}^{\gamma e} (y) \bigg] =
ik_2^{\gamma} e^{ik_2\cdot y} \epsilon^{bef} {\bf S} \pi^f (y) \nonumber\\&&
+\frac 1{f_{\pi}} e^{ik_2\cdot y} \epsilon^{bef} {\bf S}{ A}^{\gamma f} (y)
-\frac 1{f_{\pi}} k_{2\beta} \int d^4 y_2 e^{ik_2\cdot y_2}
\frac{\delta^2\hat{\cal S}}{\delta a_{\beta}^b (y_2) \delta v_{\gamma}^e (y)}.
\label{C29}
\ee
Thus, (\ref{C24}) is equal to
\be
&&+\frac 1{f_{\pi}^2} k_{2\beta}\epsilon^{cbe}\int d^4y e^{i(k_2-p_1)\cdot y}
\bigg[ {\bf S} {\bf V}_{\beta}^e (y), a_{\rm in}^{a\dagger} (k_1) \bigg]
\nonumber\\&&
+\frac i{f_{\pi}^2} k_1\cdot k_2 \epsilon^{cbe}\epsilon^{aef}
\int d^4y e^{-i(k_1-k_2 +p_1)\cdot y}
{\bf S} \pi^f (y)\nonumber\\&&
-\frac i{f_{\pi}^3} k_{2\beta}  \epsilon^{cbe}\epsilon^{aef}
\int d^4y e^{-i(k_1-k_2 +p_1)\cdot y}
{\bf S} {\bf A}_{\beta}^f (y)\nonumber\\&&
-\frac 1{f_{\pi}^2} k_{1\alpha} \epsilon^{cae}
\int d^4y e^{-i(k_1+p_1)\cdot y}
\bigg[ a_{\rm in}^{b} (k_2), {\bf S} {\bf V}_{\alpha}^e (y) \bigg]\nonumber\\&&
+\frac i{f_{\pi}^2} k_1\cdot k_2 \epsilon^{cae}\epsilon^{bef}
\int d^4y e^{-i(k_1-k_2+p_1)\cdot y}
{\bf S} \pi^f (y)\nonumber\\&&
+\frac i{f_{\pi}^3} k_{1\alpha} \epsilon^{cae}\epsilon^{bef}
\int d^4y e^{-i(k_1-k_2+p_1)\cdot y}
{\bf S} {\bf A}^f_{\alpha} (y).
\label{C30}
\ee
The pion piece in (\ref{C30}) cancels against (\ref{C21}) to give the following
result for the triple commutator (\ref{C18})
\be
&&-\frac i{f_{\pi}} m_{\pi}^2 \delta^{ab}\int d^4y e^{-i(k_1-k_2)\cdot y}
\bigg[{\bf S}\hat{\sigma} (y) , a_{\rm in}^{c\dagger } (p_1) \bigg]
+ {2\,\,\rm{perm.}} \nonumber\\&&
+\frac 1{f_{\pi}^2} k_{1}^{\alpha} \int d^4 y e^{-i(k_1-k_2)\cdot
y}\epsilon^{abe} \bigg[ {\bf S}{\bf V}_{\alpha}^e (y)  ,
a_{\rm in}^{c\dagger } (p_1) \bigg] +{2\,\,\rm{perm.}}\nonumber\\ &&
-\frac i{f_{\pi}^2} m_{\pi}^2 \delta^{ac}
\int d^4y e^{-i(k_1-k_2+p_1)\cdot y} {\bf S} \pi^b (y) -
\frac i{f_{\pi}^2} m_{\pi}^2 \delta^{bc} \int d^4y e^{-i(k_1-k_2+p_1)\cdot y}
{\bf S} \pi^a (y) \nonumber\\&&
+\frac i{f_{\pi}^3} k_{1\alpha} k_{2\beta} p_{1\gamma}
\int d^4 y_1 d^4 y_2 d^4 y_3 e^{-ik_1\cdot y_1 + ik_2 \cdot y_2 -ip_1\cdot y_3}
\frac{\delta^3\hat{\cal S}}{\delta a_{\alpha}^a (y_1) \delta a_{\beta}^b (y_2)
\delta a_{\gamma}^c (y_3)} \nonumber\\&&
+\frac 1{f_{\pi}^3} k_1^{\alpha} \epsilon^{cae}\epsilon^{bef}
\int d^4y e^{-i (k_1-k_2+p_1)\cdot y} {\bf S} {\bf A}_{\alpha}^f (y)
\nonumber\\&&
-\frac 1{f_{\pi}^3} k_2^{\beta} \epsilon^{cbe}\epsilon^{aef}
\int d^4y e^{-i (k_1-k_2+p_1)\cdot y} {\bf S} {\bf A}_{\beta}^f (y).
\label{C31}
\ee
It is now easy to see that the amplitude is symmetric under permutations
$(a\,\,k_1)$, $(b \,\,(-k_2))$ and $(c\,\,p_1)$. We remark that most of the
expressions derived above are quoted to leading order in the external fields.

\vskip .15cm
{\it 13.2. Weinberg Result}

To relate the various terms of the amplitude to physical observables, we
introduce the pion scalar form factor
\be
<0| a_{\rm in}^d (p_2 ) \bigg[ {\bf S}\hat{\sigma} (y) ,
a_{\rm in}^{c\dagger} (p_1) \bigg] | 0> = \delta^{cd} {\bf F}_S (t) e^{-i
(p_1-p_2)\cdot y}
\label{C32}
\ee
and the pion electromagnetic form factor
\be
<0| a_{\rm in}^d (p_2 ) \bigg[{\bf S V}_{\alpha}^e (y),
a_{\rm in}^{c\dagger} (p_1) \bigg] | 0> =
i\epsilon^{dec} (p_1+p_2)_{\alpha} {\bf F}_V (t) e^{-i(p_1-p_2)\cdot y}
\label{C33}
\ee
as well as the Mandelstam variables
\be
&&s= (k_1+p_1)^2 = (k_2+p_2)^2\nonumber\\&&
t= (k_1-k_2)^2 = (p_1-p_2)^2\nonumber\\&&
u= (k_1-p_2)^2 = (p_1-k_2)^2.
\label{C34}
\ee
Inserting (\ref{C31}) into (\ref{C17}), we find that
$i{\cal T}$ is a sum of
\be
-\frac i{f_{\pi}} m_{\pi}^2 \delta^{ab} \delta^{cd} {\bf F}_S (t)
+\frac i{2 f_{\pi}^2} (s-u) {\bf F}_V (t) \epsilon^{abe} \epsilon^{cde} -
\frac i{2f_{\pi}^2} t \delta^{ab} \delta^{cd} +{\rm {2\,\, perm}}
\label{C35}
\ee
and
\be
\frac i{f_{\pi}^3} k_{1\alpha} k_{2\beta } p_{1\gamma} \int d^4 y_1 d^4 y_2 d^4
y_3 e^{-ik_1\cdot y_1 + ik_2\cdot y_2 -ip_1\cdot y_3}
<0| \bigg[ a_{\rm in}^d (p_2) , \frac{\delta^3\hat{\cal S}}{\delta a_{\alpha}^a
(y_1) \delta a_{\beta}^b (y_2) \delta a_{\gamma}^c (y_3) }\bigg] |0>\nonumber\\
\label{C36}
\ee
modulo $(2\pi )^4 \delta ( k_1+ p_1 -k_2 -p_2 )$.

As a check on (\ref{C35}-\ref{C36}) we may evaluate $i{\cal T}$ at tree level
${\cal O} (1/f_{\pi}^2 )$. In this case, (\ref{C36}) does not contribute,
${\bf F}_S \rightarrow -1/f_{\pi}$ from (\ref{C25}), and ${\bf F}_V \rightarrow
{\bf F}_V (0) = 1$ by charge normalization. Thus
\be
i{\cal T}_{\rm tree} = \frac i{f_{\pi}^2} (t-m_{\pi}^2 ) \delta^{ab}
\delta^{cd}
+\frac i{f_{\pi}^2} (s-m_{\pi}^2 ) \delta^{ac} \delta^{bd} +
\frac i{f_{\pi}^2} (u-m_{\pi}^2 ) \delta^{ad} \delta^{bc}.
\label{C37}
\ee
which is Weinberg's standard result.

\vskip .15cm
{\it 13.3. General Result}

Having checked this, we return to the general result. For that, we
need to evaluate
\be
\frac{\delta^3}{\delta a_{\alpha}^a (y_1) \delta a_{\beta}^b (y_2) \delta
a_{\gamma}^c (y_3)} \bigg[ a_{\rm in}^d (p_2) , \hat{\cal S} \bigg]
\label{C38}
\ee
This functional derivative consists of six
terms
\be
-\frac 1{f_{\pi}} e^{ip_2 \cdot y_1} \bigg( \epsilon^{dae} \epsilon^{ebc} +
\epsilon^{dbe}\epsilon^{eac}\bigg) (-g^{\alpha\beta} )
\delta^4 (y_1-y_2) \nonumber\\ \times
\Delta_R (y_1-y_3 )
\bigg(-\frac{\partial}{\partial y_{3\gamma}} -
\frac{\stackrel\leftarrow\partial}{\partial
y_{3\gamma}}\bigg) \frac {\delta\hat{\cal S}}{\delta s (y_3) } +
\,\,\,{\rm {2 \,\,perm.}}
\label{C39}
\ee
\be
-\frac 1{f_{\pi}} e^{ip_2 \cdot y_3} \delta^{cd}
\bigg(-\frac{\partial}{\partial y_{3\gamma}} -\frac{\stackrel\leftarrow
\partial}{\partial
y_{3\gamma}}\bigg) \frac {\delta^3\hat{\cal S}}
{\delta s (y_3) \delta a_{\alpha}^a (y_1) \delta a_{\beta}^b (y_2)} +
\,\,\,{\rm {2 \,\,perm.}}
\label{C40}
\ee
\be
+\frac 1{f_{\pi}} e^{ip_2 \cdot y_1} \bigg( \epsilon^{dae} \epsilon^{ebf} +
\epsilon^{dbe}\epsilon^{eaf}\bigg) (-g^{\alpha\beta} )
\delta^4 (y_1-y_2) \nonumber\\ \times
\int d^4z \Delta_R (y_1- z ) \frac {\partial}{\partial
z^{\mu}} \frac {\delta^2\hat{\cal S}}{\delta a_{\mu}^f (z) \delta a_{\gamma}^c
 (y_3) } + {\rm {2 \,\,perm.}}
\label{C41}
\ee
\be
+\frac 1{f_{\pi}} e^{ip_2 \cdot y_1} \bigg( \epsilon^{dae} \epsilon^{ebf} +
\epsilon^{dbe}\epsilon^{eaf}\bigg) (-g^{\alpha\beta} )
\delta^4 (y_1-y_2) \Delta_R (y_1-y_3 ) \epsilon^{fcg}
\frac {\delta\hat{\cal S}}{\delta v_{\gamma}^g (y_3) } +
\,\,\,{\rm {2 \,\,perm.}}
\label{C42}
\ee
\be
+\frac 1{f_{\pi}} \int d^4 y e^{ip_2 \cdot y} \frac{\partial}{\partial y^{\mu}}
\frac {\delta^4\hat{\cal S}}
{\delta a_{\mu}^d (y)  \delta a_{\alpha}^a (y_1) \delta a_{\beta}^b (y_2)
\delta a_{\gamma}^c (y_3) }
\label{C43}
\ee
\be
&&+\frac 1{f_{\pi}} e^{ip_2 \cdot y_3} \epsilon^{edc}
\frac {\delta^3\hat{\cal S}}
{\delta v_{\gamma}^e (y_3)  \delta a_{\alpha}^a (y_1) \delta a_{\beta}^b (y_2)}
+ \,\,\,{\rm {2 \,\,perm.}}\nonumber\\&&
-\bigg(\epsilon^{daf}\epsilon^{fbe} +\epsilon^{dbf}\epsilon^{fae}\bigg)
\delta^4 (y_1-y_2) \, e^{ip_2\cdot y_1} \, g^{\alpha\beta}
{\bf S} {\bf j}_{A}^{\gamma \,c} (y_3) \pi_{\rm in}^e (y_1)
+ \,\,\,{\rm {2 \,\,perm.}}
\label{C44}
\ee
Applying
\be
\frac i{f_{\pi}^3} k_{1\alpha} k_{2\beta} p_{1\gamma} \int d^4 y_1 d^4 y_2 d^4
y_3 e^{-ik_1 \cdot y_1 + i k_2 \cdot y_2 -i p_1\cdot y_3}
\label{C45}
\ee
and using (\ref{C9}), we find that (\ref{C39}, \ref{C41}, \ref{C42}) sum up to
\be
&&-\frac i{f_{\pi}^3} k_1\cdot k_2 \int d^4 y_1 d^4 z e^{-i(k_1-k_2-p_2)\cdot
y_1}\bigg(\epsilon^{dae}\epsilon^{ebf} + \epsilon^{dbe}\epsilon^{eaf}\bigg)
\Delta_R (y_1-z) \partial^{\mu}
\bigg[ {\bf S} {\bf j}_{A\mu}^f (z) , a_{\rm in}^{c\dagger}(p_1) \bigg]
\nonumber\\&&
-\frac 1{f_{\pi}^4} k_1\cdot k_2 \int d^4 y_1 e^{-i(k_1-k_2-p_2+ p_1)\cdot
y_1}\bigg(\epsilon^{dae}\epsilon^{ebc} + \epsilon^{dbe}\epsilon^{eac}\bigg)
\frac {\delta\hat{\cal S}}{\delta s (y_1) }
+\,\,\,{\rm {2 \,\,perm}}.
\label{C46}
\ee
Hence the contribution to $i{\cal T}$ is
\be
-\frac i{f_{\pi}^3} k_1\cdot k_2
\bigg(\epsilon^{dae}\epsilon^{ebc} + \epsilon^{dbe}\epsilon^{eac}\bigg)
<0|\hat{\sigma} |0>+\,\,\,{\rm {2 \,\,perm}}
\label{C47}
\ee

The contribution to $i{\cal T}$ from (\ref{C40}) is
\be
+\frac i{f_{\pi}^3} m_{\pi}^2 k_1^{\alpha} k_2^{\beta}\delta^{cd}
\int d^4 y_1 d^4 y_2 e^{-ik_1\cdot y_1 +ik_2\cdot y_2}
<0| T^*\bigg( {\bf j}_{A\alpha}^a (y_1) {\bf j}_{A\beta}^b (y_2) \hat{\sigma}
(0) \bigg) |0> +{2\,\,\rm { perm}}
\label{C48}
\ee
whereas the contribution from (\ref{C43}) is
\be
&&+\frac 1{f_{\pi}^4} k_1^{\alpha} k_2^{\beta} p_1^{\gamma} p_2^{\delta}
\int d^4 y_1 d^4 y_2 d^4 y_3 e^{-ik_1\cdot y_1 +ik_2\cdot y_2 -ip_1\cdot y_3}
\nonumber\\&& \times
<0| T^*\bigg( {\bf j}_{A\alpha}^a (y_1) {\bf j}_{A\beta}^b (y_2)
{\bf j}_{A\gamma}^c (y_3) {\bf j}_{A\delta}^d (0)\bigg) |0> .
\label{C49}
\ee
We observe that the disconnected pieces in (\ref{C48}-\ref{C49})
\be
&&+\frac i{f_{\pi}^3} m_{\pi}^2 k_1^{\alpha} k_2^{\beta}\delta^{cd}
\int d^4 y_1 d^4 y_2 e^{-ik_1\cdot y_1 +ik_2\cdot y_2}
<0| T^*\bigg( {\bf j}_{A\alpha}^a (y_1) {\bf j}_{A\beta}^b (y_2)\bigg)|0>
<0|  \hat{\sigma} |0> \nonumber\\&&
+\frac 1{f_{\pi}^4} k_1^{\alpha} k_2^{\beta} p_1^{\gamma} p_2^{\delta}
\int d^4 y_1 d^4 y_2 d^4 y_3 e^{-ik_1\cdot y_1 +ik_2\cdot y_2 -ip_1\cdot y_3}
\nonumber\\&& \times
<0| T^*\bigg( {\bf j}_{A\alpha}^a (y_1) {\bf j}_{A\beta}^b (y_2) \bigg) |0>
<0| T^*\bigg( {\bf j}_{A\gamma}^c (y_3) {\bf j}_{A\delta}^d (0)\bigg) |0>
+\,\,{\rm {2 \,\,perm}}
\label{C50}
\ee
cancel out by (\ref{C10}-\ref{C11}).

As for (\ref{C44}) it is convenient to relate it back to the electromagnetic
form factor (\ref{C33}). From (\ref{C28}) we have
\be
&&\bigg[ a_{\rm in}^b (k_2 ), \bigg[ {\bf S} {\bf V}^{\gamma e} (y) ,
a_{\rm in}^{a\dagger} (k_1 ) \bigg]\bigg] =\nonumber\\&&
-ik_1^{\gamma} e^{-ik_1\cdot y} \epsilon^{aef} \bigg[ a_{\rm in}^b (k_2) ,
{\bf S}\pi^f (y) \bigg]\nonumber\\&&
+\frac 1{f_{\pi}} e^{-ik_1\cdot y} \epsilon^{aef} \bigg[ a_{\rm in}^b (k_2) ,
{\bf S}{\bf A}^{\gamma f} (y) \bigg]\nonumber\\&&
+\frac 1{f_{\pi}} k_{1\alpha} \int d^4y_1
e^{-ik_1\cdot y_1} \bigg[ a_{\rm in}^b (k_2) ,
\frac{\delta^2\hat{\cal S}}{\delta a_{\alpha}^a (y_1) \delta v_{\gamma}^e (y)}
\bigg].
\label{C51}
\ee
Hence we need,
\be
&&\frac{\delta^2}{\delta a_{\alpha}^a (y_1) \delta v_{\gamma}^e (y)}
\bigg[ a_{\rm in}^b (k_2) , \hat{\cal S} \bigg] =\nonumber\\&&
+e^{ik_2\cdot y} \epsilon^{bef} {\bf S}{\bf j}_{A}^{\alpha\, a} (y_1)
\bigg(\frac {\partial}{\partial y_{\gamma}}
-ik_2^{\gamma} \bigg) \pi_{\rm in}^f (y)\nonumber\\&&
+\frac 1{f_{\pi}} e^{ik_2\cdot y}\epsilon^{bea}
\bigg(\frac{\partial}{\partial y_{\gamma}}
-ik_2^{\gamma}\bigg) \Delta_R (y-y_1)
\bigg(\frac{\partial}{\partial  y_{1\alpha}} +
\frac{\stackrel\leftarrow\partial}{\partial  y_{1\alpha}}\bigg)
\frac{\delta\hat{\cal S}}{\delta s (y_1)}\nonumber\\&&
-\frac 1{f_{\pi}} e^{ik_2\cdot y} \epsilon^{bea} g^{\alpha \gamma} \delta^4
(y-y_1) \frac{\delta\hat{\cal S}}{\delta s (y_1)} \nonumber\\&&
+\frac 1{f_{\pi}} e^{ik_2\cdot y_1} \delta^{ab}
\bigg(\frac{\partial}{\partial y_{1\alpha}} +ik_2^{\alpha}\bigg)
\frac{\delta^2\hat{\cal S}}{\delta s (y_1)\delta v_{\gamma}^e (y)}\nonumber\\&&
+\frac 1{f_{\pi}} e^{ik_2\cdot y}\epsilon^{bef}
\bigg(\frac{\partial}{\partial
y_{\gamma}} -ik_2^{\gamma}\bigg) \int d^4z \Delta_R (y- z)
\frac{\partial}{\partial z^{\mu}} \frac{\delta^2\hat{\cal S}}
{\delta a_{\mu}^f (z) \delta a_{\alpha}^a (y_1) }\nonumber\\&&
+\frac 1{f_{\pi}} e^{ik_2\cdot y}\epsilon^{bef}
\bigg(\frac{\partial}{\partial
y_{\gamma}} -ik_2^{\gamma}\bigg) \Delta_R (y- y_1) \epsilon^{fag}
\frac{\delta\hat{\cal S}}{\delta v_{\alpha}^g (y_1)}\nonumber\\&&
+\frac 1{f_{\pi}} e^{ik_2\cdot y}\epsilon^{bef}
\frac{\delta^2\hat{\cal S}}
{\delta a_{\gamma}^f (y) \delta a_{\alpha}^a (y_1) }\nonumber\\&&
-\frac i{f_{\pi}} k_{2\beta} \int d^4y_2 e^{ik_2\cdot y_2}
\frac{\delta^3\hat{\cal S}}{\delta a_{\alpha}^a (y_1) \delta a_{\beta}^b (y_2)
\delta v_{\gamma}^e (y)}\nonumber\\&&
+\frac 1{f_{\pi}} e^{ik_2\cdot y_1}\epsilon^{baf}
\frac{\delta^2\hat{\cal S}}
{\delta v_{\alpha}^f (y_1) \delta v_{\gamma}^e (y) }
{}.
\label{C52}
\ee
Substituting this back into  (\ref{C51}) and using (\ref{C9}), we find that
(\ref{C51}) is equal to
\be
&&-ik_{1}^{\gamma} e^{-ik_1\cdot y} \epsilon^{aef} \bigg[ a_{\rm in}^b (k_2) ,
{\bf S}\pi^f (y) \bigg]\nonumber\\&&
+\frac 1{f_{\pi}} e^{-ik_1\cdot y} \epsilon^{aef} \bigg[ a_{\rm in}^b (k_2) ,
{\bf S}{\bf A}^{\gamma f} (y) \bigg]\nonumber\\&&
+\frac 1{f_{\pi}} \epsilon^{bef} e^{ik_2\cdot y} k_{1\alpha}
\int d^4y_1 e^{-ik_1\cdot y_1}
{\bf S} {\bf j}_A^{\alpha a} (y_1)
\bigg(\frac{\partial}{y_{\gamma}} -ik_2^{\gamma} \bigg)
\pi_{\rm in}^f (y)\nonumber\\&&
+\frac i{f^2_{\pi}} m_{\pi}^2 \delta^{ab}\int d^4y_1
e^{-i(k_1-k_2)\cdot y_1}
\frac{\delta^2\hat{\cal S}}{\delta s (y_1)\delta v_{\gamma}^e (y)}\nonumber\\&&
+\frac 1{f_{\pi}} e^{ik_2\cdot y}\epsilon^{bef}
\bigg(\frac{\partial}{\partial y_{\gamma}} -ik_2^{\gamma}\bigg)
\int d^4z \Delta_R (y-z)
\partial^{\mu} \bigg[ {\bf S}{\bf j}_{A\mu}^f (z) , a_{\rm in}^{a\dagger} (k_1)
\bigg] \nonumber\\&&
-\frac 1{f_{\pi}^2} e^{-i(k_1-k_2)\cdot y} \epsilon^{bea} (k_1+k_2)^{\gamma}
\frac{\delta\hat{\cal S}}{\delta s (y)}\nonumber\\&&
+\frac 1{f_{\pi}} e^{ik_2\cdot y} \epsilon^{bef}
\bigg[ {\bf S}{\bf j}_{A}^{\gamma f} (y), a_{\rm in}^{a\dagger} (k_1)
\bigg]\nonumber\\&&
+\frac i{f_{\pi}^2} e^{-i(k_1-k_2)\cdot y} \epsilon^{bef} \epsilon^{afg}
\frac{\delta\hat{\cal S}}{\delta v_{\gamma}^g (y)}\nonumber\\&&
-\frac i{f^2_{\pi}} k_{1\alpha} k_{2\beta}\int d^4 y_1 d^4y_2
e^{-ik_1\cdot y_1 +ik_2\cdot y_2}
\frac{\delta^3\hat{\cal S}}
{\delta a_{\alpha}^a (y_1) \delta a_{\beta}^b (y_2) \delta v_{\gamma}^e (y)
}\nonumber\\&&
+\frac 1{f^2_{\pi}} k_{1\alpha} \epsilon^{baf}
\int d^4y_1 e^{-i(k_1-k_2)\cdot y_1}\frac{\delta^2\hat{\cal S}}
{\delta v_{\alpha}^f (y_1) \delta v_{\gamma}^e (y) }.
\label{C53}
\ee
Taking the vacuum expectation value gives
\be
&&+i\epsilon^{bea} (k_1+k_2)_{\gamma} {\bf F}_V (t) e^{-i (k_1-k_2)\cdot y}
=\nonumber \\&&
+i\epsilon^{bea} (k_1+k_2)_{\gamma} e^{-i(k_1-k_2)\cdot y}
-\frac i{f_{\pi}} e^{-i(k_1-k_2)\cdot y}\epsilon^{bea} (k_1+k_2)_{\gamma}
<0|\hat{\sigma} |0>\nonumber\\&&
-\frac 1{f^2_{\pi}} k_1^{\alpha}
k_2^{\beta} \int d^4 y_1 d^4 y_2 e^{-ik_1\cdot
y_1 + ik_2\cdot y_2}
<0|T^*\bigg( {\bf j}_{A\alpha}^a (y_1)
{\bf j}_{A\beta}^b (y_2) {\bf V}_{\gamma}^e (y) \bigg) |0>\nonumber\\&&
-\frac 1{f^2_{\pi}} \epsilon^{baf} k_1^{\alpha} \int d^4 y_1
e^{-i(k_1-k_2)\cdot y_1}
<0|T^*\bigg( {\bf V}_{\alpha}^f (y_1) {\bf V}_{\gamma}^e (y) \bigg) |0>.
\label{C54}
\ee
Introducing the vector correlation function
\be
\int d^4 x e^{iq\cdot x} <0| T^*\bigg( {\bf V}_{\alpha}^a (x) {\bf V}_{\beta}^b
(0) \bigg) |0> = -i\delta^{ab} \bigg( -g_{\alpha\beta} q^2 + q_{\alpha}
q_{\beta} \bigg) {\bf \Pi}_V (q^2)
\label{C55}
\ee
we find that the contribution of (\ref{C44}) to $i{\cal T}$ is
\be
\frac i{2f_{\pi}^2} \epsilon^{abe}\epsilon^{cde} (s-u)
\bigg( {\bf F}_V (t) - 1 + \frac 1{f_{\pi}} <0|\hat\sigma |0>
-\frac 1{2f_{\pi}^2} t {\bf \Pi}_V (t) \bigg)  + {\rm {2 \,\, perm}}
\label{C56}
\ee

Collecting all the terms (\ref{C35}, \ref{C47}, \ref{C48}, \ref{C49},
\ref{C56}), we find the desired result for the $\pi\pi$ scattering amplitude
\be
&&i{\cal T} \bigg( p_2 d, k_2 b \leftarrow k_1 a, p_1 c \bigg) =
-\frac i{f_{\pi}} m_{\pi}^2 \delta^{ab}\delta^{cd} {\bf F}_S (t) -\frac
i{2f_{\pi}^2} t \delta^{ab}\delta^{cd} + {\rm {2 \,\, perm.}}\nonumber\\&&
+\frac i{f_{\pi}^2} \epsilon^{abe}\epsilon^{cde} (s-u)
\bigg( {\bf F}_V (t) - \frac 12 - \frac 1{4f^2_{\pi}} t {\bf \Pi}_V (t) \bigg)
+ {\rm {2 \,\, perm.}}\nonumber\\&&
+\frac i{f_{\pi}^3} m_{\pi}^2 k_1^{\alpha} k_2^{\beta} \delta^{cd} \int d^4y_1
d^4 y_2 e^{-ik_1\cdot y_1 + ik_2\cdot y_2}
<0| T^*\bigg( {\bf j}_{A\alpha}^a (y_1) {\bf j}_{A\beta}^b (y_2)
\hat\sigma (0) \bigg) |0>_{\rm conn}  + {\rm {2 \,\, perm}} \nonumber\\&&
+\frac 1{f_{\pi}^4} k_1^{\alpha} k_2^{\beta} p_1^{\gamma} p_2^{\delta}
\int d^4y_1  d^4 y_2 d^4 y_3 e^{-ik_1\cdot y_1 + ik_2\cdot y_2-ip_1\cdot y_3}
\nonumber\\&& \times
<0| T^*\bigg( {\bf j}_{A\alpha}^a (y_1) {\bf j}_{A\beta}^b (y_2)
{\bf j}_{A\gamma}^c (y_3) {\bf j}_{A\delta}^d (0)\bigg) |0>_{\rm conn}.
\label{C57}
\ee
In the chiral limit, this result reduces to
\be
&&i{\cal T} \bigg( p_2 d, k_2 b \leftarrow k_1 a, p_1 c \bigg) =
+\frac i{f_{\pi}^2} \epsilon^{abe}\epsilon^{cde} (s-u)
\bigg( {\bf F}_V (t) - \frac 23 - \frac 1{4f^2_{\pi}} t {\bf \Pi}_V (t) \bigg)
+ {\rm 2 \,\,{perm.}}\nonumber\\&&
+\frac 1{f_{\pi}^4} k_1^{\alpha} k_2^{\beta} p_1^{\gamma} p_2^{\delta}
\int d^4y_1  d^4 y_2 d^4 y_3 e^{-ik_1\cdot y_1 + ik_2\cdot y_2-ip_1\cdot y_3}
\nonumber\\&& \times
<0| T^*\bigg( {\bf j}_{A\alpha}^a (y_1) {\bf j}_{A\beta}^b (y_2)
{\bf j}_{A\gamma}^c (y_3) {\bf j}_{A\delta}^d (0)\bigg) |0>_{\rm conn}
\label{C58}
\ee
a result which was obtained in an earlier version of this paper \cite{zahed}.

The pion electromagnetic form factor ${\bf F}_V (q^2)$
is measurable from electroproduction $e^+ e^-\rightarrow \pi\pi$ for $q^2 > 4
m_{\pi}^2$ and from $\pi^+ e^- \rightarrow \pi^+ e^- $ (hydrogen target) for
$q^2<0$, whereas the vector correlation function ${\bf \Pi}_V (q^2)$ is
measurable from electroproduction for $q^2 >4 m_{\pi}^2$
or semileptonic $\tau$-decays. The first quantity
is well described by $\rho$ dominance, whereas the second requires
the $\rho$ and the
$\rho'$. It follows that the $\pi\pi$ scattering amplitude beyond tree level
should display effects of the $\rho$, in agreement with previous
estimates in the context of chiral perturbation theory \cite{leutwyler}.
In particular, the $\rho$ contribution through ${\bf F}_V$ and ${\bf \Pi}_V$
is model-independent, a question that has attracted some attention in the
literature \cite{raphael}.
It may also be interesting to compare these results with bootstrap approaches.

There is one subtlety in (\ref{C57}). The absence of the one-pion
component implies that ${\bf j}_{A\mu}^a$ is ${\cal O} (\pi^3)$ so that the
last term in (\ref{C57}) is nominally three-loop as shown in Fig. 5a.
However, divergences in the subdiagrams require the subtraction of terms such
as Fig. 5b, so that effectively the contributions start at one loop.

\vskip .15cm
{\it 13.4. Simplification}

For later use, we will record the reduction of the scalar form factor ${\bf
F}_S$, which proceeds in a similar manner to (\ref{C51}-\ref{C54}). From
(\ref{C25})
\be
\bigg[ a_{\rm in}^{b} (k_2) ,  \bigg[ {\bf S} \hat\sigma (y) , a_{\rm
in}^{a\dagger}
(k_1)\bigg]\bigg] =&&
-\frac 1{f_{\pi}} e^{-ik\cdot y} \bigg[ a_{\rm in}^b (k_2) , {\bf S}\pi^a (y)
\bigg] \nonumber\\&&
+\frac 1{f_{\pi}^2} k_{1\alpha}
\int d^4y_1 e^{-ik_1\cdot y_1}
\bigg[ a_{\rm in}^b (k_2) , \frac{\delta^2\hat{\cal S}}{\delta a_{\alpha}^a
(y_1)\delta s(y)}\bigg].
\label{C59}
\ee
Hence we need
\be
&&\frac{\delta^2}{\delta a_{\alpha}^a (y_1) \delta s (y) }
\bigg[ a_{\rm in}^b (k_2) , \hat{\cal S} \bigg] =
+e^{ik_2\cdot y} {\bf S} {\bf j}_A^{\alpha a} (y_1) \pi_{\rm in}^b (y)
\nonumber\\&&
+\frac 1{f_{\pi}} e^{ik_2\cdot y} \Delta_R (y-y_1) \delta^{ab}
\bigg(\frac {\partial}{\partial y_{1\alpha}} +
\frac {\stackrel\leftarrow\partial}{\partial y_{1\alpha}}\bigg)
\frac{\delta\hat{\cal S}}
{\delta s(y_1)} \nonumber\\&&
+\frac 1{f_{\pi}} e^{ik_2\cdot y_1} \delta^{ab}
\bigg(\frac {\partial}{\partial y_{1\alpha}} + ik_2^{\alpha} \bigg)
\frac{\delta^2\hat{\cal S}} {\delta s(y_1)\delta s (y)} \nonumber\\&&
+\frac 1{f_{\pi}} e^{ik_2\cdot y} \int d^4 z \Delta_R (y- z)
\frac{\partial}{\partial z^{\mu}}
\frac{\delta^2\hat{\cal S}}
{\delta a_{\mu}^b (z) \delta a_{\alpha}^a (y_1)} \nonumber\\&&
+\frac 1{f_{\pi}} e^{ik_2\cdot y} \Delta_R (y-y_1) \epsilon^{bae}
\frac{\delta\hat{\cal S}}
{\delta v_{\alpha}^e (y_1)} \nonumber\\&&
+\frac 1{f_{\pi}} \int d^4y_2e^{ik_2\cdot y_2}
\frac {\partial}{\partial y_2^{\beta} }\frac{\delta^3\hat{\cal S}}
{\delta a_{\alpha}^a (y_1) \delta a_{\beta}^b (y_2) \delta s(y)} \nonumber\\&&
+\frac 1{f_{\pi}} e^{ik_2\cdot y_1} \epsilon^{bae}
\frac{\delta^2\hat{\cal S}}{\delta v_{\alpha}^e (y_1) \delta s(y)}
\,\,\,.
\label{C60}
\ee
Substituting this
into (\ref{C59}) and using (\ref{C9}), we find that (\ref{C59})
is equal to
\be
&&-\frac 1{f_{\pi}} e^{-ik_1\cdot y}
\bigg[ a_{\rm in}^b (k_2) , {\bf S} \pi^a (y) \bigg] \nonumber\\&&
+\frac 1{f_{\pi}^2} k_1^{\alpha} \int d^4x e^{-ik_1\cdot x +i k_2\cdot y}
{\bf S} {\bf j}_A^{\alpha a} (x) \pi_{\rm in}^b (y)
\frac {\delta\hat{\cal S}}{\delta v^{\beta e} (x) } \nonumber\\&&
-\frac i{f^3_{\pi}} e^{-i(k_1-k_2)\cdot y}  \delta^{ab}
\frac{\delta\hat{\cal S}}
{\delta s(y)} \nonumber\\&&
+\frac i{f^3_{\pi}} m_{\pi}^2 \delta^{ab}\int d^4 y_1 e^{-i(k_1-k_2)\cdot y_1}
\frac{\delta^2\hat{\cal S}}{\delta s(y_1)\delta s (y)} \nonumber\\&&
+\frac 1{f^2_{\pi}} e^{ik_2\cdot y} \int d^4 z \Delta_R (y- z)
\partial^{\beta}
\bigg[ {\bf S} {\bf j}_{A\beta}^b (z) , a_{\rm in}^{a\dagger} (k_1) \bigg]
\nonumber\\&&
-\frac i{f^3_{\pi}} k_{1\alpha} k_{2\beta}
\int d^4 y_1 d^4y_2 e^{-ik_1\cdot y_1 +ik_2\cdot y_2}
\frac{\delta^3\hat{\cal S}}
{\delta a_{\alpha}^a (y_1) \delta a_{\beta}^b (y_2) \delta s(y)} \nonumber\\&&
+\frac 1{f^3_{\pi}} k_{1\alpha} \epsilon^{bae}\int d^4y_1
e^{-i(k_1-k_2)\cdot y_1}
\frac{\delta^2\hat{\cal S}}{\delta v_{\alpha}^e (y_1) \delta s(y)} .
\label{C61}
\ee
again, to leading order in the external fields.
Taking the vacuum expectation values gives
\be
\delta^{ab} {\bf F}_S (t) =&& -\frac 1{f_{\pi}} \delta^{ab} +\frac
1{f_{\pi}^2}\delta^{ab} <0|\hat{\sigma} |0> \nonumber\\&&
-\frac i{f_{\pi}} m_{\pi}^2 \delta^{ab} \int d^4y e^{-i(k_1-k_2)\cdot y}
<0| T^*\bigg(\hat\sigma (y) \hat\sigma (0)\bigg) |0>_{\rm conn}\nonumber\\&&
-\frac 1{f_{\pi}^2} k_1^{\alpha} k_2^{\beta} \int d^4y_1 d^4 y_2
e^{-ik_1\cdot y_1 + ik_2\cdot y_2}
<0| T^*\bigg( {\bf j}_{A\alpha}^a (y_1) {\bf j}_{A\beta}^b (y_2)
\hat\sigma (0) \bigg)|0>_{\rm conn}
\label{C62}
\ee
where we have used the fact that the disconnected pieces cancel as usual by
(\ref{C10}-\ref{C11}).

Finally, we may split the $\pi\pi$ scattering amplitude (\ref{C57}) into the
tree contribution (\ref{C37}), the $\rho$ contribution
\be
i{\cal T}_{\rm rho} = \frac i{f_{\pi}^2}\epsilon^{abe}\epsilon^{cde} (s-u)
\bigg( {\bf F}_V (t) - 1 -\frac t{4f_{\pi}^2} {\bf \Pi}_V (t) \bigg) +
{\rm {2\,\,perm.}}
\label{C63}
\ee
and the rest
\be
i{\cal T}_{\rm rest} =&& -\frac {2im_{\pi}^2}{f_{\pi}} \delta^{ab}\delta^{cd}
\bigg( {\bf F}_S (t) +\frac 1{f_{\pi}} -\frac 1{2f_{\pi}^2}
<0|\hat{\sigma} |0> \bigg) + {\rm {2 \,\, perm.}}\nonumber\\&&
+\frac 1{f_{\pi}^2} m_{\pi}^4 \delta^{ab} \delta^{cd} \int d^4 y e^{-i
(k_1-k_2)\cdot y}
<0| T^*\bigg( \hat\sigma (y) \hat\sigma (0) \bigg) |0>_{\rm conn.} +
{\rm {2\,\, perm.}} \nonumber\\&&
+\frac 1{f_{\pi}^4} k_1^{\alpha} k_2^{\beta} p_1^{\gamma} p_2^{\delta}
\int d^4 y_1 d^4 y_2 d^4 y_3 e^{-ik_1\cdot y_2 + ik_2\cdot y_2 -ip_1\cdot y_3}
\nonumber\\&& \times
<0| T^*\bigg( {\bf j}_{A\alpha}^a (y_1) {\bf j}_{A\beta}^b (y_2)
{\bf j}_{A\gamma}^c (y_3) {\bf j}_{A\delta}^d (0)\bigg) |0>_{\rm conn.}
\label{C64}
\ee
We observe that the rest contains the scalar contribution. We stress that
the full $\pi\pi$ scattering amplitude (\ref{C37},\ref{C63},\ref{C64})
is exact, and does not rely on a model approximation. It follows from chiral
symmetry and unitarity. Our result shows explicitly the importance of the
$\rho$ and $\rho '$ in the P-wave amplitude, and the relevance of
scalar correlations in the S-wave amplitude.

The combination of one-loop chiral
perturbation theory and dispersion relations as advocated by Truong and his
collaborators \cite{truong} to enforce elastic unitarity to all orders,
reproduces part of the effects in the P-wave channel (the $\rho$ pole
for instance). In the
S-channel, the Pade resummation used by these authors leads to spurious poles
which are neither expected from our result nor the data. Also, it is important
to note that all the S-wave effects given by our relation follows from the
explicit breaking of chiral symmetry and vanishes in the chiral limit.
This point is absent from the unitarized amplitudes \cite{truong}. The latter
follow from a theoretical construct that in general tends to blur the role
of chiral symmetry.

\vskip .5cm
{\bf 14. Off-Shell Pion Radiative Decay : $\pi\rightarrow e\nu\gamma^*$}
\vskip .3cm

Besides $\pi N$ and $\pi\pi$ scattering, there are other amplitudes of some
interest which may be discussed here. One is $\pi\rightarrow e\nu\gamma^*$
through the amplitude
\be
i\int d^4x e^{iq\cdot x +ik\cdot y}
<0|T^*\bigg({\bf V}_{\mu}^a (x) {\bf A}_{\nu}^b (y)\bigg)
a_{\rm in}^{c\dagger} (p)|0>
\label{ADD11}
\ee
where $k=p-q$. From (\ref{B84}) we have
\be
&&T^*\bigg( {\bf V}_{\mu}^a (x) {\bf A}_{\nu}^b (y) \bigg) =
T^*\bigg( {\bf V}_{\mu}^a (x) {\bf j}_{A\nu}^b (y) \bigg) \nonumber\\
&&+ f_{\pi} g_{\mu\nu}\epsilon^{bad} \delta^4 (x-y) \pi^d (y)
-f_{\pi}\frac{\partial}{\partial y^{\nu}}
T^*\bigg( {\bf V}_{\mu}^a (x) \pi^b (y)\bigg)
\label{ADD12}
\ee
and from (\ref{B97})
\be
&&T^*\bigg( {\bf V}_{\mu}^a (x) \pi^b (y) \bigg) = i\Delta_R (y-x)
\frac{\stackrel\leftarrow\partial}{\partial x^{\mu}}\epsilon^{bad}\pi^d
(x)\nonumber\\
&&+\frac i{f_{\pi}}\Delta_R (y-x) \epsilon^{bad} {\bf A}_{\mu}^d (x) +
T^*\bigg( {\bf V}_{\mu}^a (x) \pi_{\rm in}^b (y)\bigg)\nonumber\\
&&-\frac 1{f_{\pi}}\int d^4z \Delta_R (y-z) \frac{\partial}{\partial z_{\beta}}
T^*\bigg({\bf V}_{\mu}^a (x) {\bf j}_{A\beta}^b (z) \bigg)
\label{ADD13}
\ee
Therefore the amplitude (\ref{ADD11}) is
\be
&&+f_{\pi} g_{\mu\nu}\epsilon^{abc} -f_{\pi} (2p_{\mu}-q_{\mu}) k_{\nu}
\epsilon^{abc} \frac 1{k^2-m_{\pi}^2}\nonumber\\
&&+i\int d^4x e^{iq\cdot x +ik\cdot y}
<0|T^*\bigg({\bf V}_{\mu}^a (x) {\bf j}_{A\nu}^b (y)\bigg) a_{\rm
in}^{c\dagger} |0>\nonumber\\
&&+ie^{ik\cdot y}\int d^4z \frac{\partial}{\partial y^{\nu}}\Delta_R (y-z)
\frac{\partial}{\partial z^{\beta}}\int d^4x e^{iq\cdot x}\nonumber\\
&&\,\,\,\times
<0|T^*\bigg({\bf V}_{\mu}^a (x) {\bf j}_{A\beta}^b (z)\bigg)
a^{c\dagger}_{\rm in} |0>
\label{ADD14}
\ee
The tree terms proportional to the pion decay constant $f_{\pi}$ in
(\ref{ADD14}) are standard \cite{pidecay}.
{}From (\ref{C52}), the remainder of (\ref{ADD14}) is
\be
&&-\bigg(g_{\mu\nu} -\frac {k_{\mu}k_{\nu}}{k^2-m_{\pi}^2}\bigg)
\epsilon^{abc} <0|\hat\sigma |0>\nonumber\\
&&-(2p_{\mu}-q_{\mu}) k_{\nu} \epsilon^{abc}
\frac{m_{\pi}^2}{(k^2-m_{\pi}^2)^2} <0|\hat\sigma |0>\nonumber\\
&&+\frac i{f_{\pi}}
\bigg(\delta_{\nu}^{\beta} -\frac{k_{\nu}k^{\beta}}{k^2-m_{\pi}^2}\bigg)
\epsilon^{caf}
\int d^4x e^{-ik\cdot x}<0|T^*\bigg({\bf j}_{A\mu}^f (x) {\bf j}_{A\beta}^b (0)
\bigg) |0>\nonumber\\
&&-\frac i{f_{\pi}} (2p_{\mu} -q_{\mu}) k^{\alpha}\epsilon^{caf}
\bigg(\delta_{\nu}^{\beta} -\frac{k_{\nu}k^{\beta}}{k^2-m_{\pi}^2}\bigg)
\frac 1{k^2-m_{\pi}^2}\nonumber\\&&\,\,\,\times
\int d^4x e^{-ik\cdot x}<0|T^*\bigg({\bf j}_{A\alpha}^f (x) {\bf j}_{A\beta}^b
(0)
\bigg) |0>\nonumber\\
&&-\frac i{f_{\pi}} p^{\gamma}
\bigg(\delta_{\nu}^{\beta} -\frac{k_{\nu}k^{\beta}}{k^2-m_{\pi}^2}\bigg)
\int d^4x d^4z e^{+iq\cdot x -ip\cdot z}\nonumber\\
&&\,\,\,\times
<0|T^*\bigg({\bf V}_{\mu}^a (x) {\bf j}_{A\gamma}^c (z) {\bf j}_{A\beta}^b (0)
\bigg) |0>\nonumber\\
&&+\frac 1{f_{\pi}}\epsilon^{abc}
\bigg(\delta_{\nu}^{\beta} -\frac{k_{\nu}k^{\beta}}{k^2-m_{\pi}^2}\bigg)
\bigg(-g_{\mu\beta} q^2 +q_{\mu} q_{\beta}\bigg) \,\,\Pi_V (q^2)
\label{ADD15}
\ee
The decay amplitude $\pi\rightarrow e\nu\gamma^*$ is sensitive to the
way chiral symmetry is broken in the vacuum through $<0|\hat\sigma |0>$.

\vskip .5cm
{\bf 15. On-Shell Pion Radiative Decay : $\pi\rightarrow e\nu\gamma$}
\vskip .3cm

The on-shell $\pi\rightarrow e\nu\gamma$ decay follows from
\be
i\epsilon^{\mu} (q) \int d^4x e^{iq\cdot (x-y) + ip\cdot y}
<0| T^* \bigg( {\bf V}_{\mu}^a (x) {\bf A}_{\nu}^b (y)\bigg) a_{\rm
in}^{c\dagger}
(p)  |0>
\label{C65}
\ee
The amplitude (\ref{C65}) may be obtained from (\ref{ADD14},\ref{ADD15})
by contracting with the photon polarization vector $\epsilon^{\mu} (q)$
obeying $\epsilon\cdot q =0$ and setting the photon on shell $q^2=0$.
The result is
\be
&&+\epsilon_{\nu} (q) \epsilon^{abc} f_{\pi} +
\frac {\epsilon (q)\cdot p}{p\cdot q} (p-q)_{\nu} \epsilon^{abc}
f_{\pi}\nonumber\\&&
-\bigg( \delta^{\mu}_{\nu} -\frac {p^{\mu}k_{\nu}}{k^2-m_{\pi}^2}
+2m_{\pi}^2 \frac{p^{\mu} k_{\nu}}{(k^2-m_{\pi}^2)^2}\bigg)\,\,
\epsilon_{\mu} (q) \epsilon^{abc} <0|\hat\sigma |0>\nonumber\\
&&+\frac i{f_{\pi}} \epsilon^{\mu} (q) \epsilon^{cad}
\bigg( \delta_{\mu}^{\alpha} - 2\frac {k^{\alpha} k_{\mu}}{k^2-m_{\pi}^2}\bigg)
\bigg( \delta_{\nu}^{\beta} -\frac {k^{\beta} k_{\nu}}{k^2-m_{\pi}^2}\bigg)
\int d^4y e^{-ik\cdot y}
<0| T^*\bigg( {\bf j}_{A\alpha}^d (y) {\bf j}_{A\beta}^b (0)\bigg) |0>
\nonumber\\&&
-\frac i{f_{\pi}} \epsilon^{\alpha} (q) p^{\gamma}
\bigg( \delta_{\nu}^{\beta} -\frac{k^{\beta} k_{\nu}}{k^2-m_{\pi}^2}\bigg)
\int d^4x d^4z e^{iq\cdot x -ip\cdot z}
<0| T^* \bigg( {\bf V}_{\alpha}^a (x) {\bf j}_{A\gamma}^c (z) {\bf
j}_{A\beta}^b (0) \bigg) |0>
\label{C67}
\ee
As a check, we note that the apparent double
pole cancels by (\ref{C10}-\ref{C11}). Again,
we note that the tree contribution in (\ref{C67}) is sensitive
to the way chiral symmetry is broken in the vacuum
through $<0|\hat{\sigma}|0>$.

\vskip .5cm
{{\bf 16. Photon-Photon Collision :} $\gamma\gamma\rightarrow \pi\pi$}
\vskip .3cm

The collision of two photons at threshold has attracted considerable attention
recently in the context of chiral perturbation theory
\cite{photophoto} and dispersion techniques \cite{morgan}. These calculations
are interesting in light of the experiments carried out by the MARK II
collaboration at SLAC \cite{mark} and the Crystal Ball collaboration
at DESY \cite{ball}. Future high-energy experiments are planned at DA$\Phi$NE
at Frascati with energies up to 600 MeV.

While chiral perturbation theory implements unitarity perturbatively, thus
giving up on resonant behaviour, dispersion techniques give up
on the consequences of chiral dynamics.
In this section, we will  address this issue from our point of view.
The advantage of our formalism is that our amplitudes are unexpanded.
They are rearranged using the master equation, thus chiral dynamics,
with full consistency with analyticity and unitarity. The contributions of the
various resonances appear in the form of various vacuum correlators.

The amplitude for $\gamma\gamma\rightarrow \pi\pi$ is given by
\be
&&+\int d^4y d^4z e^{-iq_1\cdot y -iq_2\cdot z}
<0|a_{\rm in}^a (k_1) a_{\rm in}^b (k_2) {\bf S} T^*
\bigg({\bf V}_{\mu}^c (y) {\bf V}_{\nu}^d (z)\bigg) |0> \nonumber\\=
&&-\int d^4y d^4z e^{-iq_1\cdot y -iq_2\cdot z}
<0|\frac{\delta^2}{\delta v^{\mu c} (y)\delta v^{\nu d} (z)}
\bigg[ a_{\rm in}^a (k_1), \bigg[  a_{\rm in}^b (k_2) , \hat{\cal
S}\bigg]\bigg] |0>
\label{ADD30}
\ee
The first term of (\ref{T3}) gives
\be
&&+{\bf S}\int d^4y d^4z e^{-iq_1\cdot y -iq_2\cdot z}
\bigg( e^{ik_1\cdot y} \bigg(\frac{\partial}{\partial y^{\mu}} -
\frac{\stackrel\leftarrow\partial}{\partial y^{\mu}}\bigg)
\epsilon^{ace} \pi_{\rm in}^e (y) \bigg)\nonumber\\
&&\times
\bigg( e^{ik_2\cdot z} \bigg(\frac{\partial}{\partial z^{\nu}} -
\frac{\stackrel\leftarrow\partial}{\partial z^{\nu}}\bigg)
\epsilon^{bdf} \pi_{\rm in}^f (z) \bigg)\nonumber\\
&&+\bigg(q_1\,\,\mu\,\,c\leftrightarrow q_2\,\,\nu\,\,d\bigg)\nonumber\\=
&&-{\bf S} (2k_{1\mu}-q_{1\mu})(2k_{2\nu}-q_{2\nu})\epsilon^{ace}\epsilon^{bdf}
\nonumber\\&&\times
\int d^4y d^4z e^{i(k_1-q_1)\cdot y +i (k_2-q_2)\cdot z}
\pi_{\rm in}^e (y) \pi_{\rm in}^f (z) \nonumber\\
&&+\bigg(q_1\,\,\mu\,\,c\leftrightarrow q_2\,\,\nu\,\,d\bigg)
\label{ADD31}
\ee
The second term in (\ref{T3}) gives
\be
&&+\frac i{f_{\pi}}\int d^4y d^4z e^{-iq_1\cdot y -iq_2\cdot z}
\int d^4z_1 e^{ik_1\cdot z_1}\frac{\partial}{\partial z_{1\alpha}}
\frac{\delta^2}{\delta a^{\alpha a} (z_1) \delta v^{\mu c} (y) }\nonumber\\&&
\times \bigg( e^{ik_2\cdot z} \bigg(\frac{\partial}{\partial z^{\nu}} -
\frac{\stackrel\leftarrow\partial}{\partial z^{\nu}}\bigg)
\epsilon^{bdf} \pi_{\rm in}^f (z) \bigg)\nonumber\\&&
+\bigg(q_1\,\,\mu\,\,c\leftrightarrow q_2\,\,\nu\,\,d\bigg)+ ...\nonumber\\=
&&+\frac i{f_{\pi}} (2k_{2\nu}-q_{2\nu}) k_1^{\alpha} \epsilon^{bdf}
\int d^4y d^4z d^4z_1 e^{-iq_1\cdot y +i(k_2-q_2)\cdot z + ik_1\cdot z_1}
\nonumber\\
&&\times \bigg({\bf S} T^*\bigg( {\bf j}_{A\alpha}^a (z_1) {\bf V}_{\mu}^c (y)
\bigg)\bigg) \pi_{\rm in}^f (z) \nonumber\\&&
+\bigg(q_1\,\,\mu\,\,c\leftrightarrow q_2\,\,\nu\,\,d\bigg) + ...
\label{ADD32}
\ee
where $...$ stands for terms with zero vacuum expectation values. Note that
(\ref{ADD31}-\ref{ADD32}) also vanish for spacelike photons $q_1^2<0$ and
$q_2^2<0$ after taking the vacuum expectation value, since only the
annihilation part of $\pi_{\rm in} (y)$ is picked up as before.

The third term of (\ref{T3}) gives
\be
&&+i{\bf S}\int d^4y d^4z e^{-iq_1\cdot y -iq_2\cdot z}\nonumber\\
&&\times
\bigg( e^{ik_2\cdot y} \bigg(\frac{\partial}{\partial y^{\mu}} -
\frac{\stackrel\leftarrow\partial}{\partial y^{\mu}}\bigg)
\epsilon^{bcf} \Delta_R (y-z) \epsilon^{fda}
\bigg(\frac{\partial}{\partial z^{\nu}} -
\frac{\stackrel\leftarrow\partial}{\partial z^{\nu}}\bigg) e^{ik_1\cdot
z}\bigg)\nonumber\\
&&+{\bf S}\int d^4y d^4z e^{-iq_1\cdot y -iq_2\cdot z}
\bigg(e^{ik_2\cdot y} \epsilon^{bce}\epsilon^{eda}\delta^4 (y-z) e^{ik_1\cdot
z}\bigg)\nonumber\\
&&+\bigg(q_1\,\,\mu\,\,c\leftrightarrow q_2\,\,\nu\,\,d\bigg) +
....\nonumber\\=
&&+ (2\pi )^4 i \delta^4 (k_1+k_2-q_1-q_2)
{\bf S} \,\,(2k_{2\mu}-q_{1\mu})(2k_{1\nu}-q_{2\nu} )
\epsilon^{bcf}\epsilon^{fda} \frac 1{(k_2-q_1)^2-m_{\pi}^2}\nonumber\\
&&+ (2\pi )^4 i \delta^4 (k_1+k_2-q_1-q_2)
{\bf S} \epsilon^{bcf}\epsilon^{fda} g_{\mu\nu}\nonumber\\
&&+\bigg(q_1\,\,\mu\,\,c\leftrightarrow q_2\,\,\nu\,\,d\bigg) + ....
\label{ADD33}
\ee
The fourth term gives (\ref{ADD32}) with $k_1,a$ and $k_2, b$ exchanged, and
hence zero after taking the vacuum expectation value.

The term (\ref{T5}) gives
\be
&&+\frac 1{f_{\pi}^2} m_{\pi}^2 \,g_{\mu\nu}
\int d^4y d^4z e^{-iq_1\cdot y -iq_2\cdot z +ik_1\cdot y}
\epsilon^{ace}\epsilon^{ebd}\delta^4 (y-z)\nonumber\\
&&\times \int d^4z_1 \Delta_R (z-z_1) if_{\pi} {\bf S} \hat\sigma (z_1)
e^{ik_2\cdot z_1}\nonumber\\
&&+\frac 1{f_{\pi}^2} m_{\pi}^2
\int d^4y d^4z e^{-iq_1\cdot y -iq_2\cdot z}\nonumber\\
&&\times \bigg( e^{ik_1\cdot y} \bigg(\frac{\partial}{\partial y^{\mu}} -
\frac{\stackrel\leftarrow\partial}{\partial y^{\mu}}\bigg)
\epsilon^{ace} \Delta_R (y-z) \epsilon^{edb}
\bigg(\frac{\partial}{\partial z^{\nu}} -
\frac{\stackrel\leftarrow\partial}{\partial z^{\nu}}\bigg) \bigg)\nonumber\\
&&\times \int d^4z_1 \Delta_R (z-z_1) if_{\pi} {\bf S} \hat\sigma (z_1)
e^{ik_2\cdot z_1}\nonumber\\
&&+{\rm 3 \,\,perm.}\nonumber\\
&&+\frac 1{f_{\pi}^2} m_{\pi}^2
\int d^4y d^4z e^{-iq_1\cdot y -iq_2\cdot z}\nonumber\\
&&\times\int d^4z_1
\bigg( e^{ik_1\cdot y} \bigg(\frac{\partial}{\partial y^{\mu}} -
\frac{\stackrel\leftarrow\partial}{\partial y^{\mu}}\bigg)
\epsilon^{ace} \Delta_R (y-z_1) \bigg)\nonumber\\
&&\times \bigg( e^{ik_2\cdot z}
\bigg(\frac{\partial}{\partial z^{\nu}} -
\frac{\stackrel\leftarrow\partial}{\partial z^{\nu}}\bigg)
\epsilon^{bde} \Delta_R (z-z_1)
if_{\pi} {\bf S} \hat\sigma (z_1)\bigg)\nonumber\\
&&+ \bigg( k_1\,\,\, a\leftrightarrow k_2\,\,\, b\bigg)\nonumber\\
&&-\frac i{f_{\pi}} m_{\pi}^2 \delta^{ab}
\int d^4y d^4z e^{-iq_1\cdot y -iq_2\cdot z}
\int d^4z_1 e^{i(k_1+k_2)\cdot z_1}
{\bf S}T^*\bigg({\bf V}_{\mu}^c (y) {\bf V}_{\nu}^d (z) \hat\sigma (z_1)\bigg)
\nonumber\\=
&&+\frac i{f_{\pi}} m_{\pi}^2 \epsilon^{ace}\epsilon^{edb}
\frac 1{(k_1-q_1-q_2)^2 -m_{\pi}^2}
\int d^4z_1 e^{i(k_1+k_2-q_1-q_2)\cdot z_1}
{\bf S} \hat\sigma (z_1)\nonumber\\
&&-\frac i{f_{\pi}} m_{\pi}^2 \epsilon^{ace}\epsilon^{edb}
(2k_{1\mu}-q_{1\mu})(2k_{1\nu}-2q_{1\nu}-q_{2\nu})
\frac 1{(k_1-q_1)^2 -m_{\pi}^2}\nonumber\\
&&\times \frac 1{(k_1-q_1-q_2)^2 -m_{\pi}^2}
\int d^4z_1 e^{i(k_1+k_2-q_1-q_2)\cdot z_1}
{\bf S} \hat\sigma (z_1)\nonumber\\
&&+{\rm 3\,\,\,perm.}\nonumber\\
&&-\frac i{f_{\pi}} m_{\pi}^2 \epsilon^{ace}\epsilon^{bde}
(2k_{1\mu}-q_{1\mu})(2k_{2\nu}-q_{2\nu})
\frac 1{(k_1-q_1)^2 -m_{\pi}^2}\nonumber\\
&&\times \frac 1{(k_1-q_2)^2 -m_{\pi}^2}
\int d^4z_1 e^{i(k_1+k_2-q_1-q_2)\cdot z_1}
{\bf S} \hat\sigma (z_1)\nonumber\\
&&+\bigg( k_1\,\,\, a\leftrightarrow k_2\,\,\, b\bigg)\nonumber\\
&&-\frac i{f_{\pi}} m_{\pi}^2 \delta^{ab}
\int d^4y d^4z d^4z_1 e^{-iq_1\cdot y-iq_2\cdot z+ i(k_1+k_2)\cdot z_1}
{\bf S} T^*\bigg({\bf V}_{\mu}^c (y) {\bf V}_{\nu}^d (z)
\hat\sigma (z_1)\bigg)
\label{ADD34}
\ee
Taking the vacuum expectation value at this point leads to ambiguities, and
will be postponed till later.

{}From (\ref{T6}) we have
\be
&&+\frac 1{f_{\pi}^2}
\int d^4y d^4z e^{-iq_1\cdot y -iq_2\cdot z }\nonumber\\
&&\times \bigg( e^{ik_2\cdot y} \bigg(\frac{\partial}{\partial y^{\mu}} -
\frac{\stackrel\leftarrow\partial}{\partial y^{\mu}}\bigg)\epsilon^{bcf}
\int d^4z_2 \Delta_R (y-z_2) \frac{\partial}{\partial z_{2\beta}}
\nonumber\\
&&\times\bigg( e^{ik_1\cdot z_2}\epsilon^{afg}
{\bf S}T^*\bigg({\bf V}_{\beta}^g (z_2) {\bf V}_{\nu}^d (z)\bigg)
\bigg)\bigg)\nonumber\\
&&+\frac 1{f_{\pi}^2}
\int d^4y d^4z e^{-iq_1\cdot y -iq_2\cdot z +ik_2\cdot y +ik_1\cdot y}
\epsilon^{bch}\epsilon^{ahg}
{\bf S}T^*\bigg({\bf V}_{\mu}^g (y) {\bf V}_{\nu}^d (z)\bigg)\nonumber\\
&&+\frac 1{f_{\pi}^2}
\int d^4y d^4z e^{-iq_1\cdot y -iq_2\cdot z }
\int d^4z_2 e^{ik_2\cdot z_2}\frac  {\partial}{\partial z_{2\beta}}\nonumber\\
&&\times \bigg( e^{ik_1\cdot y}
\bigg(\frac{\partial}{\partial y^{\mu}} -
\frac{\stackrel\leftarrow\partial}{\partial y^{\mu}}\bigg)\epsilon^{ace}
\Delta_R (y-z_2) \epsilon^{ebg}
{\bf S}T^*\bigg({\bf V}_{\beta}^g (z_2) {\bf V}_{\nu}^d (z)\bigg)
\bigg)\nonumber\\
&&+\bigg( q_1\,\,\,\mu\,\,\, c\leftrightarrow q_2\,\,\,\nu\,\,\, d\bigg)
\nonumber\\
&&+\frac i{f^2_{\pi}}
\int d^4y d^4z  e^{-iq_1\cdot y-iq_2\cdot z}
\int d^4z_2 e^{ik_1\cdot z_2}\frac {\partial}{\partial z_{2\beta}}\nonumber\\
&&\bigg( e^{ik_2\cdot z_2}\epsilon^{abg}
{\bf S} T^*\bigg({\bf V}_{\beta}^g (z_2) {\bf V}_{\mu}^c (y)
{\bf V}_{\nu}^d (z) \bigg)\bigg)\nonumber\\
&&+\,\,\,...\nonumber\\ =
&&-\frac 1{f_{\pi}^2} (2k_{1\mu} -q_{1\mu})k_2^{\beta}
\epsilon^{ace}\epsilon^{ebg} \frac 1{(k_1-q_1)^2-m_{\pi}^2}\nonumber\\
&&\times\int d^4z d^4z_2 e^{-iq_2\cdot z+i (k_1+k_2-q_1)\cdot z_2}
{\bf S}T^*\bigg({\bf V}_{\mu}^g (y) {\bf V}_{\nu}^d (z)\bigg)\nonumber\\
&&+\frac 1{f_{\pi}^2} \epsilon^{bch}\epsilon^{ahg}\int d^4y d^4z
e^{i(k_1+k_2-q_1)\cdot y-iq_2\cdot z} {\bf S} T^*\bigg(
{\bf V}_{\mu}^g (y) {\bf V}_{\nu}^d (z) \bigg)\nonumber\\
&&+{\rm 3\,\,\,perm.}\nonumber\\
&&+\frac 1{2f_{\pi}^2} (k_2^{\beta}-k_1^{\beta})
\epsilon^{abg}
\int d^4y d^4z d^4z_2 e^{-q_1\cdot y-iq_2\cdot z+i (k_1+k_2)\cdot z_2}
\nonumber\\
&&\times {\bf S} T^*\bigg({\bf V}_{\beta}^g (z_2) {\bf V}_{\mu}^c (y)
{\bf V}_{\nu}^d (z) \bigg) + \,\,\,...
\label{ADD35}
\ee
where we have used the Ward identities (\ref{T13}) and
\be
&&+\frac{\partial}{\partial z_{2\beta}}
T^*\bigg({\bf V}_{\beta}^g (z_2) {\bf V}_{\mu}^c (y)
{\bf V}_{\nu}^d (z) \bigg)\nonumber\\ =
&&+i\epsilon^{gce}\delta^4 (z_2-y)
T^*\bigg({\bf V}_{\mu}^e (y) {\bf V}_{\nu}^d (z) \bigg)
+i\epsilon^{gde}\delta^4 (z_2-z)
T^*\bigg({\bf V}_{\mu}^c (y) {\bf V}_{\nu}^e (z) \bigg)
\label{ADD36}
\ee

{}From (\ref{T7}) we have
\be
&&-\frac 1{f_{\pi}^2} \,g_{\mu\nu}
\int d^4y d^4z e^{-iq_1\cdot y -iq_2\cdot z +ik_1\cdot y}
\epsilon^{ace}\epsilon^{edg}\delta^4 (y-z)\nonumber\\
&&\times \int d^4z_1 \Delta_R (z-z_1)
\int d^4z_2 e^{ik_2\cdot z_2}
\frac{\partial}{\partial z_{1\alpha}} \frac{\partial}{\partial z_{2\beta}}
\nonumber\\
&&\times {\bf S} T^*\bigg( {\bf j}_{A\alpha}^g (z_1)
{\bf j}_{A\beta}^b (z_2)\bigg)\nonumber\\
&&-\frac 1{f_{\pi}^2}
\int d^4y d^4z e^{-iq_1\cdot y -iq_2\cdot z}\nonumber\\
&&\times \bigg( e^{ik_1\cdot y} \bigg(\frac{\partial}{\partial y^{\mu}} -
\frac{\stackrel\leftarrow\partial}{\partial y^{\mu}}\bigg)
\epsilon^{ace} \Delta_R (y-z) \epsilon^{edg}
\bigg(\frac{\partial}{\partial z^{\nu}} -
\frac{\stackrel\leftarrow\partial}{\partial z^{\nu}}\bigg) \bigg)\nonumber\\
&&\times \int d^4z_1 \Delta_R (z-z_1) \int d^4z_2 e^{ik_2\cdot z_2}
\nonumber\\
&&\times \frac{\partial}{\partial z_{1\alpha}}
\frac{\partial}{\partial z_{2\beta}}
(-){\bf S} T^*\bigg( {\bf j}_{A\alpha}^g (z_1)
{\bf j}_{A\beta}^b (z_2)\bigg)\nonumber\\
&&+{\rm 3\,\,\,perm.}\nonumber\\
&&+\frac 1{f_{\pi}^2}
\int d^4y d^4z e^{-iq_1\cdot y -iq_2\cdot z}\nonumber\\
&&\times\bigg[ e^{ik_1\cdot y} \bigg(\frac{\partial}{\partial y^{\mu}} -
\frac{\stackrel\leftarrow\partial}{\partial y^{\mu}}\bigg)
\epsilon^{ace} \int d^4z_1 \Delta_R (y-z_1) \nonumber\\
&&\times \bigg( e^{ik_2\cdot z}
\bigg(\frac{\partial}{\partial z^{\nu}} -
\frac{\stackrel\leftarrow\partial}{\partial z^{\nu}}\bigg)
\epsilon^{bdf} \int d^4z_2 \Delta_R (z-z_2)\bigg)\nonumber\\
&&\times \frac{\partial}{\partial z_{1\alpha}}
\frac{\partial}{\partial z_{2\beta}}
{\bf S} T^*\bigg( {\bf j}_{A\alpha}^g (z_1)
{\bf j}_{A\beta}^b (z_2)\bigg)\bigg]\nonumber\\
&&+ \bigg( k_1\,\,\, a\leftrightarrow k_2\,\,\, b\bigg)\nonumber\\=
&&-\frac 1{f^2_{\pi}} \epsilon^{ace}\epsilon^{edb}
(k_1^{\alpha}-q_1^{\alpha}-q_2^{\alpha})k^{\beta}_2
\frac 1{(k_1-q_1-q_2)^2 -m_{\pi}^2}\nonumber\\
&&\int d^4z_1 d^4z_2 e^{i(k_1-q_1-q_2)\cdot z_1 +ik_2\cdot z_2}
{\bf S} T^*\bigg( {\bf j}_{A\alpha}^g (z_1) {\bf j}_{A\beta}^b (z_2)\bigg)
\nonumber\\
&&+\frac 1{f_{\pi}^2} (2k_{1\mu}-q_{1\mu})(2k_{1\nu}-2q_{1\nu}-q_{2\nu})
(k_1^{\alpha}-q_1^{\alpha} -q_2^{\alpha})k_2^{\beta}
\epsilon^{ace}\epsilon^{edg}\nonumber\\
&&\times \frac 1{(k_1-q_1)^2-m_{\pi}^2}\frac 1{(k_1-q_1-q_2)^2-m_{\pi}^2}
\int d^4z_1 d^4z_2 e^{i(k_1-q_1-q_2)\cdot z_1 +ik_2\cdot z_2}\nonumber\\
&&\times {\bf S} T^* \bigg( {\bf j}_{A\alpha}^g (z_1) {\bf j}_{A\beta}^b (z_2)
\nonumber\\
&&+{\rm 3 \,\,\, perm.}\nonumber\\
&&+\frac 1{f_{\pi}^2} \epsilon^{ace}\epsilon^{bdf}
(2k_{1\mu}-q_{1\mu})(2k_{2\nu}-q_{2\nu})(k_1^{\alpha}-q_1^{\alpha})
(k_2^{\beta}-q_{2\beta})\nonumber\\
&&\times\frac 1{(k_1-q_1)^2 -m_{\pi}^2}\frac 1{(k_2-q_2)^2 -m_{\pi}^2}
\nonumber\\
&&\int d^4z_1 d^4z_2 e^{i(k_1-q_1)\cdot z_1+i(k_2-q_2)\cdot z_2}
{\bf S} T^*\bigg( {\bf j}_{A\alpha}^e (z_1) {\bf j}_{A\beta}^f (z_2)\bigg)
\nonumber\\
&&+\bigg(k_1\,\,\, a\leftrightarrow k_2\,\,\, b\bigg)
\label{ADD37}
\ee
and
\be
&&+\frac 1{f_{\pi}^2}
\int d^4y d^4z e^{-iq_1\cdot y -iq_2\cdot z }\nonumber\\
&&\times \bigg( e^{ik_1\cdot y} \bigg(\frac{\partial}{\partial y^{\mu}} -
\frac{\stackrel\leftarrow\partial}{\partial y^{\mu}}\bigg)\epsilon^{ace}
\Delta_R (y-z) \epsilon^{edg}\bigg)\nonumber\\
&&\times\int d^4z_2  e^{ik_2\cdot z_2}
\frac{\partial}{\partial z_{2\beta}}
{\bf S}T^*\bigg({\bf j}_{A\nu}^g (z) {\bf j}_{A\beta}^b (z_2)\bigg)\nonumber\\
&&+\frac 1{f_{\pi}^2}
\int d^4y d^4z e^{-iq_1\cdot y -iq_2\cdot z }\nonumber\\
&&\times \bigg[ e^{ik_1\cdot y} \bigg(\frac{\partial}{\partial y^{\mu}} -
\frac{\stackrel\leftarrow\partial}{\partial y^{\mu}}\bigg)\epsilon^{ace}
\int d^4z_1 \Delta_R (y-z_1) \epsilon^{bdh}\nonumber\\
&&\times e^{ik_2\cdot z}\frac{\partial}{\partial z_{1\alpha}}
{\bf S}T^*\bigg({\bf j}_{A\alpha}^e (z_1) {\bf j}_{A\nu}^h (z)\bigg)\bigg]
\nonumber\\
&&+\frac i{f_{\pi}^2}
\int d^4y d^4z e^{-iq_1\cdot y -iq_2\cdot z }
\bigg[ e^{ik_1\cdot y}
\bigg(\frac{\partial}{\partial y^{\mu}} -
\frac{\stackrel\leftarrow\partial}{\partial y^{\mu}}\bigg)\epsilon^{ace}
\nonumber\\
&&\times \int d^4z_1 \Delta_R (y-z_1) \int d^4z_2 e^{ik_2\cdot z_2}
\frac {\partial}{\partial z_{1\alpha}}
\frac {\partial}{\partial z_{2\beta}}\nonumber\\
&&\times
{\bf S}T^*\bigg({\bf j}_{A\alpha}^e (z_1) {\bf j}_{A\beta}^b (z_2)
{\bf V}_{\nu}^d (z)\bigg)\bigg]  \nonumber\\
&&+\frac 1{f_{\pi}^2}
\int d^4y d^4z  e^{-iq_1\cdot y-iq_2\cdot z +ik_1\cdot y +ik_2\cdot z}
\epsilon^{acg}\epsilon^{bdh}\nonumber\\
&&\times {\bf S} T^*\bigg({\bf j}_{A\mu}^g (y) {\bf j}_{A\nu}^h (z) \bigg)
\nonumber\\
&&+\frac i{f_{\pi}^2}
\int d^4y d^4z  e^{-iq_1\cdot y-iq_2\cdot z +ik_1\cdot y}
\epsilon^{acg}\int d^4z_2 e^{ik_2\cdot z_2}\nonumber\\
&&\times \frac {\partial}{\partial z_{2\beta}}
{\bf S}T^*\bigg({\bf j}_{A\mu}^g (y) {\bf j}_{A\beta}^b (z_2)
{\bf V}_{\nu}^d (z)\bigg)  \nonumber\\
&&+\bigg( k_1\,\,\, a\leftrightarrow k_2\,\,\, b\bigg)\nonumber\\
&&-\frac 1{f_{\pi}^2}
\int d^4y d^4z  e^{-iq_1\cdot y-iq_2\cdot z}
\int d^4z_1 d^4z_2 e^{ik_1\cdot z_1+ ik_2\cdot z_2}\nonumber\\
&&\times \frac {\partial}{\partial z_{1\alpha}}
\frac {\partial}{\partial z_{2\beta}}
{\bf S}T^*\bigg({\bf j}_{A\alpha}^a (z_1) {\bf j}_{A\beta}^b (z_2)
{\bf V}_{\mu}^c (y) {\bf V}_{\nu}^d (z)\bigg)  \nonumber\\=
&&-\frac 1{f_{\pi}^2} (2k_{1\mu} -q_{1\mu})k_2^{\beta}
\epsilon^{ace}\epsilon^{edg} \frac 1{(k_1-q_1)^2-m_{\pi}^2}\nonumber\\
&&\times\int d^4z d^4z_2 e^{i(k_1-q_1-q_2)\cdot z+ ik_2\cdot z_2}
{\bf S}T^*\bigg({\bf j}_{A\nu}^g (z) {\bf j}_{A\beta}^b (z_2)\bigg)\nonumber\\
&&-\frac 1{f_{\pi}^2} (2k_{1\mu}-q_{1\mu})(k_1^{\alpha}-q_1^{\alpha})
\epsilon^{ace}\epsilon^{bdh} \frac 1{(k_1-q_1)^2-m_{\pi}^2}\nonumber\\
&&\times \int d^4z_1 d^4z e^{i(k_1-q_1)\cdot z_1 +i (k_2-q_2)\cdot z}
{\bf S} T^*\bigg({\bf j}_{A\alpha}^e (z) {\bf j}_{A\nu}^h (z) \bigg)\nonumber\\
&&-\frac 1{f_{\pi}^2}
(2k_{1\mu}-q_{1\mu})(k_1^{\alpha}-q_1^{\alpha})k_2^{\beta}
\epsilon^{ace}\frac 1{(k_1-q_1)^2-m_{\pi}^2}\nonumber\\
&&\times \int d^4z d^4z_1 d^4z_2
e^{-iq_2\cdot z +i(k_1-q_1)\cdot z_1 +ik_2\cdot z_2}
{\bf S} T^*\bigg({\bf j}_{A\alpha}^e (z) {\bf j}_{A\beta}^b (z_2)
{\bf V}_{\nu}^d (z)\bigg) \nonumber\\
&&+\frac 1{f_{\pi}^2}
\epsilon^{acg}\epsilon^{bdh}
\int d^4y d^4z e^{+i(k_1-q_1)\cdot z_1 +i(k_2-q_2)\cdot z_2}
{\bf S} T^*\bigg({\bf j}_{A\mu}^g (y) {\bf j}_{A\nu}^h (z) \bigg)\nonumber\\
&&+\frac 1{f_{\pi}^2} k_2^{\beta}\epsilon^{acg}
\int d^4y d^4z d^4z_2 e^{+i(k_1-q_1)\cdot y-iq_2\cdot z+ik_2\cdot z_2}
{\bf S} T^*\bigg({\bf j}_{A\mu}^g (y) {\bf j}_{A\beta}^b (z_2)
{\bf V}_{\nu}^d (z)\bigg)\nonumber\\
&&+\bigg( k_1\,\,\, a\leftrightarrow k_2\,\,\, b \bigg)\nonumber\\
&&+\frac 1{f_{\pi}^2}k_1^{\alpha} k_2^{\beta}
\int d^4y d^4z d^4z_1 d^4z_2 e^{-q_1\cdot y-iq_2\cdot z +ik_1\cdot
z_1+ik_2\cdot z_2}\nonumber\\
&&\times {\bf S} T^*\bigg({\bf j}_{A\alpha}^a (z_1) {\bf j}_{A\beta}^b (z_2)
{\bf V}_{\mu}^c (y) {\bf V}_{\nu}^d (z) \bigg)
\label{ADD38}
\ee
The sum of (\ref{ADD34}) and (\ref{ADD37}) gives
\be
&&-\frac i{f_{\pi}} \epsilon^{ace}\epsilon^{edb}\,g_{\mu\nu}\,\int d^4z_1
e^{i(k_1+k_2-q_1-q_2)\cdot z_1}
{\bf S} \hat\sigma (z_1)\nonumber\\
&&+\frac i{f_{\pi}} \epsilon^{ace}\epsilon^{edb}
(2k_{1\mu}- q_{1\mu})(2k_{2\nu}-2q_{1\nu} -q_{2\nu})
\frac 1{(k_1-q_1)^2-m_{\pi}^2} \nonumber\\
&&\times\int d^4z_1 e^{i(k_1+k_2-q_1-q_2)\cdot z_1}
{\bf S} \hat\sigma (z_1)\nonumber\\
&&+{\rm 3\,\,\,perm.}\nonumber\\
&&-\frac i{f_{\pi}} m_{\pi}^2\epsilon^{ace}\epsilon^{bde}
(2k_{1\mu}- q_{1\mu})(2k_{2\nu}-q_{2\nu})
\frac 1{(k_1-q_1)^2-m_{\pi}^2} \frac 1{(k_2-q_2)^2-m_{\pi}^2} \nonumber\\
&&\times\int d^4z_1 e^{i(k_1+k_2-q_1-q_2)\cdot z_1}
{\bf S} \hat\sigma (z_1)\nonumber\\
&&+\frac 1{f^2_{\pi}} \epsilon^{ace}\epsilon^{bde}
(2k_{1\mu}- q_{1\mu})(2k_{2\nu}-q_{2\nu})(k_1^{\alpha} -q_1^{\alpha})
(k_2^{\beta} -q_2^{\beta})\nonumber\\
&&\times \frac 1{(k_1-q_1)^2-m_{\pi}^2} \frac 1{(k_2-q_2)^2-m_{\pi}^2}
\nonumber\\
&&\times\int d^4z_1 d^4z_2 e^{i(k_1-q_1)\cdot z_1 +i(k_2-q_2)\cdot z_2}
{\bf S} T^*\bigg( {\bf j}_{A\alpha}^e (z_1) {\bf j}_{A\beta}^f (z_2)\bigg)
\nonumber\\
&&+\bigg( k_1\,\,\, a\leftrightarrow k_2 \,\,\, b\bigg)\nonumber\\
&&-\frac i{f_{\pi}} m_{\pi}^2 \delta^{ab}
\int d^4y d^4z d^4z_1 e^{-iq_1\cdot y-iq_2\cdot z +i (k_1+k_2)\cdot z_1}
{\bf S} T^*\bigg( {\bf V}_{\mu}^c (y) {\bf V}_{\nu}^d (z)
\hat\sigma (z_1)\bigg)
\nonumber\\&& + ....
\label{ADD39}
\ee

There is now no ambiguity in taking the vacuum expectation values. Dividing out
$(2\pi)^4\delta^4 (k_1+k_2-q_1-q_2)$, we obtain from
(\ref{ADD33},\ref{ADD34},\ref{ADD38},\ref{ADD39}) respectively,
\be
&&+i(2k_{2\mu}-q_{1\mu})(2k_{1\nu}-q_{2\nu})\epsilon^{bcf}\epsilon^{fda}
\frac 1{(k_2-q_1)^2-m_{\pi}^2}\nonumber\\
&&+i\epsilon^{bce}\epsilon^{eda} g_{\mu\nu} +
\bigg( q_1\,\,\mu\,\, c\leftrightarrow q_2\,\, \nu\,\, d \bigg)
\label{ADD40}
\ee
\be
&&+\frac i{f_{\pi}^2} (2k_{1\mu}-q_{1\mu})
(-k_{2\nu} q_2^2 +k_2\cdot q_2 q_{2\nu})\epsilon^{ace}\epsilon^{ebd}
\nonumber\\
&&\times\frac 1{(k_1-q_1)^2-m_{\pi}^2} \Pi_V (q_2^2 )\nonumber\\
&&-\frac i{2f_{\pi}^2} \epsilon^{bch}\epsilon^{ahd} (-g_{\mu\nu} q_2^2 +
q_{2\mu} q_{2\nu} )\Pi_V (q_2^2) \nonumber\\
&&+{\rm 3\,\,\, perm.}\nonumber\\
&&+\frac 1{2f_{\pi}^2} (k_2^{\beta}-k_1^{\beta})\epsilon^{abg}
\int d^4y d^4z e^{-iq_1\cdot y-iq_2\cdot z}\nonumber\\
&&\times <0|T^*\bigg( {\bf V}_{\mu}^c (y) {\bf V}_{\nu}^d (z)
{\bf V}_{\beta}^g (0) \bigg) |0>
\label{ADD41}
\ee
\be
&&-\frac 1{f_{\pi}^2} (2(k_{1\mu} -q_{1\mu})k_2^{\beta}
\epsilon^{ace}\epsilon^{edg} \frac 1{(k_1-q_1)^2-m_{\pi}^2}\nonumber\\
&&\times\int d^4z_2 e^{+ ik_2\cdot z_2}
<0|T^*\bigg({\bf j}_{A\beta}^b (z_2) {\bf j}_{A\nu}^g (0)\bigg) |0>\nonumber\\
&&-\frac 1{f_{\pi}^2} (2k_{1\mu}-q_{1\mu})(k_1^{\alpha}-q_1^{\alpha})
\epsilon^{ace}\epsilon^{bdh} \frac 1{(k_1-q_1)^2-m_{\pi}^2}\nonumber\\
&&\times \int d^4z_1 e^{i(k_1-q_1)\cdot z_1 }
<0|T^*\bigg({\bf j}_{A\alpha}^e (z_1) {\bf j}_{A\nu}^h (0) \bigg)
|0>\nonumber\\
&&-\frac 1{f_{\pi}^2}
(2k_{1\mu}-q_{1\mu})(k_1^{\alpha}-q_1^{\alpha})k_2^{\beta}
\epsilon^{ace}\frac 1{(k_1-q_1)^2-m_{\pi}^2}\nonumber\\
&&\times \int d^4z_1 d^4z_2
e^{+i(k_1-q_1)\cdot z_1 +i(k_2-q_2)\cdot z_2}
<0|T^*\bigg({\bf j}_{A\alpha}^e (z_1) {\bf j}_{A\beta}^b (z_2)
{\bf V}_{\nu}^d (0)\bigg) |0> \nonumber\\
&&+\frac 1{f_{\pi}^2}
\epsilon^{acg}\epsilon^{bdh}
\int d^4y  e^{+i(k_1-q_1)\cdot y }
<0|T^*\bigg({\bf j}_{A\mu}^g (y) {\bf j}_{A\nu}^h (0) \bigg) |0>\nonumber\\
&&+\frac 1{f_{\pi}^2} k_2^{\beta}\epsilon^{acg}
\int d^4y  d^4z_2 e^{+i(k_1-q_1)\cdot y +ik_2\cdot z_2}
<0| T^*\bigg({\bf j}_{A\mu}^g (y) {\bf j}_{A\beta}^b (z_2)
{\bf V}_{\nu}^d (0)\bigg) |0>\nonumber\\
&&+\bigg( k_1\,\,\, a\leftrightarrow k_2\,\,\, b \bigg)\nonumber\\
&&+\frac 1{f_{\pi}^2}k_1^{\alpha} k_2^{\beta}
\int d^4y d^4z_1 d^4z_2 e^{-iq_1\cdot y +ik_1\cdot z_1+ik_2\cdot
z_2}\nonumber\\
&&\times
<0|T^*\bigg({\bf j}_{A\alpha}^a (z_1) {\bf j}_{A\beta}^b (z_2)
{\bf V}_{\mu}^c (y) {\bf V}_{\nu}^d (0) \bigg) |0>
\label{ADD42}
\ee
\be
&&-\frac i{f_{\pi}} \epsilon^{ace}\epsilon^{edb} g_{\mu\nu}
<0| \hat\sigma |0>\nonumber\\
&&+\frac i{f_{\pi}} \epsilon^{ace}\epsilon^{edb}
(2k_{1\mu}- q_{1\mu})(2k_{2\nu} -q_{2\nu})
\frac 1{(k_1-q_1)^2-m_{\pi}^2}
<0|\hat\sigma |0>\nonumber\\
&&+{\rm 3\,\,\,perm.}\nonumber\\
&&-\frac i{f_{\pi}} {m_{\pi}^2}\epsilon^{ace}\epsilon^{bde}
(2k_{1\mu}- q_{1\mu})(2k_{2\nu}-q_{2\nu})\nonumber\\
&&\times
\frac 1{(k_1-q_1)^2-m_{\pi}^2} \frac 1{(k_2-q_2)^2-m_{\pi}^2}
<0|\hat\sigma |0>\nonumber\\
&&+\frac 1{f^2_{\pi}} \epsilon^{ace}\epsilon^{bdf}
(2k_{1\mu}- q_{1\mu})(2k_{2\nu}-q_{2\nu})(k_1^{\alpha} -q_1^{\alpha})
(k_2^{\beta} -q_2^{\beta})\nonumber\\
&&\times \frac 1{(k_1-q_1)^2-m_{\pi}^2} \frac 1{(k_2-q_2)^2-m_{\pi}^2}
\nonumber\\
&&\times\int d^4z_1 e^{i(k_1-q_1)\cdot z_1}
<0| T^*\bigg( {\bf j}_{A\alpha}^e (z_1) {\bf j}_{A\beta}^f (0)\bigg) |0>
\nonumber\\
&&+\bigg( k_1\,\,\, a\leftrightarrow k_2 \,\,\, b\bigg)\nonumber\\
&&-\frac i{f_{\pi}} m_{\pi}^2 \delta^{ab}
\int d^4y d^4z  e^{-iq_1\cdot y-iq_2\cdot z }
<0|T^*\bigg( {\bf V}_{\mu}^c (y) {\bf V}_{\nu}^d (z)
\hat\sigma (0)\bigg) |0>
\label{ADD43}
\ee
The photon-photon collision process is sensitive to the way chiral symmetry
is broken in the vacuum through $<0|\hat\sigma |0>$.
The expressions (\ref{ADD40}-\ref{ADD43}) are our exact result for the
$\gamma\gamma\rightarrow \pi\pi$ process. It is important to stress that
this result follows only from the dictates of chiral symmetry and unitarity.

At low energy, we expect the
two-point correlation function ${\bf \Pi}_V$ to be $\rho$ and $\rho$'
dominated, while the two-point correlators $<{\bf j}_A {\bf j}_A>$
to be $a_1$ dominated. The three- and higher-point correlation functions
involve more complex structure.
We note that the isoscalar part of the electromagnetic current contributes to
both the isospin $I=0,1$ channels of $\gamma\gamma\rightarrow \pi\pi$. Thus
an exact  assessment of the various correlation functions entering
(\ref{ADD40}-\ref{ADD43}) require a full
$SU(3)\times SU(3)$ treatment. This goes beyond the scope of the present
analysis. An estimation  of the various correlators in
(\ref{ADD40}-\ref{ADD43}) using $SU(2)\times SU(2)$  chiral
counting arguments will be given below.

For the process
$\gamma\gamma \rightarrow \pi^+\pi^-$, the nearness of the pion pole to the
physical region means that the effects of the high mass resonances
$\rho\,\, , \rho '\,\,, \omega\,\,, a_1$ ..., are small. Thus, the process
is dominated by the Born terms. Since our result accounts properly for the Born
terms, we expect overall agreement with the data much like the Born
approximation with some radiative corrections \cite{terentev}, or one-loop
chiral perturbation theory \cite{photophoto}. However, the large discrepancy
noted between one-loop chiral perturbation theory and the Crystal Ball data
\cite{ball} for $\gamma\gamma\rightarrow \pi^0\pi^0$ shows some evidence
of correlations, since the Born terms vanish in this case.
This has been partly confirmed by the inclusion of vector
mesons to one-loop chiral
perturbation theory by Ko \cite{ko}, and the use of dispersion analysis
by Morgan and Pennington and others \cite{morgan}.
At low energy, the $\omega$, the $\rho$ and the $a_1$ play an
important role in the fusion process to neutral pions. These effects are
all accounted for in (\ref{ADD40}-\ref{ADD43}) in a model independent way.
We also note that some of these correlations do appear in a
recent two-loop calculation by Belluci, Gasser and Sainio in the context of
chiral perturbation theory \cite{belluci}, through pion rescattering and
unspecified counter terms. Since our result
(\ref{ADD40}-\ref{ADD43}) contains the unexpanded correlation functions
as dictated by chiral symmetry and unitarity, it would be interesting to
carry a more systematic comparison between our results,
the approximate estimations and the available data within the 1 GeV range.

\vskip .5cm
{{\bf 17. Compton Scattering : } $\gamma\pi\rightarrow\gamma\pi$}
\vskip .3cm

When atoms are immersed in an electromagnetic cavity, their interaction with
light induces a polarisation of the electric charge distributions. These
electromagnetic deformations of an extended object can be characterized by
electric and magnetic polarisabilities. Compton scattering of light off pions
through atomic targets via Primakoff effect \cite{lebedev,serp} and
photon-photon collision \cite{mark,ball} have led to electric polarisabilities
for charged pions
ranging from about $2$ $10^{-4}$ fm$^{3}$ to $20$ $10^{-4}$ fm$^{3}$. The
order of magnitude discrepancy between the quoted data illustrates the
experimental difficulties encountered in the threshold measurements as well as
the systematic uncertainties associated to the way the various extrapolations
are performed.

This notwithstanding, there have appeared a number of estimations
of the pion polarisabilities using various chiral models \cite{polamodel} and
chiral perturbation theory \cite{polacpt}. In the latter
the reaction $\gamma\gamma\rightarrow \pi\pi$ and its crossed version
$\gamma\pi\rightarrow \gamma\pi$ were used. While the
$\gamma\gamma\rightarrow \pi^+\pi^-$ seems to be well described by one loop
chiral perturbation theory (mostly the Born amplitude though),
the reaction $\gamma\gamma\rightarrow \pi^0\pi^0$ is totally off at one loop.
Dispersion techniques and two-loop calculations have been carried out recently
to try to account for the deficiency of the one-loop calculation
\cite{donoghue,belluci}, leading to new estimates for the pion
polarisabilities.

In our approach, Compton scattering off pions will be treated exactly to first
order in the electromagnetic charge. Unitarity will be maintained to all
orders, and  the strong correlations in the neutral channel will be expressed
in terms of vacuum correlators. The Compton scattering amplitude
follows from the matrix element
\be
\int d^4y d^4z e^{-iq_1\cdot y+iq_2\cdot z}
<0|a_{\rm in}^b (k_2) {\bf S}T^*
\bigg({\bf V}_{\mu}^c (y) {\bf V}_{\nu}^d (z)\bigg) a_{\rm in}^{a \dagger}
(k_1) |0>
\label{ADD44}
\ee
This amplitude follows from the result (\ref{ADD15},\ref{ADD43}) for
$\gamma\gamma\rightarrow \pi\pi$ by crossing $q_2\leftrightarrow -q_2$
and $k_1\leftrightarrow -k_1$. Modulo the overall factor
$(2\pi)^4 \delta^4 (k_1+q_1-k_2-q_2)$ for energy momentum conservation,
the result is
\be
&&-i(2k_{2\mu}-q_{1\mu})(2k_{1\nu}-q_{2\nu})\epsilon^{bcf}\epsilon^{fda}
\frac 1{(k_2-q_1)^2-m_{\pi}^2}\nonumber\\
&&+i\epsilon^{bce}\epsilon^{eda} g_{\mu\nu} +
\bigg( q_1\,\,\mu\,\, c\leftrightarrow - q_2\,\, \nu\,\, d \bigg)
\label{ADD45}
\ee
\be
&&-\frac i{f_{\pi}^2} (2k_{1\mu}+q_{1\mu})
(-k_{2\nu} q_2^2 +k_{2}\cdot q_2 q_{2\nu})\epsilon^{ace}\epsilon^{ebd}
\nonumber\\
&&\times\frac 1{(k_1+q_1)^2-m_{\pi}^2} \Pi_V (q_2^2 )\nonumber\\
&&-\frac i{2f_{\pi}^2} \epsilon^{bch}\epsilon^{ahd} (-g_{\mu\nu} q_2^2 +
q_{2\mu} q_{2\nu} )\Pi_V (q_2^2) \nonumber\\
&&+{\rm 3\,\,\, perm.}\nonumber\\
&&+\frac 1{2f_{\pi}^2} (k_2^{\beta}+k_1^{\beta})\epsilon^{abg}
\int d^4y d^4z e^{-iq_1\cdot y+iq_2\cdot z}\nonumber\\
&&\times <0|T^*\bigg( {\bf V}_{\mu}^c (y) {\bf V}_{\nu}^d (z)
{\bf V}_{\beta}^g (0) \bigg) |0>
\label{ADD46}
\ee
\be
&&+\frac 1{f_{\pi}^2} (2k_{1\mu} +q_{1\mu})k_2^{\beta}
\epsilon^{ace}\epsilon^{edg} \frac 1{(k_1+q_1)^2-m_{\pi}^2}\nonumber\\
&&\times\int d^4z_2 e^{+ ik_2\cdot z_2}
<0|T^*\bigg({\bf j}_{A\beta}^b (z_2) {\bf j}_{A\nu}^g (0)\bigg) |0>\nonumber\\
&&-\frac 1{f_{\pi}^2}
(2k_{1\mu}+q_{1\mu})(k_1^{\alpha}+q_1^{\alpha})
\epsilon^{ace}\epsilon^{bdh} \frac 1{(k_1+q_1)^2-m_{\pi}^2}\nonumber\\
&&\times \int d^4z_1 e^{-i(k_1+q_1)\cdot z_1  }
<0|T^*\bigg({\bf j}_{A\alpha}^e (z_1) {\bf j}_{A\nu}^h (0)\bigg) |0>\nonumber\\
&&-\frac 1{f_{\pi}^2}
(2k_{1\mu}+q_{1\mu})(k_1^{\alpha}+q_1^{\alpha})k_2^{\beta}
\epsilon^{ace}\frac 1{(k_1+q_1)^2-m_{\pi}^2}\nonumber\\
&&\times \int d^4z_1 d^4z_2
e^{-i(k_1+q_1)\cdot z_1 +i(k_2+q_2)\cdot z_2}
<0|T^*\bigg({\bf j}_{A\alpha}^e (z_1) {\bf j}_{A\beta}^b (z_2)
{\bf V}_{\nu}^d (0)\bigg) |0> \nonumber\\
&&+\frac 1{f_{\pi}^2}
\epsilon^{acg}\epsilon^{bdh}
\int d^4y  e^{-i(k_1+q_1)\cdot y }
<0|T^*\bigg({\bf j}_{A\mu}^g (y) {\bf j}_{A\nu}^h (0) \bigg) |0>\nonumber\\
&&+\frac 1{f_{\pi}^2} k_2^{\beta}\epsilon^{acg}
\int d^4y  d^4z_2 e^{-i(k_1+q_1)\cdot y +ik_2\cdot z_2}
<0| T^*\bigg({\bf j}_{A\mu}^g (y) {\bf j}_{A\beta}^b (z_2)
{\bf V}_{\nu}^d (0)\bigg) |0>\nonumber\\
&&+\bigg(- k_1\,\,\, a\leftrightarrow k_2\,\,\, b \bigg)\nonumber\\
&&-\frac 1{f_{\pi}^2}k_1^{\alpha} k_2^{\beta}
\int d^4y d^4z_1 d^4z_2 e^{-q_1\cdot y -ik_1\cdot z_1+ik_2\cdot z_2}\nonumber\\
&&\times
<0|T^*\bigg({\bf j}_{A\alpha}^a (z_1) {\bf j}_{A\beta}^b (z_2)
{\bf V}_{\mu}^c (y) {\bf V}_{\nu}^d (0) \bigg) |0>
\label{ADD47}
\ee
\be
&&-\frac i{f_{\pi}} \epsilon^{ace}\epsilon^{edb}
<0| \hat\sigma |0>\nonumber\\
&&-\frac i{f_{\pi}} \epsilon^{ace}\epsilon^{edb}
(2k_{1\mu}+ q_{1\mu})(2k_{2\nu}-2q_{1\nu} -q_{2\nu})
\frac 1{(k_1+q_1)^2-m_{\pi}^2}
<0|\hat\sigma |0>\nonumber\\
&&+{\rm 3\,\,\,perm.}\nonumber\\
&&+\frac i{f_{\pi}} {m_{\pi}^2}\epsilon^{ace}\epsilon^{bde}
(2k_{1\mu}+q_{1\mu})(2k_{2\nu}+q_{2\nu})\nonumber\\
&&\times
\frac 1{(k_1+q_1)^2-m_{\pi}^2} \frac 1{(k_2+q_2)^2-m_{\pi}^2}
<0|\hat\sigma |0>\nonumber\\
&&+\frac 1{f^2_{\pi}} \epsilon^{ace}\epsilon^{bdf}
(2k_{1\mu}+ q_{1\mu})(2k_{2\nu}+q_{2\nu})(k_1^{\alpha} +q_1^{\alpha})
(k_2^{\beta} +q_2^{\beta})\nonumber\\
&&\times \frac 1{(k_1+q_1)^2-m_{\pi}^2} \frac 1{(k_2+q_2)^2-m_{\pi}^2}
\nonumber\\
&&\times\int d^4z_1 e^{-i(k_1+q_1)\cdot z_1}
<0| T^*\bigg( {\bf j}_{A\alpha}^e (z_1) {\bf j}_{A\beta}^f (0)\bigg) |0>
\nonumber\\
&&+\bigg(- k_1\,\,\, a\leftrightarrow k_2 \,\,\, b\bigg)\nonumber\\
&&-\frac i{f_{\pi}} m_{\pi}^2 \delta^{ab}
\int d^4y d^4z  e^{-iq_1\cdot y+iq_2\cdot z }
<0|T^*\bigg( {\bf V}_{\mu}^c (y) {\bf V}_{\nu}^d (z)
\hat\sigma (0)\bigg) |0>
\label{ADD48}
\ee
which is valid for isospin $0$ and $2$ channels.
Again, we note that the Compton scattering process is sensitive to the
chiral symmetry breaking term. The expressions(\ref{ADD45}-\ref{ADD48})
are our full result for the Compton process $\gamma\pi\rightarrow\gamma\pi$.
Most of the remarks we made above for the fusion process
$\gamma\gamma\rightarrow \pi\pi$ apply $verbatim$ to $\gamma\pi\rightarrow
\gamma\pi$ by crossing. Below we will give an estimation for
(\ref{ADD45}-\ref{ADD48}) using $SU(2)\times SU(2)$ chiral counting arguments.

The Compton amplitudes for neutral and charged pions can be used for a precise
definition of the pion polarisabilities. Since (\ref{ADD30}) by
crossing yields (\ref{ADD44}), we define the neutral ($a=b=3$) and the charged
amplitude ($a=b=1$) to be
\be
{\cal V}_{\mu\nu}^{ab}= &&+\int d^4y d^4z e^{-iq_1\cdot y -iq_2\cdot z}
<0|a_{\rm in}^a (k_1) a_{\rm in}^b (k_2) {\bf S} T^*
\bigg({\bf V}_{\mu}^3 (y) {\bf V}_{\nu}^3 (z)\bigg) |0>
\label{X100}
\ee
Factoring out $(2\pi )^4 \delta^4 (q_1+q_2-k_1-k_2)$, using covariance and
gauge invariance yield  \cite{belluci}
\be
{\cal V}_{\mu\nu} = A (s,t,u) {\bf T}_{1\mu\nu} + B (s,t,u) {\bf T}_{2\mu\nu}
\label{X101}
\ee
with generically
\be
{\bf T}_{1\mu\nu} =&&\frac s2 g_{\mu\nu} -q_{1\mu} q_{2\nu}\nonumber\\
{\bf T}_{1\mu\nu} =&& 2s \Delta_{\mu}\Delta_{\nu} -\nu^2 g_{\mu\nu}
-2\nu \bigg( q_{1\nu} \Delta_{\mu} -q_{2\mu} \Delta_{\nu}\bigg)
\label{X102}
\ee
and $\Delta_{\mu} = (k_1-k_2)_{\mu}$ and $\nu = t-u$. The amplitudes
$A$ and $B$  are analytic in $s,t,u$ and symmetric under crossing
$(t,u)\leftrightarrow (u,t)$. For $s\geq 4m_{\pi}^2$ they describe the
fusion process, and for $s\leq 0$ they describe the Compton process.
In terms of (\ref{X102}), the neutral pion polarisabilities read
\cite{belluci}
\be
\alpha_{\pi}^0 = &&+\frac {\alpha}{2m_{\pi}} \bigg( A^0 (0,m_{\pi}^2,m_{\pi}^2)
+16 m_{\pi}^2 B^0 (0,m_{\pi}^2,m_{\pi}^2 )\bigg)\nonumber\\
\beta_{\pi}^0 = &&-\frac {\alpha}{2m_{\pi}} A^0 (0, m_{\pi}^2, m_{\pi}^2 )
\label{X103}
\ee
with $\alpha =e^2/4\pi =1/137$.
The charged pion polarisabilities follow from (\ref{X103})
by exchanging the upper script $0\rightarrow \pm$, and multiplying the overall
result by a minus sign. From (\ref{ADD40}-\ref{ADD43}) we have
\be
\alpha_{\pi} = &&-\frac {\alpha m_{\pi}}2
\lim_{s\to0}\lim_{t\to{m_{\pi}^2}} \frac 1{s^2} \bigg( V (s,t,u) +
\frac {24}{\epsilon_1\cdot \epsilon_2} \hat{V} (s,t,u) \bigg)\nonumber\\
\beta_{\pi} = &&+\frac {\alpha m_{\pi}}2
\lim_{s\to0}\lim_{t\to{m_{\pi}^2}} \frac 1{s^2}  V (s,t,u)
\label{X104}
\ee
with
\be
&&V=V^{33\mu}_{\mu} \qquad{\rm and}\qquad
\hat V = \epsilon_1^{\mu}\epsilon_2^{\nu} V_{\mu\nu}^{33}
\qquad : \gamma\gamma\rightarrow \pi^0\pi^0\nonumber\\
&&V=V^{11\mu}_{\mu} \qquad{\rm and}\qquad
\hat V = \epsilon_1^{\mu}\epsilon_2^{\nu} V_{\mu\nu}^{11}
\qquad : \gamma\gamma\rightarrow \pi^+\pi^-\nonumber
\ee
and
\be
V_{\mu\nu}^{ab} =
&&+\frac 1{f_{\pi}^2}k_1^{\alpha} k_2^{\beta}
\int d^4y d^4z_1 d^4z_2 e^{-iq_1\cdot y +ik_1\cdot z_1+ik_2\cdot
z_2}\nonumber\\
&&\times
<0|T^*\bigg({\bf j}_{A\alpha}^a (z_1) {\bf j}_{A\beta}^b (z_2)
{\bf V}_{\mu}^3 (y) {\bf V}_{\nu}^3 (0) \bigg) |0>\nonumber\\
&&-\frac i{f_{\pi}} m_{\pi}^2 \delta^{ab}
\int d^4y d^4z  e^{-iq_1\cdot y-iq_2\cdot z }
<0|T^*\bigg( {\bf V}_{\mu}^3 (y) {\bf V}_{\nu}^3 (z)
\hat\sigma (0)\bigg) |0>
\label{X105}
\ee
The relation between $A,B$ in (\ref{X103}) and $V,\hat V$ in (\ref{X104})
is straightforward. An assessment of the pion
polarisabilities in $SU(2)\times SU(2)$ will be given below using a new
one-loop effective action as we now discuss.

\vskip .5cm
{\bf 18. One-Loop Effective Action}
\vskip .3cm

To give a quantitative estimate for the unknown terms in $\pi\pi\rightarrow
\pi\pi$ (\ref{C63}-\ref{C64}), $\pi\rightarrow e\nu\gamma$ (\ref{C67}),
$\pi\rightarrow e\nu\gamma^*$ (\ref{ADD15}),
$\gamma\gamma\rightarrow \pi\pi$ (\ref{ADD41}-\ref{ADD43})
and $\gamma\pi\rightarrow \gamma\pi$ (\ref{ADD45}-\ref{ADD48}),
we will expand the exact results
of the preceding section in $1/f_{\pi}$. For that we will need power counting.
We recall that for the gauged nonlinear sigma model,
\be
\hat{\bf I}= f_{\pi}^2 m_{\pi}^2 \int d^4 x +  \int d^4x \,\,\pi (x) \cdot
J(x) + {\bf I}_Q + {\cal O} (\pi^3 /f_{\pi} )
\label{D1}
\ee
where the quadratic action ${\bf I}_Q$ has been defined in (\ref{B91}).
If we count $\phi = (v, a, s, J)$ as order $f_{\pi}^{0}$, then $\pi\sim
f_{\pi}^0$, so that ${\bf j}_{V\mu}\sim f_{\pi}^0$, and ${\bf j}_{A\mu}^a \sim
f_{\pi}^0$, whereas $\hat\sigma\sim f_{\pi}^{-1}$. We will $assume$ that the
same is true for QCD (or more appropriately the real world). As a check, we
note that this assumption is consistent with (\ref{C25}-\ref{C26}) for the
q-number piece, while $<0|\hat\sigma |0>\sim f_{\pi}^{-1}$ reproduces the
Gell-Mann-Oakes-Renner (GOR) relation
\be
\hat{m} <0| \overline qq |0 > =-m_{\pi}^2 f_{\pi}^2 + {\cal O} ( 1 )
\label{D2}
\ee
by (\ref{B68}-\ref{B81}).

To leading order, the master formulas are then the strong version of
(\ref{B57}) and
\be
\frac{\delta\hat{\cal S}_0}{\delta J} =&&
i\hat{\cal S}_0 \bigg( 1 +G_R {\bf K} \bigg) \pi_{\rm in} -i G_R J
\hat{\cal S}_0 \nonumber \\ =&&
i \bigg( 1 +G_A {\bf K} \bigg) \pi_{\rm in}\hat{\cal S}_0 -i G_A J
\hat{\cal S}_0.
\label{D3}
\ee
The pionic part of the S-matrix to this order $\hat{\cal S}_0$ is completely
fixed by (\ref{B57}) and (\ref{D3}) up to a phase factor which may depend on
the external fields. The phase factor may be determined from the vacuum
persistence amplitude
\be
e^{iZ_0 [v,a,s,J]} \equiv &&<0\,\,{\rm in} | \hat{\cal S}_0 | 0\,\,{\rm in} >
\nonumber\\ = &&
\int [d\pi ] {\rm exp}\bigg(
{i{\bf I}_Q + i  \int d^4x \pi (x)\cdot J (x)}\bigg)
\nonumber\\ = &&
e^{iZ_0 [v,a,s,0]}
\,\,\,{\rm exp}\bigg({-i\frac 12 \int d^4x d^4 y J^a (x)
G_F^{ab} (x,y) J^b (y)}\bigg)
\label{D4}
\ee
with
\be
Z_0 [v,a,s,0] = &&-i\,\, {\rm ln} (
\frac{{\rm det} (-\Box -m_{\pi}^2 +i0 )}
     {{\rm det} (-\Box -m_{\pi}^2 -{\bf K} +i0 )} \bigg)^{\frac 12}
\nonumber\\= &&
+\frac i2 {\rm Tr}{\rm ln} \bigg( 1-{\bf K} \Delta_F \bigg) =
-\frac i2 \sum_{n=1}^{\infty} \frac 1{n} {\rm Tr} ({\bf K}\Delta_F )^n
\label{D5}
\ee
where $G_F$ is the Green's function
\be
\bigg(
-\nabla^{\mu}\nabla_{\mu} + {\ub a}^{\mu} {\ub a}_{\mu} - m_{\pi}^2 -s \bigg)
G_F (x, y)  = \delta^4 (x-y) {\bf 1}
\label{D6}
\ee
with Steuckelberg-Feynman boundary conditions.

Equations (\ref{D4}-\ref{D5}) imply that $Z_0 [v,a,s, J]$ gives the Green's
functions of ${\bf j}_{V\mu}^a$, ${\bf j}_{A\mu}^a$, and $\hat\sigma$ to
one-loop, while Green's functions involving $\pi$ are given to tree level. This
is awkward, but since Green's functions for $\pi$ may be expressed in terms of
${\bf j}_{V\mu}^a$, ${\bf j}_{A\mu}^a$, and $\hat\sigma$ as in section 4, this
is not a fundamental difficulty. $Z_0 [v,a,s , 0]$ is then the desired one-loop
effective action.

In (\ref{D4}-\ref{D5})  $s$ and $a^a_{\mu}$ appear only in the combination
$\hat s = s{\bf 1} -{\ub a}_{\mu}{\ub a}^{\mu}$. We assume the same is true
after
renormalization. A general analysis without this assumption is given in
Appendix E. With this in mind, we note that local isospin invariance
\be
{\bf X}_V Z_0 [v,a,s,0] =0
\label{D7}
\ee
implies that the sum of graphs with external $v$-legs only is fixed up to one
constant
\be
-\frac {c_1}4 \int d^4x \,\,
{ v}_{\mu\nu} (x)\cdot { v}^{\mu\nu} (x)
\label{D8}
\ee
where $v_{\mu\nu}$ is the field strength
\be
{\ub v}_{\mu\nu} = \bigg[\nabla_{\mu} , \nabla_{\nu} \bigg]  =
\partial_{\mu}{\ub v}_{\nu} -\partial_{\nu} {\ub v}_{\mu} +
[{\ub v}_{\mu} , {\ub v}_{\nu} ].
\label{D9}
\ee
For mixed $v-\hat s$ graphs,  Fig. 6a is proportional to ${\rm Tr} \,\,{\ub
v}_{\mu}(x)\hat s (y) = 0$, whereas the divergences in Fig. 6b and Fig. 6c
cancel. This corresponds to the fact that there are no contact terms which obey
(\ref{D7}) and have the correct index structure. For the remaining graphs, the
possible tadpole of Fig. 6d is eliminated by (\ref{C10}-\ref{C11}), whereas
Fig. 6e gives another constant
\be
\frac {\hat{c}_1}4 \int d^4x\,\, {\rm Tr} \bigg(
\hat{s} (x)\hat{s} (x)\bigg) \,\,\,.
\label{D10}
\ee
The consistency of the  above assumptions may be checked in $several$
ways as we now discuss.

\vskip .3cm
{\it 18.1. Gell-Mann-Oakes-Renner (GOR)}
\vskip .15cm

First, we note that $<0|\hat\sigma
|0>=0$ to one loop is compatible with chiral perturbation theory to one loop.
Indeed, using the notation of Gasser and Leutwyler \cite{leutwyler}
we have \cite{novikov}
\be
\hat{m} <0|\overline u u |0> =
\hat{m} <0|\overline u u |0>_0
\bigg(1-\frac {3M_{\pi}^2}{32\pi^2 F_{\pi}^2} \,\,{\rm ln} M_{\pi}^2\bigg) =
-\frac 12 F^2 M^2
\bigg(1-\frac {3M_{\pi}^2}{32\pi^2 F_{\pi}^2} \,\,{\rm ln} M_{\pi}^2\bigg)
\nonumber\\
\label{D11}
\ee
with \cite{pagels2}
\be
M_{\pi}^2 =  M^2 \bigg( 1+ \frac {M^2}{32\pi^2 F^2} {\rm ln} M^2 + ...\bigg)
\label{D12}
\ee
\be
F_{\pi} =  F \bigg( 1- \frac {M^2}{16\pi^2 F^2} {\rm ln} M^2 + ...\bigg).
\label{D13}
\ee
The chiral logarithms are seen to cancel, leaving to one-loop
\be
{\hat m} <0| \overline u u |0> = -\frac 12 F_{\pi}^2 M_{\pi}^2
\label{D14}
\ee
which is the GOR relation.
At this stage, we should mention that alternatives to the GOR relation have
been advocated by some authors \cite{heretics}. A way to distinguish between
these
various schemes is to increase the accuracy of the S-wave $\pi\pi$ scattering
lengths, or to improve on the empirical analysis of
$\pi\rightarrow e\nu\gamma$, $\pi\rightarrow e\nu\gamma^*$,
$\gamma\gamma\rightarrow \pi\pi$ or $\gamma\pi\rightarrow\gamma\pi$
at threshold.
We note that the GOR relation cannot be tested in  pion-nucleon or
in photon-nucleon processes (to leading order in the charge)
since $<0|\hat\sigma |0>$ drops out from the connected part of the
pertinent amplitudes.

\vskip .3cm
{\it 18.2. KSFR-Relation}
\vskip .15cm

The vanishing of the three point function
$<0|T^*\bigg({\bf j}_{A\alpha}^a (x) {\bf j}_{A\beta}^b (y)
{\bf V}_{\gamma}^c (z) \bigg)|0>$, as well as $<0|\hat\sigma |0>$ to one loop
gives
\be
{\bf F}_V (q^2) = 1+ \frac {q^2}{2f_{\pi}^2} \Pi_V (q^2) + {\cal O}
\bigg(\frac 1{f_{\pi}^4} \bigg)
\label{D15}
\ee
by (\ref{C54}-\ref{C55}). For space-like momenta $q^2<0$, the vector
form-factor of the pion ${\bf F}_V (q^2)$ and the vector-isovector correlator
${\bf \Pi}_V (q^2)$ may be parameterized as
\be
{\bf F}_V (q^2) = \frac {-m_{\rho}^2}{q^2-m_{\rho}^2 +i0}
\label{D16}
\ee
\be
{\bf \Pi}_V (q^2) =-f_{\rho}^2 \frac 1{q^2-m_{\rho}^2 +i0}
\label{D17}
\ee
where
\be
<0 | {\bf V}_{\mu}^a (x) |\rho^b (p) > =
\delta^{ab} \epsilon_{\mu} (p) f_{\rho} m_{\rho} e^{-ip\cdot x}.
\label{D18}
\ee
The result (\ref{D15}) then implies
\be
1 = \frac {f_{\rho}^2}{2f_{\pi}^2} +{\cal O}\bigg( \frac 1{f_{\pi}^4} \bigg)
\label{D19}
\ee
which is essentially the KSFR relation \cite{KSFR}.
With $f_{\pi}=93 $ MeV and $f_{\rho} =144$ MeV, we find
\be
1=1.20 +{\cal O}\bigg( \frac 1{f_{\pi}^4} \bigg).
\label{D20}
\ee
The $20\%$ discrepancy is somewhat large and may be due to the finite width of
the $\rho$ as well as the strong presence of the $\rho$' resonance in
$\Pi_V (q^2)$.

To bring the above result to a more conventional form, we may introduce the
transition amplitude $i{\cal T}$ for the
$\rho^c (p) \rightarrow \pi^a (k)\pi^b
(q)$ decay. By the mass-shell conditions $k^2=q^2=m_{\pi}^2$, $p^2=m_{\rho}^2$,
and energy-momentum conservation $p=k+q$, all inner products of the four
vectors $p, k, q$ are fixed. Isospin invariance implies that $i{\cal T}$ is
proportional to $\epsilon^{abc}$, so it is antisymmetric under
$k\leftrightarrow q$ by Bose statistics. Since the amplitude is linear in the
polarization vector $\epsilon_{\mu} (q)$ of the $\rho$, it follows that
\be
i{\cal T} \bigg( \pi^a (k) \pi^b (q) \leftarrow \rho^c (p) \bigg) =
g_{\rho\pi\pi} \epsilon^{abc} (k-q)\cdot \epsilon (p)
\label{D21}
\ee
which defines the $\rho\pi\pi$ coupling constant $g_{\rho \pi\pi}$.
By (\ref{D18}) ${\bf
V}_{\mu}^a /m_{\rho} f_{\rho}$ may be used as an interpolating field for the
the $\rho$. The WLSZ reduction formula and current conservation imply
\be
&&<0| a_{\rm in}^a (k) a_{\rm in}^b (q) {\bf S} {\bf V}_{\mu}^c (y) |0> =
\nonumber\\&&
f_{\rho} m_{\rho} g_{\rho\pi\pi} i\epsilon^{acb} (k-q)_{\mu}
\frac 1{(k+q)^2 -m_{\rho}^2 +i0} e^{i (k+q)\cdot y} + ...\,.
\label{D22}
\ee
Comparison with the crossed version of (\ref{C33}) and (\ref{D16}) yields
\be
f_{\rho} g_{\rho\pi\pi} = -m_{\rho}.
\label{D23}
\ee
Hence (\ref{D19}) is equivalent to
\be
m_{\rho}^2 = 2 g_{\rho\pi\pi}^2 f_{\pi}^2 +
{\cal O}\bigg( \frac 1{f_{\pi}^2} \bigg)
\label{D24}
\ee
which is the conventional form of the KSFR relation.

\vskip .3cm
{{\it 18.3. On-Shell Radiative Decay :} $\pi\rightarrow e\nu\gamma$}
\vskip .15cm

A third consequence of our assignment is that the one-loop contribution to the
radiative decay of the pion $\pi\rightarrow e\nu\gamma$ vanishes.
It follows that to the same order the structure dependent axial-vector form
factor vanishes. To see this
consider the uncontracted amplitude (\ref{C65}) with $a=3$
\be
{\cal T}_{\nu  \mu }^{bc} (p,q) =
\int d^4x e^{iq\cdot x}
<0| T^* \bigg( {\bf V}_{\mu}^3 (x) {\bf A}_{\nu}^b (0)\bigg)
a_{\rm in}^{c\dagger} (p)  |0>
\label{ZZ1}
\ee
Covariance, implies the general decomposition
\be
{\cal T}_{\nu \mu }^{bc} (p,q) =
\epsilon^{3bc} \bigg( A (\nu ) g_{\mu\nu} + B (\nu ) q_{\nu}p_{\mu}
+ C(\nu ) q_{\nu} q_{\mu} + D (\nu ) q_{\mu} p_{\nu} + E(\nu ) p_{\nu}
p_{\mu}\bigg)
\label{ZZ2}
\ee
with $\nu= p\cdot q$, $p^2=m_{\pi}^2$ and $q^2=0$. The structure dependent form
factors $A,B, ...,E$ are constrained by gauge invariance $q^{\mu}{\cal
T}^{bc}_{\nu\mu} =0$.  Because of the Adler theorem, there is no pion-pole
contribution to $A$. The non-pole contribution to (\ref{ZZ1}) follows from
(\ref{C67}) in the form
\be
{\overline{\cal T}}_{\nu\mu}^{bc} (p,q) = &&+
\epsilon^{3bc} g_{\mu\nu} \bigg( f_{\pi} -<\hat\sigma > \bigg)\nonumber\\
&&+\epsilon^{3dc} \frac i{f_{\pi}}\int d^4y e^{-ik\cdot y} <0 |T^*\bigg( {\bf
j}_{A\mu}^d (y) {\bf j}_{A\nu}^b (0) \bigg) |0> \nonumber\\
&&-\frac i{f_{\pi}} p^{\gamma} \int d^4x d^4z e^{iq\cdot x-ip\cdot z}
<0|T^*\bigg( {\bf V}_{\mu}^3 (x) {\bf j}_{A\gamma}^c (z) {\bf j}_{A\beta }^b
(0) \bigg) |0>
\label{ZZ3}
\ee
Setting
\be
i\int d^4y e^{-ik\cdot y} <0 |T^*\bigg( {\bf
j}_{A\mu}^d (y) {\bf j}_{A\nu}^b (0) \bigg) |0> =
\delta^{bd} ( -g_{\mu\nu} k^2 +k_{\mu}k_{\nu} ) {\bf \Pi}_A (k^2)
\label{ZZ4}
\ee
with $k=p-q$, and using (\ref{ZZ2}-\ref{ZZ4}) we obtain
\be
{A} (\nu ) = f_{\pi} -<\hat\sigma > -\frac {k^2}{f_{\pi}}
{\bf\Pi}_A(k^2)
\label{ZZ5}
\ee
Empirically, the axial structure function (\ref{ZZ5}) is fitted to
$A(\nu) = f_{\pi} -\nu a (\nu )$ with finite $a (0)$. This implies first that
\be
\frac{<\hat\sigma >}{f_{\pi}} = -\frac{m_{\pi}^2}{f_{\pi}^2} {\bf \Pi}_A
(m_{\pi}^2 ) \sim -\frac {m_{\pi}^2}{m_A^2} \sim - 0.01
\label{ZZ6}
\ee
suggesting a 1 \% deviation from the GOR result,
and second that
\be
a(0)= -\frac 2{f_{\pi}} {\bf \Pi}_A (0)
\label{ZZ7}
\ee
in agreement with the result derived by Terentev  using
the soft pion limit \cite{terentev}.
Since to one-loop the one-pion subtracted correlator
(\ref{ZZ4}) does not acquire a transverse part, we conclude that to the same
order the structure dependent axial-vector form factor vanishes, $i.e.$
$a(0)=0$.

We note that $\rho$ Vector Meson Dominance
(VMD) as implied by our counting rules, does not necessarily mean $a_1$
VMD. Indeed, we can write down effective lagrangians
which realize chiral symmetry by $\rho\rightarrow \rho\pi$ rather than
$\rho\rightarrow a_1$. Consistency with the chiral counting arguments
presented above implies that the $a_1$ will only appear at two-loops
or higher since $a_1\sim 3\pi$. In this way, we think that the chiral duality
arguments as used in chiral perturbation theory \cite{raphael}
should only make use of the rho effects to one-loop for consistency.
This is more so since the axial-vector correlation function does not
generate a transverse part to one-loop.

We also note that in the soft pion limit,
the structure-dependent axial form factor obeys the Das-Mathur-Okubo
theorem \cite{das}
\be
a( 0 ) = \frac 1{f_{\pi}} \int_0^{\infty} \frac {ds}{s}
\bigg( \rho_A (s) -\rho_V (s) \bigg) + \frac 13 f_{\pi} <r^2>_V
\label{R4}
\ee
where $<r^2>_V =6{\bf F}_V' (0)$ is the pion-isovector charge radius.
The spectral densities in (\ref{R4}) are understood in the chiral limit.
They are given by
\be
\rho_V (s) =&& \frac 1{\pi} {\rm Im} {\Pi}_V (s)\nonumber\\
\rho_A (s) =&& \frac 1{\pi} {\rm Im} {\Pi}_A (s)
\label{RX4}
\ee
Using resonance saturation in the zero width approximation
\be
&&\rho_V (s) = f_{\rho}^2 \delta (s -m_{\rho}^2 )\nonumber\\&&
\rho_A (s) = f_{A}^2 \delta (s -m_{A}^2 )
\label{R5}
\ee
along with the two Weinberg sum rules \cite{weinberg1}
\be
\int_0^{\infty} ds \bigg(\rho_V (s) -\rho_A (s) \bigg) =&&
f_{\pi}^2 +{\cal O} (m_{\pi}^2 )\nonumber\\
\int_0^{\infty} ds\,\,s \bigg(\rho_V (s) -\rho_A (s) \bigg) =&& {\cal O}
(m_{\pi^2})
\label{RX5}
\ee
that is $m_A^2=2m_{\rho}^2$ and $f_A^2=f_{\pi}^2$ respectively,
and the KSFR relation $f_{\rho}^2=2f_{\pi}^2$, we have
\be
f_{\pi} a(0) = \bigg( \frac{f_A^2}{m_A^2} -\frac{f_{\rho}^2}{m_{\rho}^2}\bigg)
+\frac 13 f_{\pi}^2 <r^2>_V = \frac{f_A^2}{m_A^2}
\label{RY5}
\ee
We have used $<r^2>_V=6/m_{\rho}^2$ in the VDM limit (see below).
This clearly shows that if the axial spectral density is set to zero at
one-loop as it should, then $a(0) =0$ in agreement with (\ref{ZZ7}).

Dominguez and Sola \cite{sola} have estimated $a(\nu =0)$ using (\ref{R4}),
duality and the empirical vector and axial spectral functions as measured from
semileptonic decays $\tau\rightarrow \nu_{\tau} +n\pi$ with $n=\,\, even$
(vector) and $n=\,\,odd$ (axial) up to the kinematical phase space limit
$t\sim 3$ GeV$^2$. Using $<r^2>_V = 0.44\pm 0.03$ fm$^2$, they have
found $2f_{\pi} a(0) = 0.017\pm 0.001\pm 0.004$. Since the $a_1$
contributes empirically to the isovector-axial correlator, this result
provides an upper bound to ours. Finally, the ratio
$\gamma  = {a (\nu =0 )}/{F (\nu = 0)}$
of the structure dependent axial-form factor to the structure-dependent vector
form factor has been measured, with $F(\nu =0) = 0.0265$ following from
$\pi^0\rightarrow \gamma\gamma$ decay  \cite{pidecay}.
Two values of $\gamma$ have been reported :
$\gamma = 0.44 \pm 0.12$  and $\gamma =-2.36\pm 0.12$ \cite{pidecay}.
Our one-loop result $\gamma =0$ is incompatible with the second result.
The possible relationship between this result and the charged pion
polarisability will be discussed below.

\vskip .3cm
{{\it 18.4. Off-Shell Radiative Decay :} $\pi\rightarrow e\nu\gamma^*$}
\vskip .15cm

A fourth consequence of our assumption is that the off-shell radiative decay
described by (\ref{ADD15}) simplifies to

\be
-\frac i{f_{\pi}}\epsilon^{abc}
\bigg(g_{\nu}^{\beta} -\frac{k_{\nu}k^{\beta}}{k^2-m_{\pi}^2}\bigg)
\bigg(-g_{\mu\beta} q^2 +q_{\mu} q_{\beta}\bigg) \,\,\Pi_V (q^2)
\label{ADD16}
\ee
and indeed vanishes for $q^2=0$. This result is related to the second
axial-vector structure form factor $R$ as measured at SIN
\cite{sin}. From (\ref{ADD16}) we have $R= \sqrt{2} m_{\pi} \Pi_V (0)/f_{\pi}$.
Using $\Pi_V (0)= 2f_{\pi}^2 F_V ' (0)$ with $F_V ' (0) = <r^2>_V/6$, we
have $R=\sqrt{2}m_{\pi}f_{\pi} <r^2>_V/3$. With the empirical value for the
isovector radius of the pion quoted above, we obtain
$R = 0.069\pm 0.005$ in very good agreement with
$R({\rm exp}) = 0.06\pm 0.02$ as measured at SIN
\cite{sin}. Our result  agrees with  the soft pion  result \cite{rpcac}.

\vskip .3cm
{{\it 18.5. Pion-Pion Scattering :} $\pi\pi\rightarrow \pi\pi$}
\vskip .15cm

A fifth consequence of our assumption is the structure of the $\pi\pi$
scattering amplitude. With our assignments, the one-loop
contribution to (\ref{C64}) is contained entirely in
\be
Z_0\rightarrow \frac 14 \int d^4x d^4y {\rm Tr} \bigg(\hat{s} (x)\hat{s}
(y)\bigg)
\,\,\int \frac {d^4q}{(2\pi )^4} e^{-iq\cdot(x-y)} \bigg(\hat{c}_1 +{\cal J}
(q^2)\bigg)
\label{D25}
\ee
where
\be
{\cal J} (q^2) =&&
-i\int \frac {d^4k}{(2\pi )^4}
\bigg( \frac 1{k^2-m_{\pi}^2 +i0} \frac 1{(k-q)^2-m_{\pi}^2 +i0} -
(q=p=0) \bigg)
\label{D26}
\ee
{}From Appendix F, we have ($q^2 \geq 4m_{\pi}^2$)
\be
16\pi^2 {\cal J} (q^2) = 2+
\sqrt{1-\frac {4m_{\pi}^2}{q^2}} \,\,\bigg( {\rm ln}
\bigg(\frac{\sqrt{1-4m_{\pi}^2/q^2}-1}{\sqrt{1-4m_{\pi}^2/q^2}+1}\bigg)
+i \pi \bigg)
\label{D261}
\ee
The factor $1/16\pi^2$ is worth some comment, since it suggests that the
expansion parameter is $q^2/16\pi^2 f_{\pi}^2\sim q^2/ 1{\rm GeV}^2$
\cite{georgi}, rather than $q^2/f_{\pi}^2\sim q^2/ 0.01{\rm GeV}^2$, which
differs by a factor of 100.

It follows from (\ref{D25}) that
\be
<0| T^*\bigg(\hat\sigma (x)\hat\sigma (y)\bigg) |0>_{\rm conn} =
-\frac {3i}{2f_{\pi}^2} \int \frac {d^4q}{(2\pi )^4} e^{-iq\cdot (x-y)}
\,\,\bigg( \hat{c}_1 + {\cal J} (q^2) \bigg)
\label{D27}
\ee
\be
&&<0|T^* \bigg( {\bf j}_{A\alpha}^a (y_1) {\bf j}_{A\beta}^b (y_2)\hat\sigma
(y_3)\bigg) |0>_{\rm conn} = \nonumber\\&&
-\frac 2{f_{\pi}} g_{\alpha\beta}\delta^{ab}
 \delta^4 (y_1-y_2) \int \frac{d^4q}{(2\pi)^4}
e^{-iq\cdot(y_1-y_3)} \bigg( \hat{c}_1 +{\cal J} (q^2 )\bigg)
\label{D28}
\ee
\be
&&<0|T^* \bigg( {\bf j}_{A\alpha}^a (y_1) {\bf j}_{A\beta}^b (y_2)
{\bf j}_{A\gamma}^c(y_3) {\bf j}_{A\delta}^d (y_4)\bigg)|0>_{\rm conn} =
i\bigg( 2\delta^{ab}\delta^{cd} +\delta^{ac}\delta^{bd} +\delta^{ad}\delta^{bc}
\bigg) g_{\alpha\beta} g_{\gamma\delta} \nonumber\\&&
\qquad\times\delta^4 (y_1-y_2) \delta^4 (y_3-y_4)
\int \frac{d^4q}{(2\pi)^4}
e^{-iq\cdot(y_1-y_3)} \bigg( \hat{c}_1 +{\cal J} (q^2 )\bigg) +{\rm
{2\,\,perm}}.
\label{D29}
\ee
\be
{\bf F}_S (t) +\frac 1{f_{\pi}} = - \frac 1{f_{\pi}^3}
\bigg( t-\frac {m_{\pi}^2}2\bigg) \bigg( \hat{c}_1 +{\cal J} (t) \bigg)
\label{D30}
\ee
\be
i{\cal T}_{\rm rest} =&&+\bigg( \frac i{f_{\pi}^4} m_{\pi}^2
\delta^{ab}\delta^{cd}
\bigg( 2t -\frac 52 m_{\pi}^2\bigg) \bigg(\hat{c}_1 +{\cal J} (t)\bigg)
\nonumber\\&&
\qquad +\frac i{4f_{\pi}^4}
\bigg( 2\delta^{ab}\delta^{cd} +\delta^{ac}\delta^{bd} +\delta^{ab}\delta^{bc}
\bigg) (t-2m_{\pi}^2)^2 \bigg(\hat{c}_1 +{\cal J} (t)\bigg)\bigg) \nonumber\\&&
+{\rm {2\,\,perm.}} + {\cal O} \bigg(\frac 1{f_{\pi}^6}\bigg)
\label{D31}
\ee
to one loop.
As a check, let us introduce the amplitude $A(s,t,u)$
\be
i{\cal T} = i\delta^{ac}\delta^{bd} A(s,t,u) +
i\delta^{ab}\delta^{cd} A(t,u,s) +
i\delta^{ad}\delta^{bc} A(u,t,s)
\label{D32}
\ee
so that
\be
A_{\rm tree} (s,t,u) = \frac 1{f_{\pi}^2} (s-m_{\pi}^2)
\label{D33}
\ee
\be
A_{\rm rho} (s,t,u) = &&+\frac 1{f_{\pi}^2} (s-u)
\bigg( {\bf F}_V (t) - 1 -\frac{t}{4f_{\pi}^2} {\bf \Pi}_V (t) \bigg)
\nonumber\\&&+\frac 1{f_{\pi}^2} (s-t)
\bigg( {\bf F}_V (u) - 1 -\frac{u}{4f_{\pi}^2} {\bf \Pi}_V (u) \bigg)
\label{D34}
\ee
\be
A_{\rm rest} (s,t,u) =&& +\frac 1{2f_{\pi}^4} (s^2-m_{\pi}^4)
\bigg(\hat{c}_1 +{\cal J} (s) \bigg)
+\frac 1{4f_{\pi}^4} (t-2m_{\pi}^2)^2
\bigg(\hat{c}_1 +{\cal J} (t) \bigg)\nonumber\\&&
+\frac 1{4f_{\pi}^4} (u-2m_{\pi}^2)^2
\bigg(\hat{c}_1 +{\cal J} (u) \bigg) +{\cal O} \bigg(\frac 1{f_{\pi}^6}\bigg)
\label{D35}
\ee

To one loop
\be
{\bf \Pi}_V (q^2) = c_1 +\frac 1{72\pi^2} + \frac 13 \bigg(
1-\frac{4m_{\pi}^2}{q^2}\bigg) {\cal J} (q^2)
\label{D36}
\ee
where we have adopted the convention that $c_1=\hat{c}_1=0$ with the BPHZ
subtraction scheme. Then
\be
{\bf F}_V (q^2) = 1 + \frac 1{2f_{\pi}^2}
\bigg( c_1 q^2 + \frac {q^2}{72\pi^2} + \frac 13 \bigg(
q^2-{4m_{\pi}^2}\bigg) {\cal J} (q^2)\bigg)
\label{D37}
\ee
by (\ref{D15}). It follows that
\be
A_{\rm rho} (s,t,u) + A_{\rm rest} (s,t,u) = B(s,t,u) + C(s,t,u)
\label{D38}
\ee
to one loop, where
\be
B(s,t,u) =&&+ \frac 1{2f_{\pi}^4} (s^2-m_{\pi}^4) {\cal J} (s) \nonumber\\&&+
\frac 1{6f_{\pi}^4}\bigg( t(t-u) -2m_{\pi}^2 t + 4m_{\pi}^2 u -2m_{\pi}^4
\bigg) {\cal J} (t) \nonumber\\&&
+\frac 1{6f_{\pi}^4}\bigg( u(u-t) -2m_{\pi}^2 u + 4m_{\pi}^2 t -2m_{\pi}^4
\bigg) {\cal J} (u)
\label{D39}
\ee
\be
C(s,t,u) =&& +\frac 1{4f_{\pi}^4}\bigg( c_1 +\frac 1{72\pi^2} \bigg)
\bigg( (s-u)t + (s-t)u\bigg) \nonumber\\&&
+\frac 1{2f_{\pi}^4}\hat{c}_1 (s^2-m_{\pi}^4) +\frac 1{4f_{\pi}^4}\hat{c}_1
\bigg( (t-2m_{\pi}^2)^2 + (u-2m_{\pi}^2)^2\bigg) \nonumber\\
=&&-\frac 1{2f_{\pi}^4} \bigg( c_1 +\frac 1{72\pi^2}\bigg) (s-2m_{\pi}^2)^2
+\frac 1{2f_{\pi}^4} \hat{c}_1 (s^2-m_{\pi}^4) \nonumber\\&&
+\frac 1{8f_{\pi}^4} \bigg( c_1 + \frac 1{72\pi^2} + \hat{c}_1\bigg)
\bigg( (t-2m_{\pi}^2)^2 + (u-2m_{\pi}^2)^2 \bigg).
\label{D40}
\ee
The structure of $B(s,t,u)$ is dictated by unitarity and is in agreement with
previous analyses by various authors.

Amplitudes with fixed isospin $I$ in the $s$ channel are given by
\be
&&T^0 (s,t) = 3 A(s,t,u) + A(t,u,s) + A (u,s,t)\nonumber\\&&
T^1 (s,t) = A(t,u,s) - A (u,s,t)\nonumber\\&&
T^2 (s,t) = A(t,u,s) + A (u,s,t).
\label{D41}
\ee
In the region of elastic unitarity $4m_{\pi}^2<s<16m_{\pi}^2$, $T^I(s,t)$
can be expressed in terms of phase shifts $\delta_l^I (s)$ as
\be
T^I(s,t) =32\pi \sum_{l=0}^{\infty} (2l+1) P_l ({\rm cos} \theta )
\bigg( 1+ \frac {m_{\pi}^2}{Q^2}\bigg)^{\frac 12}
\,\,e^{i\delta_l^I (s)} {\rm sin} (\delta_l^I (s) )
\label{D42}
\ee
where $P_l ({\rm cos}\theta )$ are the Legendre polynomials, $\theta$
is the scattering angle, and $Q$ is the momentum in the center of mass frame
\be
&&
s= + 4(Q^2+m_{\pi}^2 )\nonumber\\&&
t= -2 Q^2(1-{\rm cos}\theta )\nonumber\\&&
u= -2 Q^2(1+{\rm cos} \theta ).
\label{D43}
\ee
Bose symmetry implies that the $I=0,2$ channels contain only even $l$, while
the $I=1$ channel contains only odd $l$.

The phase shifts at threshold are parameterized as follows
\be
\bigg( 1+ \frac {m_{\pi}^2}{Q^2}\bigg)^{\frac 12}
\,\,{\rm Re}\bigg( e^{i\delta_l^I (s)} {\rm sin} (\delta_l^I (s) )\bigg) =
Q^{2l} \bigg( a_l^I + b_l^I Q^2 + {\cal O} (Q^4) \bigg).
\label{D44}
\ee
We will refer to the $a$'s as the scattering lengths, and $b$'s as the range
parameters.
It follows that the tree term (\ref{D33}) contributes only to S and P waves,
\be
&&
a_0^0 ({\rm tree} ) =+ \frac {7m_{\pi}^2}{32\pi f_{\pi}^2} =+ 0.16\nonumber\\&&
b_0^0 ({\rm tree} ) =+ \frac {1}{4\pi f_{\pi}^2} =+ 0.18
\,\,m_{\pi}^{-2}\nonumber\\&&
a_1^1 ({\rm tree} ) =+ \frac {1}{24\pi f_{\pi}^2} =+ 0.030
\,\,m_{\pi}^{-2}\nonumber\\&&
a_0^2 ({\rm tree} ) = -\frac {m_{\pi}^2}{16\pi f_{\pi}^2} = -0.045
\nonumber\\&&
b_0^2 ({\rm tree} ) = -\frac {1}{8\pi f_{\pi}^2} = -0.089\,\, m_{\pi}^{-2}.
\label{D45}
\ee
Experimentally, the phase shifts $\delta_l^I (s)$ are measurable from the final
state interaction in $K^-\rightarrow \pi^+\pi^- e^- \overline{\nu}$ decay
and also from $\pi p\rightarrow \pi\pi p$ (see above)
as well as $\pi p\rightarrow \pi\pi\Delta$
decays by extrapolation to the pion  pole in the $t$-channel.
The results are \cite{petersen}
\be
&&
a_0^0 ({\rm exp })= 0.26 \pm 0.05\nonumber\\&&
b_0^0 ({\rm exp}) = ( 0.25 \pm 0.03 )\,\, m_{\pi}^{-2}\nonumber\\&&
a_1^1 ({\rm exp}) = ( 0.038 \pm 0.002 ) \,\,m_{\pi}^{-2}\nonumber\\&&
(2a_0^0 -5a_0^2) ({\rm exp}) = 0.614 \pm 0.0028  \nonumber\\&&
b_0^2 ({\rm exp}) = (- 0.082 \pm 0.008 ) \,\,m_{\pi}^{-2}\nonumber\\&&
a_2^0 ({\rm exp}) = ( 17 \pm 3  )10^{-4} \,\,m_{\pi}^{-4}\nonumber\\&&
a_2^2 ({\rm exp}) = ( 1.3 \pm 3 ) 10^{-4} \,\,m_{\pi}^{-4}.
\label{D46}
\ee
where $m_{\pi}$ is taken to be the charged pion mass $139.6$ MeV.
The tree level predictions are rather off, for instance $40\%$ for $a_0^0$.
This means that the role of the scalar correlation function
following from (\ref{C64}) may be important.
This point is presently under investigation \cite{jim}.

For the $\rho$ contribution the expressions
(\ref{D16}-\ref{D17}) are appropriate for $q^2=t$
or $u$, but not for $q^2=s$, since they do not have the proper threshold
behavior. This point may be corrected. Since we are working only to one-loop,
we may simply use (\ref{D36}-\ref{D37}), so that
\be
A_{\rm rho} (s,t,u) = \frac 1{4 f_{\pi}^4} (s-u) t {\bf \Pi}_V (t) +
\frac 1{4 f_{\pi}^4} (s-t) u {\bf \Pi}_V (u) .
\label{D47}
\ee

The necessary parameter is then $c_1 ={\bf \Pi}_V (0) =
f_{\rho}^2/m_{\rho}^2 = 0.035$. Alternativaly, we may take $c_1=2f_{\pi}^2 {\bf
F}'_V (0) = 0.033 \pm 0.002$ where we have used $<r^2>_V = 6{\bf F}'_V (0)
=0.439 \pm 0.03$ fm$^2$. For definiteness we choose the former.
Expanding in $Q$ gives
\be
A_{\rm rho} (s,t,u) = &&+\frac 1{4f_{\pi}^4} (s-u) t {\bf \Pi}_V (0) +
\frac 1{4f_{\pi}^4} (s-t) u {\bf \Pi}_V (0) \nonumber\\&&
+\frac 1{4f_{\pi}^4} 4m_{\pi}^2 {\bf \Pi}'_V (0) t^2 +
\frac 1{4f_{\pi}^4} 4m_{\pi}^2 {\bf \Pi}'_V (0) u^2 + {\cal O} (Q^6
)\nonumber \\ =&&+
\frac 1{f_{\pi}^4} \frac {f_{\rho}^2}{m_{\rho}^2}
\bigg( -6Q^4 + 2 Q^4 {\rm cos}^2 \theta -4m_{\pi}^2 Q^2 \bigg) \nonumber\\ &&
+\frac 1{60\pi^2 f_{\pi}^4} Q^4 (1+{\rm cos}^2 \theta ) +{\cal O} (Q^6 )
\label{D48}
\ee
\be
A_{\rm rho} (t,u,s) = &&+\frac 1{4f_{\pi}^4} (t-u) s {\bf \Pi}_V (4m_{\pi}^2) +
\frac 1{4f_{\pi}^4} 4m_{\pi}^2 {\bf \Pi}'_V (4m_{\pi}^2) (s-4m_{\pi}^2) (t-u)
\nonumber\\&&
+\frac 1{4f_{\pi}^4} u (t-s) {\bf \Pi}_V (0)  +
\frac 1{4f_{\pi}^4} u^2 (-4m_{\pi}^2 ){\bf \Pi}'_V (0)  + {\cal O} (Q^6
)\nonumber \\ =&& +
\frac 1{f_{\pi}^4} \frac {f_{\rho}^2}{m_{\rho}^2}
\bigg( +3Q^4 + 6Q^4 {\rm cos} \theta -Q^4 {\rm cos}^2 \theta +2m_{\pi}^2 Q^2
+ 6m_{\pi}^2 Q^2 {\rm cos} \theta \bigg) \nonumber\\ &&
+\frac 1{18\pi^2 f_{\pi}^4} (Q^4 +m_{\pi}^2 Q^2) {\rm cos} \theta
+\frac 1{6\pi^2f_{\pi}^4} Q^4 {\rm cos}\theta \nonumber\\&&
-\frac 1{120\pi^2 f_{\pi}^4} Q^4
(1+{\rm cos}\theta )^2 + {\cal O} (Q^6 )
\nonumber\\
\label{D49}
\ee
\be
A_{\rm rho} (u,t,s) = && +
\frac 1{f_{\pi}^4} \frac {f_{\rho}^2}{m_{\rho}^2}
\bigg( +3Q^4 - 6Q^4 {\rm cos} \theta -Q^4 {\rm cos}^2 \theta +2m_{\pi}^2 Q^2
- 6m_{\pi}^2 Q^2 {\rm cos} \theta \bigg) \nonumber\\ &&
-\frac 1{18\pi^2 f_{\pi}^4} (Q^4 +m_{\pi}^2 Q^2) {\rm cos} \theta
-\frac 1{6\pi^2f_{\pi}^4} Q^4 {\rm cos}\theta \nonumber\\&&+
\frac 1{120\pi^2 f_{\pi}^4} Q^4
(1-{\rm cos}\theta )^2 + {\cal O} (Q^6 ).
\nonumber\\
\label{D50}
\ee
Therefore, the $\rho$ contribution to the scattering lengths is given by
\be
&&
a_0^0 ({\rm rho })= 0 \nonumber\\&&
b_0^0 ({\rm rho} ) = -\frac {m_{\pi}^2}{4\pi f_{\pi}^4} \frac{f_{\rho
}^2}{m_{\rm \rho}^2} = -0.014 \,m_{\pi}^{-2} \nonumber\\&&
a_2^0 ({\rm rho} ) = +\frac {1}{60\pi f_{\pi}^4} \frac{f_{\rho
}^2}{m_{\rho}^2} + \frac {1}{7200\pi^3 f_{\pi}^4}
 = +9.3 \,\,10^{-4}\, m_{\pi}^{-4} \nonumber\\&&
a_1^1 ({\rm rho} ) = +\frac {m_{\pi}^2}{8\pi f_{\pi}^4} \frac{f_{\rho
}^2}{m_{\rho}^2} + \frac {m_{\pi}^2}{864\pi^3 f_{\pi}^4}
 = +0.0070 \,m_{\pi}^{-2} \nonumber\\&&
a_0^2 ({\rm rho} ) = 0\nonumber\\&&
b_0^2 ({\rm rho} ) = +\frac {m_{\pi}^2}{8\pi f_{\pi}^4} \frac{f_{\rho
}^2}{m_{\rho}^2}  = +0.0072 \,m_{\pi}^{-2} \nonumber\\&&
a_2^2 ({\rm rho} ) = -\frac {1}{120\pi f_{\pi}^4} \frac{f_{\rho
}^2}{m_{\rho}^2} -\frac {1}{14400\pi^3 f_{\pi}^4}
 = -4.7 \,\,10^{-4}\, m_{\pi}^{-4}
\label{D51}
\ee
Also
\be
{\rm Re}\bigg( A_{\rm rest} (s,t,u) \bigg) = &&+
\frac 1{f_{\pi}^4} Q^4 \bigg(\frac {32}3 \hat{c}_1 -\frac{19}{40\pi^2}\bigg) +
\frac 1{f_{\pi}^4} Q^4 \bigg( {\rm cos}^2 \theta -\frac 13 \bigg)
\bigg( 2\hat{c}_1 -\frac 3{40\pi^2} \bigg)\nonumber\\&&
+\frac 1{f_{\pi}^4} m_{\pi}^2 Q^2 \bigg(20 \hat{c}_1 +\frac {49}{48\pi^2}\bigg)
+\frac 1{f_{\pi}^4} m_{\pi}^4  \bigg( \frac{19}{2} \hat{c}_1 +
\frac {15}{16\pi^2}\bigg) +{\cal O} (Q^6)
\nonumber\\
\label{D52}
\ee
\be
{\rm Re}\bigg( A_{\rm rest} (t,u,s) \bigg) = &&+
\frac 1{f_{\pi}^4} Q^4 \bigg( 8 \hat{c}_1 +\frac{11}{360\pi^2}\bigg) +
\frac 1{f_{\pi}^4} Q^4 {\rm cos}\theta
\bigg(- 2\hat{c}_1 -\frac {17}{240\pi^2} \bigg)\nonumber\\&&
+\frac 1{f_{\pi}^4} Q^4 \bigg({\rm cos}^2\theta -\frac 13\bigg)
\bigg( 3\hat{c}_1 -\frac {19}{480\pi^2} \bigg)
+\frac 1{f_{\pi}^4} m_{\pi}^2 Q^2 \bigg(6 \hat{c}_1 +\frac {35}{96\pi^2}\bigg)
\nonumber\\&&
+\frac 1{f_{\pi}^4} m_{\pi}^2 Q^2 {\rm cos}\theta
\bigg( 2 \hat{c}_1 -\frac {1}{32\pi^2}\bigg)
+\frac 1{f_{\pi}^4} m_{\pi}^4 \bigg( \frac 32 \hat{c}_1 +
\frac {1}{8\pi^2}\bigg)
+{\cal O} (Q^6)
\nonumber\\
\label{D53}
\ee
\be
{\rm Re}\bigg( A_{\rm rest} (u,t,s) \bigg) = &&+
\frac 1{f_{\pi}^4} Q^4 \bigg( 8 \hat{c}_1 +\frac{11}{360\pi^2}\bigg) +
\frac 1{f_{\pi}^4} Q^4 {\rm cos}\theta
\bigg( 2\hat{c}_1 +\frac {17}{240\pi^2} \bigg)\nonumber\\&&
+\frac 1{f_{\pi}^4} Q^4 \bigg({\rm cos}^2\theta -\frac 13\bigg)
\bigg( 3\hat{c}_1 -\frac {19}{480\pi^2} \bigg)
+\frac 1{f_{\pi}^4} m_{\pi}^2 Q^2 \bigg(6 \hat{c}_1 +\frac {35}{96\pi^2}\bigg)
\nonumber\\&&
+\frac 1{f_{\pi}^4} m_{\pi}^2 Q^2 {\rm cos}\theta
\bigg( -2 \hat{c}_1 +\frac {1}{32\pi^2}\bigg)
+\frac 1{f_{\pi}^4} m_{\pi}^4 \bigg( \frac 32 \hat{c}_1 +
\frac {1}{8\pi^2}\bigg)
+{\cal O} (Q^6)
\nonumber\\
\label{D54}
\ee
so that
\be
&&
a_0^0 ({\rm rest }) = \frac {63m_{\pi}^4}{64\pi f_{\pi}^4} \hat{c}_1 +
\frac {49 m_{\pi}^4}{512 \pi^3 f_{\pi}^4} \nonumber\\&&
b_0^0 ({\rm rest }) = \frac {9m_{\pi}^2}{4\pi f_{\pi}^4} \hat{c}_1 +
\frac {91 m_{\pi}^2}{768 \pi^3 f_{\pi}^4} \nonumber\\&&
a_2^0 ({\rm rest }) = \frac {1}{20\pi f_{\pi}^4} \hat{c}_1
-\frac {73 }{57600 \pi^3 f_{\pi}^4} \nonumber\\&&
a_1^1 ({\rm rest }) = \frac {m_{\pi}^2}{24\pi f_{\pi}^4} \hat{c}_1 -
\frac {m_{\pi}^2}{1536 \pi^3 f_{\pi}^4} \nonumber\\&&
a_0^2 ({\rm rest }) = \frac {3m_{\pi}^4}{32\pi f_{\pi}^4} \hat{c}_1 +
\frac {m_{\pi}^4}{128 \pi^3 f_{\pi}^4} \nonumber\\&&
b_0^2 ({\rm rest }) = \frac {3m_{\pi}^2}{8\pi f_{\pi}^4} \hat{c}_1 +
\frac {35m_{\pi}^2}{1536 \pi^3 f_{\pi}^4} \nonumber\\&&
a_2^2 ({\rm rest }) = \frac {1}{40\pi f_{\pi}^4} \hat{c}_1 -
\frac {19}{57600 \pi^3 f_{\pi}^4} .
\label{D55}
\ee
It follows that $\hat{c}_1$ can be determined in seven ways
\be
&&
16\pi^2 \hat{c}_1 = 16\pi^2 \frac{64\pi f_{\pi}^4}{63 m_{\pi}^4}
\bigg( a_0^0 ({\rm exp}) -a_0^0 ({\rm tree})\bigg) -\frac {14}{9} =
8\pm 5 \nonumber\\&&
16\pi^2 \hat{c}_1 = 16\pi^2 \frac{4\pi f_{\pi}^4}{9 m_{\pi}^2}
\bigg( b_0^0 ({\rm exp}) -b_0^0 ({\rm tree}) -b_0^0 (\rm rho )\bigg)
-\frac {91}{108} = 3\pm 1 \nonumber\\&&
16\pi^2 \hat{c}_1 = 16\pi^2 {20\pi f_{\pi}^4}
\bigg( a_2^0 ({\rm exp}) -a_2^0 (\rm rho )\bigg) +\frac {73}{180} =
2\pm 1 \nonumber\\&&
16\pi^2 \hat{c}_1 = 16\pi^2 \frac{24\pi f_{\pi}^4}{m_{\pi}^2}
\bigg( a_1^1 ({\rm exp}) -a_1^1 ({\rm tree}) -a_1^1 (\rm rho )\bigg)
+\frac 14 =
2\pm 5 \nonumber\\&&
16\pi^2 \hat{c}_1 = 16\pi^2 \frac{32\pi f_{\pi}^4}{3 m_{\pi}^4}
\bigg( a_0^2 ({\rm exp}) -a_0^2 ({\rm tree})\bigg) -\frac 43 =
26\pm 21 \nonumber\\&&
16\pi^2 \hat{c}_1 = 16\pi^2 \frac{8\pi f_{\pi}^4}{3 m_{\pi}^2}
\bigg( b_0^2 ({\rm exp}) -b_0^2 ({\rm tree}) -b_0^2 (\rm rho )\bigg)
-\frac {35}{36} =
-1\pm 2 \nonumber\\&&
16\pi^2 \hat{c}_1 = 16\pi^2 {40\pi f_{\pi}^4}
\bigg( a_2^2 ({\rm exp}) -a_2^2 (\rm rho)\bigg) +\frac {19}{90} =
3\pm 1 \nonumber\\&&
\label{D56}
\ee
which is seen to be consistent, to the the possible
exception of $16\pi^2\hat{c}_1= -1\pm 2$.
The value $16\pi^2\hat{c}_1= 26\pm 21$ involves a large error bar and should
not be taken seriously.  We note, however, that overall consistency can be
achieved with $1.1$ standard deviation. Alternatively, one can input either a
scattering length or a range parameter to fix $\hat{c}_1$, and predict the
remaining scattering lengths and range parameters.

Finally, we note that at low energies the partial phase shifts have been
usually extracted from the high statistics experiments carried out both at
CERN and Saclay using $K_{e4}$ decays \cite{cernsaclay}. The near future
experiments planned at DA$\Phi$NE are expected to give much better accuracy
\cite{franzini} in the threshold region, allowing for a detailed comparison
between various theoretical proposals. In light of this, we will give
elsewhere a  comprehensive
analysis of the phase shifts near threshold as they follow
from our exact (expanded) results, in comparison with
one-loop chiral perturbation theory, dispersion methods and the data
\cite{jim}.

\vskip .3cm
{{\it 18.6. Fusion Process :} $\gamma\gamma\rightarrow \pi\pi$}
\vskip .15cm

A sixth consequence of our assumptions is the fusion process
$\gamma\gamma \rightarrow \pi\pi$. Specifically, consider
$\gamma (q_1) \gamma (q_2)\rightarrow
\pi (k_1) \pi (k_2)$ in the gauge where the photon polarisabilities satisfy the
condition $\epsilon_{\mu}(q_i ) q^{\mu}_j = 0$, with $i,j=1,2$. The
Mandelstam variables are
\be
&&s = (q_1+q_2)^2 = 2q_1\cdot q_2 \nonumber\\
&&t = (q_1-k_1)^2 = m_{\pi}^2 -2q_1\cdot k_1\nonumber\\
&&u = (q_1-k_2)^2 = m_{\pi}^2 -2q_1\cdot k_2
\label{FIN1}
\ee
Throughout $q_1^2=q_2^2=0$ and $p_1^2=p_2^2=m_{\pi}^2$.
The case of one or two tagged photons will not be discussed here.
Contracting the photon polarisations with (\ref{ADD40}-\ref{ADD43})
and using the mass shell conditions give
\be
&&+i\epsilon_1\cdot \epsilon_2 \bigg( \epsilon^{bce}\epsilon^{eda} +
\epsilon^{bde} \epsilon^{eca} \bigg) \nonumber\\
&&+4i \epsilon_1\cdot\epsilon_2 \bigg(
\frac 1{u-m_{\pi}^2} \epsilon^{bcf}\epsilon^{fda} +
\frac 1{t-m_{\pi}^2} \epsilon^{bdf}\epsilon^{fca} \bigg)\nonumber\\
&&+\frac 1{2f_{\pi}^2} \epsilon_1^{\mu}\epsilon^{\nu}_2 (k_2-k_1)^{\beta}
\epsilon^{abg}\int d^4 y d^4 z e^{-iq_1\cdot y -iq_2\cdot z}
<0| T^*\bigg( {\bf V}_{\mu}^c (y) {\bf V}_{\nu}^d (z)
{\bf V}_{\beta}^g (0) \bigg) |0>
\nonumber\\
&&+\frac 1{f_{\pi}^2} \epsilon_1^{\mu}\epsilon^{\nu}_2 k_1^{\alpha} k_2^{\beta}
\int d^4y d^4z_1 d^4 z_2 e^{-iq_1\cdot y +ik_1\cdot z_1 + ik_2\cdot z_2}
\nonumber\\
&&\times <0| T^*\bigg( {\bf j}_{A\alpha}^a (z_1) {\bf j}_{A\beta}^b (z_2)
{\bf V}_{\mu}^c (y) {\bf V}_{\nu}^d (0) \bigg) |0>
\nonumber\\
&&-\frac 1{f_{\pi}} m_{\pi}^2 \delta^{ab} \epsilon_1^{\mu}\epsilon_2^{\nu}
\int d^4 y d^4 z e^{-iq_1\cdot y -iq_2\cdot z}
<0| T^*\bigg( {\bf V}_{\mu}^c (y) {\bf V}_{\nu}^d (z) \hat\sigma (0)\bigg) |0>
\label{FIN2}
\ee
We observe that on-shell the $a_1$ contribution through $<{\bf j}_A{\bf j}_A>$
drops. The first and second term in (\ref{FIN2}) are the seagull and Born terms
respectively. Typical contributions to the remaining correlators in
(\ref{FIN2}) are shown in Fig. 7.
To estimate the various vacuum correlators entering (\ref{FIN2}) we will
use our one loop effective action.

To one-loop, the contribution to the three-point vector correlator
$<{\bf V} {\bf V} {\bf V} >$ contains a divergent part given by $c_1$ and a
finite part. Using (\ref{D8}) we have
\be
<0| T^*\bigg( {\bf V}_{\mu}^c (y) {\bf V}_{\nu}^d (y)
{\bf V}_{\rho}^e (z) \bigg) |0> =
&& -c_1 g_{\mu\nu}\epsilon^{cde} \delta^4
(x-z) \partial_{\rho} \delta^4 (y-z) \nonumber\\
&& +c_1 g_{\mu\nu}\epsilon^{cde} \delta^4
(y-z) \partial_{\rho} \delta^4 (x-z) \nonumber\\
&&+\epsilon^{cde} \int \frac{d^4p}{(2\pi)^4}\int \frac{d^4q}{(2\pi )^4}
e^{iq\cdot x-ip\cdot y-i(q-p)\cdot z}
{\cal J}_{\mu\nu\rho} (q,p)\nonumber\\
\label{FIN3}
\ee
with
\be
{\cal J}_{\mu\nu\rho} (q,p) =
i\int \frac{d^4k}{(2\pi )^4}
\bigg(&& \frac 1{k^2-m_{\pi}^2 +i0} \frac 1{(k+q)^2-m_{\pi}^2+i0}
          \frac 1{(k+p)^2-m_{\pi}^2+i0}\nonumber\\
&&\times (2k_{\mu}+q_{\mu})(2k_{\nu} +p_{\nu})(2k_{\rho}+(q-p)_{\rho})
+{\rm counterterms} \bigg)\nonumber\\
\label{FIN4}
\ee

The one-loop contribution to the three-point correlator
$<{\bf V}{\bf V} \hat\sigma >$ in (\ref{FIN2}) is finite since the divergences
in Fig. 6b and Fig. 6c cancel out. Thus
\be
&&<0|T^*\bigg( {\bf V}_{\mu}^c (x) {\bf V}_{\nu}^d (y) \hat\sigma (z) \bigg)
|0>_{\rm conn.} =\nonumber\\
&&+\frac 2{f_{\pi}} \delta^{cd} g_{\mu\nu} \delta^4 (x-y) \int
\frac{d^4q}{(2\pi )^4} e^{-iq\cdot (y-z)} {\cal J} (q^2) \nonumber\\
&&+\frac 2{f_{\pi}} \delta^{cd} \int \frac{d^4p}{(2\pi )^4}\int
\frac{d^4q}{(2\pi )^4}
e^{iq\cdot x -ip\cdot y -i (q-p)\cdot z} {\cal J}_{\mu\nu} (q, p)
\label{FIN5}
\ee
where ${\cal J} (q^2)$ is given by (\ref{D26}-\ref{D261})
and ${\cal J}_{\mu\nu} (q, p)$ is defined by
\be
{\cal J}_{\mu\nu} (q,p) =
i\int \frac{d^4k}{(2\pi )^4}
\bigg(&& \frac 1{k^2-m_{\pi}^2+i0 } \frac 1{(k+q)^2-m_{\pi}^2 +i0}
          \frac 1{(k+p)^2-m_{\pi}^2 +i0} \nonumber\\
&&\times (2k_{\mu}+q_{\mu})(2k_{\nu} +p_{\nu})
 - (q=p=0) \bigg)
\label{FIN6}
\ee

Finally, the four-point function $<{\bf j}_A {\bf j}_A {\bf V} {\bf V}>$
is also finite by the same argument as above, since $\hat s = s{\bf
1}- {\underline a}_{\mu} {\underline a}^{\mu}$. Thus, to one-loop
\be
&&<0|T^*\bigg( {\bf j}_{A\alpha}^a (z_1) {\bf j}_{A\beta}^b (z_2)
{\bf V}_{\mu}^c (x) {\bf V}_{\nu}^d (y)  \bigg) |0>_{\rm conn.} =\nonumber\\
&&-i\bigg( 2\delta^{ab}\delta^{cd} +\delta^{ac} \delta^{bd}
+\delta^{ad}\delta^{bc} \bigg) g_{\mu\nu} g_{\alpha\beta}\nonumber\\
&&\times \delta^4 (x-y) \delta^4 (z_1-z_2) \int \frac{d^4q}{(2\pi )^4}
e^{-iq\cdot (y-z_1)} {\cal J} (q^2 ) \nonumber\\
&&-i\bigg( 2\delta^{ab}\delta^{cd} +\delta^{ac} \delta^{bd}
+\delta^{ad}\delta^{bc} \bigg) g_{\alpha\beta} \delta^4 (z_1-z_2)\nonumber\\
&&\times \int \frac{d^4p}{(2\pi )^4}\int \frac{d^4q}{(2\pi )^4}
e^{+iq\cdot x-ip\cdot y-i(q-p)\cdot z_1} {\cal J}_{\mu \nu} (q,p )
\label{FIN7}
\ee

Inserting (\ref{FIN3},\ref{FIN5},\ref{FIN7}) into (\ref{FIN2}) and performing
the Fourier transforms yield
\be
&&+i\epsilon_1\cdot \epsilon_2 \bigg( \epsilon^{bce}\epsilon^{eda} +
\epsilon^{bde} \epsilon^{eca} \bigg) \nonumber\\
&&+4i \epsilon_1\cdot k_1 \epsilon_2\cdot k_2 \bigg(
\frac 1{u-m_{\pi}^2} \epsilon^{bcf}\epsilon^{fda} +
\frac 1{t-m_{\pi}^2} \epsilon^{bdf}\epsilon^{fca} \bigg)\nonumber\\
&&+\frac 1{2f_{\pi}^2} \epsilon^{abg} \epsilon^{cdg}
\bigg( ic_1 \epsilon_1\cdot \epsilon_2  (t-u)
+ \epsilon_1^{\mu}\epsilon_2^{\nu}
(k_2-k_1)^{\beta} {\cal J}_{\mu\nu\beta} (q_1, -q_2) \bigg) \nonumber\\
&&-\frac i{2f_{\pi}^2} (s-2m_{\pi}^2)
\bigg(2\delta^{ab}\delta^{cd} +\delta^{ac}\delta^{bd} +\delta^{ad}\delta^{bc}
\bigg) \bigg( \epsilon_1\cdot \epsilon_2 {\cal J} (s) +
\epsilon_1^{\mu} \epsilon_2^{\nu} {\cal J}_{\mu\nu} (q_1, -q_2 ) \bigg)
\nonumber\\
&&-\frac {2i}{f_{\pi}^2} m_{\pi}^2 \delta^{ab} \delta^{cd}
\bigg( \epsilon_1\cdot \epsilon_2 {\cal J} (s) +
\epsilon_1^{\mu} \epsilon_2^{\nu} {\cal J}_{\mu\nu} (q_1, -q_2 ) \bigg)
\label{FIN8}
\ee
Using the results of Appendix F, we have
\be
&&+\epsilon_1^{\mu} \epsilon_2^{\nu} {\cal J}_{\mu\nu} (q_1, -q_2 ) =
2\epsilon_1\cdot \epsilon_2 \,\,{\cal K} (s)
\label{FIN9}
\ee
and
\be
&&+\epsilon^{\mu}_1\epsilon^{\nu}_2 (k_2-k_1)^{\beta} {\cal J}_{\mu\nu\beta}
(q_1, -q_2) =-2i \epsilon_1\cdot \epsilon_2 (t-u){\cal H} (s)
\label{FIN10}
\ee
Thus,
\be
&&+i\epsilon_1\cdot \epsilon_2 \bigg( \epsilon^{bce}\epsilon^{eda} +
\epsilon^{bde} \epsilon^{eca} \bigg) \nonumber\\
&&+4i \epsilon_1\cdot k_1 \epsilon_2\cdot k_2 \bigg(
\frac 1{u-m_{\pi}^2} \epsilon^{bcf}\epsilon^{fda} +
\frac 1{t-m_{\pi}^2} \epsilon^{bdf}\epsilon^{fca} \bigg)\nonumber\\
&&+\frac 1{2f_{\pi}^2} \epsilon_1\cdot \epsilon_2 (t-u)\epsilon^{abg}
\epsilon^{cdg} \bigg( c_1  -2 {\cal H} (s)  \bigg)\nonumber\\
&&-\frac i{2f_{\pi}^2} \epsilon_1\cdot\epsilon_2 (s-2m_{\pi}^2)
\bigg(2\delta^{ab}\delta^{cd} +\delta^{ac}\delta^{bd} +\delta^{ad}\delta^{bc}
\bigg) \bigg( {\cal J} (s) + 2{\cal K} (s) \bigg)
\nonumber\\
&&-\frac {2i}{f_{\pi}^2} \epsilon_1\cdot\epsilon_2 m_{\pi}^2
\delta^{ab} \delta^{cd}
\bigg( {\cal J} (s) + 2 {\cal K} (s) \bigg)
\label{FIN11}
\ee
The explicit form of ${\cal J} (s)$, ${\cal K} (s)$ and ${\cal H} (s)$ are
given in Appendix F.

\vskip 0.5cm
{$\bullet$ $\,\,\,\gamma\gamma\rightarrow\pi^0\pi^0$}

For the neutral fusion process
$\gamma\gamma \rightarrow \pi^0\pi^0$ the first (seagull), second (Born)
and third terms in
(\ref{FIN11}) drop and the contracted amplitude reduces to
\be
{\cal T}_{\gamma\gamma\rightarrow \pi^0\pi^0} =
-\frac {2ie^2}{f_{\pi}^2} \epsilon_1\cdot \epsilon_2 \bigg( {\cal J} (s) +
2{\cal
K} (s) \bigg) (s-m_{\pi}^2 )
\label{FIN12}
\ee
Using $s\geq 4m_{\pi}^2$
\be
-16\pi^2 \bigg( {\cal J} (s) + 2{\cal K} (s) \bigg) =
1+\frac {m_{\pi}^2}s \bigg( {\rm ln}\,\,
\bigg( \frac{\sqrt{s} -\sqrt{s-4m_{\pi}^2}}{\sqrt{s}
+\sqrt{s-4m_{\pi}^2}}\bigg) + i \pi \bigg)^2
\label{FIN13}
\ee
we finally obtain
\be
{\cal T}_{\gamma\gamma\rightarrow \pi^0\pi^0} =
-\frac {2ie^2}{16\pi^2f_{\pi}^2}
\epsilon_1\cdot \epsilon_2 (s-m_{\pi}^2)
\bigg( 1+\frac {m_{\pi}^2}s \bigg( {\rm ln}\,\,
\bigg( \frac{\sqrt{s} -\sqrt{s-4m_{\pi}^2}}{\sqrt{s}
+\sqrt{s-4m_{\pi}^2}}\bigg) + i \pi \bigg)^2 \bigg)
\label{FIN14}
\ee
in agreement with the result derived using one-loop chiral perturbation theory
without kaon loops \cite{photophoto}.

The differential cross section for the neutral fusion process follows from
(\ref{FIN14}) in the following form
\be
\bigg(\frac{d\sigma}{d\Omega}\bigg)_{\gamma\gamma\rightarrow \pi^0\pi^0}=
\frac{\alpha^2\beta_V}{4s} \vert \frac{m_{\pi}}{2\alpha} {\bf \alpha}_{\pi}^0
(s) s \vert^2
\label{FIN15}
\ee
with a neutral polarisation function for the fusion process given by
\be
{\bf \alpha}_{\pi}^0 (s) = \frac{\alpha}{8\pi^2 m_{\pi} f_{\pi}^2}
\bigg(\frac{m_{\pi}^2}s -1\bigg)\,
\bigg( 1+\frac {m_{\pi}^2}s \bigg( {\rm ln}\,\,
\bigg( \frac{\sqrt{s} -\sqrt{s-4m_{\pi}^2}}{\sqrt{s}
+\sqrt{s-4m_{\pi}^2}}\bigg) + i \pi \bigg)^2 \bigg)
\label{FIN16}
\ee
Here $\beta_V$ $\beta_V$ is the pion velocity in the CM frame.

\vskip 0.5cm
$\bullet$ $\,\,\,\gamma\gamma\rightarrow\pi^+\pi^-$

For the charged fusion process
$\gamma\gamma \rightarrow \pi^+\pi^-$ the seagull and Born terms in
(\ref{FIN11}) contribute. The final contracted amplitude is
\be
{\cal T}_{\gamma\gamma\rightarrow \pi^+\pi^-} = &&
- {2ie^2}\epsilon_1\cdot \epsilon_2
\bigg( 1-\frac{s}{32\pi^2f_{\pi}^2}
\bigg( 1+\frac {m_{\pi}^2}s \bigg( {\rm ln}\,\,
\bigg( \frac{\sqrt{s} -\sqrt{s-4m_{\pi}^2}}{\sqrt{s}
+\sqrt{s-4m_{\pi}^2}}\bigg) + i \pi \bigg)^2 \bigg)\bigg)\nonumber\\
&&-4ie^2\epsilon_1\cdot k_1\epsilon_2\cdot k_2
\bigg( \frac 1{t-m_{\pi}^2} + \frac 1{u-m_{\pi}^2} \bigg)
\label{FIN17}
\ee
This result is different from the result obtained from one-loop chiral
perturbation theory \cite{photophoto}.

The differential cross section for the charged fusion process follows from
(\ref{FIN17}) in the following form
\be
\bigg(\frac{d\sigma}{d\Omega}\bigg)_{\gamma\gamma\rightarrow \pi^+\pi^-}=
\frac{\alpha^2\beta_V}{4s}
\bigg(&& +\vert 1+ \frac{m_{\pi}}{2\alpha} {\bf \alpha}_{\pi}^{\pm }
(s) s \vert^2 \nonumber\\
&& +\vert {\bf B} + \frac{m_{\pi}}{2\alpha} {\bf \alpha}_{\pi}^{\pm }
(s) s \vert^2 \bigg)
\label{FIN18}
\ee
with a charged polarisation function for the fusion process given by
\be
{\bf \alpha}_{\pi}^{\pm} (s) = -\frac{\alpha}{16\pi^2 m_{\pi} f_{\pi}^2}
\bigg(\frac{m_{\pi}^2}s -1\bigg)\,
\bigg( 1+\frac {m_{\pi}^2}s \bigg( {\rm ln}\,\,
\bigg( \frac{\sqrt{s} -\sqrt{s-4m_{\pi}^2}}{\sqrt{s}
+\sqrt{s-4m_{\pi}^2}}\bigg) + i \pi \bigg)^2 \bigg)
\label{FIN19}
\ee
and a Born contribution
\be
{\bf B} = -1 + \frac{2sm_{\pi}^2}{{(t-m_{\pi}^2)}{(u-m_{\pi}^2)}}
\label{FIN20}
\ee

\vskip .3cm
{{\it 18.7. Compton Scattering :} $\gamma\pi\rightarrow \gamma\pi$}
\vskip .15cm

The one-loop Compton scattering amplitudes follow from the one-loop
fusion amplitudes by crossing $s\rightarrow t$ in the region $t\leq 0$.

\vskip 0.5cm
$\bullet$  $\,\,\, \gamma\pi^0\rightarrow \gamma\pi^0$

The differential cross section for the neutral Compton process is
\be
\bigg(\frac{d\sigma}{d\Omega}\bigg)_{\gamma\pi^0\rightarrow \gamma\pi^0}=
\frac{m_{\pi}^2}{4s} \vert {\bf \alpha}_{\pi}^0 (t) t \vert^2
\label{FIN21}
\ee
with again
\be
{\bf \alpha}_{\pi}^0 (t) = \frac{\alpha}{8\pi^2 m_{\pi} f_{\pi}^2}
\bigg(\frac{m_{\pi}^2}t -1\bigg)\,
\bigg( 1+\frac {m_{\pi}^2}t\,\, {\rm ln}^2\,\,
\bigg( \frac{\sqrt{t} -\sqrt{t-4m_{\pi}^2}}{\sqrt{t}
+\sqrt{t-4m_{\pi}^2}}\bigg) \bigg)
\label{FIN22}
\ee
For $t\leq 0$ the logarithm in (\ref{FIN22}) may be Taylor expanded
\be
{\rm ln}^2\,\,
\bigg( \frac{\sqrt{t} -\sqrt{t-4m_{\pi}^2}}{\sqrt{t}
+\sqrt{t-4m_{\pi}^2}}\bigg) =
-\frac{t}{m_{\pi}^2} -\frac 1{12} \bigg(\frac t{m_{\pi}^2}\bigg)
+ {\cal O} \bigg(\bigg( \frac t{m_{\pi}^2}\bigg)^3\bigg)
\label{FIN23}
\ee
Inserting (\ref{FIN23}) into (\ref{FIN22}) yields the neutral pion
polarisabilities at $t=0$,
\be
\alpha_{\pi}^0 =-\beta_{\pi}^0=
{\bf \alpha}_{\pi}^0 (0) = -\frac{\alpha}{96\pi^2 m_{\pi} f_{\pi}^2}
=-0.49 \,\, 10^{-4}\,\,{\rm fm}^3
\label{FIN24}
\ee
in agreement with the result obtained in the context of one-loop chiral
perturbation theory \cite{babusci}. In the latter, the kaon loops
do not not contribute at threshold.  The result (\ref{FIN24}) is to be
compared with two existing measurements
$|{\bf \alpha}_{\pi}^0 ({\rm exp})| =  (0.69\pm 0.07\pm 0.04)$ 10$^{-4}$ fm$^3$
\cite{babusci} and $|{\bf \alpha}_{\pi}^0 ({\rm exp})| =
(0.8\pm 2.0)$ 10$^{-4}$ fm$^3$ \cite{EX0}. The first empirical value
relies on the high statistics data from MARK II at SLAC \cite{mark},
and appears to be more reliable than the second one.

\vskip 0.5cm
$\bullet$ $\,\,\, \gamma\pi^{\pm}\rightarrow \gamma\pi^{\pm}$

The differential cross section for the charged Compton process is
\be
\bigg(\frac{d\sigma}
{d\Omega}\bigg)_{\gamma\pi^{\pm}\rightarrow \gamma\pi^{\pm}}=
\frac{\alpha^2}{2s}
\bigg(&& +\vert 1+ \frac{m_{\pi}}{2\alpha} {\bf \alpha}_{\pi}^{\pm }
(t) t \vert^2 \nonumber\\
&& +\vert \overline{\bf B} + \frac{m_{\pi}}{2\alpha} {\bf \alpha}_{\pi}^{\pm }
(t) t \vert^2 \bigg)
\label{FIN25}
\ee
with a charged polarisation function for the charged Compton process given by
(\ref{FIN19}) with $s\rightarrow t$, and a Born contribution
\be
\overline{\bf B} = -1 + \frac{2tm_{\pi}^2}{{(s-m_{\pi}^2)}{(u-m_{\pi}^2)}}
\label{FIN26}
\ee
{}From (\ref{FIN19}) with $s\rightarrow t$ and the Taylor expansion
(\ref{FIN23})
we conclude that the one-loop pion charge polarisabilities vanish
\be
\alpha_{\pi}^{\pm}=- \beta_{\pi}^{\pm}={\bf \alpha}_{\pi}^{\pm} ( t= 0) = 0
\label{FIN27}
\ee
This result is to be compared with the one-loop chiral perturbation theory
result ${\bf \alpha}^{\pm} = 2.7$ 10$^{-4}$ fm$^{3}$
\cite{terentev,donoghue}, and the following three empirical results
\be
{\bf \alpha}_{\pi}^{+} (\rm exp ) = && (20\pm 12 )\,\,\,10^{-4}
\,\,{\rm fm}^3\nonumber\\
{\bf \alpha}_{\pi}^{+} (\rm exp ) = && (6.8\pm 1.4\pm 1.2 )\,\,\,10^{-4}
\,\,{\rm fm}^3\nonumber\\
{\bf \alpha}_{\pi}^{+} (\rm exp ) = && (2.2\pm 1.6 )\,\,\,10^{-4}
\,\,{\rm fm}^3
\label{FIN28}
\ee
from \cite{aibergenov}, \cite{antipov} and \cite{babusci}
respectively. The last measurement follows from the high statistics data
of the MARK II collaboration at SLAC using $\gamma\gamma\rightarrow
\pi^+\pi^-$.

The result (\ref{FIN27}) appears to be compatible with the vanishing of
the axial structure form factor (\ref{ZZ7}) in $\pi\rightarrow e\nu\gamma$
to one-loop, and thus $\gamma =0$. The possible relationship between $\gamma$
and the electric pion polarisability was first suggested by Terentev using
the soft pion limit \cite{terentev}. From (\ref{ZZ7}) and (\ref{X104}) it
appears that the correspondence implied by the soft pion limit does not
hold in general.

We note that the non-vanishing of the axial form factor in one-loop chiral
perturbation theory through the combination of parameters $L_9+L_{10}\neq 0$
\cite{babusci,donoghue}, is also what makes the charged pion electric
polarisability nonzero. We have already indicated above that to one-loop
the axial form factor (\ref{ZZ7}) vanishes. The axial correlator does not
acquire a transverse part, and $a$ $priori$ does not need to be renormalized,
thus the consistency of our two-parameter effective action.
The finite $a_1\sim 3\pi$ contribution is expected at two-loops and higher.

Given the large discrepancies in the extracted polarisabilities quoted
above, we cannot conclude whether our estimates for the neutral or charged
pion electric polarisabilities, rule in or out the two-parameter one-loop
effective action we have suggested. The forthcoming measurements of the pion
electric polarisabilities at FERMILAB are therefore welcome.

\vskip .5cm
{\bf 19. Discussion}

We have set up a general framework for analyzing the consequences of
chiral $SU(2) \times SU(2) $ for both QCD and
the nonlinear sigma model.
Our derivation in section 4 shows that the main results
(\ref{B57},\ref{C2},\ref{C3})
are simply restatements of the the symmetry conditions
(\ref{B72}-\ref{B73}) subject to the boundary conditions (\ref{B60}-\ref{B61}).
The latter in turn are equivalent to
\be
<0| {\bf A}_{\mu}^a (x) |\pi^b (p) > = i f_{\pi} p_{\mu} \delta^{ab}
e^{-ip\cdot x}
\label{E1}
\ee
given the absence of stable axial vector mesons or other stable pseudoscalars.
Hence any result which is a consequence of symmetry and (\ref{E1}) as well as
general principles such as unitarity and causality is  contained
in (\ref{B57},\ref{C2},\ref{C3}) plus general principles.
We are then justified in calling them master formulas.
The subsequent exact formulas we have derived prior to any expansion,
encompass both the results of few-loops chiral perturbation theory, dispersion
relations or models. They provide important insights to low energy processes
at and beyond threshold. They form the core results of this paper.

The only other
non-trivial assumption in our work is power counting in $1/f_{\pi}$.
This counting is not without
problems. Indeed, it is known that the GOR relation
(\ref{D2}) gives rise to the $U_A(1)$ problem. It also gives a non-zero
contribution to the cosmological constant, which must be subtracted one way or
another. However, if a subtraction is made, there is no guarantee that the
relation will remain operationally meaningful.

The various analyses of sections 5-12 show that the standard results
of current algebra (Goldberger-Treiman, Tomozawa-Weinberg, Adler-Weisberger)
as well as results on $\pi N\rightarrow \pi\pi N$,
$\gamma N\rightarrow \pi\pi N$, $\pi N\rightarrow \pi\gamma N$,
$\gamma N \rightarrow \pi N$  and others
do not provide tests of (\ref{D2}), since $<0|\hat{\sigma}
|0>$ appears only in disconnected pieces and gets canceled. The
Weinberg sum rules or other high-energy results are also insensitive to
(\ref{D2}), since free quark theory or perturbative QCD do not give rise
to symmetry breaking in the chiral limit. Therefore, a test of (\ref{D2}) must
rely on the on- and off-shell
radiative  decay $\pi \rightarrow e\nu\gamma$ of the pion,
$\gamma\gamma\rightarrow \pi\pi$ fusion, $\gamma\pi\rightarrow \gamma\pi$, or
$\pi\pi$ scattering within the present
framework \footnote{A possible alternative
framework would be the QCD sum rules. However, they suffer from precision
statements and also face the same problem with the cosmological
constant.}.

We may note that a few authors \cite{heretics} have advocated an alternative
scenario of chiral symmetry breaking without the GOR relation (\ref{D2}) and a
corresponding generalization of chiral perturbation theory \cite{stern}.
However, we have been unable so far to find an expansion scheme within our
framework that is compatible with this alternative.

In any event,
there is no doubt that the $1/f_{\pi}$ counting we have assumed is valid for
the non-linear sigma model, so we may compare our results with the  work
of Gasser and Leutwyler \cite{leutwyler}, who also start with the nonlinear
sigma model coupled to external fields. At one-loop level ten parameters :
$l_1, ..., l_7$ and $h_1, ..., h_3$ are then introduced. Among these, $l_7$ and
$h_3$ refer to the densities $\overline q i\gamma_5 q$ and $\overline q \tau^a
q$
, both of which are associated with isospin breaking that we have
ignored. The parameters $h_1$ and $h_2$ do not enter the expression for
physical quantities. Overall, only six parameters are left to compare with. The
general analysis of one-loop effects in Appendix E is overall
compatible with this result. However, we arrive at some different conclusions.
In particular, we find that chiral symmetry does not impose any constraints on
$\pi\pi$ scattering at one-loop level other than those already given by
unitarity, causality and the tree result (\ref{C37}), and that all six
parameters can be extracted from pionic data without recourse to $SU(3)$ or
other considerations. Also, there are unresolved issues in the
approach suggested by Gasser and Leutwyler, as we discuss in Appendix G.
Finally, since the approach we suggested in section 18 makes use of only
two parameters, a discussion is now needed.

The point is that we are not necessarily forced to entertain the most
general possibility that is compatible with unitarity, causality and (broken)
chiral symmetry. A good example is conventional renormalization theory.
Nothing in the general principles of causality, unitarity and so on forbids the
introduction of non-renormalizable counterterms in a
renormalizable theory (at least within a perturbative context), but this is not
ordinarily done.

Similarly, while we pay due respect to unitarity, causality and
chiral symmetry, we may attempt to do so with a minimal number of free
parameters. In particular, while the master formulas and the results that
follow from it are quite general as we have seen, the one-loop effective action
of section 18 was subjected to the maximum number of constraints that could
reasonably be imposed. These assumptions were theoretically motivated by simple
power counting, and simplified solutions. The major outcome of these
assumptions
was the GOR relation and the KSFR relation, giving a motivation
$a$ $posteriori$. Also we found that in the on-shell radiative decay
$\pi\rightarrow e\nu\gamma$, the ratio $\gamma$ (of the two structure
dependent axial and vector currents) vanishes, that in the off-shell
radiative decay the axial
structure form-factor $R$ is consistent with the soft pion result, and that
the pion polarisabilities are $\alpha_{\pi}^0 = -\beta^0_{\pi}=
-0.49$ 10$^4$ fm$^3$
and $\alpha_{\pi}^{\pm} =-\beta_{\pi}^{\pm}=0$. The overall uncertainties in
the present
measurements of these low-energy parameters, do not allow for a definite
assessment of our new one-loop effective action. Better measurements are
therefore called for.

A feature of our approach which deserves attention is that processes
involving nucleons
could be discussed as well as purely pionic processes, in
contrast with some other approaches. Moreover, the effects of the vector mesons
are unambiguously accounted for. Also, unlike chiral perturbation
theory, it is unnecessary to start from the chiral limit, once the explicit
symmetry breaking term is given. This has allowed us to bypass the issue of
chiral logarithms associated with vanishing energy denominators. While such
terms are theoretically interesting and serve as useful checks, their
appearance is cumbersome if the goal is to derive relations between
experimentally measurable quantities. Finally,
one aspect of current algebra which
is clear from our derivation but not chiral perturbation theory is that the
exact formulas of current algebra alone actually do not constitute a test of
the idea that pions are Nambu-Goldstone bosons in the chiral limit. Indeed, the
formulas are valid even if $f_{\pi}\rightarrow 0$ or $\infty$ as
$m_{\pi}\rightarrow 0$. It is only when extra assumptions such as the
possibility of a loop expansion are introduced that the nature of the chiral
limit becomes relevant.

We may also note that in our approach, we could work with finite equations,
once it is known that renormalization does not spoil the current conservation
equations through anomalies. In section 18, we have adopted a conventional
analysis in terms of divergent diagrams, but this could also be avoided
in principle since (\ref{D3}) is a finite equation.
In particular, for ordinary scalar theories our
approach correctly handles overlapping divergences at the two-loop level
\cite{yamagishi}. Thus, a two-loop calculation in our scheme is feasible.
In fact, it is not difficult to write down the required equations as shown in
Appendix H. We hope to give a detailed account of these equations in the
future.

Finally, we would like to point out that there is in principle no difficulty in
incorporating isospin breaking effects into our framework. However, a
quantitative analysis requires the consideration of competing electromagnetic
effects, so we  will not pursue this point here. Given the interest in
threshold low energy processes beyond leading order, it is desirable that
the present formalism be extended so as to include baryon dynamics.
This and related issues are currently under investigation.

\vglue 0.6cm
{\bf \noindent  Acknowledgements \hfil}
\vglue 0.4cm
This work was supported in part  by the US DOE grant DE-FG-88ER40388.

\newpage
{\bf Appendix A : Quantum Field Theory minus Feynman Streamlined}
\vskip .1cm

The advantages of working with the Peierls formula for the commutator
(\ref{B19}-\ref{B21}) has been emphasized by DeWitt \cite{dewitt}. Here,
we shall list some further applications of the Schwinger-Peierls-Bogoliubov
relations (\ref{B7}-\ref{B8},\ref{B15}-\ref{B25}).

One case is when the system is in some pure or mixed state in the remote past,
in which case we  immediately obtain the generalized Green-Kubo formula
\be
\frac{\delta}{\delta v^{\nu b} ( y) } \langle{\bf V}_{\mu}^a (x) \rangle =
i \langle R^*\bigg[ {\bf V}_{\mu}^a (x) , {\bf V}_{\nu}^b (y) \bigg] \rangle
\label{G1}
\ee
where $\langle \,\,\rangle$ denotes the ensemble average.

Another case arises if we introduce a c-number source following
Schwinger\footnote{Schwinger has subsequently criticised the use of  field
operators. However, unitarity and causality of the S-matrix in the presence of
a c-number source imply the existence of local field operators, as we have
seen.}
\be
{\bf I} [\phi ] \rightarrow {\bf I} [\phi ] +\int d^4x \,J (x) \phi (x)
\label{G2}
\ee
We then have
\be
\phi (x)  = -i{\cal S}^{\dagger} [J]
\bigg( \frac{\delta{\cal S} [J]}{\delta J (x)}\bigg)\,\,.
\label{G3}
\ee
We further assume that
\be
\phi (x)\rightarrow \phi_{\rm in, out} (x)
\qquad\qquad x^0\rightarrow \mp \infty
\label{G4}
\ee
so that we have the Yang-Feldman equations
\be
\phi (x) =  &&\phi_{\rm in} (x) +\int d^4y \Delta_R (x-y) j(y) \nonumber\\ =&&
{\cal S}^{\dagger} [J] \phi_{\rm in} (x) {\cal S} [J] +
\int d^4y \Delta_A (x-y) j(y)
\label{G5}
\ee
\be
j(x) = \bigg(-\Box -m^2 \bigg) \phi (x) .
\label{G6}
\ee
Combining (\ref{G3}-\ref{G6}) as in (\ref{B93}-\ref{B94}) and
(\ref{C1}-\ref{C3}), yields
\be
\bigg[a_{\rm in} (k) , {\cal S} [J] \bigg] =
-\int d^4y e^{ik\cdot y}\bigg(-\Box -m^2 \bigg)
\frac {\delta {\cal S}}{\delta J (y)}
\label{G7}
\ee
\be
\bigg[{\cal S} [J], a_{\rm in}^{\dagger} (k) \bigg]= -
\int d^4y e^{-ik\cdot y} \bigg(-\Box-m^2\bigg)
\frac{\delta{\cal S}}{\delta J (y)}.
\label{G8}
\ee
Iteration then gives the generalized WLSZ formula
\be
&&
\bigg[ ...\bigg[\bigg[{\cal S} [J], a_{\rm in}^{\dagger} (k_1)\bigg]
a_{\rm in}^{\dagger} (k_2)\bigg] ...a_{\rm in}^{\dagger} (k_n) \bigg]
\nonumber\\ =&&
(-1)^n \int d^4y_1 ... d^4y_n e^{-ik_1\cdot y_1 -...-ik_n\cdot y_n}
\bigg(-\Box_1-m^2\bigg) ...\bigg(-\Box_n -m^2\bigg)
\frac{\delta^n {\cal S} [J]}{\delta J(y_1) ...\delta J(y_n)}\nonumber\\ =&&
(-i)^n \int d^4y_1...d^4y_n e^{-ik_1\cdot y_1 ...-ik_n\cdot y_n}
\bigg(-\Box_1-m^2\bigg) ...\bigg(-\Box_n -m^2\bigg)
{\cal S} [J] T^*\bigg( \phi (y_1) ...\phi (y_n)\bigg).
\nonumber\\
\label{G9}
\ee

We may also consider the case when the action splits into a free part and an
interacting part
\be
{\bf I} [\phi ] ={\bf I}_{\rm free} [\phi ] +
\int d^4x  \,g\,{\cal L}_{\rm int} (\phi (x)).
\label{G10}
\ee
Following Gell-Mann and Low we take $g$ to be variable
\be
{\bf I} [\phi ]\rightarrow {\bf I}_{\rm free} [\phi ] +
\int d^4x \, g(x)\, {\cal L}_{\rm int} (\phi (x)).
\label{G11}
\ee
We then have
\be
\frac{\delta^n{\cal S} [g]}{\delta g (x_1) ...\delta g(x_n)} =
(+i)^n {\cal S} [g] T^*\bigg({\cal L}_{\rm int}(\phi(x_1))  ...
{\cal L}_{\rm int}
(\phi (x_n))\bigg).
\label{G12}
\ee
Expansion in a formal power series gives the Lagrangian version of Dyson's
formula
\be
{\cal S} [g] =\sum_{n=0}^{\infty} \frac{i^n}{n!}\,
\int d^4x_1 ...d^4x_n \,g(x_1)...g(x_n)\,
T^*\bigg({\cal L}_{\rm int} (\phi_{\rm in} (x_1)) ....
         {\cal L}_{\rm int} (\phi_{\rm in} (x_n))\bigg).
\label{G13}
\ee
We may further combine (\ref{G2}) and (\ref{G11}) to obtain the Gell-Mann-Low
formula
\be
T^*\bigg(\phi (x_1) ...\phi(x_m)\bigg)_{J=0} = &&
{\cal S}^{\dagger}[g]\sum_{n=0}^{\infty} \frac{i^n}{n!}
\int d^4y_1 ...d^4y_n \, g(y_1) ... g(y_n)\, \nonumber\\&&
T^*\bigg(\phi_{\rm in} (x_1)
...\phi_{\rm in} (x_m) \,{\cal L_{\rm int}} (\phi_{\rm in} (y_1)) ...
{\cal L_{\rm int}} (\phi_{\rm in} (y_n)\bigg).
\label{G14}
\ee
Another possibility is to use DeWitt's background field method as discussed by
one of us \cite{yamagishi}.

\vskip 1.5cm
{\bf Appendix B : Dirac's Terminology}
\vskip .1cm

In this Appendix we introduce some terminology in relation to the
constraint problem discussed in section 3. Let ${\cal M}$
be the subspace of the functions $(v, a, J )$ defined by (\ref{B29}). Following
Dirac, we will call an equation weak if it holds on ${\cal M}$ but not
necessarily in its neighborhood, and denote it by $\approx$.
An equation will be
called strong if it holds in some neighborhood of ${\cal M}$, and will be
denoted by $=$. Since the gradient of the constraint
\be
\bigg( \frac{\delta}{\delta v_{\mu}^a (x)} (f_{\pi}\nabla^{\nu} a_{\nu} - J)^b
(y) ,
       \frac{\delta}{\delta a_{\mu}^a (x)} (f_{\pi}\nabla^{\nu} a_{\nu} - J)^b
(y) ,
       \frac{\delta}{\delta J^a (x)} (f_{\pi}\nabla^{\nu} a_{\nu} - J)^b (y)
\bigg) =\nonumber\\
\bigg( -f_{\pi}\epsilon^{abc} a^{\mu c} (x) \delta^4 (x-y) ,
       -f_{\pi}\delta^{ab} \partial^{\mu} \delta^4 (x-y ) + f_{\pi}
       \epsilon^{abc} v^{\mu c} (x) \delta^4 (x-y) ,
        -\delta^{ab}  \delta^4 (x-y )
\bigg)\nonumber\\
\label{B30}
\ee
is nonvanishing, its components define a set of normals on ${\cal M}$. A vector
with the components $(\xi_{\mu}^a , \eta_{\mu}^a , \zeta^a )$ is tangent to
${\cal M}$ if and only if
\be
\int d^4x \bigg(
\xi_{\mu}^a (x) \frac{\delta}{\delta v_{\mu}^a (x)} +
\eta_{\mu}^a  (x) \frac{\delta}{\delta a_{\mu}^a (x)} +
\zeta^a (x) \frac{\delta}{\delta J^a (x)}\bigg)
(f_{\pi}\nabla^{\mu} a_{\mu} - J)^b (y) =\nonumber
\ee
\be
-f_{\pi}({\ub a}^{\mu} \xi_{\mu} )^b (y) + f_{\pi}(\nabla^{\mu}\eta_{\mu} )^b
(y) -\zeta^b (y)
\label{B31}
\ee
vanishes at least
weakly. Following the mathematical literature, we will identify a
vector with the first order linear differential operator
\be
{\bf X} = \int d^4x \bigg(
\xi_{\mu}^a (x) \frac{\delta}{\delta v_{\mu}^a (x)} +
\eta_{\mu}^a  (x) \frac{\delta}{\delta a_{\mu}^a (x)} +
\zeta^a (x) \frac{\delta}{\delta J^a (x)}\bigg) .
\label{B32}
\ee
The condition for ${\bf X}$ to be a tangent vector is then simply given by
\be
{\bf X} (f_{\pi}\nabla^{\mu} a_{\mu} - J) \approx 0.
\label{B33}
\ee
An example of a tangent vector is
\be
{\bf T}_V^a (x) =
{\bf X}_V^a (x) + {\ub J}^{ac} (x) \frac{\delta}{\delta J^c (x )}
\label{B34}
\ee
which is the generator of local isospin transformations in the extended space
$(v, a, J )$
\be
\bigg[ {\bf T}_V^a (x) , {\bf T}_V^b (y ) \bigg] =
-\epsilon^{abc} {\bf T}_V^c (y) \delta^4 (x-y) \,\,\,.
\label{B35}
\ee

Now suppose that $f$ is a functional defined on ${\cal M}$, and
$f^{(1)}$ and $f^{(2)}$ two smooth extensions to its neighborhood. Since
(\ref{B30}) is nonvanishing, we may write
\be
f^{(1)} -f^{(2)} =\int d^4y \lambda^b (y )
(f_{\pi}\nabla^{\nu} a_{\nu} - J)^b (y)
\label{B36}
\ee
so that
\be
{\bf X}f^{(1)} -{\bf X}f^{(2)}  \approx
\int d^4 y \lambda^b (y) {\bf X} (f_{\pi}\nabla^{\nu} a_{\nu} - J)^b (y).
\label{B37}
\ee
It follows that ${\bf X} f$ is well defined if ${\bf X}$ is a tangent vector.
On the other hand, individual partial derivatives are only defined up to the
following shifts
\be
\frac{\delta f}{\delta v_{\mu}^a (x)} &&\rightarrow
\frac{\delta f}{\delta v_{\mu}^a (x)} + f_{\pi}({\ub a}^{\mu} \lambda )^a (x )
\nonumber\\
\frac{\delta f}{\delta a_{\mu}^a (x)} &&\rightarrow
\frac{\delta f}{\delta a_{\mu}^a (x)} - f_{\pi}(\nabla^{\mu} \lambda )^a (x )
\nonumber\\
\frac{\delta f}{\delta J^a (x)} &&\rightarrow
\frac{\delta f}{\delta J^a (x)} -\lambda^a (x ).
\label{B38}
\ee
We observe that $\lambda$ plays the role of a Lagrange multiplier associated
with the constraint (\ref{B29}).
Time-ordered and retarded products of the original currents
${\bf V}_{\mu}^a$ and ${\bf A}_{\mu}^a$ may be obtained by applying the tangent
vectors
\be
&&\frac{\delta}{\delta v_{\mu}^a (x)} + f_{\pi} {\ub a}^{\mu ac} (x)
\frac{\delta}{\delta J^c (x)} \nonumber\\
&&\frac{\delta}{\delta a_{\mu}^a (x)} - f_{\pi} {\nabla}^{\mu ac} (x)
\frac{\delta}{\delta J^c (x)}
\label{BA38}
\ee
to (\ref{B40}-\ref{B42}).

\vskip  1.5cm
{\bf Appendix C : Four Dimensional Gell-Mann Algebra}
\vskip .10cm
The case for which the Dirac's constraint problem
can be solved explicitly is of some interest. Consider the case of
only left handed external fields, $v_{\mu}^a = - a_{\mu}^a = w_{\mu}^a/2$
and $J^a=0$. The Veltman-Bell equation reads
\be
\nabla^{\mu}_L {\bf L}_{\mu} =0 \qquad\qquad
{\bf L}_{\mu} = \frac 12 \bigg( {\bf V}_{\mu} -{\bf A}_{\mu} \bigg)
\label{K1}
\ee
where $\nabla_L^{\mu} = \partial^{\mu} +{\ub w}^{\mu}$. For the nonlinear sigma
model, ${\bf L}_{\mu}^a$ is the Sugawara current \cite{sugawara}. The
constraint $\partial^{\alpha} w_{\alpha} =0$ may be solved as
\be
w_{\alpha}^a =\partial^{\mu} W_{\mu\alpha}^a \qquad\qquad
W_{\mu\alpha}^a = -W_{\alpha\mu}^a
\label{K2}
\ee
The appearance of antisymmetric tensor fields is natural within the context of
the nonlinear sigma model \cite{Freedman}. This point was the starting point
of our early investigation \cite{zahed}\footnote{However,
the treatment of ${\bf L}_{\mu}^a$ given there is incorrect.}. We have
\be
\partial_{\mu} {\bf L}_{\nu} -\partial_{\nu} {\bf L}_{\mu} =
-2\frac{\delta {\bf I}}{\delta W^{\mu\nu}} =
+2i {\cal S}^{\dagger} \frac{\delta{\cal S}}{\delta W^{\mu\nu}}
\label{K3}
\ee
with the asymptotic conditions
\be
{\bf L}_{\mu} (x) \rightarrow \frac {f_{\pi}}2
         \partial_{\mu} \pi_{\rm in, out}(x)
\qquad\qquad (x_0\rightarrow \mp \infty )
\nonumber
\ee
\be
\partial_{\mu} {\bf L}_{\nu}(x) -\partial_{\nu} {\bf L}_{\mu}(x) \rightarrow 0
\qquad\qquad (x_0\rightarrow \mp \infty )
\label{K4}
\ee

Eqn. (\ref{K3}) may be solved for ${\bf L}_{\nu}^a$ by first imposing the
Fock-Schwinger condition $x^{\nu} {\bf L}_{\nu}^a =0$. Multiplying by $x^{\mu}$
then gives
\be
x^{\mu}\partial_{\mu} {\bf L}_{\nu}^a  +{\bf L}_{\nu}^a = 2i{\cal S}^{\dagger}
\,x^{\mu}\frac{\delta{\cal S}}{\delta W^{a \mu\nu}}
\label{K5}
\ee
Alternativaly, one may employ the Poincare lemma construction. Either way, the
result is
\be
{\bf L}^a_{\nu} (x) = +\frac {f_{\pi}}2 \partial_{\nu} \phi^a (x) +
2i {\cal S}^{\dagger} \int_0^x d\xi^{\mu} \frac{\delta {\cal S}}
{\delta W^{a\mu\nu}(\xi )}
\label{K6}
\ee
where the line integral is over a straight path from $0$ to $x$. Substituting
into the Veltman-Bell equation (\ref{K1}) gives
\be
-\nabla^{\mu}_L \partial_{\mu} \phi^a (x) =
+\frac{4i}{f_{\pi}} {\cal S}^{\dagger} \nabla_L^{\nu}
\int_0^x d\xi^{\mu} \frac{\delta {\cal S}}
{\delta W^{a\mu\nu}(\xi )}
\label{K7}
\ee
Introducing the retarded and advanced Green's functions
\be
-\nabla^{\mu}_L \partial_{\mu} {\bf G}_R (x,y) = \delta^4 (x-y) \,{\bf 1}
\label{K8}
\ee
and solving under (\ref{K4}) yield
\be
\phi^a (x) =&&+\int d^4z \bigg( 1 +{\bf G}_{R, A} {\ub w}^{\mu}\partial_{\mu}
\bigg)^{ac} (x,z) \,\pi_{\rm in\, ,\, out}^c (z)\nonumber\\&& +
\frac{4i}{f_{\pi}}{\cal S}^{\dagger}\int d^4z {\bf G}_{R, A}^{ac} (x,z)
\nabla_L^{\kappa \, ce} (z)
\int_0^z d\xi^{\alpha} \frac{\delta {\cal S}}
{\delta W^{e\alpha\kappa}(\xi )}
\label{K9}
\ee
Hence
\be
{\cal S}^{\dagger} \pi_{\rm in}^a (x) {\cal S} = && \pi_{\rm out}^a (x)
\nonumber\\= &&+\int d^4z
\bigg( \bigg( 1 +{\bf G}_A {\ub w}^{\mu}\partial_{\mu} \bigg)^{-1}
\bigg( 1 +{\bf G}_R {\ub w}^{\nu}\partial_{\nu} \bigg)\bigg)^{ac} (x,z)
\pi_{\rm in}^c (z) \nonumber\\&&
+\frac {4i}{f_{\pi}} {\cal S}^{\dagger}
\int d^4z \bigg(\bigg(  1 +{\bf G}_A {\ub w}^{\mu}\partial_{\mu} \bigg)^{-1}
{\bf G}\bigg)^{ac} (x, z) \nabla_L^{\kappa\, ce} (z)
\int_0^z d\xi^{\alpha} \frac{\delta {\cal S}}
{\delta W^{e\alpha\kappa}(\xi )}\nonumber\\
\label{K10}
\ee
where ${\bf G} = {\bf G}_R -{\bf G}_A$, and
\be
{\bf L}_{\mu}^a (x) = &&+\frac {f_{\pi}}2 \frac{\partial}{\partial x^{\mu}}
\int d^4z \bigg( 1 + {\bf G}_A {\ub w}^{\mu} \partial_{\mu} \bigg)^{ac} (x, z)
\pi_{\rm out}^c (z) \nonumber\\&&
+2i{\cal S}^{\dagger} \int_0^x d\xi^{\alpha} \frac{\delta {\cal S}}
{\delta W^{a\alpha\mu} (\xi )}\nonumber\\&&
+2i{\cal S}^{\dagger} \frac{\partial}{\partial x^{\mu}}
\int d^4z {\bf G}^{ac}_A (x,z) \nabla_L^{\kappa\, ce} (z)
\int_0^z d\xi^{\alpha} \frac{\delta {\cal S}}
{\delta W^{e\alpha\kappa} (\xi )}
\label{K11}
\ee

The commutator of two currents may be computed from (\ref{K10}, \ref{K11}) by
noting that
\be
\delta {\bf G}_{R, A} = {\bf G}_{R, A} \,\delta{\ub w}^{\mu} \partial_{\mu}
{\bf G}_{R, A}
\label{K12}
\ee
and
\be
\bigg[ \pi_{\rm out} , {\cal S}^{\dagger} \delta {\cal S} \bigg] =
&&\delta\pi_{\rm out}\nonumber\\ =&&+
\frac 2{f_{\pi}} \bigg(1+ {\bf G}_A {\ub w}^{\mu}\partial_{\mu} \bigg)^{-1}
{\bf G} {\delta\ub w}^{\alpha} {\bf L}_{\alpha} \nonumber\\&&
+\frac {4i}{f_{\pi}} \bigg(1+ {\bf G}_A
{\ub w}^{\mu}\partial_{\mu} \bigg)^{-1}
{\bf G} \nabla^{\kappa} \int d\xi^{\mu} \delta
\bigg( {\cal S}^{\dagger} \frac{\delta{\cal S}}{\delta W^{\mu \kappa} (\xi )}
\bigg)
\label{K13}
\ee
We find
\be
\bigg[ {\bf L}_{\mu}^a (x) , {\bf L}_{\nu}^b (y) \bigg] = &&
-i\frac{\partial}{\partial x^{\mu}}\int_0^y d\eta^{\beta} {\bf G}^{ac} (x, \eta
) \frac{\stackrel\leftarrow\partial}{\partial\eta^{\beta}}
\epsilon^{cbe} {\bf L}_{\nu}^e (\eta )\nonumber\\&&
+i\frac{\partial}{\partial x^{\mu}}\int_0^y d\eta^{\beta} {\bf G}^{ac} (x, \eta
) \frac{\stackrel\leftarrow\partial}{\partial\eta^{\nu}}
\epsilon^{cbe} {\bf L}_{\beta}^e (\eta )\nonumber\\&&
-i\frac{\partial}{\partial x^{\mu}}\int_0^y d\eta^{\beta} {\bf G}^{ac} (x, \eta
) \epsilon^{cbe}
\bigg(\partial_{\beta} {\bf L}_{\nu}^e -\partial_{\nu} {\bf L}_{\beta}^e \bigg)
 (\eta) \nonumber\\&&
-4\frac{\partial}{\partial x^{\mu}}\int d^4z {\bf G}_R^{ac} (x, z)
\nabla_L^{\kappa ce} (z) \int_0^z d\xi^{\alpha} \int_0^y
d\eta^{\beta}\nonumber\\&&
\times \frac{\delta}{\delta W^{b\beta\nu} (\eta )}
\bigg( {\cal S}^{\dagger} \frac{\delta{\cal S}}{\delta W^{e\alpha\kappa}
(\xi )}\bigg) \nonumber\\&&
+4\frac{\partial}{\partial x^{\mu}}\int d^4z {\bf G}_A^{ac} (x, z)
\nabla_L^{\kappa ce} (z) \int_0^z d\xi^{\alpha} \int_0^y
d\eta^{\beta}\nonumber\\&&
\times \frac{\delta}{\delta W^{e\alpha\kappa} (\eta )}
\bigg( {\cal S}^{\dagger} \frac{\delta{\cal S}}{\delta W^{b\beta\nu}
(\xi )}\bigg) \nonumber\\&&
-i\frac{\partial}{\partial x^{\mu}}\frac{\partial}{\partial y^{\nu}}
\int d^4z' {\bf G}_A^{bd} (y, z' ) \nabla_L^{\lambda df} (z' )
\int_0^{z'} d\eta^{\beta} \nonumber\\&&
\times {\bf G}^{ac} (x , \eta )
\frac{\stackrel\leftarrow\partial}{\partial\eta^{\beta}}\epsilon^{cfg}
{\bf L}_{\lambda}^g (\eta )\nonumber\\&&
+i\frac{\partial}{\partial x^{\mu}}\frac{\partial}{\partial y^{\nu}}
\int d^4z' {\bf G}_A^{bd} (y, z' ) \nabla_L^{\lambda df} (z' )
\int_0^{z'} d\eta^{\beta} \nonumber\\&&
\times {\bf G}^{ac} (x , \eta )
\frac{\stackrel\leftarrow\partial}{\partial\eta^{\lambda}}\epsilon^{cfg}
{\bf L}_{\beta}^g (\eta )\nonumber\\&&
-i\frac{\partial}{\partial x^{\mu}}\frac{\partial}{\partial y^{\nu}}
\int d^4z' {\bf G}_A^{bd} (y, z' ) \nabla_L^{\lambda df} (z' )
\int_0^{z'} d\eta^{\beta} \nonumber\\&&
\times {\bf G}^{ac} (x , \eta )\epsilon^{cfg}
\bigg(\partial_{\beta} {\bf L}_{\lambda}^g - \partial_{\lambda}{\bf
L}_{\beta}^g \bigg) (\eta) \nonumber\\&&
-4\frac{\partial}{\partial x^{\mu}}\int d^4z {\bf G}_R^{ac} (x, z)
\nabla_L^{\kappa ce} (z) \int_0^z d\xi^{\alpha} \nonumber\\&&
\times
\frac{\partial}{\partial y^{\nu}}\int d^4z' {\bf G}_A^{bd} (y, z')
\nabla_L^{\lambda df} (z') \int_0^{z'} d\eta^{\beta} \nonumber\\&&
\times \frac{\delta}{\delta W^{b\lambda\eta} (\eta )}
\bigg( {\cal S}^{\dagger} \frac{\delta{\cal S}}{\delta W^{e\alpha\kappa}
(\xi )}\bigg) \nonumber\\&&
-\bigg( x\,\,\, \mu\,\,\, a \leftrightarrow y\,\,\, \nu\,\,\, b \bigg)
\nonumber\\&&
+ \frac {f^2_{\pi}}4
\frac{\partial}{\partial x^{\mu}}\frac{\partial}{\partial y^{\nu}}
{\bf G}^{ab} (x, y)\nonumber\\&&
-4\int_0^x d\xi^{\alpha} \int_0^y d\eta^{\beta}
\bigg[{\cal S}^{\dagger} \frac{\delta{\cal S}}{\delta W^{a\alpha\mu} (\xi )} ,
{\cal S}^{\dagger} \frac{\delta{\cal S}}{\delta W^{b\beta\nu} (\eta )} \bigg]
\label{K14}
\ee
This may be regarded as a four dimensional extension of the Gell-Mann algebra.
Unfortunately, locality is not manifest in (\ref{K14}).

\vskip 1.5cm
{\bf Appendix D : Commutation Relations and Characteristic Curves}
\vskip .1cm

This Appendix describes some consistency checks and reformulations of the
master equations. We first note that the retarded version of (\ref{B93})
allows us to calculate the retarded commutators involving $\pi$. Using
(\ref{B95}), we find
\be
\delta \pi =&&+ G_R \bigg(\delta {\bf K}\bigg) \pi - G_R \delta J +
G_R \bigg( \nabla^{\mu} a_{\mu} - J/f_{\pi}\bigg) \nonumber\\&&
+G_R\bigg( \nabla^{\mu} \delta a_{\mu} +\delta {\ub v}^{\mu} a_{\mu} -
\frac 1{f_{\pi}}\delta J
\bigg) \hat{\sigma}\nonumber\\&&
-\frac 1{f_{\pi}} G_R \bigg( \delta{\ub v}^{\mu}{\bf j}_{A\mu}
+\delta{\ub a}^{\mu} {\bf j}_{V\mu} +\nabla^{\mu}\delta{\bf j}_{A\mu} +
{\ub a}^{\mu}\delta{\bf j}_{V\mu} \bigg).
\label{J1}
\ee
Hence
\be
R^*\bigg[ \pi^a (x) , {\bf j}_{A\beta}^b (y) \bigg] = &&
-i\frac{\delta \pi^a (x)}{\delta a^{\beta b} (y) }  \nonumber\\ =&&
+iG_R^{ac} (x, y) {\ub a}_{\beta}^{cd} (y) \epsilon^{dbe} \pi^e (y) +
i\nabla_{\beta}^{bc} (y) \bigg( G_R^{ac} (x, y) \hat\sigma (y) \bigg)
\nonumber\\&&
+\int d^4x' G_R^{ac} (x, x') \bigg(\nabla^{\mu} a_{\mu} -J/f_{\pi}\bigg)^c (x')
\,\,R^* \bigg[ \hat\sigma (x') , {\bf j}_{A\beta}^b (y) \bigg] \nonumber\\&&
+\frac i{f_{\pi}} G_A^{ac} (x,y)\epsilon^{cbd}
\bigg({\bf j}_{V\beta}^b (y) + f_{\pi} {\ub a}_{\beta}^{de} (y) \pi^e (y)
\bigg)
\nonumber\\&&
-\frac 1{f_{\pi}}\int d^4x' G_R^{ac} (x, x')
\bigg(\nabla^{\alpha cd} (x') R^* \bigg[ {\bf j}_{A\alpha}^d (x') ,
{\bf j}_{A\beta}^b (y) \bigg] \nonumber\\&&
+{\ub a}^{\alpha cd} (x' ) R^* \bigg[{\bf j}_{V\alpha}^d (x') ,
{\bf j}_{A\beta}^b (y) \bigg] \bigg)
\label{J2}
\ee
\be
R^*\bigg[ \pi^a (x) , {\bf j}_{V\beta}^b (y) \bigg] = &&
-i\frac{\delta \pi^a (x)}{\delta v^{\beta b} (y) }  \nonumber\\ =&&
+iG_R^{ac} (x, y) \bigg(\frac{\stackrel\leftarrow\partial}
{\partial y^{\beta}}\epsilon^{cbd}
-{\ub v}_{\beta}^{ce} (y) \epsilon^{ebd} \bigg) \pi^d (y) \nonumber\\&&
+iG_R^{ac} (x, y) {\ub a}_{\beta}^{cb} (y) \hat\sigma (y)\nonumber\\&&
+\int d^4x' G_R^{ac} (x, x') \bigg(\nabla^{\mu} a_{\mu} -J/f_{\pi}\bigg)^c (x')
\,\,R^* \bigg[ \hat\sigma (x') , {\bf j}_{V\beta}^b (y) \bigg] \nonumber\\&&
+\frac i{f_{\pi}} G_R^{ac} (x,y) \epsilon^{cbd}
\bigg({\bf j}_{A\beta}^d (y) -f_{\pi}\nabla_{\beta}^{de} (y) \pi^e (y) \bigg)
\nonumber\\&&
-\frac 1{f_{\pi}}\int d^4x' G_R^{ac} (x, x')
\bigg(\nabla^{\alpha cd} (x') R^* \bigg[ {\bf j}_{A\alpha}^d (x') ,
{\bf j}_{V\beta}^b (y) \bigg] \nonumber\\&&+
{\ub a}^{\alpha cd} (x' ) R^* \bigg[{\bf j}_{V\alpha}^d (x') ,
{\bf j}_{V\beta}^b (y) \bigg] \bigg)\nonumber\\
\label{J3}
\ee
\be
R^*\bigg[ \pi^a (x) , \hat\sigma (y) \bigg] = &&
-\frac i{f_{\pi}}
\frac{\delta \pi^a (x)}{\delta s (y) }\nonumber\\ =&&
-\frac{i}{f_{\pi}} G_R^{ac} (x, y) \pi^c (y)
\nonumber\\&&
+\int d^4x' G_R^{ac} (x, x') \bigg(\nabla^{\mu} a_{\mu} -J/f_{\pi}\bigg)^c (x')
\,\,R^* \bigg[ \hat\sigma (x') , \hat\sigma (y) \bigg] \nonumber\\&&
-\frac 1{f_{\pi}}\int d^4x' G_R^{ac} (x, x')
\bigg(\nabla^{\alpha cd} (x') R^* \bigg[ {\bf j}_{A\alpha}^d (x') ,
\hat\sigma (y) \bigg] \nonumber\\&&+
{\ub a}^{\alpha cd} (x' ) R^* \bigg[{\bf j}_{V\alpha}^d (x') ,
\hat\sigma (y) \bigg] \bigg)\nonumber\\
\label{J4}
\ee
\be
R^*\bigg[ \pi^a (x) , \pi^b (y) \bigg] = &&
-i \frac{\delta \pi^a (x)}{\delta J^b (y) }  \nonumber\\ =&&
+iG_R^{ac} (x, y)  + \frac i{f_{\pi}} G_R^{ab} (x,y) \hat\sigma (y)
\nonumber\\&&
+\int d^4x' G_R^{ac} (x, x') \bigg(\nabla^{\mu} a_{\mu} -J/f_{\pi}\bigg)^c (x')
\,\,R^* \bigg[ \hat\sigma (x') , \pi^b (y) \bigg] \nonumber\\&&
-\frac 1{f_{\pi}}\int d^4x' G_R^{ac} (x, x')
\bigg(\nabla^{\alpha cd} (x') R^* \bigg[ {\bf j}_{A\alpha}^d (x') ,
\pi^b (y) \bigg] \nonumber\\&&
+{\ub a}^{\alpha cd} (x' ) R^* \bigg[{\bf j}_{V\alpha}^d (x') ,
\pi^b (y) \bigg] \bigg)\nonumber\\
\label{J5}
\ee
which are consistent with causality.

On the other hand, the advanced version of (\ref{B93}) and (\ref{C1}) allow us
to
calculate the full commutators. We may write (\ref{C1}) as
\be
\pi_{\rm out} = &&+\pi_{\rm in} + \tilde G {\bf K} \pi_{\rm in}\nonumber\\&&
-\tilde G J +\tilde G \bigg(\nabla^{\mu}a_{\mu}-J/f_{\pi}\bigg)\hat\sigma
\nonumber\\&&
-\frac 1{f_{\pi}}\tilde{G}\bigg(
\nabla^{\mu}{\bf j}_{A\mu} +{\ub a}^{\mu}{\bf j}_{V\mu}\bigg)
\label{J6}
\ee
where
\be
\tilde{G} = \bigg( 1+ G_A {\bf K}\bigg)^{-1} G =
\Delta \bigg( 1+ {\bf K} G_R \bigg)
\label{J7}
\ee
\be
G = G_R -G_A.
\label{J8}
\ee
Noting that
\be
\delta\tilde{G} = \tilde{G} (\delta {\bf K}) G_R
\label{J9}
\ee
we find
\be
\bigg[ \pi_{\rm out} , \hat{\cal S}^{\dagger} \delta\hat{\cal S}\bigg] = &&
\delta\pi_{\rm out}  \nonumber\\ =&&
+\tilde{G}\bigg(\delta{\bf K}\bigg)\pi -  \tilde{G} \delta J +
\tilde{G}\bigg(\nabla^{\mu}\delta a_{\mu} +\delta{\ub v}^{\mu} a_{\mu} -\delta
J/f_{\pi}\bigg)\hat\sigma\nonumber\\&&
+\tilde{G}\bigg(\nabla^{\mu}a_{\mu} -J\bigg) \delta\hat\sigma
\nonumber\\&&
-\frac 1{f_{\pi}}\tilde{G}
\bigg(\delta{\ub v}^{\mu}{\bf j}_{A\mu} +\delta{\ub a}^{\mu}{\bf j}_{V\mu} +
\nabla^{\mu}\delta{\bf j}_{A\mu} +{\ub a}^{\mu}\delta {\bf j}_{V\mu} \bigg).
\label{J10}
\ee
Hence
\be
\bigg[ \pi^a (x) , {\bf j}_{A\beta}^b (y) \bigg] = &&
+iG^{ac} (x, y) {\ub a}_{\beta}^{cd} (y) \epsilon^{dbe} \pi^e (y) +
i\nabla_{\beta}^{bc} (y) \bigg( G^{ac} (x, y) \hat\sigma (y) \bigg)
\nonumber\\&&
+\int d^4x' G_R^{ac} (x, x') \bigg(\nabla^{\mu} a_{\mu} -J/f_{\pi}\bigg)^c (x')
\,\,R^* \bigg[ \hat\sigma (x') , {\bf j}_{A\beta}^b (y) \bigg] \nonumber\\&&
-\int d^4x' G_A^{ac} (x, x') \bigg(\nabla^{\mu} a_{\mu} -J/f_{\pi}\bigg)^c (x')
\,\,R^* \bigg[ {\bf j}_{A\beta}^b (y), \hat\sigma (x') \bigg] \nonumber\\&&
+\frac i{f_{\pi}} G^{ac} (x,y) \epsilon^{cbd}
\bigg({\bf j}_{V\beta}^d (y) + f_{\pi} {\ub a}_{\beta}^{de} (y) \pi^e (y)
\bigg)\nonumber\\&&
-\frac 1{f_{\pi}}\int d^4x' G_R^{ac} (x, x')
\bigg(\nabla^{\alpha cd} (x') R^* \bigg[ {\bf j}_{A\alpha}^d (x') ,
{\bf j}_{A\beta}^b (y) \bigg] \nonumber\\&&+
{\ub a}^{\alpha cd} (x' ) R^* \bigg[{\bf j}_{V\alpha}^d (x') ,
{\bf j}_{A\beta}^b (y) \bigg] \bigg)\nonumber\\&&
+\frac 1{f_{\pi}}\int d^4x' G_A^{ac} (x, x')
\bigg(\nabla^{\alpha cd} (x') R^* \bigg[ {\bf j}_{A\beta}^b (y) ,
{\bf j}_{A\alpha}^d (x') \bigg] \nonumber\\&&+
{\ub a}^{\alpha cd} (x' ) R^* \bigg[{\bf j}_{A\beta}^b (y) ,
{\bf j}_{V\alpha}^d (x') \bigg] \bigg)\nonumber\\
\label{J11}
\ee
\be
\bigg[ \pi^a (x) , {\bf j}_{V\beta}^b (y) \bigg] = &&
+iG^{ac} (x, y) \bigg(\frac{\stackrel\leftarrow\partial}
{\partial y^{\beta}}\epsilon^{cbd}
-{\ub v}_{\beta}^{ce} (y) \epsilon^{ebd} \bigg) \pi^d (y) \nonumber\\&&
+iG^{ac} (x, y) {\ub a}_{\beta}^{cb} (y) \hat\sigma (y)\nonumber\\&&
+\int d^4x' G_R^{ac} (x, x') \bigg(\nabla^{\mu} a_{\mu} -J/f_{\pi}\bigg)^c (x')
\,\,R^* \bigg[ \hat\sigma (x') , {\bf j}_{V\beta}^b (y) \bigg] \nonumber\\&&
-\int d^4x' G_A^{ac} (x, x') \bigg(\nabla^{\mu} a_{\mu} -J/f_{\pi}\bigg)^c (x')
\,\,R^* \bigg[ {\bf j}_{V\beta}^b (y) ,\hat\sigma (x')   \bigg] \nonumber\\&&
+\frac i{f_{\pi}} G^{ac} (x,y) \epsilon^{cbd}
\bigg({\bf j}_{A\beta}^d (y) -f_{\pi}\nabla_{\beta}^{de} (y) \pi^e (y) \bigg)
\nonumber\\&&
-\frac 1{f_{\pi}}\int d^4x' G_R^{ac} (x, x')
\bigg(\nabla^{\alpha cd} (x') R^* \bigg[ {\bf j}_{A\alpha}^d (x') ,
{\bf j}_{V\beta}^b (y) \bigg] \nonumber\\&&+
{\ub a}^{\alpha cd} (x' ) R^* \bigg[{\bf j}_{V\alpha}^d (x') ,
{\bf j}_{V\beta}^b (y) \bigg] \bigg)\nonumber\\ &&
+\frac 1{f_{\pi}}\int d^4x' G_A^{ac} (x, x')
\bigg(\nabla^{\alpha cd} (x') R^* \bigg[ {\bf j}_{V\beta}^b (y) ,
{\bf j}_{A\alpha} (x') \bigg] \nonumber\\&&+
{\ub a}^{\alpha cd} (x' ) R^* \bigg[{\bf j}_{V\beta}^b (y) ,
{\bf j}_{V\alpha}^d (x') \bigg] \bigg)\nonumber\\
\label{J12}
\ee
\be
\bigg[ \pi^a (x) , \hat\sigma (y) \bigg] = &&
-\frac{i}{f_{\pi}} G^{ac} (x, y) \pi^c (y)
\nonumber\\&&
+\int d^4x' G_R^{ac} (x, x') \bigg(\nabla^{\mu} a_{\mu} -J/f_{\pi}\bigg)^c (x')
\,\,R^* \bigg[ \hat\sigma (x') , \hat\sigma (y) \bigg] \nonumber\\&&
-\int d^4x' G_A^{ac} (x, x') \bigg(\nabla^{\mu} a_{\mu} -J/f_{\pi}\bigg)^c (x')
\,\,R^* \bigg[ \hat\sigma (y), \hat\sigma (x') \bigg] \nonumber\\&&
-\frac 1{f_{\pi}}\int d^4x' G_R^{ac} (x, x')
\bigg(\nabla^{\alpha cd} (x') R^* \bigg[ {\bf j}_{A\alpha}^d (x') ,
\hat\sigma (y) \bigg] \nonumber\\&&+
{\ub a}^{\alpha cd} (x' ) R^* \bigg[{\bf j}_{V\alpha}^d (x') ,
\hat\sigma (y) \bigg] \bigg)\nonumber\\&&
+\frac 1{f_{\pi}}\int d^4x' G_A^{ac} (x, x')
\bigg(\nabla^{\alpha cd} (x') R^* \bigg[ \hat\sigma (y) ,
{\bf j}_{A\alpha}^d (x')  \bigg] \nonumber\\&&+
{\ub a}^{\alpha cd} (x' ) R^* \bigg[\hat\sigma (y),
{\bf j}_{V\alpha}^d (x')  \bigg] \bigg)\nonumber\\
\label{J13}
\ee
which are consistent with (\ref{J2}-\ref{J4}) as well as locality. The
comparison between the two also gives
\be
R^*\bigg[ {\bf j}_{A\beta}^b (y), \pi^a (x) \bigg] = &&
+iG_A^{ac} (x, y) {\ub a}_{\beta}^{cd} (y) \epsilon^{dbe} \pi^e (y) +
i\nabla_{\beta}^{bc} (y) \bigg( G_A^{ac} (x, y) \hat\sigma (y) \bigg)
\nonumber\\&&
+\int d^4x' G_A^{ac} (x, x') \bigg(\nabla^{\mu} a_{\mu} -J/f_{\pi}\bigg)^c (x')
\,\,R^* \bigg[ {\bf j}_{A\beta}^b (y), \hat\sigma (x') \bigg] \nonumber\\&&
+\frac i{f_{\pi}} G_A^{ac} (x,y)\epsilon^{cbd}
\bigg({\bf j}_{V\beta}^b (y) + f_{\pi} {\ub a}_{\beta}^{de} (y) \pi^e (y)
\bigg)
\nonumber\\&&
-\frac 1{f_{\pi}}\int d^4x' G_A^{ac} (x, x')
\bigg(\nabla^{\alpha cd} (x') R^* \bigg[ {\bf j}_{A\beta}^b (y) ,
{\bf j}_{A\alpha}^a (x') \bigg] \nonumber\\&&+
{\ub a}^{\alpha cd} (x' ) R^* \bigg[{\bf j}_{A\beta}^b (y) ,
{\bf j}_{V\alpha}^d (x') \bigg] \bigg)\nonumber\\
\label{J14}
\ee
\be
R^*\bigg[ {\bf j}_{V\beta}^b (y), \pi^a (x) \bigg] = &&
+iG_A^{ac} (x, y) \bigg(\frac{\stackrel\leftarrow\partial}
{\partial y^{\beta}}\epsilon^{cbd}
-{\ub v}_{\beta}^{ce} (y) \epsilon^{ebd} \bigg) \pi^d (y) \nonumber\\&&
+iG_A^{ac} (x, y) {\ub a}_{\beta}^{cb} (y) \hat\sigma (y)\nonumber\\&&
+\int d^4x' G_A^{ac} (x, x') \bigg(\nabla^{\mu} a_{\mu} -J/f_{\pi}\bigg)^c (x')
\,\,R^* \bigg[ {\bf j}_{V\beta}^b (y) ,\hat\sigma (x') \bigg] \nonumber\\&&
+\frac i{f_{\pi}} G_A^{ac} (x,y) \epsilon^{cbd}
\bigg({\bf j}_{A\beta}^d (y) -f_{\pi}\nabla_{\beta}^{de} (y) \pi^e (y) \bigg)
\nonumber\\&&
-\frac 1{f_{\pi}}\int d^4x' G_A^{ac} (x, x')
\bigg(\nabla^{\alpha cd} (x') R^* \bigg[ {\bf j}_{V\beta}^b (y) ,
{\bf j}_{A\alpha}^d (x')  \bigg] \nonumber\\&&+
{\ub a}^{\alpha cd} (x' ) R^* \bigg[{\bf j}_{V\beta}^b (y),
{\bf j}_{V\alpha}^d (x') \bigg] \bigg)\nonumber\\
\label{J15}
\ee
\be
\bigg[ \hat\sigma (y) ,\pi^a (x) \bigg] = &&
-\frac{i}{f_{\pi}} G_A^{ac} (x, y) \pi^c (y)
\nonumber\\&&
+\int d^4x' G_A^{ac} (x, x') \bigg(\nabla^{\mu} a_{\mu} -J/f_{\pi}\bigg)^c (x')
\,\,R^* \bigg[ \hat\sigma (y), \hat\sigma (x') \bigg] \nonumber\\&&
-\frac 1{f_{\pi}}\int d^4x' G_A^{ac} (x, x')
\bigg(\nabla^{\alpha cd} (x') R^* \bigg[ \hat\sigma (y) ,
{\bf j}_{A\alpha}^d (x')  \bigg] \nonumber\\&&+
{\ub a}^{\alpha cd} (x' ) R^* \bigg[ \hat\sigma (y) ,
{\bf j}_{V\alpha}^d (x') \bigg] \bigg).\nonumber\\
\label{J16}
\ee

With some further work, we also find
\be
\bigg[ \pi^a (x) , \pi^b (y) \bigg] = &&
+iG_R^{ac} (x, y)  - \frac i{f_{\pi}}
\int d^4z G_R^{ac} (x,z) \pi^c (z) G_A^{bd} (y, z)
\bigg(\nabla^{\mu} a_{\mu} - J/f_{\pi}\bigg)^d (z)\nonumber\\&&
+\int d^4x' d^4y' G_R^{ac} (x, x' )
\bigg(\nabla^{\mu} a_{\mu} -J/f_{\pi} \bigg)^c (x')\nonumber\\&&
G_A^{bd} (y,y') \bigg(\nabla^{\mu} a_{\mu} - J/f_{\pi} \bigg)^d (y') R^*\bigg[
\hat\sigma (x ') , \hat\sigma (y ' )\bigg]\nonumber\\&&
-\frac 1{f_{\pi}}\int d^4 x' d^4 y' G_R^{ac} G_A^{bd} (y,y')
\bigg(\nabla^{\mu} a_{\mu} - J/f_{\pi} \bigg)^d (y') \nonumber\\&&
\times\bigg(\nabla^{\alpha ce} R^* \bigg[{\bf j}_{A\alpha}^e (x'), \hat{\sigma}
(y')\bigg] +{\ub a}_{\alpha}^{ce} (x') R^* \bigg[{\bf j}_{V\alpha}^e (x'),
\hat{\sigma} (y')\bigg] \bigg)\nonumber\\&&
-\frac i{f_{\pi}} \int d^4 z G_A^{bd} (y, z) \nabla^{\alpha de} (z)
\bigg( G_R^{ac} (x, z ) {\ub a}_{\alpha}^{cf} (z) \epsilon^{feg} \pi^g (z)
\bigg)
\nonumber\\&&
-\frac i{f_{\pi}} \int d^4 z G_A^{bd} (y, z)
(\nabla^{\alpha}\nabla_{\alpha})^{dc} (z)
\bigg( G_R^{ac} (x, z ) \hat\sigma (z) \bigg)
\nonumber\\&&
-\frac 1{f_{\pi}}\int d^4 x' d^4 y' G_R^{ac} \bigg(\nabla^{\mu} a_{\mu} -
J/f_{\pi}\bigg)^c (x') G_A^{bd} (y, y')\nonumber\\&&
\times\bigg(\nabla^{\alpha de} R^* \bigg[\hat{\sigma} (x') , {\bf
j}_{A\alpha}^e (y')
\bigg] +{\ub a}_{\alpha}^{de} (x') R^* \bigg[\hat{\sigma} (y') ,
{\bf j}_{V\alpha}^e (y') \bigg] \bigg)\nonumber\\&&
-\frac i{f^2_{\pi}} \int d^4z G_A^{bd} (y, z) \nabla^{\alpha de} (z)
\bigg( G_R^{ac} (x, z) \epsilon^{cef}
\bigg( {\bf j}_{V\alpha}^f (z) + f_{\pi} {\ub a}_{\alpha}^{fg} (z) \pi^g (z)
\bigg)\bigg)\nonumber\\&&
+\frac 1{f_{\pi}^2} \int d^4x' d^4 y' G_R^{ac} (x, x') G_A^{bd} (y,y')
\nonumber\\&&
\bigg(\nabla^{\alpha ce} (x') \nabla^{\beta df} (y')
R^*\bigg[{\bf j}_{A\alpha}^e (x') , {\bf j}_{A\beta}^f (y')\bigg]\nonumber\\&&
+{\ub a}^{\alpha ce} (x') \nabla^{\beta df} (y')
R^*\bigg[{\bf j}_{A\alpha}^e (x') , {\bf j}_{V\beta}^f (y')\bigg]\nonumber\\&&
+\nabla^{\alpha ce} (x') {\ub a}^{\beta df} (y')
R^*\bigg[{\bf j}_{A\alpha}^e (x') , {\bf j}_{V\beta}^f (y')\bigg]\nonumber\\&&
+{\ub a}^{\alpha ce} (x') {\ub a}^{\beta df} (y')
R^*\bigg[{\bf j}_{V\alpha}^e (x') , {\bf j}_{V\beta}^f (y')\bigg]\bigg)
\nonumber\\&&
-\frac i{f_{\pi}}\int d^4z G_A^{bd} (y, z) {\ub a}^{\alpha de} (z)
\bigg( G_R^{ac} (x, z)
\bigg(\frac{\stackrel\leftarrow\partial}{\partial z^{\alpha}}\epsilon^{cef} -
{\ub v}_{\alpha}^{cg} (z) \epsilon^{gef} \bigg) \pi^f (z) \bigg)\nonumber\\&&
+\frac i{f_{\pi}} \int d^4z G_R^{ac} (x, z) G_A^{bd} (y, z)
({\ub a}^{\alpha}{\ub a}_{\alpha} )^{cd} (z) \hat\sigma (z) \nonumber\\&&
-\frac i{f_{\pi}^2} \int d^4z G_R^{ac} (x, z) G_A^{bd} (y, z)
{\ub a}^{\alpha de} (z) \epsilon^{cef}
\bigg({\bf j}_{A\alpha}^f (z) - f_{\pi} \nabla_{\alpha}^{fg} (z) \pi^g (z)
\bigg)\nonumber\\&&
+ \frac i{f_{\pi}} G_R^{ab} (x, y) \hat\sigma (x) -
\bigg( x\,\, a \leftrightarrow y\,\, b\bigg)
\label{J17}
\ee
from (\ref{B93}) and (\ref{J10}), which may be checked to be consistent with
(\ref{J5}) and (\ref{J14}-\ref{J16}). Time-ordered products may be also
computed from
\be
T^*\bigg( \pi^a (x) {\bf j}_{A\beta}^b (y) \bigg) =
R^* \bigg[ \pi^a (x) , {\bf j}_{A\beta}^b (y) \bigg] +
{\bf j}_{A\beta}^b (y) \pi^a (x)
\label{J18}
\ee
and so on. The results coincide with (\ref{B96}-\ref{B99}).

Another type of commutator appears in connection with integrability. We may
define
\be
\Upsilon_R {\cal O} = &&\frac{\delta{\cal O}}{\delta J} -i{\cal O}
\bigg( 1+G_R {\bf K}\bigg) \pi_{\rm in} + i G_R J {\cal O}
\nonumber\\&&
-\frac 1{f_{\pi}} G_R \bigg(\nabla^{\mu} a_{\mu} - J/f_{\pi}\bigg)
\frac{\delta{\cal O}}{\delta s}
+\frac 1{f_{\pi}} G_R {\bf X}_A {\cal O}
\label{J19}
\ee
\be
\Upsilon_A {\cal O} = &&\frac{\delta{\cal O}}{\delta J} -i
\bigg( 1+G_R {\bf K}\bigg) \pi_{\rm in}{\cal O} + i G_A J {\cal O}
\nonumber\\&&
-\frac 1{f_{\pi}} G_A \bigg(\nabla^{\mu} a_{\mu} - J/f_{\pi}\bigg)
\frac{\delta{\cal O}}{\delta s}
+\frac 1{f_{\pi}}G_A {\bf X}_A {\cal O}
\label{J20}
\ee
for a general operator ${\cal O}$.
With some further work, we find
\be
\bigg[\Upsilon_R^a (x) , \Upsilon_R^b (y) \bigg] \,\,{\cal O} = &&
i
\bigg(\frac{\delta}{\delta J^a (x) }\int d^4y' G_R^{bd} (y, y') J^d (y') \bigg)
{\cal O}\nonumber\\&&
-\frac 1{f_{\pi}}
\bigg(\frac{\delta}{\delta J^a (x) }\int d^4y' G_R^{bd} (y, y')
\bigg(\nabla^{\mu} a_{\mu} - J/f_{\pi}\bigg)^d (y') \bigg)
\frac{\delta{\cal O}}{\delta s (y')}\nonumber\\&&
+ i\frac 1{f_{\pi}} {\cal O}
\int d^4 x' G_R^{ac} (x, x')
\bigg(\nabla^{\mu} a_{\mu} - J/f_{\pi}\bigg)^c (x')\nonumber\\&&
\frac {\delta}{\delta s (x')} \int d^4y' \,G_R^{bd} (y,y') {\bf K}^{de} (y')
\pi_{\rm in}^e (y')\nonumber\\&&
- \frac i{f_{\pi}}
\int d^4 x' G_R^{ac} (x, x') \bigg(\nabla^{\mu} a_{\mu} - J/f_{\pi}\bigg)^c
(x')
\nonumber\\&&\times\bigg(
\frac{\delta}{\delta s (x')}\int d^4y' G_R^{bd} (y, y') J^d (y')\bigg)
{\cal O}\nonumber\\&&
+\frac 1{f_{\pi}^4}
\int d^4 x' G_R^{ac} (x, x') \bigg(\nabla^{\mu} a_{\mu} - J/f_{\pi}\bigg)^c
(x')
\nonumber\\&&\times\bigg(
\frac{\delta}{\delta s (x')}\int d^4y' G_R^{bd} (y, y')
\bigg(\nabla^{\mu} a_{\mu} -J/f_{\pi}\bigg)^d (y')\bigg)
\frac{\delta{\cal O}}{\delta s (y')}\nonumber\\&&
-\frac 1{f_{\pi}^2}
\int d^4 x' G_R^{ac} (x, x') \bigg(\nabla^{\mu} a_{\mu} - J/f_{\pi}\bigg)^c
(x')
\nonumber\\&&\times\bigg(
\frac{\delta}{\delta s (x')}\int d^4y' G_R^{bd} (y, y') {\bf X}_A^{d} (y')
\bigg) {\cal O}\nonumber\\&&
-\frac i{f_{\pi}} {\cal O}
\int d^4 x' G_R^{ac} (x, x') {\bf X}_A^c (x')
\int d^4 y' G_R^{bd} (y, y') {\bf K}^{de} (y')\pi^e_{\rm in} (y')\nonumber\\&&
+\frac i{f_{\pi}}
\int d^4 x' G_R^{ac} (x, x') \bigg({\bf X}_A^c (x')
\int d^4 y' G_R^{bd} (y, y') J^d (y')\bigg) {\cal O} \nonumber\\&&
-\frac 1{f_{\pi}^2}
\int d^4 x' G_R^{ac} (x, x') \bigg({\bf X}_A^c (x')
\int d^4 y' G_R^{bd} (y, y') \nonumber\\&&
\times\bigg(\nabla^{\mu} a_{\mu} -J/f_{\pi}\bigg)^d (y')
\bigg) \frac{\delta{\cal O}}{\delta s (y')} \nonumber\\&&
+\frac 1{f_{\pi}^2}\int d^4 x' G_R^{ac} (x, x') \bigg({\bf X}_A^c (x')
\int d^4 y' G_R^{bd} (y, y') \bigg) {\bf X}_A^d (y') {\cal O} \nonumber\\&&
-\bigg( x\,\, a \leftrightarrow y\,\, b\bigg)\nonumber\\&&
+\frac 1{f_{\pi}^2}
\int d^4 x' d^4y' G_R^{ac} (x, x') G_R^{bd} (y, y')
\bigg[ {\bf X}_A^c (x') , {\bf X}_A^d (y')\bigg]
{\cal O} \nonumber\\&&
-{\cal O}
\bigg[ \bigg( 1 +G_R {\bf K} \bigg)^{bd} \pi_{\rm in}^d (y) ,
\bigg( 1 +G_R {\bf K} \bigg)^{ac} \pi_{\rm in}^c (y) \bigg] \nonumber\\ =&&
\frac 1{f_{\pi}^2}\int d^4z G_R^{ac} (x, z) G_R^{bd} (y, z)
\bigg( J^c (z) \Upsilon_R^d (z) {\cal O} \nonumber\\&&
-J^d (z) \Upsilon^c_R (z) {\cal O} -\epsilon^{cde}
\bigg({\bf X}_V^e (z) +
{\ub J}^{ef} (z) \frac {\delta}{\delta J^f (z)}\bigg) {\cal O}\bigg).
\label{J21}
\ee
Similarly,
\be
\bigg[ \Upsilon_A^a (x), \Upsilon_A^b (y) \bigg] =&&
+\frac 1{f_{\pi}^2}\int d^4z G_A^{ac} (x, z) G_A^{bd} (y, z)
\bigg( J^c (z) \Upsilon_A^d (z) {\cal O} \nonumber\\&&
-J^d (z) \Upsilon^c_A (z) {\cal O} -\epsilon^{cde}
\bigg({\bf X}_V^e (z) +
{\ub J}^{ef} (z) \frac {\delta}{\delta J^f (z)}\bigg) {\cal O}\bigg)
\label{J22}
\ee
\be
\bigg[ \Upsilon_R^a (x), \Upsilon_A^b (y) \bigg] =&&
-\frac 1{f_{\pi}}\int d^4z G_R^{ac} (x, z) G_A^{bd} (y, z)
\bigg(\nabla^{\mu} a_{\mu} - J/f_{\pi} \bigg)^c (z) \Upsilon^d_A (z) {\cal
O}\nonumber\\&&
-\frac 1{f_{\pi}}\int d^4z G_R^{ac} (x, z) \nabla_{\mu}^{cf} (z)
\nonumber\\&&
\times\bigg( G_A^{bd} (y, z) \bigg(\epsilon^{dfg} {\ub a}^{\mu ge} +
{\ub a}^{\mu dg} (z) \epsilon^{gfe} \bigg) \Upsilon_A^e (z) {\cal O}\bigg)
\nonumber\\&&
+\frac 1{f_{\pi}}\int d^4z G_R^{ac} (x, z) {\ub a}_{\mu}^{cf} (z)
G_A^{bd} (y, z)\epsilon^{dfg} \nabla^{\mu ge} (z) \Upsilon^e_A (z) {\cal O}
\nonumber\\&&
+\frac 1{f_{\pi}}
\int d^4z G_R^{ac} (x, z) {\ub a}_{\mu}^{cf} (z)
\Upsilon^e_A (z) {\cal O} \epsilon^{efg} \nabla^{\mu gd} (z) G_A^{bd} (y, z)
\nonumber\\&&
-\bigg(x\,\,, a\,\,, R\leftrightarrow y\,\,, b\,\,, A \bigg)\nonumber\\&&
-\frac 1{f_{\pi}^2}
\int d^4z G_R^{ac} (x, z) G_A^{bd} (y, z) \epsilon^{cde}
\nonumber\\&&
\times\bigg({\bf X}_V^e (z) +
{\ub J}^{ef} (z) \frac {\delta}{\delta J^f (z)}\bigg) {\cal O}.
\label{J23}
\ee
We may also check that
\be
\bigg[ {\bf X}_V^a (x) +{\ub J}^{ac} (x) \frac{\delta}{\delta J^c (x) },
\Upsilon^b_{R, A} (y) \bigg] =
-\epsilon^{abc} \delta^4 (x-y) \Upsilon^c_{R, A} (y).
\label{J24}
\ee

The eqns. (\ref{B35}, \ref{J21}-\ref{J24}) imply that the master equations
(\ref{B57}, \ref{B94}) constitute an integrable system. It follows that the
solutions of (\ref{B57},\ref{C1}) can be consistently extended to solutions of
(\ref{B57},\ref{B94}). Since the space of external fields and Fock space of
$\pi_{\rm in}$ is topologically trivial, there should be no global obstruction.
Eqns. (\ref{J2}-\ref{J5}, \ref{J11}-\ref{J13}) then guarantee that causality
and
locality are duly maintained.

The linear character of the master formula is further emphasized if we
introduce coherent states,
\be
a_{\rm in}^a (k) |\alpha \,\,{\rm in} > = &&
\alpha^a (k) |\alpha \,\,{\rm in} > \nonumber\\
a^{a\dagger }_{\rm in} (k) |\alpha \,\,{\rm in} > = &&
(2\pi)^3 2k^0 \frac{\delta}{\delta\alpha^a (k)}
|\alpha \,\,{\rm in} > \nonumber\\
<\beta \,\,{\rm in} | a^{\dagger a}_{\rm in} (k)  = &&
 <\beta \,\,{\rm in} | \beta^{a *} (k)  \nonumber\\
<\beta \,\,{\rm in} | a_{\rm in}^a (k)  = &&
(2\pi)^3 2k^0 \frac{\delta}{\delta\beta^{a *} (k)}
<\beta \,\,{\rm in} |.
\label{J25}
\ee
The master formula (\ref{B94}) then reads
\be
&&\frac {\delta}{\delta J^a (x)}
<\beta\,\, {\rm in} | \hat{\cal S} |\alpha \,\, {\rm in} > \nonumber\\&&
-i\int d^3 k \int d^4y e^{ik\cdot y} \bigg( 1 +G_R {\bf K}\bigg)^{ab}
(x, y) \frac{\delta}{\delta\alpha^a (k)}
<\beta\,\, {\rm in} | \hat{\cal S} |\alpha \,\, {\rm in} > \nonumber\\&&
-\frac 1{f_{\pi}}\int d^4 y G_R^{ab} (x, y)
\bigg(\nabla^{\mu} a_{\mu} - J/f_{\pi} \bigg)^b (y)
\frac {\delta}{\delta s (x)}
<\beta\,\, {\rm in} | \hat{\cal S} |\alpha \,\, {\rm in} > \nonumber\\&&
+\frac 1{f_{\pi}}\int d^4 y G_R^{ab} (x, y)
{\bf X}_A^b (y)
<\beta\,\, {\rm in} | \hat{\cal S} |\alpha \,\, {\rm in} > -\theta_R^a (x)
<\beta\,\, {\rm in} | \hat{\cal S} |\alpha \,\, {\rm in} >  = 0
\label{J26}
\ee
\be
\theta_R^a (x) =&&+ i \int d^4y
\bigg( 1 + G_R {\bf K}\bigg)^{ab} (x, y)\bigg(
\frac 1{(2\pi)^3}\int \frac {d^3k}{2k^0} \alpha^b (k) e^{-ik\cdot y}\bigg)
\nonumber\\&&
-i  \int d^4 y G_R^{ab} (x, y) J^b (y)
\label{J27}
\ee
\be
&&\frac {\delta}{\delta J^a (x)}
<\beta\,\, {\rm in} | \hat{\cal S} |\alpha \,\, {\rm in} > \nonumber\\&&
-i \int d^3 k \int d^4y e^{-ik\cdot y} \bigg( 1 +G_A {\bf K}\bigg)^{ab}
(x, y) \frac{\delta}{\delta\beta^{a *} (k)}
<\beta\,\, {\rm in} | \hat{\cal S} |\alpha \,\, {\rm in} > \nonumber\\&&
-\frac 1{f_{\pi}}\int d^4 y\,\, G_A^{ab} (x, y)
\bigg(\nabla^{\mu} a_{\mu} - J /f_{\pi}\bigg)^b (y)
\frac {\delta}{\delta s (x)}
<\beta\,\, {\rm in} | \hat{\cal S} |\alpha \,\, {\rm in} > \nonumber\\&&
+\frac 1{f_{\pi}}\int d^4 y\,\, G_A^{ab} (x, y)
{\bf X}_A^b (y)
<\beta\,\, {\rm in} | \hat{\cal S} |\alpha \,\, {\rm in} > -\theta_A^a (x)
<\beta\,\, {\rm in} | \hat{\cal S} |\alpha \,\, {\rm in} >  = 0
\label{J28}
\ee
\be
\theta_A^a (x) =&&+ i\int d^4y
\bigg( 1 + G_A {\bf K}\bigg)^{ab} (x, y)\bigg(
\frac 1{(2\pi)^3}\int \frac {d^3k}{2k^0} \beta^{b *} (k) e^{+ik\cdot y}
\bigg)\nonumber\\&&
-i  \int d^4 y G_A^{ab} (x, y) J^b (y).
\label{J29}
\ee

These equations can be solved as usual by introducing characteristics.
For the retarded equations (\ref{J26}-\ref{J27})
\be
\frac{d J^b (y, \tau )}{d\tau} = && +\delta^4 (x-y) \delta^{ab}\nonumber\\
\frac{d \alpha^b (k, \tau )}{d\tau} = && -i \int d^4y e^{+ik\cdot y}
\bigg( 1+ G_R {\bf K} \bigg)^{ab} (x, y, \tau )\nonumber\\
\frac{d s (y, \tau )}{d\tau} = && -\frac 1{f_{\pi}}G_R^{ab} (x, y, \tau )
\bigg(\nabla^{\mu} a_{\mu} - J/f_{\pi}\bigg)^b (y, \tau) \nonumber\\
\frac{d a_{\mu}^c (y, \tau )}{d\tau} = && +\frac 1{f_{\pi}}
G_R^{ab} (x, y, \tau )
\bigg(-\frac{\stackrel\leftarrow\partial}{\partial y^{\mu}}\delta^{bc} +
{\ub v}_{\mu}^{bc} (y, \tau )\bigg) \nonumber\\
\frac{d v_{\mu}^c (y, \tau )}{d\tau} = && +G_R^{ab} (x, y, \tau )
{\ub a}_{\mu}^{bc} (y, \tau )
\label{J30}
\ee
where $x$ and $a$ are fixed, and $G_R (x, y, \tau )$ means that $s$,
$a_{\mu}^a$ and $v_{\mu}^a$ in $G_R (x, y)$ are replaced by their running
values. The master formula then simplifies to
\be
<\beta\,\, {\rm in} | \hat{\cal S} [\tau_1 ] |\alpha [\tau_1 ]
\,\, {\rm in} >  =
<\beta\,\, {\rm in} | \hat{\cal S} [\tau_0 ] |\alpha [\tau_0 ]
\,\, {\rm in} >
{\rm exp}\bigg( \int_{\tau_0}^{\tau_1} d\tau
\theta_R^a (x, \tau ) \bigg)
\label{J31}
\ee
and a similar relation for the advanced version.
So far, we have been unable to extract practical information from
(\ref{J30}), even though it is a system of ordinary differential equations.
This is to be contrasted with the nonlinear partial differential equation
for operator-valued distributions we started from. Another open issue regarding
this formulation is whether any useful information could be extracted in the
classical limit.

\vskip 1.5cm
{\bf Appendix E : General Analysis of One-Loop Effects}
\vskip .10 cm

Let us consider the phase change in the S-matrix $\hat{\cal S }\rightarrow
\hat{\cal S} e^{i\alpha}$, where $\alpha$ may depend on the external fields
$\phi = (v,a,s,J)$. Substitution into the causality condition
\be
\frac{\delta}{\delta \phi (y)}\bigg( \hat{\cal S}^{\dagger}
\frac{\delta\hat{\cal S}}{\delta\phi (x)}\bigg) = 0\qquad
( x\preceq y )
\label{I1}
\ee
gives
\be
\frac{\delta^2\alpha}{\delta\phi (y) \delta\phi (x)} =0
\qquad ( x\preceq y ).
\label{I2}
\ee
By symmetry
\be
\frac{\delta^2\alpha}{\delta\phi (y)\delta\phi (x)} =0
\qquad x\neq y
\label{I3}
\ee
There are two types of solutions to (\ref{I3}). One is a constant plus linear
terms in $\phi$. The constant may  be eliminated by the normalization
$<0| {\bf S} |0> = 1$. Terms linear in $v,a,J$ are forbidden by isospin,
whereas
translation invariance requires the term linear in $s$ to be of the form
\be
\int d^4x \, s(x).
\label{I4}
\ee
The other type of solution is
\be
\int d^4x \, P(x)
\label{I5}
\ee
where $P(x)$ is a polynomial in $\phi (x)$ and its derivatives. (\ref{I4}) may
then be absorbed into this case. We also note that the space of $\phi$ is
topologically trivial, so there should be no Wess-Zumino type terms which
require extra coordinates.

The $J$ dependence is fixed by the master equation (\ref{B94}), so we may set
$J=0$. We further assume that the theory contains only two scales $f_{\pi}$
and $m_{\pi}$, which should be the case for QCD with two flavors. Then the
$P$'s with mass dimension 5 or higher in $\phi$ and its derivatives will be
accompanied by powers of $1/f_{\pi}$ times some dimensionless function
$h(m_{\pi}/f_{\pi})$. The existence of the chiral limit requires $h(0)$ to be
finite, so such terms belong to higher order in the loop expansion.

The analysis must be modified for QCD with three flavors or the linear sigma
model, which contain an extra mass scale $m_K$ or $m_{\sigma}$. However, if we
assume that the theory is finite as $m_K/f_{\pi}$ or
$g=m_{\sigma}^2/2f_{\pi}^2$ go to zero, the same conclusion applies.
It is also possible that the theory contains a heavy scale $M$. In that case,
we assume that the amplitudes can be expanded in $p/f_{\pi}$,
$f_{\pi}/M$ and $p/M$, where $p$ stands for a typical momentum or a light mass.
If the latter two terms can be ignored, we are back to the same result.

Let us apply our results to $\hat{\cal S}_0$. Substitution of $\hat{\cal
S}_0\rightarrow \hat{\cal S}_0 e^{i\alpha}$ into (\ref{D3}), gives
$\delta\alpha /\delta J =0$, in agreement with the explicit representation
(\ref{D4}). As we have seen,
it is then sufficient to consider $P$'s with dimension 4 or less in
$\phi =(v,a,s)$ and its derivatives. Besides local isospin invariance ${\bf
X}_V \alpha =0$, G-parity requires $a_{\mu}^a$ to appear in even powers,
and parity forbids the use of the Levi-Civita $\epsilon_{\mu\nu\rho\sigma}$.

There are no terms with three or four derivatives. Ignoring total divergences,
there are three terms with two derivatives
\be
&&-\frac {c_1}{4} v^{\mu\nu} \cdot v_{\mu \nu}\nonumber\\&&+
\frac {c_2}2 (\nabla^{\mu} a^{\nu} )\cdot (\nabla_{\mu} a_{\nu} )\nonumber\\&&
+\frac {c_3}2 (\nabla^{\mu} a_{\mu} )^2
\label{I6}
\ee
since the other combination
\be
+(\nabla^{\mu} a^{\nu} ) (\nabla_{\nu} a_{\mu})
\label{I7}
\ee
reduces to the last term of (\ref{I6}) and (\ref{I8}) below up to a total
divergence. There is essentially one term with one derivative
\be
+\frac {c_4}2 \epsilon^{abc} v_{\mu\nu}^a a^{\mu b} a^{\nu c}
\label{I8}
\ee
and six terms with no derivatives
\be
&&+\frac{c_5}4 (a_{\mu}\cdot a^{\mu} )^2 \qquad ,\nonumber\\&&
+\frac{c_6}4 (a_{\mu}\cdot a_{\nu} )(a^{\mu}\cdot a^{\nu} ) \qquad ,
\nonumber\\&&
+\frac {c_7}2 s a_{\mu}\cdot a^{\mu} \qquad ,\nonumber\\&&
+\frac {c_8}2 s^2 \qquad ,\nonumber\\&&
+\frac {c_9}2 m_{\pi}^2 a_{\mu}\cdot a^{\mu}\qquad ,\nonumber\\&&
+c_{10} m_{\pi}^2 s\qquad .
\label{I9}
\ee

The results of section 18 imply that ,
$<0|\hat{\sigma} |0>$, $<0|T^*({\bf j}_{A\alpha}^a (x) {\bf j}_{A\beta }^b (y)
) |0>$ and
$<0|T^*({\bf j}_{A\alpha}^a (x) {\bf j}_{A\beta }^b (y) {\bf V}_{\gamma}^c (z)
) |0>$ at one-loop level are given entirely by the contact terms above.
Specifically,
\be
<0| \hat\sigma  |0> = c_{10} \frac {m_{\pi}^2}{f_{\pi}}
\label{I10}
\ee
\be
<0|T^*\bigg({\bf j}_{A\alpha}^a (x) {\bf j}_{A\beta }^b (y)\bigg) |0> = &&
+ic_2 \delta^{ab}\,g_{\alpha \beta} \Box \delta^4 (x-y) \nonumber\\&&+ ic_3
\delta^{ab}
\partial_{\alpha}\partial_{\beta}\delta^4 (x-y) \nonumber\\&&
-ic_9 \delta^{ab} g_{\alpha\beta} m_{\pi}^2 \delta^4 (x-y)
\label{I11}
\ee
\be
<0|T^*\bigg({\bf j}_{A\alpha}^a (x) {\bf j}_{A\beta }^b (y) {\bf V}_{\gamma}^c
(z)
\bigg) |0> = &&
+c_2 \epsilon^{abc} g_{\alpha\beta} \delta^4 (x-z) \partial_{\gamma} \delta^4
(y-z)\nonumber\\ &&
+c_3 \epsilon^{abc} g_{\alpha\gamma} \delta^4 (x-z) \partial_{\beta} \delta^4
(y-z)\nonumber\\ &&
+c_4 \epsilon^{abc} g_{\alpha\gamma} \delta^4 (x-y) \partial_{\beta} \delta^4
(y-z)\nonumber\\ &&
+ \bigg( x, \alpha , a \rightarrow y, \beta, b \bigg).
\label{I12}
\ee
Substitution of (\ref{I10}-\ref{I11}) into (\ref{C10}-\ref{C11}) gives
\be
c_{10} = c_2 +c_3 +c_9
\label{I13}
\ee
while substitution into (\ref{C54}-\ref{C55}) gives
\be
{\bf F}_V (t) = 1 -\frac {m_{\pi}^2}{f_{\pi}^2} c_9 +
\frac t{2f_{\pi}^2} \bigg( c_4 -c_2 +{\bf \Pi}_V (t) \bigg)
+{\cal O} \bigg(\frac 1 {f_{\pi}^4}\bigg).
\label{I14}
\ee
The normalization condition ${\bf F}_V (0) = 1$ requires\footnote{It is not
clear to us, however, why $c_9$ manages to violate ${\bf F}_V (0) =1$.}
\be
c_9 = 0.
\label{I15}
\ee
Combining with (\ref{D36}) then gives
\be
{\bf F}_V (t) = 1 + \frac t{2f_{\pi}^2}
\bigg( (c_4 +2c_1)- (c_2+c_1)
+\frac 13 (1-\frac {4m_{\pi}^2}{t} ) {\cal J} (t) +
\frac{1}{72\pi^2}\bigg) + {\cal O} \bigg( \frac {1}{f_{\pi}^4}\bigg).
\label{I16}
\ee
Substitution of (\ref{I10}-\ref{I12}) into (\ref{C67}) also yields
\be
\frac 1{f_{\pi}} \bigg( (c_4+2c_1) - 2(c_2+c_1)\bigg)
\bigg( p\cdot q \epsilon_{\nu} (q) -\epsilon (q) \cdot p q_{\nu} \bigg)
+{\cal O}\bigg( \frac {1}{f_{\pi}^3}\bigg).
\label{I17}
\ee

We may also compute other quantities. From (\ref{B85})
\be
<0| T^*\bigg( {\bf A}_{\mu}^a (x) {\bf A}_{\nu}^b (y) \bigg) |0> =
&&-if_{\pi}^2
\delta^4 (x-y) g_{\mu\nu} \delta^{ab} \nonumber\\&&+
<0| T^*\bigg( {\bf j}_{A\mu}^a (x) {\bf j}_{A\nu}^b (y) \bigg) |0>\nonumber\\&&
-f_{\pi} \frac{\partial}{\partial x^{\mu}}
<0| T^*\bigg( {\pi }^a (x) {\bf j}_{A\nu}^b (y) \bigg) |0>\nonumber\\&&
-f_{\pi} \frac{\partial}{\partial y^{\nu}}
<0| T^*\bigg( {\bf j}_{A\mu}^a (x) {\pi}^b (y) \bigg) |0>\nonumber\\&&
+f_{\pi}^2 \frac{\partial}{\partial x^{\mu}}\frac{\partial}{\partial y^{\nu}}
<0| T^*\bigg( {\pi }^a (x) {\pi }^b (y) \bigg) |0>
\label{I18}
\ee
and from (\ref{B96})
\be
&&<0| T^*\bigg( {\pi}^a (x) {\bf j}_{A\nu}^b (y) \bigg) |0> =
i\delta^{ab} \frac{\partial}{\partial y^{\nu}}\Delta_R (x-y) <0|\hat{\sigma}
|0> \nonumber\\&&
-\frac 1{f_{\pi}} \int d^4x' \Delta_R (x-x')
\frac{\partial}{\partial x_{\alpha}'}
<0| T^*\bigg( {\bf j}_{A\mu}^a (x') {\bf j}_{A\nu}^b (y) \bigg) |0>.
\label{I19}
\ee
Also from (\ref{B96},\ref{B99}) we have
\be
<0| T^* \bigg(\pi^a (x) \pi^b (y) \bigg) |0>  =&&+ i\delta^{ab} \Delta_F (x-y)
+
\frac {2i}{f_{\pi}} \delta^{ab} \Delta_R (x-y)  <0| \hat{\sigma}  | 0>
\nonumber\\&&
+\frac i{f_{\pi}} m_{\pi}^2 \delta^{ab} \int d^4 x' \Delta_R (x-x')
\Delta_A (y-x') <0|\hat{\sigma }  |0> \nonumber\\&&
+\frac i{f^2_{\pi}} \int d^4 x' d^4 y' \Delta_R (x-x')
\Delta_A (y-y') \nonumber\\&&\times \frac{\partial}{\partial x_{\alpha}'}
\frac{\partial}{\partial y_{\beta}'}
<0|T^*\bigg( {\bf j}_{A\alpha}^a (x') {\bf j}_{A\beta }^b (y')\bigg) |0>.
\label{I20}
\ee
Hence
\be
&&i\int d^4 x e^{ip\cdot (x-y)} <0| T^* \bigg( {\pi}^a (x) {\pi}^b (y)
\bigg) |0>  =\nonumber\\&& -\delta^{ab} \frac 1{p^2-m_{\pi}^2 +i0}
+\frac 1{f_{\pi}^2} \delta^{ab} (c_2+c_3) +{\cal O} \bigg( \frac
1{f_{\pi}^4}\bigg)
\label{I21}
\ee
\be
&&i\int d^4 x e^{ip\cdot (x-y)} <0| T^* \bigg( {\bf A}_{\mu}^a (x) {\bf
A}_{\nu}^b (y) \bigg) |0>  = +f_{\pi}^2 g_{\mu\nu} \delta^{ab} \nonumber\\&&
-f_{\pi}^2 \delta^{ab}\, p_{\mu} p_{\nu}\, \frac 1{p^2-m_{\pi}^2 +i0}
-c_2 \delta^{ab}\bigg( -g_{\mu\nu} p^2 + p_{\mu} p_{\nu} \bigg)  +
{\cal O}\bigg( \frac 1{f_{\pi}^2}\bigg) .
\label{I22}
\ee

The contact term proportional to $f^2_{\pi} g_{\mu\nu} \delta^{ab}$ in
(\ref{I18},\ref{I22}) deserves some comment. The appearance of such a term is
natural in the gauged  nonlinear sigma model where the action is quadratic in
$a_{\mu}^a$, so that ${\bf A}_{\mu}^a $ contains a term proportional to
$a_{\mu}^a$,
\be
{\bf A}_{\mu}^a (x) = f_{\pi}^2 a_{\mu}^a (x) + ...\,.
\label{I23}
\ee
However, the appearance of such a term is rather surprising for QCD, where the
action is nominally linear in $a_{\mu}^a$. Nevertheless such a term must appear
if QCD does what it is supposed to do, namely generate pions. In particular,
the
Ward identity in the chiral limit (\ref{B26}) implies
\be
i\int d^4x e^{ip\cdot (x-y)}<0|T^*\bigg( {\bf A}_{\mu}^a (x) {\bf A}_{\nu}^b
(y) \bigg) |0> = f_{\pi}^2 \delta^{ab}\bigg( g_{\mu\nu} -\frac {p_{\mu}
p_{\nu}}{p^2+i0}\bigg) + ...\,.
\label{I24}
\ee
We observe that while the problem is related to the short distance behavior of
the time-ordered product, it cannot be seen in perturbative QCD, since
dimensional transmutation implies that $f_{\pi}\sim \mu e^{-1/g^2}$, where
$\mu$ is the renormalization scale and $g$ is the QCD coupling constant.

To come back to the main issue, we note that two parameters are $off$ $shell$
in the following sense. The master formulas (\ref{B57}) and (\ref{B94}) are
invariant under $\hat{\cal S}\rightarrow {\hat {\cal S}}e^{i\alpha}$, if
$\alpha$ is a linear combination of
\be
&&-\frac{c_1}8 \int d^4x \bigg( F_{L\mu\nu}\cdot F_{L}^{\mu\nu} +
F_{R\mu\nu}\cdot F_{R}^{\mu\nu} \bigg) =\nonumber\\&&
-\frac {c_1}4 \int d^4x \bigg(
v_{\mu\nu}\cdot v^{\mu\nu} +
(\nabla^{\mu} a^{\nu} -\nabla^{\nu} a^{\mu} )\cdot
(\nabla_{\mu} a_{\nu} -\nabla_{\nu} a_{\mu} )\nonumber\\&&
+\epsilon^{abc}\epsilon^{ade} a_{\mu}^b a_{\nu}^c a^{\mu d} a^{\nu e} +
2\epsilon^{abc} v_{\mu\nu}^a a^{\mu b} a^{\nu c}\bigg)=\nonumber\\&&
-\frac {c_1}4 \int d^4x \bigg(
v_{\mu\nu}\cdot v^{\mu\nu} +
2(\nabla^{\mu} a^{\nu}) \cdot
(\nabla_{\mu} a_{\nu}) -2 (\nabla^{\mu} a_{\mu} )^2\nonumber\\&&
+\epsilon^{abc}\epsilon^{ade} a_{\mu}^b a_{\nu}^c a^{\mu d} a^{\nu e} +
4\epsilon^{abc} v_{\mu\nu}^a a^{\mu b} a^{\nu c}\bigg)
\label{I25}
\ee
and
\be
\frac {c_3-c_1}2 \int d^4x \bigg(
s^2+ 2m_{\pi}^2 s + (\nabla^{\mu} a_{\mu} - J/f_{\pi} )^2 \bigg)
\label{I26}
\ee
where $F_{R,L}^{\mu\nu}$ are the field strengths associated with $v_{\mu}\pm
a_{\mu}$ respectively. Therefore, such a phase redefinition will change the
Green's functions only by some polynomials in the momenta, and will not affect
on-shell quantities which are given by pole terms. Therefore on-shell
quantities can depend only on six parameters, which may be chosen to be
\be
c_2+c_1,\,\, c_4+2c_1,\,\, c_5+c_1,\,\, c_6-c_1,\,\,c_7,\,\,
c_8-(c_3-c_1).
\label{I27}
\ee
It is easy to see that (\ref{I16}-\ref{I17}) involve only linear combinations
of the above.  We may note that the result is essentially independent of the
high energy behavior of QCD or any other model. The only constraint on the high
energy behavior is that it should not destroy the Veltman-Bell
equations in the first place.

Comparison of (\ref{C55}, \ref{D36}) and
(\ref{I13},\ref{I16},\ref{I17},\ref{I21},\ref{I22}) with the corresponding
expressions in the work of Gasser and Leutwyler \cite{leutwyler} suggests the
following identification
\be
c_1 +\frac 1{72\pi^2} &&\leftrightarrow \frac 1{48\pi^2} (\overline{h}_2-\frac
13)\nonumber\\
c_{10} && \leftrightarrow \frac 1{8\pi^2} (\overline{h}_1-\overline{l}_4)
\nonumber\\
c_4-c_2 +c_1 +\frac 1{72\pi^2} &&\leftrightarrow \frac 1{48\pi^2}
(\overline{l}_6-\frac 13)\nonumber\\
c_4-2c_2 &&\leftrightarrow \frac 1{48\pi^2} (\overline{l}_6-\overline{l}_5)
\nonumber\\
c_2+c_3 &&\leftrightarrow \frac 1{8\pi^2} (\overline{h}_1-\overline{l}_4)
\nonumber\\
c_2 &&\leftrightarrow \frac 1{48\pi^2} (\overline{l}_5-\overline{h}_2).
\label{I28}
\ee
The above results are consistent. In particular, we recover a relation between
the pion electromagnetic form factor (\ref{I16}), the pion radiative decay
$\pi\rightarrow e\nu\gamma$ (\ref{I17}) and the difference between the vector
and axial-vector correlator
\be
<0|T^*\bigg({\bf V}_{\mu}^a (x){\bf V}_{\nu}^b (y)
-{\bf A}_{\mu}^a (x){\bf A}_{\nu}^b (y) \bigg) |0>
\nonumber
\ee
as given by (\ref{C55}, \ref{D36}) and (\ref{I22}). This relation is not
necessarily equivalent to the Das-Mathur-Okubo relation \cite{das}, since the
spectral function for the axial-vector correlator does not have a transverse
part at one loop as discussed in 18.3.
We believe that this remark applies also to the analysis
performed by Gasser and Leutwyler,  although it appears to have been stated
otherwise \cite{leutwyler}. The correspondence also implies
that the parameters $c_2+c_4+c_1$ and $c_2+c_1$ involving the vector current
${\bf V}_{\mu}^a (x)$ are relatively large, but the difference is rather small.
Numerically,
\be
16\pi^2 (c_4-c_2+c_1) = 32\pi^2 f_{\pi}^2 {\bf F}_V' (0) =
5.17\pm 0.35
\label{I29}
\ee
The radiative decay $\pi\rightarrow e\nu\gamma$ gives
\be
16\pi^2 (c_4-2c_2) = 0.88 \pm 0.24
\label{I30}
\ee
so that
\be
16\pi^2 (c_2+c_1) = 4.29 \pm 0.42
\label{I31}
\ee

Let us now consider $\pi\pi$ scattering again. Equations (\ref{D27}-\ref{D31})
are modified to
\be
<0| T^*\bigg(\hat\sigma (x)\hat\sigma (y)\bigg) |0>_{\rm conn} =
-\frac {3i}{2f_{\pi}^2} \int \frac {d^4q}{(2\pi )^4} e^{-iq\cdot (x-y)}
\,\,{\cal J} (q^2) -\frac {i}{f_{\pi}^2} c_8 \delta^4 (x-y)
\label{I32}
\ee
\be
<0|T^* \bigg( {\bf j}_{A\alpha}^a (y_1) {\bf j}_{A\beta}^b (y_2)\hat\sigma
(y_3)\bigg) |0>_{\rm conn} =&&
-\frac 2{f_{\pi}} g_{\alpha\beta} \delta^4 (y_1-y_2) \int \frac{d^4q}{(2\pi)^4}
e^{-iq\cdot(y_1-y_3)} {\cal J} (q^2 ) \nonumber\\&&
-\frac 1{f_{\pi}} c_7 \delta^4 (y_1-y_3) \delta^4 (y_2-y_3) g_{\alpha\beta}
\delta^{ab}
\label{I33}
\ee
\be
&&<0|T^* \bigg( {\bf j}_{A\alpha}^a (y_1) {\bf j}_{A\beta}^b (y_2)
{\bf j}_{A\gamma}^c(y_3) {\bf j}_{A\delta}^d (y_4)\bigg)|0>_{\rm conn}
=\nonumber\\&&
i\bigg( 2\delta^{ab}\delta^{cd} +\delta^{ac}\delta^{bd} +\delta^{ab}\delta^{bc}
\bigg) g_{\alpha\beta} g_{\gamma\delta} \nonumber\\&&
\times \delta^4 (y_1-y_2) \delta^4 (y_3-y_4)
\int \frac{d^4q}{(2\pi)^4}
e^{-iq\cdot(y_3-y_1)} {\cal J} (q^2 ) \nonumber\\&&
+ \bigg( 2ic_5 g_{\alpha\beta}g_{\gamma\delta} +ic_6
g_{\alpha\gamma}g_{\beta\delta} + ic_6
g_{\alpha\delta}g_{\beta\gamma}\bigg)\delta^{ab}\delta^{cd}
\nonumber\\&&
\times\delta^4 (y_1-y_2) \delta^4 (y_3-y_4) \delta^4 (y_1-y_3) +{\rm
{2\,\,perm}}.
\label{I34}
\ee
\be
{\bf F}_S (t) +\frac 1{f_{\pi}} =&& - \frac 1{f_{\pi}^3}
\bigg( t-\frac {m_{\pi}^2}2\bigg) {\cal J} (t) \nonumber\\&&
+\frac {m_{\pi}^2}{f_{\pi}^3} (c_2+c_3-c_8) -\frac 1{2f_{\pi}^3} c_7
(t-2m_{\pi}^2)
\label{I35}
\ee
\be
i{\cal T}_{\rm rest} =&&+ \frac i{f_{\pi}^4} m_{\pi}^2 \delta^{ab}\delta^{cd}
\bigg( 2t -\frac 52 m_{\pi}^2\bigg) \,{\cal J} (t)
\nonumber\\&&
+\frac i{4f_{\pi}^4}
\bigg( 2\delta^{ab}\delta^{cd} +\delta^{ac}\delta^{bd} +\delta^{ab}\delta^{bc}
\bigg) (t-2m_{\pi}^2)^2 {\cal J} (t) \nonumber\\&&
+\frac {im_{\pi}^2}{f_{\pi}^4} \delta^{ab}\delta^{cd}
\bigg( m_{\pi}^2 (c_8-c_2-c_3) + (t-2m_{\pi}^2) c_7 \bigg) \nonumber\\&&
+\frac {ic_5}{2f_{\pi}^4} \delta^{ab}\delta^{cd} (t-2m_{\pi}^2)^2
+\frac {ic_6}{4f_{\pi}^4} \delta^{ab}\delta^{cd} (s-2m_{\pi}^2)^2 \nonumber\\&&
+\frac {ic_6}{4f_{\pi}^4} \delta^{ab}\delta^{cd} (u-2m_{\pi}^2)^2 +
{\rm {2\,\,perm}} + {\cal O} \bigg(\frac 1{f_{\pi}^6}\bigg).
\label{I36}
\ee
Similarly, equations (\ref{D35}, \ref{D37}, \ref{D40}) read
\be
A_{\rm rest} (s,t,u) = &&+\frac 1{2f_{\pi}^4} (s^2-m_{\pi}^4)
{\cal J} (s) \nonumber\\&&
+\frac 1{4f_{\pi}^4} (t-m_{\pi}^2)^2
{\cal J} (t) +\frac 1{4f_{\pi}^2} (u-2m_{\pi}^2)^2
{\cal J} (u)\nonumber\\&&
+\frac {m_{\pi}^2}{f_{\pi}^4} \bigg( m_{\pi}^2 (c_8-c_2-c_3) + (s-2m_{\pi}^2)
c_7 \bigg) \nonumber\\&&
+\frac 1{2f_{\pi}^4} c_5 (s-2m_{\pi}^2)^2
+\frac 1{4f_{\pi}^4} c_6 (t-2m_{\pi}^2)^2
+\frac 1{4f_{\pi}^4} c_6 (u-2m_{\pi}^2)^2 \nonumber\\&&
 +{\cal O} \bigg(\frac 1{f_{\pi}^6}\bigg)
\label{I37}
\ee
\be
{\bf F}_V (q^2 ) = 1 + \frac 1{2f_{\pi}^2}
\bigg( (c_1+c_4-c_2+\frac 1{72\pi^2}) q^2 +\frac 13 (q^2-4m_{\pi}^2) {\cal J}
(q^2) \bigg) +
{\cal O} \bigg(\frac 1{f_{\pi}^4}\bigg)
\label{I38}
\ee
\be
C(s,t,u) = &&
-\frac 1{f_{\pi}^4} (s-2m_{\pi}^2)^2 (c_1+c_4-c_2 +\frac 1{144\pi^2}
)\nonumber\\&&
+\frac 1{4f_{\pi}^4} (s^2 +(t-u)^2) (c_1+c_4-c_2 +\frac 1{144\pi^2}
)\nonumber\\&&
+\frac {m_{\pi}^2}{f_{\pi}^4}
\bigg( m_{\pi}^2 (c_8-c_2-c_3) + (s-2m_{\pi}^2) c_7 \bigg)
\nonumber\\&&
+\frac 1{2f_{\pi}^4} (c_5+c_1) (s-2m_{\pi}^2)^2 +
\frac 1{8f_{\pi}^4} (c_6-c_1) (s^2+ (t-u)^2).
\label{I39}
\ee
At first sight (\ref{I39}) does not agree with Ref. \cite{leutwyler} which
gives
\be
C(s,t,u) = \frac 1{96\pi^2 F^4} &&
\bigg( +2(\overline{l}_1-\frac 43) (s-2M^2)^2 \nonumber\\&&+
(\overline{l}_6 -\frac 56 ) (s^2 + (t-u)^2) - 12 M^2 s + 15 M^4\bigg).
\label{I40}
\ee
with only two parameters.
However, this is because the tree term in Ref. \cite{leutwyler} is taken as
$(s-M^2)/F^2$ rather than (\ref{D33}). If we express $M$ and $F$ in terms of
the physical values $m_{\pi}$ and $f_{\pi}$, we pick up an extra contribution
to $C(s,t,u)$
\be
\frac {M^2}{8\pi^2 F^4} (s-M^2) \overline{l}_4 -
\frac{M^4}{32\pi^2 F^4} \overline{l}_3
\label{I41}
\ee
so there are four parameters in total.

As a check, comparison of (\ref{I35}) with the corresponding expression in Ref.
\cite{leutwyler} gives
\be
c_7 &&\leftrightarrow \frac 1{8\pi^2} (\overline{l}_4 -1 )\nonumber\\
c_2 +c_3 +c_7-c_8 &&\leftrightarrow \frac 1{32\pi^2}
(\overline{l}_3 -1 )
\label{I42}
\ee
while (\ref{I39}-\ref{I41}) give
\be
-(c_1+c_4-c_2 +\frac 1{144\pi^2} ) +\frac 12 (c_5+c_1)
&&\leftrightarrow \frac 1{48\pi^2} (\overline{l}_1 -\frac 43 )\nonumber\\
+\frac 14
(c_1+c_4-c_2 +\frac 1{144\pi^2} ) +\frac 18 (c_6-c_1) &&
\leftrightarrow \frac 1{96\pi^2} (\overline{l}_2 -\frac 56 )\nonumber\\
+c_7 &&\leftrightarrow \frac 1{8\pi^2} (\overline{l}_4 -1 )\nonumber\\
+(c_8-c_2-c_3)
&&\leftrightarrow \frac 1{8\pi^2} \overline{l}_4 -\frac 1{32\pi^2}
(\overline{l}_3 +3 )
\label{I43}
\ee
which are seen to be consistent.

We therefore conclude that chiral symmetry does not yield any constraints
on $\pi\pi$ scattering at one-loop level other than those already given by
unitarity, causality and the tree term. It also follows that all six on-shell
parameters can be extracted from the pion data $without$ recourse to $SU(3)$ or
other considerations.
Specifically, equations (\ref{D47}, \ref{D51}, \ref{D55}) are modified to
\be
A_{\rm rho} (s,t,u) =&&+\frac 1{4f_{\pi}^4} (s-u) t {\bf \Pi}_V (t) +
\frac 1{4f_{\pi}^4} (s-t) u {\bf \Pi}_V (u) \nonumber\\&&
+\frac 1{f_{\pi}^2m_{\rho}^2} (s-u) t \bigg( 1-\frac
{f_{\rho}^2}{2f_{\pi}^2}\bigg)
+\frac 1{f_{\pi}^2m_{\rho}^2} (s-t) u \bigg( 1-\frac
{f_{\rho}^2}{2f_{\pi}^2}\bigg)
\label{I44}
\ee
\be
&&
a_0^0 ({\rm rho })= 0 \nonumber\\&&
b_0^0 ({\rm rho} ) = +\frac {m_{\pi}^2}{4\pi f_{\pi}^4} -
\frac{m_{\pi}^2}{\pi f_{\pi}^2 m_{\rho}^2}
= -0.010 \,\,m_{\pi}^{-2} \nonumber\\&&
a_2^0 ({\rm rho} ) = -\frac {1}{60\pi f_{\pi}^4} \frac{f_{\rho
}^2}{m_{\rho}^2} + \frac 1{15\pi f_{\pi}^2 m_{\rho}^2} +
\frac {1}{7200\pi^3 f_{\pi}^4}
 = 6.5 \,\,10^{-4} \,\,m_{\pi}^{-4} \nonumber\\&&
a_1^1 ({\rm rho} ) = -\frac {m_{\pi}^2}{8\pi f_{\pi}^4} \frac{f_{\rho
}^2}{m_{\rho}^2} + \frac {m_{\pi}^2}{2\pi f_{\pi}^2 m_{\rho}^2} +
\frac {m_{\pi}^2}{864\pi^3 f_{\pi}^4}
 = +0.0049\,\, m_{\pi}^{-2} \nonumber\\&&
a_0^2 ({\rm rho} ) = 0\nonumber\\&&
b_0^2 ({\rm rho} ) = -\frac {m_{\pi}^2}{8\pi f_{\pi}^4} \frac{f_{\rho
}^2}{m_{\rho}^2} +\frac{m_{\pi}^2}{2\pi f_{\pi}^2 m_{\rho}^2}
= +0.0047 \,\,m_{\pi}^{-2} \nonumber\\&&
a_2^2 ({\rm rho} ) = +\frac {1}{120\pi f_{\pi}^4} \frac{f_{\rho
}^2}{m_{\rho}^2}-\frac 1{30\pi f_{\pi}^2 m_{\rho}^2}
-\frac {1}{14400\pi^3 f_{\pi}^4}
 = -3.3 \,\,10^{-4} \,\,m_{\pi}^{-4}
\label{I45}
\ee
\be
&&
a_0^0 ({\rm rest }) = \frac {m_{\pi}^4}{32\pi f_{\pi}^4}
\bigg( 5(c_8-c_2-c_3) + 2c_7 +10 c_6 +10 c_5 +\frac {49}{16\pi^2}\bigg)
\nonumber\\&&
b_0^0 ({\rm rest }) = \frac {m_{\pi}^2}{32\pi f_{\pi}^4}
\bigg( 8c_7 + 24 c_6 + 32 c_5 +\frac {91}{24\pi^2}\bigg)
\nonumber\\&&
a_2^0 ({\rm rest }) = \frac {1}{240\pi f_{\pi}^4}
\bigg( 8c_6 + 4c_5 - \frac{73}{240\pi^2}\bigg)
\nonumber\\&&
a_1^1 ({\rm rest }) = \frac {m_{\pi}^2}{96\pi f_{\pi}^4}
\bigg( 4c_7 + 4 c_6 -8 c_5 -\frac {1}{16\pi^2}\bigg)
\nonumber\\&&
a_0^2 ({\rm rest }) = \frac {m_{\pi}^4}{32\pi f_{\pi}^4}
\bigg( 2(c_8-c_2-c_3) -4c_7 +4c_6 +4c_5 +
\frac {1}{4\pi^2}\bigg)
\nonumber\\&&
b_0^2 ({\rm rest }) = \frac {m_{\pi}^2}{32\pi f_{\pi}^4}
\bigg(- 4c_7 + 12 c_6 +8 c_5 +\frac {35}{48\pi^2}\bigg)
\nonumber\\&&
a_2^2 ({\rm rest }) = \frac {1}{240\pi f_{\pi}^4}
\bigg( 2 c_6 +4 c_5 -\frac {19}{240\pi^2}\bigg).
\label{I46}
\ee
These results  are consistent with the data (\ref{D46}),
but the error bars are rather large.
For $c_7$,
\be
16\pi^2 c_7 = &&+\frac{64\pi^3 f_{\pi}^4}{m_{\pi}^2}
\bigg( b_0^0 ({\rm exp} ) - b_0^0 ({\rm tree}) - b_0^0 ({\rm rho}) \bigg)
\nonumber\\&&
-640 \pi^3 f_{\pi}^4 \bigg( a_2^0 ({\rm exp} ) - a_2^0 ({\rm rho} ) \bigg)
\nonumber\\&&-
3200 \pi^3 f_{\pi}^4 \bigg( a_2^2 ({\rm exp} ) - a_2^2 ({\rm rho} ) \bigg)
-\frac {189}{20} = 9.0 \pm 13\nonumber
\ee
\be
16\pi^2 c_7 = &&+\frac{384\pi^3 f_{\pi}^4}{m_{\pi}^2}
\bigg( a_1^1 ({\rm exp} ) - a_1^1 ({\rm tree}) - a_1^1 ({\rm rho}) \bigg)
\nonumber\\&&
-1280 \pi^3 f_{\pi}^4 \bigg( a_2^0 ({\rm exp} ) - a_2^0 ({\rm rho} ) \bigg)
\nonumber\\&&+
3200 \pi^3 f_{\pi}^4 \bigg( a_2^2 ({\rm exp} ) - a_2^2 ({\rm rho} ) \bigg)
-\frac {19}{60} = 7.8 \pm 8\nonumber
\ee
\be
16\pi^2 c_7 = &&-\frac{128\pi^3 f_{\pi}^4}{m_{\pi}^2}
\bigg( b_0^2 ({\rm exp} ) - b_0^2 ({\rm tree}) - b_0^2 ({\rm rho}) \bigg)
\nonumber\\&&
+1280 \pi^3 f_{\pi}^4 \bigg( a_2^0 ({\rm exp} ) - a_2^0 ({\rm rho} ) \bigg)
\nonumber\\&& +
640 \pi^3 f_{\pi}^4 \bigg( a_2^2 ({\rm exp} ) - a_2^2 ({\rm rho} ) \bigg)
+\frac {19}{4} = 13.1 \pm 7\nonumber
\label{I47}
\ee
\be
16\pi^2 c_7 =&&+ \frac{128\pi^3 f_{\pi}^4}{3m_{\pi}^4}
\bigg( a_0^0 ({\rm exp}) -a_0^0 ({\rm tree} ) \bigg)
\nonumber\\ &&
-\frac {320\pi^3 f_{\pi}^4}{3m_{\pi}^4}
\bigg(a_0^2 ({\rm exp}) -a_0^2 ({\rm tree}) \bigg) -\frac {13}4 = 6.0 \pm 19
\label{IA47}
\ee

For the remainder we have,
\be
16\pi^2 (c_6-c_1) =
640\pi^3 f_{\pi}^4 \bigg( &&+ a_2^0 ({\rm exp} ) -
a_2^0 ({\rm rho} )
-a_2^2 ({\rm exp} ) + a_2^2 ({\rm rho} ) \bigg)
\nonumber\\&&+
\frac 35 -16\pi^2 \frac
{f_{\rho}^2}{m_{\rho}^2} = -2.6 \pm 1.7\nonumber
\ee
\be
16\pi^2 (c_5+c_1) =
320\pi^3 f_{\pi}^4 \bigg(&&+ 4a_2^2 ({\rm exp} ) - 4a_2^2 ({\rm rho} )
-a_2^0 ({\rm exp} ) + a_2^0 ({\rm rho} ) \bigg)
\nonumber\\&&+
\frac 1{60} +16\pi^2 \frac
{f_{\rho}^2}{m_{\rho}^2} = +7.1 \pm 2.5\nonumber
\ee
\be
16\pi^2 (c_8-c_2-c_3) = &&
+\frac{512\pi^3 f_{\pi}^4}{5m_{\pi}^4} \bigg(+ a_0^0 ({\rm exp} ) - a_0^0
({\rm tree} ) \bigg)\nonumber\\&&
-640 \pi^3 f_{\pi}^4 \bigg( a_2^0 ({\rm exp} ) -a_2^0 ({\rm rho} )\bigg)
\nonumber\\&&-
1280 \pi^3 f_{\pi}^4 \bigg( a_2^2 ({\rm exp} ) -a_2^2 ({\rm rho}
)\bigg)\nonumber\\&& -
\frac {32\pi^2}5 c_7 -\frac {331}{30} = 41 \pm 32
\label{I48}
\ee
where we have used the average value $c_7=0.057$ following from
(\ref{I47}) for the last term in (\ref{I48}).

We note that the pion isoscalar charge radius $<r^2>_{S}$ follows from
\be
{\bf F}_S (t) = {\bf F}_S (0) \bigg( 1 + \frac t6 <r^2>_S +{\cal O}
(t^2)\bigg).
\label{I51}
\ee
Taylor expanding (\ref{I35}) and using the average value
$c_7 = 0.057 $, we obtain the central value
\be
<r^2>_S = \frac {3}{f_{\pi}^2} \bigg( c_7 -\frac 1{96\pi^2}\bigg)\sim 0.75
\,\,{\rm fm}^2 .
\label{I52}
\ee
in agreement with the value $0.7\pm 0.2$ fm$^2$ as quoted in Ref.
\cite{leutwyler}.
A better measurement of the $\pi\pi$ scattering lengths and range parameters
are called for, to narrow the accuracy on the determination of $c_7$ and thus
the pion scalar charge radius. The isovector charge radius of the pion is used
as an input in (\ref{I29}).

Finally, we note that the one-loop correction to the
radiative decay of the pion
for the general case $\pi\rightarrow e\nu\gamma^*$ is no longer (\ref{ADD15})
but
\be
&&+\frac 1{f_{\pi}}\epsilon^{abc}
\bigg(g_{\mu\nu} (k^2-m_{\pi}^2) -k_{\mu}k_{\nu}\bigg)\nonumber\\
&&+\frac 1{f_{\pi}}\epsilon^{abc} c_2
(2p_{\mu}-q_{\mu}) k_{\nu} \frac {m_{\pi}^2}{k^2-m_{\pi}^2}\nonumber\\
&&+\frac 1{f_{\pi}}\epsilon^{abc} c_2
(2p_{\mu}-q_{\mu}) p_{\beta}
\bigg(g_{\nu}^{\beta} -\frac{k_{\nu}k^{\beta}}{k^2-m_{\pi}^2}\bigg)\nonumber\\
&&-\frac 1{f_{\pi}}\epsilon^{abc} c_4
p_{\mu} q_{\beta}
\bigg(g_{\nu}^{\beta} -\frac{k_{\nu}k^{\beta}}{k^2-m_{\pi}^2}\bigg)\nonumber\\
&&+\frac 1{f_{\pi}}\epsilon^{abc} c_4 p\cdot q
\bigg(g_{\mu\nu} -\frac{k_{\mu}k_{\nu}}{k^2-m_{\pi}^2}\bigg)\nonumber\\
&&-\frac i{f_{\pi}}\epsilon^{abc}
\bigg(g_{\nu}^{\beta} -\frac{k_{\nu}k^{\beta}}{k^2-m_{\pi}^2}\bigg)
\bigg(-g_{\mu\beta} q^2 +q_{\mu} q_{\beta}\bigg) \,\,\Pi_V (q^2)
\label{ADD17}
\ee
For on-shell photons $q^2=0$, the last term drops. The one-loop correction
depends on the two parameters $c_2$ and $c_4$.

\vskip 1.5cm
{\bf Appendix F : One-Loop Integrals}
\vskip .10 cm

In this appendix we summarise some one-loop integrals needed in some
of the calculations discussed in section 18.

$\bullet \,\,\,{\cal J} (s )$

\be
{\cal J} (s= q^2) =
-i\int \frac {d^4k}{(2\pi )^4}
\bigg( \frac 1{k^2-m_{\pi}^2 +i0} \frac 1{(k-q)^2-m_{\pi}^2 +i0} -
(q=0) \bigg)
\label{INT1}
\ee
After Feynman parametrization we have
\be
16\pi^2 {\cal J} (s=q^2) =
\int_0^1 dx \frac {x(1-2x) q^2}{x(1-x) q^2 -m_{\pi}^2 +i0}
\label{INT2}
\ee

\vskip .5cm
$\bullet \,\,\,{\cal J}_{\mu\nu} (q,p )$

\be
{\cal J}_{\mu\nu} (q,p) =
i\int \frac{d^4k}{(2\pi )^4}
\bigg(&& \frac 1{k^2-m_{\pi}^2+i0 } \frac 1{(k+q)^2-m_{\pi}^2 +i0}
          \frac 1{(k+p)^2-m_{\pi}^2 +i0} \nonumber\\
&&\times (2k_{\mu}+q_{\mu})(2k_{\nu} +p_{\nu})
+{\rm counterterms}\bigg)
\label{INT3}
\ee
After Feynman parametrisation we have
\be
16\pi^2 {\cal J}_{\mu\nu} (q,p) =
&&+2g_{\mu\nu} \int_0^1 dx\int_0^{1-x} dy\,\,{\rm ln}
\bigg( \bigg( m_{\pi}^2 + (xq+yp)^2 -xq^2-yp^2-i0 \bigg)/m_{\pi}^2\bigg)
\nonumber\\
&&+\int_0^1 dx\int_0^{1-x} dy \bigg( (1-2x) q_{\mu} -2y p_{\mu}\bigg)
\bigg( (1-2y) p_{\nu} -2x q_{\nu} \bigg) \nonumber\\
&&\times \bigg( m_{\pi}^2 + (xq+yp)^2 -xq^2-yp^2-i0 \bigg)^{-1}
\label{INT4}
\ee

\vskip .5cm
$\bullet \,\,\,{\cal J}_{\mu\nu\rho} (q,p )$

\be
{\cal J}_{\mu\nu\rho} (q,p) =
i\int \frac{d^4k}{(2\pi )^4}
\bigg(&& \frac 1{k^2-m_{\pi}^2 +i0} \frac 1{(k+q)^2-m_{\pi}^2+i0}
          \frac 1{(k+p)^2-m_{\pi}^2+i0}\nonumber\\
&&\times (2k_{\mu}+q_{\mu})(2k_{\nu} +p_{\nu})(2k_{\rho}+(q-p)_{\rho})
 - (q=p=0) \bigg)
\label{INT5}
\ee
After Feynman parametrization we have
\be
16 \pi^2 {\cal J}_{\mu\nu\rho} (q,p) =
&&-2i\int_0^1dx \int_0^{1-x} dy \bigg[  g_{\mu\nu} \bigg(
(1-2x) q_{\rho} + (1-2y) p_{\rho} \bigg)\nonumber\\&&
+ g_{\nu\rho} \bigg( (1-2x) q_{\mu} - 2y p_{\mu} \bigg) +
g_{\rho\mu} \bigg( (1-2y) p_{\nu} - 2x q_{\nu} \bigg) \bigg]
\nonumber\\&&\times {\rm ln} \bigg(
\bigg ( m_{\pi}^2 + (xq+yp)^2 -xq^2-yp^2-i0 \bigg)/ m_{\pi}^2 \bigg)
\nonumber\\
&&-i\int_0^1dx \int_0^{1-x} dy
\bigg( (1-2x) q_{\mu} - 2y p_{\mu} \bigg)\nonumber\\
&&\times\bigg( (1-2y) p_{\nu} -2x q_{\nu} \bigg)
\bigg( (1-2x) q_{\rho} + (1-2x )p_{\rho}\bigg)\nonumber\\
&&\times \bigg( m_{\pi}^2 + (xq+yp)^2 -xq^2-yp^2-i0 \bigg)^{-1}
\label{INT6}
\ee

The Feynman integrals appearing in the above can be undone by noting that
\be
{\bf I}_1 (s) =&&
\int_0^1 dx \frac {x(1-2x) s}{x(1-x) s -m_{\pi}^2 +i0}
\nonumber\\ =&&
2+ \bigg( 1-\frac {4m_{\pi}^2}{s}\bigg)^{1/2}
\bigg({\rm ln}
\bigg(\frac{\sqrt{s-4m_{\pi}^2}-\sqrt{s}}{\sqrt{s-4m_{\pi}^2}+\sqrt{s}}\bigg)
+ i\pi \bigg)
\label{INT7}
\ee
and that
\be
{\bf I}_2 (s) = && \int_0^1 \frac{dx}x \,\,{\rm ln} \bigg( 1
-\frac{s}{m_{\pi}^2}
x (1-x) \bigg) \nonumber\\ = &&
\int_0^{\frac 1{x_+}} \frac {dx}x \,\,{\rm ln} \bigg( 1-x \bigg) +
\int_0^{\frac 1{x_-}} \frac {dx}x \,\,{\rm ln} \bigg( 1-x \bigg)
\label{INT18}
\ee
with
\be
2x_{\pm} = 1\pm  \sqrt{ 1-\frac{4m_{\pi}^2}s }
\label{INT19}
\ee
Since
\be
\int_0^{z} \frac {dx}x \,\,{\rm ln} \bigg( 1-x \bigg) +
\int_0^{\frac z{z-1}} \frac {dx}x \,\,{\rm ln} \bigg( 1-x \bigg) =
\frac 12 \,\,{\rm ln}^2 \bigg( 1-z\bigg)
\label{INT20}
\ee
and choosing the branch cut of the logarithm along the Re$s \geq 0$ axis, we
have
\be
{\bf I}_2 (s) = \frac 12
\bigg({\rm ln}
\bigg(\frac{\sqrt{s-4m_{\pi}^2}-\sqrt{s}}{\sqrt{s-4m_{\pi}^2}+\sqrt{s}}\bigg)
+ i\pi \bigg)^2
\label{INT21}
\ee

With the above in mind, we have
\be
16\pi^2 {\cal J} (s) =  2 +\bigg( 1-\frac{4m_{\pi}^2}s \bigg)^{1/2}
\bigg( {\rm ln} \bigg( \frac{x_-}{x_+}\bigg) + i\pi \bigg)
\label{INT22}
\ee
\be
16\pi^2 {\cal K} (s) = &&\int_0^1 dx\int_0^{1-x} dy\,\,
{\rm ln} \bigg( 1-\frac{s}{m_{\pi}^2} xy -i0 \bigg)\nonumber\\ =
&&-\frac 32 -\frac 12
\bigg( 1-\frac{4m_{\pi}^2}s \bigg)^{1/2}
\bigg( {\rm ln} \bigg( \frac{x_-}{x_+}\bigg) + i\pi \bigg)
-\frac{m_{\pi}^2}{2s}
\bigg( {\rm ln}\,\bigg( \frac{x_-}{x_+}\bigg) + i\pi \bigg)^2
\label{INT23}
\ee
\be
16\pi^2 {\cal H} (s) = &&\int_0^1 dx\int_0^{1-x} dy\, (1-2x)\,
{\rm ln} \bigg( 1-\frac{s}{m_{\pi}^2} xy -i0 \bigg)\nonumber\\ =
&& -\bigg(\frac {11}{18} +\frac{8m_{\pi}^2}{3s}\bigg)\nonumber\\
&& +\bigg(\frac 13
\bigg( 1-\frac{4m_{\pi}^2}s \bigg)^{3/2}-\frac 12
\bigg( 1-\frac{4m_{\pi}^2}s \bigg)\bigg)
\bigg( {\rm ln} \bigg( \frac{x_-}{x_+}\bigg) + i\pi \bigg)\nonumber\\
&&-\frac{m_{\pi}^2}{2s}
\bigg( {\rm ln}\, \bigg( \frac{x_-}{x_+}\bigg) + i\pi \bigg)^2
\label{INT24}
\ee

\vskip 1.5cm
{\bf Appendix G : Unresolved Issues in Chiral Perturbation Theory}
\vskip .1cm

In this Appendix, we wish to call upon some unresolved technical
issues as they appear in the general context of chiral perturbation theory
as discussed by Gasser and Leutwyler \cite{leutwyler,fleutwyler}.
The first issue is related to the
use of the pion equations of motion in the nonlinear sigma model
\be
\bigg(-\nabla^{\mu}\nabla_{\mu} +{\ub a}^{\mu}\cdot{\ub a}_{\mu} -m_{\pi}^2 -
s\bigg) \pi +  J = {\cal O}\bigg( \frac{\pi^2}{f_{\pi}}\bigg).
\label{F2}
\ee
In the path-integral approach of Ref. \cite{leutwyler}, the equation must be
solved with Steuckelberg-Feynman boundary conditions, giving
\be
\pi = - G_F \,J + {\cal O}\bigg( \frac{\pi^2}{f_{\pi}}\bigg).
\label{F3}
\ee
The result corresponds to equation (10.21) in Ref. \cite{leutwyler}. Explicitly
\be
U^i_1 = (\Box +M^2 )^{-1} \bigg( \partial^{\mu} a_{\mu}^i +\chi^i \bigg).
\label{F4}
\ee
Since $G_F$ is complex, $\pi$ is also complex. This is expected. Indeed, as
emphasized by Faddeev \cite{faddeev}, it is in general unjustified to assume
that
the fields in the path integral are real. This circumstance does not cause any
problems for ordinary scalar theories. However, for the nonlinear sigma model,
a complex $\pi$ means that the chiral field $\overline{U}$ evaluated at the
stationary point is non-unitary, that is $\overline
U\,\overline{U}^{\dagger}\neq
{\bf 1}$, which leads to several difficulties \cite{zahed}.
The first one is that we will obtain different answers according to whether we
write the
original Lagrangian as
\be
+\frac {f_{\pi}^2}4 {\rm Tr} \bigg( \partial^{\mu}U\,\partial_{\mu}
U^{\dagger}\bigg)\,\,, \qquad
+\frac {f_{\pi}^2}4 {\rm Tr} \bigg( \partial^{\mu}U\,\partial_{\mu}
U^{-1}\bigg)\,\,, \qquad
-\frac {f_{\pi}^2}4 {\rm Tr}
\bigg( U^{\dagger}\partial^{\mu} U\, U^{\dagger}\partial_{\mu}U
\bigg).
\nonumber
\ee
The second one is that the integration measure $d\mu (U )$ must be redefined.
A third one is that there is an infinite number of terms such as
${\rm Tr} \bigg(\overline{U}\,\overline{U}^{\dagger}\bigg)^n$ which are
invariant under $SU(2)\times SU(2)$ and may be added to the Lagrangian.

All such difficulties disappear if we work in Euclidean space. However, the
problem still remain as to whether we should continue $U^{\dagger}$ or $U^{-1}$
or
something else (e.g. Green's functions) when coming to Minkowski space.

The second issue arises when multi-loop effects are considered. In his
general analysis of the foundations of chiral perturbation theory, Leutwyler
makes use of stationary points of higher derivative actions \cite{fleutwyler}.
However, it is known that higher derivative actions generally give rise to
pathological effects that are at odd with positivity and causality, thereby
jeoperdizing the main conclusions beyond one loop. The recent two-loop
calculation by Belluci, Gasser and Sainio \cite {belluci} avoids this problem
by using the stationary point of the minimal derivative action.
The minimal stationary point is pathological free and allows for simple chiral
counting. We do not know whether this analysis could be used to revise
Leutwyler's general analysis. The latter is important for
the overall consistency of chiral perturbation theory beyond one loop.

Overall, we believe that a resolution of the concerns we have raised is
desirable for further comparison between Ref. \cite{leutwyler} and our
approach.

\vskip 1.5cm
{\bf Appendix H : Elements for a Two-Loop Calculation}
\vskip .1cm

In this Appendix, we will outline the necessary elements for extending our
one-loop analysis of section 18 to two loops. To lowest order (\ref{C1})
reads as

\be
\bigg[ \pi_{\rm in}, \hat{\cal S}_0 \bigg] =\hat{\cal S}_0\tilde{G}{\bf
K}\pi_{\rm in} -\tilde{G} \,J\,\hat{\cal S}_0
\label{F5}
\ee
where $\tilde{G}$ is given in (\ref{J7}). We now write
\be
\hat{\cal S} =\hat{\cal S}_0
\bigg(1+ \frac i{f_{\pi}} \Theta_1 -\frac 1{f_{\pi}^2}\Theta_2 \bigg)
\nonumber
\ee
where unitarity requires
\be
\Theta_1^{\dagger} =\Theta_1 \qquad\qquad \Theta_2 +\Theta_2^{\dagger} =
\Theta_1^2.
\label{F6}
\ee
Substitution into (\ref{C1}) gives
\be
\bigg[ \bigg( 1+\tilde G {\bf K} \bigg) \pi_{\rm in}, \Theta_1 \bigg] =
-\hat{\cal S}_0^{\dagger} \tilde{G}\nabla^{\mu}a_{\mu}
\frac{\delta\hat{S}_0}{\delta s} + \hat{\cal S}_0^{\dagger}\tilde G {\bf X}_A
\hat{\cal S}_0
\label{F7}
\ee
\be
\bigg[ \bigg( 1+\tilde G {\bf K} \bigg) \pi_{\rm in}, \Theta_2 \bigg] = &&
-i\hat{\cal S}_0^{\dagger} \tilde{G} \,J
-\hat{\cal S}_0^{\dagger} \tilde{G}\nabla^{\mu}a_{\mu}
\frac{\delta\hat{S}_0}{\delta s} \Theta_1 \nonumber\\&&
 -\tilde{G}\nabla^{\mu}a_{\mu} \frac{\delta\Theta_1}{\delta s}
+ \hat{\cal S}_0^{\dagger}\tilde G ({\bf X}_A \hat{\cal S}_0 )\Theta_1 +
\tilde{G} {\bf X}_A \Theta_1.
\label{F8}
\ee
Multiplying by
\be
\bigg(1+\tilde{G}{\bf K}\bigg)^{-1} = 1 -\Delta{\bf K} \bigg(
1+G_A {\bf K} \bigg)
\nonumber
\ee
yields
\be
\bigg[ \pi_{\rm in} , \Theta_1 \bigg] =
-\hat{\cal S}_0^{\dagger}\Delta \bigg( 1 +{\bf K} G_A \bigg)
\nabla^{\mu}a_{\mu}
\frac{\delta\hat{S}_0}{\delta s} + \hat{\cal S}_0^{\dagger}
\Delta \bigg( 1 +{\bf K} G_A \bigg) {\bf X}_A
\hat{\cal S}_0
\label{F9}
\ee
\be
\bigg[ \pi_{\rm in} , \Theta_2 -\Theta_2^{\dagger} \bigg] = &&
\bigg[\bigg[ \pi_{\rm in} , \Theta_1\bigg],\Theta_1 \bigg] -
2i  \hat{\cal S}_0^{\dagger} \Delta \bigg( 1+ {\bf K} G_A \bigg) J
\frac{\delta \hat{\cal S}_0}{\delta s} \nonumber\\&&
-\Delta \bigg( 1 +{\bf K} G_A \bigg)
\nabla^{\mu}a_{\mu}
\frac{\delta\Theta_1}{\delta s} +
\Delta \bigg( 1 +{\bf K} G_A \bigg) {\bf X}_A \Theta_1.
\label{F10}
\ee
Equations (\ref{F5}-\ref{F10}) completely fix the pionic part of $\Theta_1$ and
$\Theta_2$ up to c-numbers. Furthermore, (\ref{J10}) to lowest order
\be
\bigg[ \pi_{\rm out} , \hat{\cal S}_0^{\dagger} \delta\hat{\cal S}_0 \bigg] =
\tilde{G} (\delta {\bf K} ) \pi - {\tilde G} \delta J
\label{F11}
\ee
implies that $\hat{\cal S}_0^{\dagger} \delta\hat{\cal S}_0$ is at most
quadratic in $\pi_{\rm in}$, so that $\Theta_1$ is at most cubic in $\pi_{\rm
in}$, and $\Theta_2-\Theta_2^{\dagger}$ is quintic in $\pi_{\rm in}$.
These relations are sufficient to extend the present analysis to two loops.
This construction will be discussed elsewhere.

\newpage
{\bf Figure Captions}

\vskip 3cm

\noindent{\bf Figure 1} : Threshold contribution to $\gamma N\rightarrow \pi N$
without baryon dynamics. (a) is the pion t-channel pole and (b) the
Kroll-Ruderman term.

\vskip .5cm

\noindent{\bf Figure 2} : Contributions to $\gamma N\rightarrow \gamma\pi N$
at threshold, to leading order in the electromagnetic charge.

\vskip .5cm

\noindent{\bf Figure 3} : Contributions to $\gamma N\rightarrow \pi\pi N$
at threshold, to leading order in the electromagnetic charge. The solid line
(V)
refers to the isovector vector-current ${\bf V}$.

\vskip .5cm

\noindent{\bf Figure 4} : Contributions to $\pi N\rightarrow \pi\gamma N$
at threshold, to leading order in the electromagnetic charge. The solid line
(V)
refers to the isovector vector-current ${\bf V}$.

\vskip .5cm

\noindent {\bf Figure 5} : (a) Three-loop contribution to the four-point
function
$<0|T^*({\bf j}_A{\bf j}_A{\bf j}_A{\bf j}_A )|0>$, where ${\bf j}_A$ is
one-pion
reduced. (b) Related divergences from the subdiagrams.

\vskip .5cm

\noindent {\bf Figure 6} : Divergent $v\hat s$-graphs (a,b,c), $\hat s$-tadpole
(d), and $\hat s \hat s$-graph (e). The divergences are at the origin of
the unspecified parameters in the effective action at one-loop.

\vskip .5cm

\noindent{\bf Figure 7} : Typical contributions to the vacuum correlators in
the process $\gamma\gamma\rightarrow \pi\pi$ as given by
(\ref{FIN2}). The solid V-lines refer to the isovector vector-current ${\bf
V}$,
the solid A-lines refers to the one-pion reduced isovector axial-current
${\bf j}_A$, and the solid S-line refers to the scalar current triggered by
$\hat\sigma$.

\newpage
\setlength{\baselineskip}{15pt}

\end{document}